\documentclass[a4paper,11pt, oneside, openright]{book}

\usepackage{amsmath}
\usepackage{amssymb}
\usepackage{amsfonts}
\usepackage{chngpage}
\usepackage{longtable}
\usepackage{amsthm}
 \usepackage[lmargin=110pt,rmargin=70pt,tmargin=127pt,bmargin=123pt]{geometry}
\usepackage{placeins}
\usepackage{times,epsfig}
\usepackage{array, multirow}
\usepackage{pdflscape}
\usepackage[]{appendix}
\usepackage{cite}
\usepackage{amscd}
\usepackage{color,graphicx}
\usepackage[T1]{fontenc}
\usepackage[utf8]{inputenc}
\usepackage[english]{babel}
\usepackage{rotating}
 \usepackage[normal]{caption2}
 \usepackage{hyphenat}
\usepackage{xcolor}
\usepackage{url}

\usepackage{hyperref}
\hypersetup{colorlinks,%
citecolor=blue,%
filecolor=red,%
linkcolor=black,%
urlcolor=blue,%
pdftex}

\usepackage{epstopdf}
\usepackage{tikz}
\usetikzlibrary{"arrows","shapes.geometric"}
\usetikzlibrary{"calc"}

\usepackage[nobottomtitles]{titlesec}
\usepackage{titletoc}
\usepackage{fancyhdr}
\newcommand{\numerogordopart}[1]{\definecolor{gris}{gray}{0.75}\fontfamily{phv}\selectfont{\fontsize{30mm}{30mm}\selectfont\color{gris}{#1}}}

\captionstyle{normal}         
\setcaptionwidth{0.9\textwidth}

\titleformat{\chapter}
	[display]
	{\rmfamily\bfseries\scshape\Large}
	{\filleft \numerogordopart{\thechapter}}
	{2ex}
	{\titlerule \vspace{2ex}\filright}
	[\vspace{2ex}\titlerule]

\titleformat{\section}{\vspace{.8ex} \rmfamily\bfseries\upshape\Large} {\bf\thesection}{.5em}{}[]
\titleformat{\subsection}{\vspace{.8ex} \rmfamily\bfseries\upshape\large} {\bf\thesubsection}{.5em}{}[]

\titlecontents{chapter}
	[0pt]
	{\addvspace{0.5pc}\scshape}
	{\contentsmargin{0pt}%
		\bfseries
		\makebox[0pt][r]{
		\thecontentslabel
		\enspace
		}
	}
	{\contentsmargin{0pt}%
		\bfseries
		\makebox[0pt][r]{
		\thecontentslabel
		\enspace
		}
	}
	{	\bf
		\hspace
		\fill
		\thecontentspage
		\hspace*{-1.4pc}
	}
        [\addvspace{0.5pc}]

\fancypagestyle{plain}{%
\fancyhf{} 
\fancyfoot[C]{\thepage} 

}
	
\newcolumntype{P}{>{\raggedright\arraybackslash}p{3cm}}
\newcolumntype{M}{>{\raggedright\arraybackslash}m{1.2cm}}

\makeatletter

\begin{document}


\thispagestyle{empty}
\pagenumbering{roman}
\clearpage
%
%
%
%
%
%
%
%
%
%
%
%
%
%

\thispagestyle{empty}
\begin{center}

~
\mbox{\parbox{3cm} {\centering\scriptsize\mbox{\includegraphics[width=3.0cm] {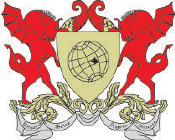}}}}

\vspace{0.5cm}

{\large UNIVERSIDADE FEDERAL DE VI\c{C}OSA}

\vspace{1.5cm}

{\Large MASTER'S THESIS} \vspace{1.5cm}


\vspace{2mm}


{\Large\sc\textbf{ KARDAR-PARISI-ZHANG UNIVERSALITY, ANOMALOUS SCALING AND CROSSOVER EFFECTS IN THE
 GROWTH OF CdTe THIN FILMS }}
\vspace{1mm}




\vspace{2cm}

{\textbf{Author: }}

\vspace{1mm}

{\textbf{Renan Augusto Lisboa Almeida}}

\vspace{3mm}

{\textbf{Supervisors:}}

\vspace{1mm}

{\textbf{Sukarno Olavo Ferreira}}

\vspace{1mm}

{\textbf{Tiago Jos\'{e} de Oliveira}}

\vspace{2.5cm}

{\large  PHYSICS DEPARTMENT} \vspace{5mm}

{\large \textrm{Vi\c{c}osa, Frebuary 2015}}
\end{center}



\newpage\thispagestyle{empty}
\newpage\thispagestyle{empty}

\begin{center}


\vspace{2mm}


{\Large MASTER'S THESIS}

\vspace{1.5cm}

{\Large\sc\textbf{ KARDAR-PARISI-ZHANG UNIVERSALITY, ANOMALOUS SCALING AND CROSSOVER EFFECTS IN THE
 GROWTH OF CdTe THIN FILMS  }}
\vspace{1mm}




\vspace{1.5cm}

\vspace{1cm}

\begin{tabular}{ll}
{\textbf{Autor: }} & {\textbf{Renan Augusto Lisboa Almeida}} \\
{\textbf{Supervisors: }} & {\textbf{Sukarno Olavo Ferreira}, \textbf{Tiago Jos\'{e} de Oliveira}}
\end{tabular}




\vspace{1.5cm}

\end{center}

{\center
{\large Qualification Exam} \vspace{1cm}

\begin{tabular}[h]{l | c | r}
	  &\phantom{aaaaaaaaaaaaaaaaaaaaaaaaaaaaaaaaaaaaaaaaaaa} &{\large}\phantom{aaaaaaaa}\\ \hline
	& &\\
	{\large President:}& Ph.D. Sukarno Olavo Ferreira & \\
	 & &\\ \hline
	  & &\\
	{\large Invited Examiner:}& Ph.D. Maximiliano Luis Munford &\\
	 & & \\ \hline
	 & &\\
	{\large Invited Examiner: }& D.Sc. Thiago Albuquerque de Assis &\\
	& &\\ \hline
\end{tabular}

\vspace{2cm}

{\large Result of the Qualification: \textbf{APPROVED} by unanimity}
}
\vspace{2cm}

\begin{center}
{Vi\c{c}osa, February, 10, 2015}
\end{center}

 \newpage\thispagestyle{empty}
 \newpage\thispagestyle{empty}
 \begin{center}
%
%
%
%
\vspace*{\fill}

 {\Large\sc\textbf{ KARDAR-PARISI-ZHANG UNIVERSALITY, ANOMALOUS SCALING AND CROSSOVER EFFECTS IN THE
 GROWTH OF CdTe THIN FILMS }}
\vspace{1.5cm}

\Large{Renan Augusto Lisboa Almeida}

\vspace{0.5cm}

February 2015
 
 \vspace*{\fill} 
 \end{center}

 \thispagestyle{empty}
 \vspace*{50mm}
 \begin{flushright}
 \textit{This work is dedicated to all those people that believe 
in the humanity and that are fighting for decreasing 
the poverty, social differences and for improving the education. 
These are the true hard and necessary works in which 
all of us should also focus on.}
\pagebreak
 \end{flushright}

\begin{center} \large{\textbf{ACKNOWLEDGMENTS}} 
\end{center}
The conclusion of this dissertation coincides with the end of a
long special period in my life, which goes beyond of physics. Thus,
I think it is appropriate to start thanking those people who have helped me at 
both directions since the beginning of this journey. My parents are my basis. An integral 
part of what I am today is why I have been learning, through examples, that true, passion and 
faith must guide our attitudes, even when all the world seems to behave conversely. Thanks for 
all support and motivation, mainly in difficult times. I also must thanks Miriam Santos for 
so many years of accomplicity. It will be impossible to forget them...

I am enormously grateful to my supervisors, Sukarno Ferreira and Tiago Oliveira. Sukarno is a great experimentalist. He knows almost everything about any growth and characterization technique, as well knows a lot 
about electronic. However, the most remarkable in Sukarno is his personality, 
always open to teach, advise, and help young and expert physicists. Fortunately, I was ``adopted'' by him since the beginning of my undergraduate and today I am proud of 
finishing this MSc under his supervision.
In the same vein, Tiago is ``just'' the most competent person that I have known. His
criticism has played a crucial role in my progress as a physicist and it is very clear that without it, 
this work could be just more one containing several meaningless exponents. Additionally, Tiago has been showing me, as matter
of example, a great ``scientific integrity'', a lack in many unprepared people that share the 
authorship in papers by convenience, instead merit. I am grateful to both for supporting my level change towards the Doctorate,
as well as my adventure in Japan.

By the way, I would like to thank Prof. S. Ferreira for sharing that adventure with me. ``Thank'' for leaving me lost in the middle of Kyoto one day before my talk, without a map... at the end
of the day (or at the beginning), I was able to give a good talk. 
Many thanks also for recommending me to an abroad PhD. The KPZ workshop was a great experience. I am glad to meet J. Krug, T. Halpin-Healy, J. Kim, K. Takeuchi, M. Myllys, P. Yunker, P. Ferrari... 
and so many inspirational physicists and mathematicians. The former two have given important insights about my work, as well as wise advices about my academic plans. It was a privilege to receive them. About P. Yunker, I hope that we can work together soon. \textit{Muchas gracias} to my friends J. Rodr\'{i}guez-Laguna, S. N. Santalla and E. Vivo, which made that time in Japan a fantastic one. We didn't meet a Japanese temple, but now Edoardo and I have known how is a Japanese party... Thank also for the help with this beautiful \LaTeX template.

About my friends from Vi\c{c}osa, I express my special thanks to Isnard Ferraz and Frederico Vasconcellos. You have transformed the lab. and the moments out of it much nicer. I also thank all those who have, sincerely, helped me along the way. It includes Pablo Lisboa, Fellipe Rufino, Reinaldo Bastos, Armand Vidal, Alberto Romani, Ismael Carrasco and many other good friends.

I'm thankful to Prof. S. G. Alves for introducing me in this field two years ago and to Prof. R. Cuerno for his availability to accept evaluate this dissertation. Unfortunately, the short time that I had for writing it, and boring borucratic problems, prevented out our plans this time.
Special thanks are deserved to Prof. Maximiliano Luis Munford and Prof. Thiago Albuquerque de Assis for accepting evaluate this dissertation and for giving valuable contributions for the final version of this text.

\vspace{0.2cm}
Thanks the \textit{Coordena\c{c}\~{a}o de Aperfei\c{c}oamento de Pessoal de N\'{\i}vel Superior} (CAPES) by one and half year of Master Scholarship.
\pagebreak

\addcontentsline{toc}{chapter}{Contents}
 \markboth{Contents}{Contents}
\vfill
\parindent3em

\tableofcontents
 \listoffigures\addcontentsline{toc}{chapter}{LIST OF FIGURES}
 \listoftables\addcontentsline{toc}{chapter}{LIST OF TABLES}
 \clearpage

 \pagebreak
 \chapter*{\begin{center}\begin{large}RESUMO\end{large} \end{center}}
 \lhead{\bfseries Resumo}
 \addcontentsline{toc}{chapter}{RESUMO}
 \begin{flushright}
\parbox[t]{14.0cm}{{\setlength{\baselineskip}%
{0.6\baselineskip}
  ALMEIDA, Renan Augusto Lisboa, Universidade Federal de Vi\c{c}osa, Fevereiro, $2015$ \\
  \textbf{KARDAR-PARISI-ZHANG UNIVERSALITY, ANOMALOUS SCALING AND \\ 
  CROSSOVER EFFECTS IN THE GROWTH OF CdTe THIN FILMS}. Orienta- \\
  dor: Sukarno Olavo Ferreira. Co-Orientador: Tiago Jos\'{e} de Oliveira.
 \par}} 
\end{flushright}

\vspace{2.0 cm}

 Neste trabalho estuda-se a din\^{a}mica de crescimento de filmes finos de Telu- 
 reto de C\'{a}dmio (CdTe) para
 temperaturas de deposi\c{c}\~{a}o (T) entre $150\,^{\circ}\mathrm{C}$ e $300\,^{\circ}\mathrm{C}$. Uma rela\c{c}\~{a}o entre a evolu\c{c}\~{a}o dos morros e flutua\c{c}\~{o}es de longos comprimentos de onda na surperf\'{i}cie do filme 
 de CdTe \'{e} estabelecida. Encontra-se que escalas de curtos comprimentos de onda s\~{a}o ditadas por uma competi\c{c}\~{a}o entre o resultado da forma\c{c}\~{a}o de defeitos na borda de gr\~{a}os vizinhos colididos 
 e entre um processo de relaxa\c{c}\~{a}o originado da difus\~{a}o e da deposi\c{c}\~{a}o de part\'{i}culas (mol\'{e}culas de CdTe) sobre essas regi\~{o}es. Um modelo de Monte Carlo Cin\'{e}tico corrobora as explica\c{c}\~{o}es.
 \`{A} medida que $T$ \'{e} elevada, essa competi\c{c}\~{a}o d\'{a} origem a diferentes cen\'{a}rios na escala de rugosidade tais como: crescimento descorrelacionado, \textit{crossover} de descorrelacionado para crescimento correlacionado e escala an\^{o}mala transiente. Em particular, para $T = 250\,^{\circ}\mathrm{C}$, mostra-se
 que flutua\c{c}\~{o}es na superf\'{i}cie de CdTe s\~{a}o descritas pela c\'{e}lebre equa\c{c}\~{a}o Kardar-Parisi-Zhang (KPZ), ao mesmo tempo que, a universalidade das distribui\c{c}\~{o}es de altura, rugosidade local e altura m\'{a}xima para a classe KPZ \'{e}, finalmente, experimentalmente demonstrada. A din\^{a}mica das flutua\c{c}\~{o}es na superf\'{i}cie de filmes 
 crescidos a outras temperaturas ainda \'{e} descrita pela equa\c{c}\~{a}o KPZ, mas com diferentes valores
 para a tens\~{a}o superficial ($\nu$) e para o excesso de velocidade ($\lambda$). A saber, para $T = 150\,^{\circ}\mathrm{C}$ encontra-se um crescimento Poissoniano que indica $\nu = \lambda = 0$. Para $T = 200\,^{\circ}\mathrm{C}$,
 entretanto, um \textit{crossover} aleat\'{o}rio-para-KPZ \'{e} encontrado, com $\lambda > 0$ neste segundo regime. A origem da escala KPZ
 para filmes crescidos a $T \in [200, 250]\,^{\circ}\mathrm{C}$ decorre da complexa din\^{a}mica de empacotamento dos gr\~{a}os durante a qual espa\c{c}os nas vizinhan\c{c}as dos mesmos n\~{a}o s\~{a}o totalmente preenchidos. Esse mecanismo de agrega\c{c}\~{a}o tem o mesmo efeito da agrega\c{c}\~{a}o lateral do modelo de deposi{c}\~{a}o bal\'{i}stica o qual leva a um excesso de velocidade ($\lambda > 0$). Finalmente, para
 filmes crescidos a $T = 300\,^{\circ}\mathrm{C}$ demonstra-se que $\lambda < 0$. Em particular, o mecanismo KPZ para filmes crescidos a esta temperatura decorre da alta taxa de recusa da deposi\c{c}\~{a}o de part\'{\i}culas, a qual \'{e} dependente das inclina\c{c}\~{o}es locais. Este f\^{e}nomeno 
 pode ser explicado em termos do coeficiente de \textit{sticking}, o qual \'{e} t\~{a}o pequeno qu\~{a}o mais localmente inclinada for a
 superf\'{i}cie. Devido a efeitos de tempo-finito (\textit{crossover} temporais e escala an\^{o}mala) ocorrendo em $T = 200\,^{\circ}\mathrm{C}$ e $300\,^{\circ}\mathrm{C}$, expoentes de escala falham em revelar a Classe de Universalidade do crescimento. Contudo, um novo m\'{e}todo desenvolvido, que avan\c{c}a sobre a simples compara\c{c}\~{a}o entre 
 expoentes e seus valores teoricamente esperados, permite-nos concluir que o 
 crescimento de CdTe, em uma ampla faixa de temperatura, pertence \`{a} classe KPZ.


 \pagebreak
 \chapter*{\begin{center}\begin{large}ABSTRACT\end{large} \end{center}}
 \lhead{\bfseries Abstract}
 \addcontentsline{toc}{chapter}{ABSTRACT}
 \begin{flushright}
\parbox[t]{14.0cm}{{\setlength{\baselineskip}%
{0.6\baselineskip}
  ALMEIDA, Renan Augusto Lisboa, Universidade Federal de Vi\c{c}osa, February, $2015$ \\
  \textbf{KARDAR-PARISI-ZHANG UNIVERSALITY, ANOMALOUS SCALING AND \\ 
  CROSSOVER EFFECTS IN THE GROWTH OF CdTe THIN FILMS}. Supervi-\\
  sor: Sukarno Olavo Ferreira. Co-Supervisor: Tiago Jos\'{e} de Oliveira.
 \par}} 
\end{flushright}

\vspace{2.0 cm}

 In this work one reports on the growth dynamic of CdTe thin films for deposition temperatures ($T$) in the range
 of $150\,^{\circ}\mathrm{C}$ to $300\,^{\circ}\mathrm{C}$. A relation between the mound evolution and 
 large-wavelength fluctuations at CdTe surface has been established. One finds that short-length scales are 
 dictated by an interplay between the effects of the formation of defects at colided boundaries of neighboring 
 grains and a relaxation process which stems from the diffusion and deposition of particles (CdTe molecules) 
 torward these regions. A Kinetic Monte Carlo model corroborates these reasonings. As $T$ is increased,
 that competition gives rise to different scenarios in the roughening scaling such as: uncorrelated growth, 
 a crossover from random to correlated growth and transient anomalous scaling. In particular, for $T = 250\,^{\circ}\mathrm{C}$, one shows that
 fluctuations of CdTe surface are described by the celebrated Kardar-Parisi-Zhang (KPZ) equation, in the meantime that, the universality of height, local roughness and maximal height distributions
 for the KPZ class is, finally, experimentally demonstrated. The dynamic of fluctuations at the CdTe surface for other
 temperatures still is described by the KPZ equation, but with different values for the superficial tension ($\nu$)
 and excess of velocity ($\lambda$). Namely, for $T = 150\,^{\circ}\mathrm{C}$ one finds a Poissonian growth that
 indicates $\nu = \lambda = 0$. For $T = 200\,^{\circ}\mathrm{C}$, however, a Random-to-KPZ crossover is found,
 with $\lambda > 0$ in the second regime. The origin of the KPZ scaling for films grown at 
 $T \in [200, 250]\,^{\circ}\mathrm{C}$ stems from a complex dynamic of grain packing during which all available space at the neighborhood of grains are not filled. This aggregation mechanism has the same effect of the lateral aggregation of the balistic deposition model which leads to an excess of velocity ($\lambda > 0$). Finally, for films grown at
 $T = 300\,^{\circ}\mathrm{C}$ one demonstrates that a KPZ growth with $\lambda < 0$ takes place. In particular, the KPZ mechanism at this $T$ comes from the high refuse rate of the deposition of particles,
 which depends on the local slopes. This phenomenon can be explained in terms of the sticking coefficient
 which is so smaller as more locally inclinated is the surface. Due to finite-time
 effects (temporal crossover and anomalous scaling) taking place in $T = 200\,^{\circ}\mathrm{C}$ and $T = 300\,^{\circ}\mathrm{C}$, scale exponents fail in reveal the Universality Class of the growth. Notwhithstanding, a new
 scheme developed, which advances over the simple comparison between exponents and their theoretically predicted values,
 allow us, surely, to conclude that the growth of CdTe, in a wide range of deposition temperature, belongs to the 
 KPZ class.

 
%



 \chapter{Introduction}

  \setcounter{chapter}{1} \pagenumbering{arabic}
 
  \lhead{\bfseries 1. Introduction}

\section{Motivation}

Semiconductor thin films are the basis of our opto-electronic technology and can be found everywhere \cite{Ohring}. Great part 
of the current thin-film stage was due the vacuum technology \textit{adventum} at 1950's and 1960's, 
which supported the developing of sophisticated growth techniques as Molecular Beam Epitaxy (MBE),
Hot Wall Epitaxy (HWE), Chemical Vapor Deposition (CVD) and others \cite{Ohring, Venables, Pimpinelli}.
Since then, the quality and control on doping, thickness, chemistry composition and structure 
of films became so accurate that the results have been touching directly our life: computers each 
time smaller and more powerful (including cell phones and others mobile devices), systems of 
global position localization (GPS) available to population, lasers of several wavelengths, 
medical equipments of high social impact as magnetic resonance, positron-electron 
tomography and so forth.

Among the most prominent compounds intended for thin film productions, stands out the 
Cadmium-Telluride (CdTe) owing to its suitable semiconductor properties such as 
direct energy gap ($E_{gap} = 1.53$ eV to $300$ K) and high optical absorption 
coefficient ($\approx 5 \times 10^5$ cm$^{-1}$) \cite{Triboulet}. Applications concerned on CdTe films
span the fabrication of solar cells of high efficiency \cite{Li} as well as of the X-ray, $\gamma$-ray and 
infrared detectors \cite{Rams}. 
There are extensive studies on \textit{CdTe growth} in branches as diverse as the controlled growth of tetrapod-branched 
nanocrystals \cite{Manna}, self-assembly of quantum dots \cite{SukarnoQD1,SukarnoQD2}, nano-wires \cite{Williams},
microcavities of ultrafast optical responses \cite{Saba}, etc. Nevertheless, there are still a 
few works on \textit{CdTe growth dynamics} itself \cite{Sukarno, Ribeiro, Mata, Fabio}, discussing issues 
as kinetic roughening, growth symmetries, as well as morphological aspects like grain size, grain shape
and their dependency on deposition temperature\footnote{This term refers to the temperature of the substrate.}, molecular flux and thickness. In fact, \textit{these 
are the most important structures and parameters affecting elasto-mechanical and electrical film features and, 
consequently, the efficiency of devices built up from them} \cite{Ohring, Venables, Assender, Siniscalco}.

From the theoretical side, the surface (or interface) growth is an interesting out-of-equilibrium Statistical Mechanics 
subject because of scale invariance and Universality emergence, as occurs in thermal fluctuations 
of equilibrium systems at criticality \cite{Kardar_book2}. The scaling invariance implies absence of 
any characteristic length in the system beyond system size itself. In turn, the Universality concept
means that systems of different microscopic nature can exhibit the same large scale behavior, 
since they are ruled by interactions sharing dimensionality, symmetries and conservation 
laws \cite{Kardar_book2}. Actually, the Universality idea goes beyond of interface studies and has been found 
in others far-from-equilibrium contexts such as epidemic spreading in random networks \cite{Barabasi_Sci}, crackling 
noise \cite{Sethna} and social dynamics \cite{Castellano}. However, most efforts have been focusing in surface 
growth due to its ubiquitousness in the nature with examples ranging from Physics, Chemistry, Biology to 
Applied Mathematics and Geology \cite{Barabasi}.

 \begin{figure}[ht]
    \vspace{0.5 cm}
    \centering
    \includegraphics[height = 5.0 cm, width = 10.0 cm]{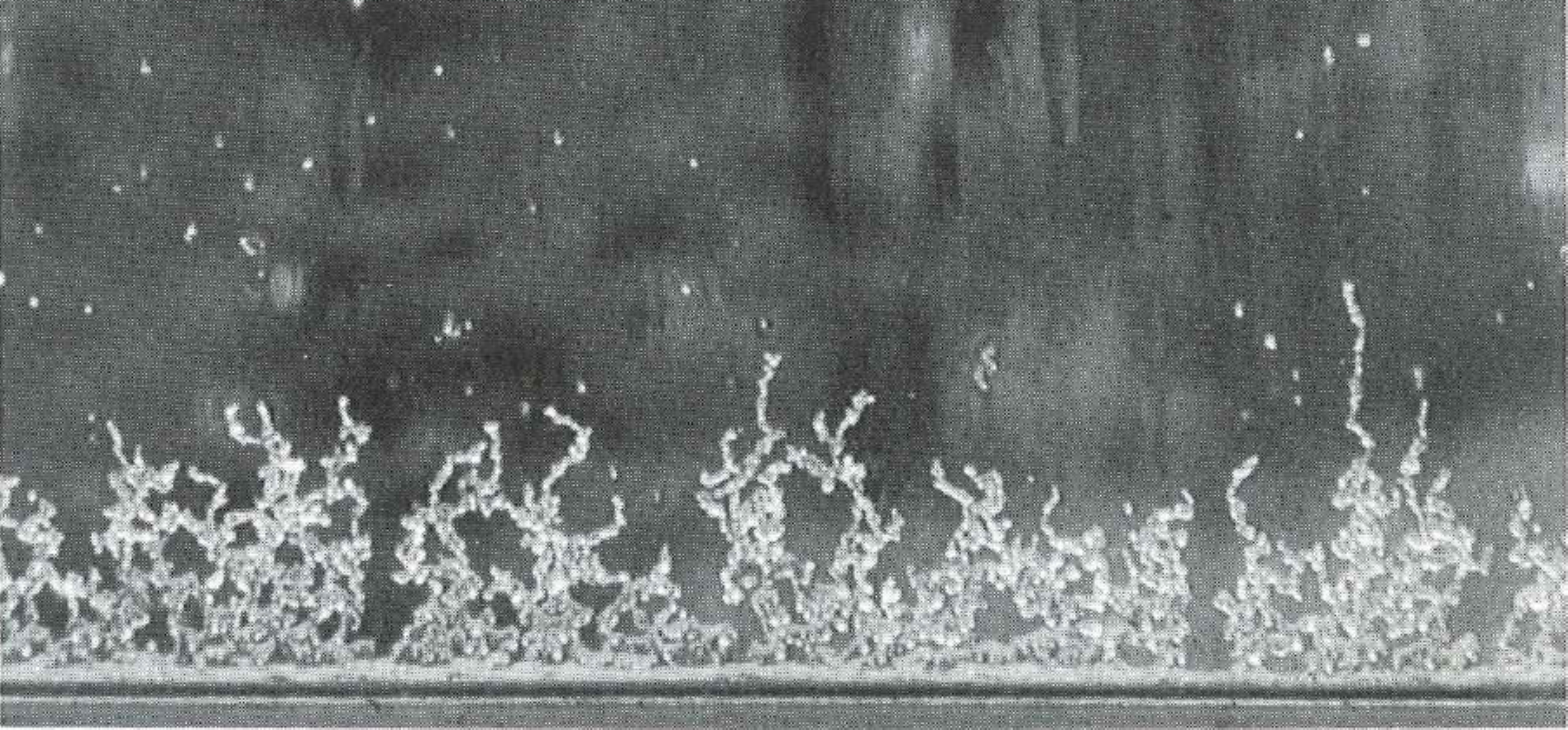}
    \caption[Fresh snow interface]{Fresh snow falling on the window and forming a \textit{rough} 
    interface. Extracted from \cite{Barabasi}.}
    \label{snow}
    \vspace{0.5 cm}
    \end{figure}

For instance, one can observe the process of snowflakes falling on the window, sliding down through it and 
sticking on the first aggregate they have met. Although the whole system is part of our daily 
experience and seems to be quite ``simple'', the generated interface is undoubtedly fascinating - see fig. \ref{snow}. 
One can notice the presence of large voids, branches and a \textit{rough} aspect.

\begin{figure}[ht]
    \vspace{0.5 cm}
    \centering
    \includegraphics[height = 5.0 cm, width = 10.0 cm]{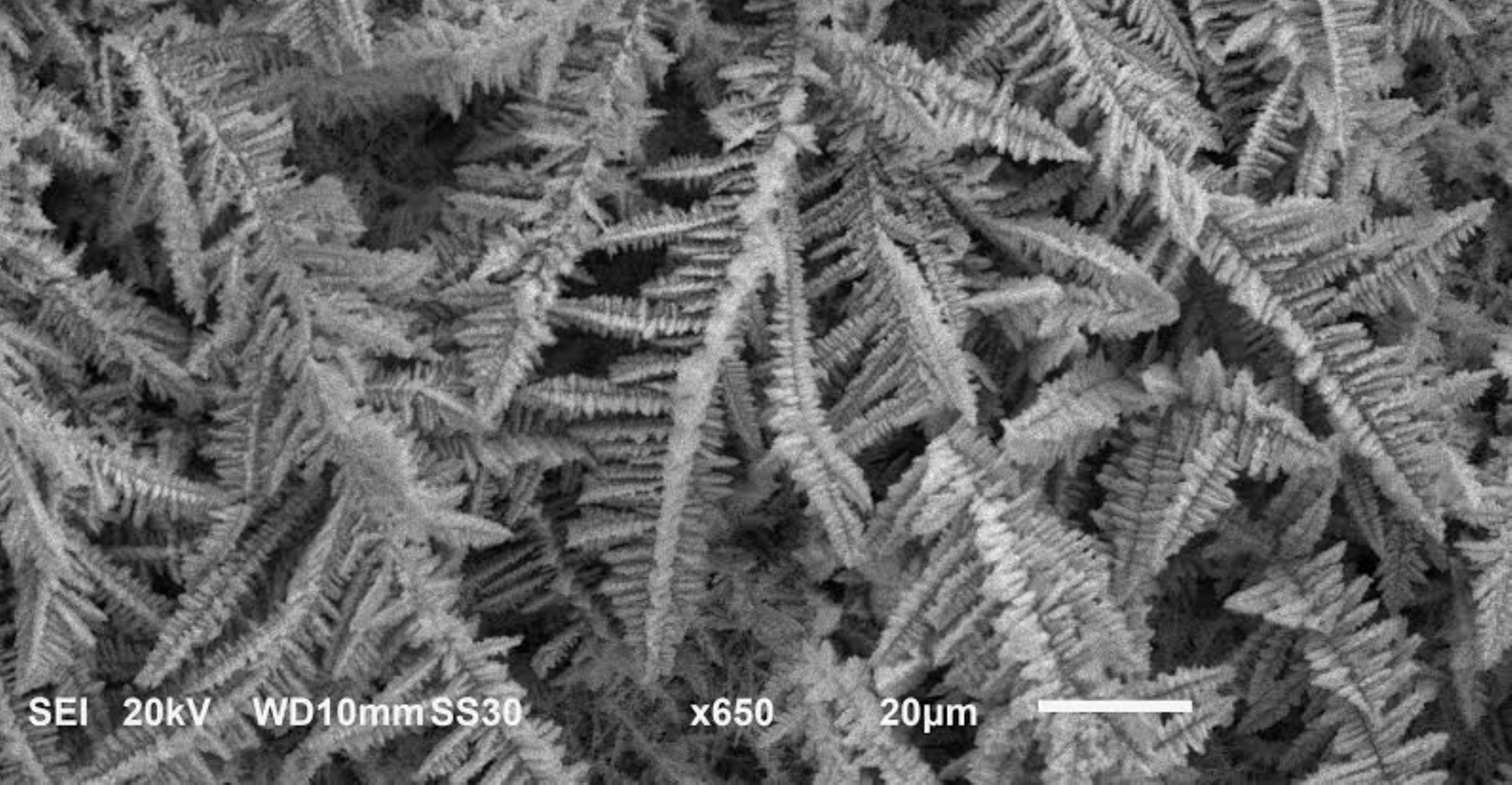}
    \caption[SEM image of Bi film grown on Si.]
    {SEM image of Bi thin film grown on Si by electrodeposition technique. 
    Courtesy of prof. Ren\^{e} C. Silva from Universidade Federal de Vi\c{c}osa, Brazil.}
    \label{BionSi}
    \vspace{0.5 cm}
    \end{figure}

Complex interfaces (in the sense of formation, growth and dynamic) are found 
in the nature from the bacterial growth, spreading of flame fronts, propagation of fluids in porous materials, etc, as well as 
surfaces made in the lab like those from MBE, eletroconvection and other thin-film deposition techniques \cite{Barabasi}. 
Figure \ref{BionSi} displays a Scanning Electron Microscopy (SEM) surface image of 
Bi film electrochemically grown on a Si substrate. This surface, grown on a two-dimensional substrate, is 
contrasted with that one formed by snowflakes and presents a rougher aspect which resembles to fern leaves, an example of 
macroscopic object exhibiting \textit{fractal properties}. Indeed, interface growth is a close topic to the fractal geometry 
by Mandelbrot \cite{Mandelbrot} and a concise link between them is established in the chapter 2.

In order to describe the dynamic of growing surfaces which are driven by \textit{local} processes, \textit{continuum growth equations} can be proposed. These equations does not take into account the microscopic nature of systems, but only the underlying symmetries and relevant relaxation processes that rule the dynamic at a coarsening-grained level and in a hydrodynamic limit\footnote{Details are found in the chapter 2.}. Under appropriate scale transformations, some statistical quantities are kept unchanged and define, this way, ``critical'' exponents. In a successful description, the set of these exponents is related to an Universality Class (UC) in which systems sharing dimensionality and large scale behavior can be grouped. 

Among a few UCs theoretically predicted, the most remarkable one is that of Kardar-Parisi-Zhang (KPZ) \cite{kpz1986}. It has became
a celebrated UC because: 1- its continuum growth equation, which is non-linear, can be mapped in classical equilibrium problems of the Statistical Physics \cite{Healy-Review} and in several mathematical motivated ones \cite{Ivan_Review}. 2- In d = 1 + 1 dimensions\footnote{This notation means one substrate topological dimension ($d_s$) plus one growth direction\cite{Barabasi}.}, important
KPZ models as the Single-Step model \cite{Johansson} and the Poly-Nuclear Growth model \cite{Prahofer} allowed uncovering a 
parallel between the (rescaled) height distribution of KPZ interfaces and the famous Tracy-Widow distributions emerging from the Random Matrix Theory \cite{TW1,TW2}. 3 - Ten years later, analytical treatments on the basis of the famous Directed Polymer in a Random Medium (DPRM) model \cite{Kardar_DPRM, Healy-Review} allowed to solve the KPZ(-DPRM) equation \cite{Sasamoto, Amir, Calabrese2, Dotsenko, Calabrese, Imamura}, in the meantime that experiments carried out by K. Takeuchi and M. Sano \cite{Takeuchi1, Takeuchi2, Takeuchi3, Takeuchi4}, regarding the growth of turbulent liquid crystals, have confirmed the analytical findings and suggested new universal KPZ features. Previous experiments concerning on the low combustion of paper \cite{Marco1, Marco2, Marco3, Marco4}, and a recent experimental realization on the deposit of colloidal particles at edge of water drops \cite{Yunker} have also enormously contributed to enoble the KPZ$_{d = 1 + 1}$ paradigm. Shortly after Takeuchi and Sano experiments \cite{
Takeuchi1, 
Takeuchi2, 
Takeuchi3, Takeuchi4}, numerical models belonging to the KPZ class have supported and, indeed, gone beyond liquid-crystal results \cite{Sidiney_EPL, Tiago1d, Prolhac, Healy_stationary} filling, finally, the last pieces toward a consistent KPZ$_{d = 1 + 1}$ \textit{triumvirate}.

In d = 2 + 1, the KPZ situation, however, is very different from its lower dimensional counterpart. There are not 
analytical results and almost one knows about the most important dimension for technological applications comes 
from numerical results: the scaling exponents \cite{Kelling, Marinari_exp} and the height \cite{Fabio2d}, squared local roughness \cite{Marinari, Racz} and maximal relative height distributions \cite{Tiago_mahd} in the \textit{steady-state} are older examples. Recently, the universality for height distributions in the \textit{growth regime} has been verified through large-scale simulations \cite{Tiago2d, Healy_2d1, Healy_2d2, Carrasco}. From the experimental side,
evidences of KPZ$_{d = 2 + 1}$ systems are extremely rare \cite{Ojeda, KPZ2d} and a long quest to find out one robust realization confirming the KPZ universality beyond scaling exponents has persisted until the beginning of 2014 \cite{Almeida, Healy_exp}. In fact, this poverty of experimental evidences has touching all UCs. Several works (see chapter 11 in the Ref. \cite{Barabasi} and section 3 in the Ref. \cite{Cuerno1}) have been reporting exponent values that do not match with anyone known UC. Inevitably, it has created a deep nuisance in the area, suggesting that the theoretical framework as yet is not completed. Advances as  the \textit{anomalous scaling} \cite{Schroeder, Lopez96, Lopez97, Lopez99, Ramasco00, Lopez05} have shown that  some systems are ruled by different exponents at local and global scales and that the analysis of interface fluctuations in the Fourier space are essentials for accessing the true form of the 
local scaling law \cite{Ramasco00}. However, in these situations, a larger number of exponents must be found in order to classify the interface dynamic. It makes the work still harder.
 
From an \textit{experimental point of view}, there are some basic aspects hampering the 
association between a two-dimensional growth and its UC. They are listed in the following:

\vspace{0.3 cm}

 \begin{itemize}
 \item[i)] \textbf{Difficulty for imaging surfaces and for growing films at long times -} Unlike one-dimensional 
 growing interfaces, it is hard recording the time evolution of two-dimensional 
 surfaces during the growth. Thus, in general, \textit{ex-situ} probe microscopic techniques are used for
 imaging the interface at discrete growth times. Each growth time corresponds to an one distinct 
 sample produced in the lab, which, depending on the growth technique and on the growth parameters, 
 can take some hours (or even days) to become ready. Because of this procedure, two-dimensional growths 
 are, usually, conditioned to the \textit{finite-time growth}, instead to the \textit{asymptotic-time growth}, 
 where the scaling regime of the true UC is expected to emerge. 
 
\item[ii)] \textbf{\textit{Finite-time} effects -} During the growth, the interface dynamic can suffer a 
 transition, i.e, a crossover. Temporal crossovers can be identified measuring the time evolution of some correlation 
 function\footnote{They are thoroughly defined in the section \ref{section_CF}.}, but even in simulations this is not an easy procedure \cite{WV, Schroeder, Smilauer, Xun}. For instance, studying the growth of SiO$_{2}$ on Si by CVD, F. Ojeda \textit{et al.}  \cite{Ojeda} have found the true asymptotic scaling regime only for samples grown at the range 
 of $\sim 10^2$ to $10^3$ min, after two initial crossovers - see fig. \ref{cross}. From this example, it is clear that if the data from an experiment corresponds to a crossover region (very difficult to be detected), the exponent value extracted from there could not match with any universal value. The same can occur when transient anomalous scaling takes place \cite{Schroeder}. Both
 facts are also reasons of why so many experimental works have found exponents which are not theoretically expected.

 \begin{figure}[!t]
    \vspace{0.5 cm}
    \centering
    \includegraphics[height = 5.5 cm]{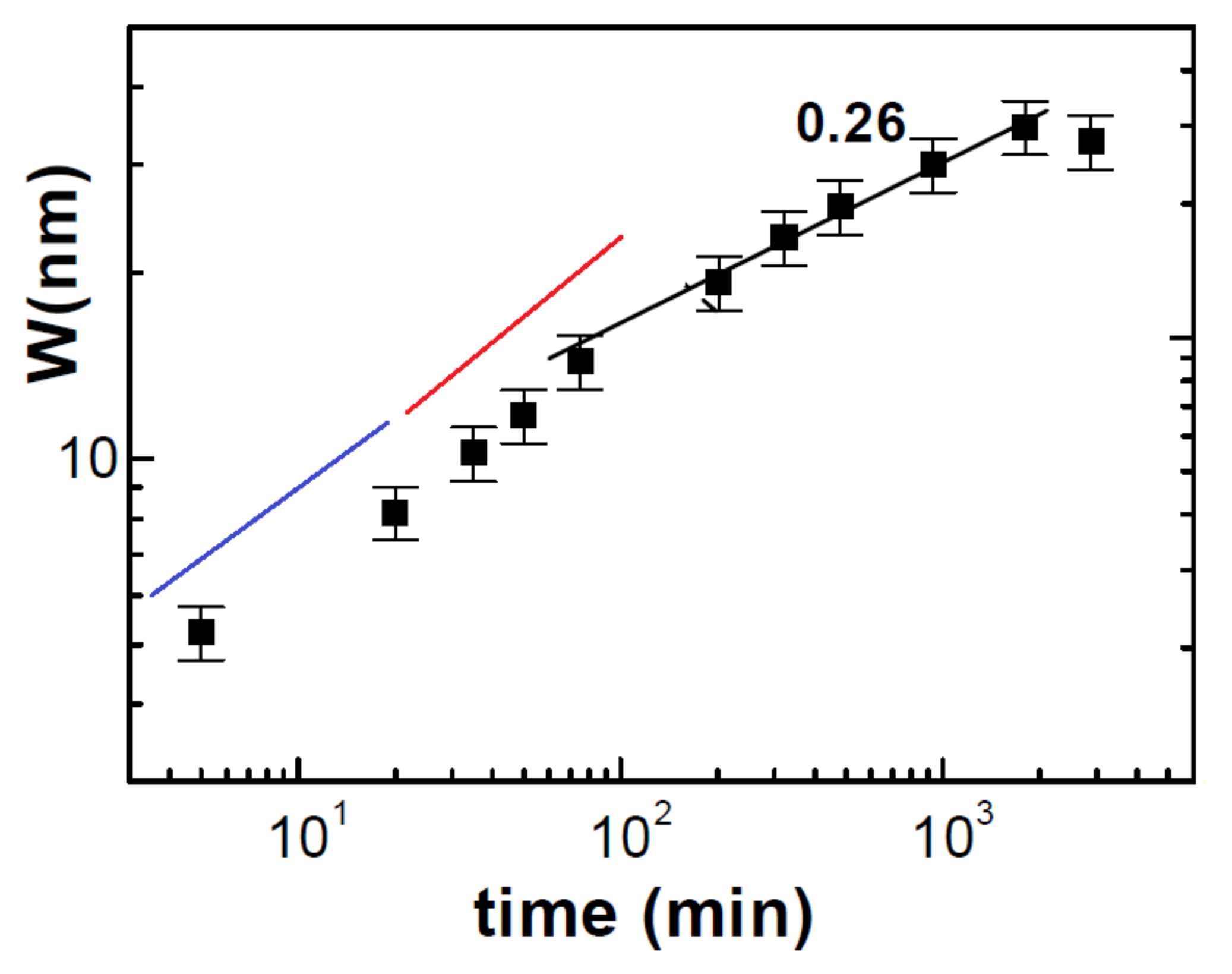}
    \caption[Example of crossovers occurring while an interface evolves.]
    {Example of crossovers ocurring in the two-dimensional growth of SiO$_2$ on 
    Si(001) substrates by CVD. The crossovers are identified 
    by different slopes (blue to red and red to black) at the W(t) scaling in a log $\times$ log plot. The short number of 
    samples makes hard to distinguish different regimes. The value 0.26 refers to the slope of the black curve and was
    related to the assymptotic scaling regime. Figure extracted and edited from \cite{Ojeda}.}
    \label{cross}
    \vspace{0.5 cm}
    \end{figure}
 
 \item[iii)] \textbf{The presence of morphological instabilities -} When the surface can be decomposed
 into an array of globules (grains, mounds, etc.), one says there is the presence 
 of morphological instabilities \cite{Machling}. This feature induces a characteristic length in the system
 $\zeta$, below which scaling invariance is awaited to be broken. The presence of $\zeta$ modifies
 local scaling of correlation functions and has led to many equivocated associations between experimental growths and UCs. In particular, non-universal local exponents have been confused with critical ones, as very well explained by Oliveira and Aar\~{a}o Reis \cite{Tiago_grain1, Tiago_grain2}. These studies are addressed in the section \ref{section_grainmodel}.

 \item[iv)] \textbf{Dimensional fragility of the KPZ equation -} Concerning on the KPZ class, as theoretically discussed in the 
 ref. \cite{Nicoli}, a same experimental KPZ$_{d = 1 + 1}$ system, when grown in d = 2 + 1 dimensions, might 
 has its exponents deviated from those expected for the KPZ class. This occurs due to the presence of morphological 
 instabilities and non-locality, which are introduced by hand as a perturbation term in the non-linear term of KPZ equation\footnote{See chapter 3.}. M. Nicoli \textit{et al.} have been calling attention for this 
 \textit{dimensional fragility} as a further obstacle in front of experimental confirmations 
 of KPZ$_{d = 2 + 1}$ growth. As extension, one can wonder whether this fragility is really a 
 particular feature of the KPZ equation or can take place in others non-linear growth equations.

\end{itemize}

\vspace{0.2cm}
 Guided by all this experience with two-dimensional experimental systems, as well as the
 deep knowledge about the KPZ$_{d = 1 + 1}$ \textit{triumvirate}, we have found out
 the first robust experimental confirmation of the KPZ class in two-dimensional systems, going beyond the \textit{standard} comparison with exponents \cite{Almeida}. Through a very systematic scheme for investigating the UC of growing films, which advances over 
 the comparison of scaling exponents and provide a reliable way for circumvent out (partially) the 
 obstacles ii-iv), we have confirmed that the growth of CdTe thin films on Si(001) for $T = 250 \,^{\circ}\mathrm{C}$ belongs to the 
 KPZ class as demonstrated by scaling exponents and universal height distributions, universal squared local roughness distributions and universal maximal relative height distributions \cite{Almeida}. Moreover, this is the first time that it has been experimentally demonstrated 
 the universality of such distributions \cite{Almeida}.
 
 As the goal of the present work is twofold (theoretically and experimentally-motivated), we have also studied
 the effect of the deposition temperature on this novel two-dimensional Kardar-Parisi-Zhang system. We have shown that
 in the presence of finite-time effects (crossovers and anomalous scaling), scaling exponents are not able
 to point, in a compelling way, the UC class of the system, mainly due to inexorable experimental obstacles. However,
 we have shown that this can be circumvented by analysing distributions, which allow us to prove that fluctuations
 of CdTe interface are described, asymptotically, by the Kardar-Parisi-Zhang equation in a broad range of temperature \cite{AlmeidaEPL}.
 Finally, one demonstrates that is possible tunning, experimentally, the KPZ non-linearity through the 
 deposition temperature \cite{AlmeidaEPL}.

 

 The rest of this dissertation is organized as follows. 
\textbf{Chapter 2} makes a link between fractals and surface fluctuations,
 introducing basic statistical tools. Up-to-dated results about the specific subjects of 
 each section can be found diluted along the text and, when indicated, in the Appendix sections. 
 The \textbf{chapter 3} is deserved for a short review on the Kardar-Parisi-Zhang equation, which includes
 the mainly ingredients and advances whithin the KPZ theory. In the \textbf{chapter 4}, experimental 
 procedures used during this work are carefully described, as well as the
 techniques and growth parameters. \textbf{Chapter 5} presents the first results of this work and exhibits a breaktrough
 in the context of the two-dimensional KPZ paradigm. On the sequence line, in the \textbf{chapter 6} one distills the 
 effect of the deposition temperature on the CdTe growth dynamic, and one explores the richnnes emerging from
 the CdTe system. Finally, in the \textbf{chapter 7}, conclusions and perspectives are drawn with a compreheensive
 overview of this work, standing out its mainly contributions to the context of kinetic roughening surfaces.
 Appendix sections cover topics as the random growth equation, the Villain-Lai-Das-Sarma (VLDS) class, etc., the origin and development 
 of anomalous scaling and, finally, an introductory material for young students concerning on the basic physics of crystal growth.

 \chapter{Fractals, Scale Invariance and Universality in Interface Growth}
 \lhead{\bfseries 2. Fractals, Scale Invariance and Universality in Non-Equilibrium Processes}
  In this chapter, a link between growing surfaces and their descriptions, built on the analogy of tools
 from Statistical-Mechanics of equilibrium phenomena at the criticality is given. Continuum growth equations
 are discussed and their critical exponents are extracted. Universality is discussed on the theoretical, numerical
 and experimental approaches.

\section{From Fractality to the Family-Vicsek \textit{Ans\"{a}tze}}

A \textit{fractal} is a complex and irregular\footnote{In the sense that Euclidean geometry can not
describe it.} object which, under an appropriated scale transformation, (any)one of 
its parts represent it as the whole. Mountains, clouds, tree leaves and coastlines are some examples of fractals
decorating the nature \cite{Mandelbrot}. In a mathematical sense, a fractal is 
said \textit{deterministic} if a ``zoom'' in the system always reproduces 
exactly the whole object. However, in the nature, one finds \textit{statistical fractals}. 
For example, if one compares two snapshots from a mountain at different magnifications they do not 
overlap but, nonetheless, their statistical properties are the same \cite{Barabasi}.

A fractal object presents a \textit{dilatation symmetry} 
(or \textit{homogeneity} property) \cite{Kardar_book2}. Statistical homogeneity is described 
by the condition:

\begin{equation}
 f(\varsigma^{\alpha_1}x_1, \varsigma^{\alpha_2}x_2, \varsigma^{\alpha_3}x_3, ...) = {\varsigma^{\alpha}}f(x_1, x_2, x_3, ...),
 \label{dilatation}
\end{equation}
where the object is formed by a set of points $f = f(\textbf{x})$, $\varsigma$ is 
a scale factor and $\alpha$ is called the H\"{o}lder exponent \cite{Barabasi}.

If ${\alpha_1} = {\alpha_2} = ... = {\alpha}$ (meaning an isotropic scale transformation) satisfies the eq. \ref{dilatation}, so one has \textit{self-similarity}. Otherwise, for anisotropic transformations, one has
\textit{self-affinity} \cite{Barabasi, Mandelbrot}.

At a \textit{critical point} of a phase transition, \textit{correlation 
functions}\footnote{A measurement of how local fluctuations in one part of the system affects 
those in the other ones. They are defined explicitly in the next section.} behave exactly 
as eq. \ref{dilatation} \cite{Kardar_book2}. In the Statistical Mechanics language, it 
implies that the \textbf{correlation length}\footnote{The length over the which one local part of the 
system affects the other ones.} ($\xi$) diverges and there is no 
characteristic length in the system beyond system size itself \cite{Kardar_book2}. 
For \textit{non-equilibrium processes} such as surface growth, it has been shown that interface
fluctuations also behave as those at critical point of a phase transition 
\cite{Barabasi, Kardar_book2, Cuerno1}. In other words, \textit{typically, interfaces provided by the
nature are fractals in the statistical sense}. Let $h(x,t)$ be the height of an one-dimensional 
interface at the position $x$ at time $t$. Assuming homogeneity, as in eq. \ref{dilatation}, then:

\begin{equation}
 h(x,t) = \varsigma^{-\alpha}h(\varsigma x, \varsigma^{z}t),
 \label{hx}
\end{equation}
where \textbf{z} is the \textbf{dynamic exponent}, and $\alpha$, in the surface growth context,
is named \textbf{roughness} exponent \cite{Barabasi}. Notice that, the dilation symmetry has also been assumed
on the temporal axis, which has lead to the z definition. Hence, $\alpha$ and $z$ are characteristic of systems
displaying spacial \textit{and} temporal scaling invariance.

A surface evolving in time is not a deterministic process. The height $h(x)$ at surface
is a \textit{stochastic variable} and have a set of possible \textit{outcomes} 
$\{h_1, h_2, h_3 ...\}$ distributed from a given
\textit{probability} $P(h)$. In general, the probability for a given height of the interface take the value $h$ is defined as 
the ratio between the number of occurrences $N_h$ of this specific event and the total number $N$ of heights sampled:

\begin{equation}
 P(h) = \lim_{N \rightarrow \infty} N_h/N.
 \label{prob}
\end{equation}

The probability depends on the range ($dh$) in which a value between $h$ and $h + dh$ is 
measured. So, is more interesting to define the \textit{probability density function} (pdf) 
$p(h) \equiv dP(h)/dh$ \cite{Kardar_book2, Kardar_book1, Callen}, which must
satisfy the normalization requirement:

\begin{equation}
 \int_{-\infty}^{\infty} dh \, p(h) = 1.
 \label{norma}
\end{equation}

A pdf is built up from its \textit{moments} or {cumulants}. The nth-\textit{moment} 
($m_n$) of a pdf is:
 
\begin{equation}
 m_n \equiv \langle h^n \rangle \equiv \int_{-\infty}^{\infty} dh \, p(h) \, (h^n),
 \label{mom}
\end{equation}
where the brackets mean to take the \textit{expectation value} of the variable $h^n$.

The moments are generated through the \textit{characteristic function} \~{p}$(k)$, which
is the Fourier transform of $p(h)$ \cite{Kardar_book1}. From the 
logarithm of \~{p}$(k)$, one defines the nth-\textit{cumulant} ($\langle h^n \rangle_c$):
 
\begin{equation}
 \ln \textrm{\~{p}}(k) = \sum_{n=1}^{\infty} \frac{(-ik)^n}{n!} \langle h^n \rangle_c.
 \label{cum}
\end{equation}

By expanding the logarithm in the \~{p}$(k)$ definition and comparing therm by therm with the expansion
of \~{p}$(k)$ in powers of k, one reaches to the relation between the moments and 
the cumulants \cite{Kardar_book1}. For the first four cumulants, one has, eq. \ref{cum2}:

\begin{equation}
\left. \begin{array}{p{10cm}}
$\langle h \rangle_c  = \langle h \rangle$,	\\
$\langle h^2 \rangle_c  =  \langle h^2 \rangle - \langle h \rangle^2$, \\
$\langle h^3 \rangle_c = \langle h^3 \rangle - 3\langle h^2 \rangle \langle h \rangle + 2\langle h \rangle^3$,	\\
$\langle h^4 \rangle_c = \langle h^4 \rangle -4\langle h^3 \rangle \langle h \rangle -  3\langle h^2 \rangle^2 + 
12\langle h^2 \rangle \langle h \rangle^2 - 6\langle h \rangle^4$.
\end{array} \right.
\label{cum2}
\end{equation}
The first cumulant is called \textit{mean} and the second one is \textit{variance}.

\begin{figure}[t]
    \vspace{0.5 cm}
    \centering
    \includegraphics[width = 8.0 cm]{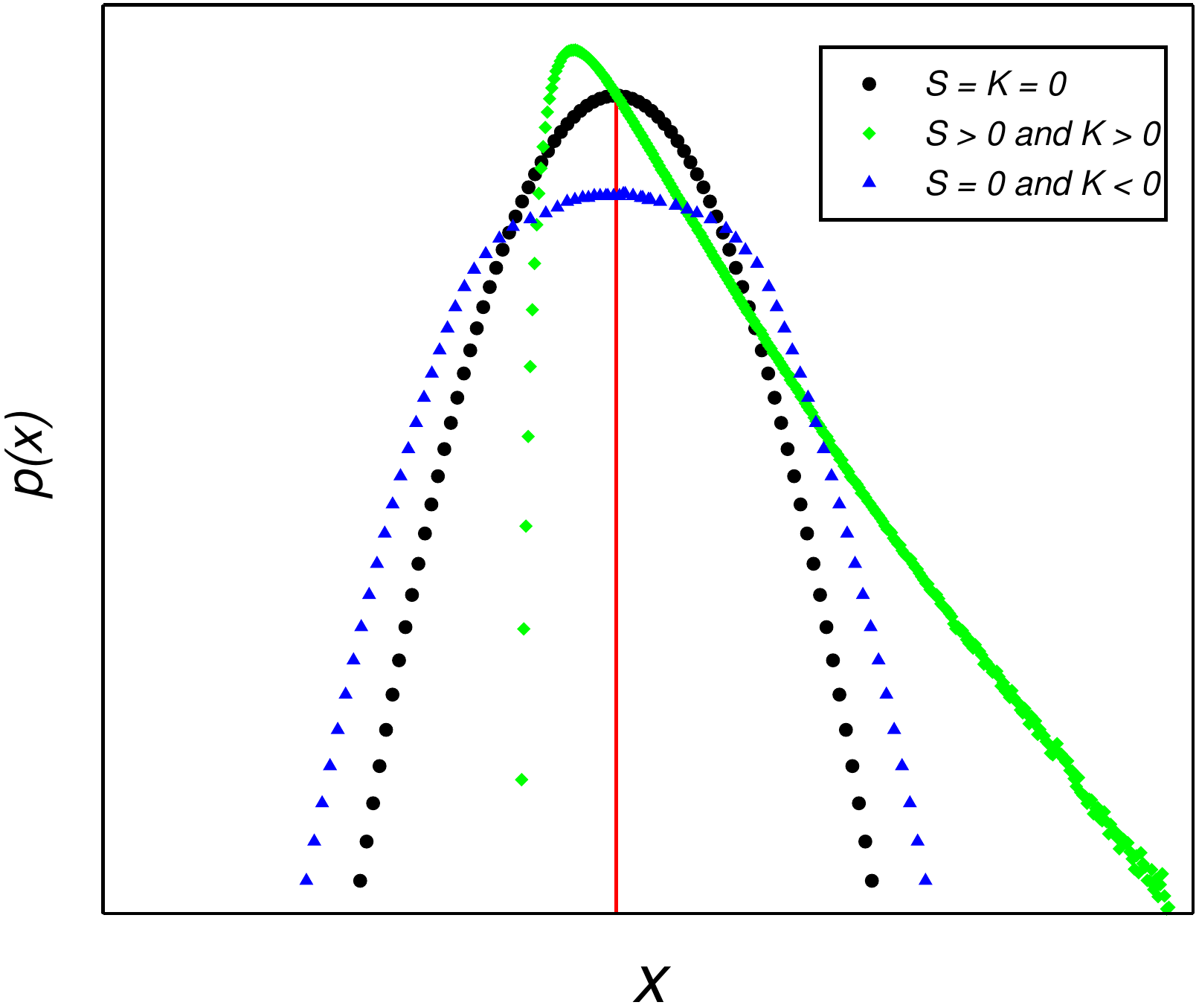}\quad
    \caption[Comparison between distributions having the same expectation value.]
    {Comparison between different distributions having the same expectation value, which is 
    indicated by the solid red line. A curve presenting S < 0 can be obtained performing a parity
    transformation on the green curve.}
    \label{skewf}
    \vspace{0.5 cm}
    \end{figure}

\vspace{0.2cm}
Other important quantities are dimensionless cumulant ratios. In particular, the
\textit{Skewness} S (eq. \ref{skew}) is an indicative of pdf asymmetry \cite{Kardar_book1}. If the expectation value is larger than 
the \textit{more probable} one, then $S > 0$; otherwise, $S < 0$. The \textit{Kurtosis} K (eq. \ref{skew}) provide 
information about the ``weight'' of pdf tails. A symmetric pdf with $K > 0$ presents a 
peak sharper than that from the Gaussian and its tails take longer to fall down. The opposite occurs for $K < 0$. By definition, the 
Gaussian pdf has $S = K = 0$. A representation comparing distributions with the same
expectation value for different S and K values is shown in the fig. \ref{skewf}.

\begin{equation}
 S = \langle h^3 \rangle_c/[\langle h^2 \rangle_c]^{3/2} \quad \textrm{and} \quad K = \langle h^4 \rangle_c/[\langle h^2 \rangle_c]^2.
 \label{skew}
\end{equation}

The global squared \textbf{roughness} ($w^2$) of an interface is defined as the variance of heights composing it\footnote{In surfaces with translational symmetry, this is also the variance of $p(h)$.}. Interestingly, the roughness (or width) is a very important variable from both experimental and theoretical point of view. The electrical conductivity of some thin films, for example, is strongly reduced as 
rougher is their surface \cite{Ohring, Venables, Pimpinelli}. The importance from the
theoretical side will be clear soon. Consider an one-dimensional interface of size $L$. The 
roughness of such interface at time $t$ reads:

\begin{equation}
 w(L,t) = [\langle h(x,t)^2 \rangle - \langle h(x,t) \rangle^2]^{1/2},
 \label{roug}
\end{equation}
where the $\langle...\rangle$ refers to a spatial average over the whole system of size $L$.

Now, we can return to the hypothesis made in eq. \ref{hx}. Inserting this equation in the
eq. \ref{roug} and performing algebraic manipulations\footnote{One must assume $\varsigma^zt = 1$. This can be done because
$\varsigma$ is just an arbitrary factor. We can change its value until obtain the result $\varsigma^zt = 1$.} one finds:

\begin{equation}
w(L,t) = t^{\alpha/z}f(L/t^{1/z}),
\label{roug2}
\end{equation}
this equation precede a canonical \textit{ans\"{a}tze} in the kinetic roughening theory. Defining
the \textbf{growth exponent} $\beta \equiv \alpha/z$, the Family-Vicsek (FV) dynamic scaling
\textit{ans\"{a}tze} predicts that $f$ asymptotically behaves as $f(u) \sim u^{\alpha}$ for $u \ll 1$ 
and $ \sim$ $const$ for $u \gg 1$ \cite{FVS}. Leading the \textit{ans\"{a}tze} for eq. \ref{roug2}, finally, 
one obtains:

\begin{equation}
 w(L,t) \sim \left\{ 
       \begin{array}{c}
       t^{\beta},\quad \textrm{for} \quad t^{1/z} \ll L,\\
       L^{\alpha},\quad \textrm{for} \quad t^{1/z} \gg L.
       \end{array} \right.
\label{roug3}
\end{equation}

 \begin{figure}[!h]
    \vspace{0.5 cm}
    \centering
    \includegraphics[width = 9.0 cm]{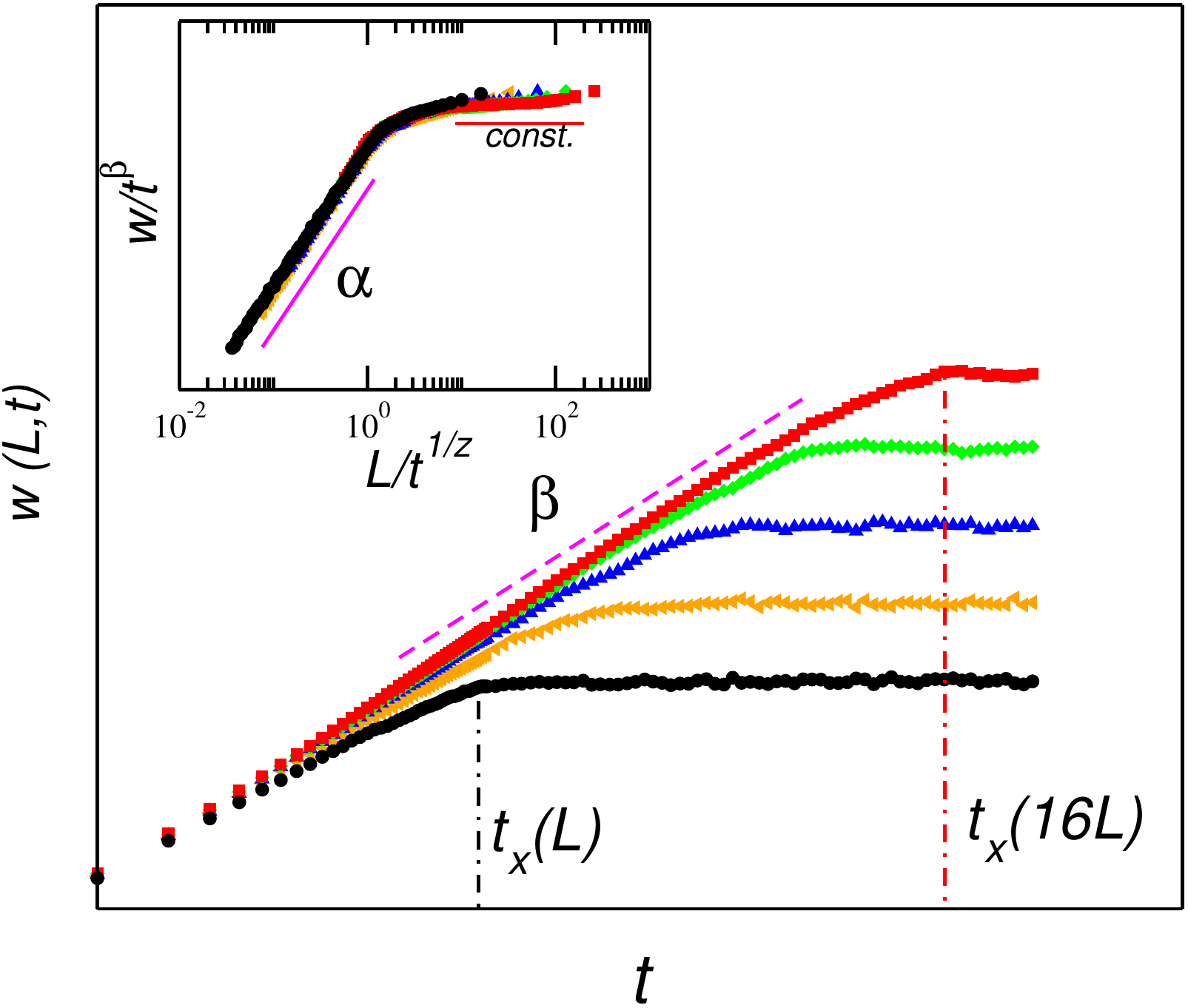}
    \caption[Typical roughness versus time plot (log $\times$ log) during an interface 
    growth.]{Typical time roughness plot (log $\times$ log) during an interface growth for
    different substrate sizes. The curves refer to substrate of size: black circles (L),
    orange left triangles (2L), blue up triangles (4L), green diamonds (8L) and
    red squares (16L). $t_x$ indicates the characteristic time in which $\xi_{||} \approx$ substrate size.
    Inset confirms the FV \textit{ans\"{a}tze}, described in eq. \ref{roug3} using the 
    appropriate values for $\beta$ and $z$.}
    \label{wmodel}
    \vspace{0.5 cm}
    \end{figure}

\vspace{0.2cm}
Roughness behavior is sketched in the fig. \ref{wmodel}. The system roughens as $t^{\beta}$ while the correlations are spreading through it. This is 
called \textit{growth regime}. At the time $t_x$, the \textit{correlation length} 
becomes of the same order of $L$ and a distinct regime is reached: roughness stops growing in time and turns to depend solely on $L$. 
This is the \textit{steady state} or \textit{saturation regime}. Comparing what eq. \ref{roug3} tell us, it is clear
to associate $\xi_{||} \sim t^{1/z}$, where $\xi_{||}$ is the \textit{parallel} 
correlation length. In inset of fig. \ref{wmodel}, the FV \textit{ans\"{a}tze} is tested, showing
remarkable collapse for the curves. We notice that the roughness presents a power-law dependence in space and time, as occurs with
the correlation functions in equilibrium critical phenomena. The parallel goes further and,
indeed, the ``critical'' exponents ($\alpha$ and $z$) \textit{do not depend on microscopic
details of the system under investigation,} i.e, there is universality in fluctuations of growing interfaces.

Once the saturation regime depends on $t^{1/z}$ be of the same order of $L$, in experimental situations where 
$L$ is much larger than the characteristic size of particles constituting the interface, the time required to the system
gets into the steady state is hardly achieved. As far as we know, only the evolution of profiles (d = 1 + 1) yielded by 
slow combustion of paper \cite{Marco1} and the growth (d = 2 + 1) of SiO$_2$ films by CVD (after 2 days of deposition) 
have experimentally achieved the stationary state \cite{Ojeda}.

The FV \textit{ans\"{a}tze} has been confirmed in several examples of surface growth 
such as the propagation of fluid flow in a porous medium \cite{Barabasi}, paper wetting \cite{Barabasi}, the 
growth of turbulent liquid crystals \cite{Takeuchi1, Takeuchi2, Takeuchi3}, the slow 
combustion of paper sheets \cite{Marco1, Marco2} and so on \cite{Krug_advances}. However,
the FV scaling is not the most general one. Indeed, FV \textit{ans\"{a}tze} fails for describing local scaling of growth process
with \textit{anomalous roughening}. A detailed description of anomalous scaling is let to the appendix section \ref{section_anomalous}.

\section{Correlation Functions}
\label{section_CF}
Correlation functions (CF) play a central role in systems exhibiting criticality (out-of or at
equilibrium) because they provide a measurement of the correlation length. They 
are formed by an operation (sum, product, etc.) involving
suitable quantities describing the \textit{microscopic} state of the 
system\footnote{Examples are the local magnetization for a spin lattice, the local height 
for an interface, etc. }, which are separated by a distance of $\boldsymbol{l}$. Taking an interface of 
linear size L for instance, one can describe it by its height field $\{h\}_L$ or 
even by its slope field $\{\nabla h\}_L$ (of course, in an appropriate 
coarsening grained level). A CF for an interface growing can, hence, be written as:

\begin{equation}
 C_h(|\boldsymbol{l}|,t) = \langle[ h(\textrm{\textbf{x}}, t) -  h(\textrm{\textbf{x}} +  \boldsymbol{l}, t) ]^2\rangle.
\label{hh_CF}
 \end{equation}

If the system presents translational symmetry, \textit{any CF will
depend solely on the magnitude of $\boldsymbol{l}$} \cite{Kardar_book2}, what is not true for anisotropic 
systems (see \cite{Edoardo_thesis} and ref. therein). Furthermore, as discussed at the beginning 
of this chapter, a CF displays a dilation symmetry at
the criticality. Hence, inserting eq. \ref{hx} in eq. \ref{hh_CF} and performing algebraic manipulations 
one reaches to:
 
\begin{equation}
 C_h(l,t) = t^{2\alpha/z}f(l/t^{1/z}),
 \label{FV_CF}
\end{equation}
where $f(u)$ is the scaling function that, for while, obeys the FV ans\"{a}tze. $C_h$ is often
called \textit{height-difference correlation function}. 

\begin{figure}[h]
    \vspace{0.5 cm}
    \centering
    \includegraphics[width = 9.0 cm]{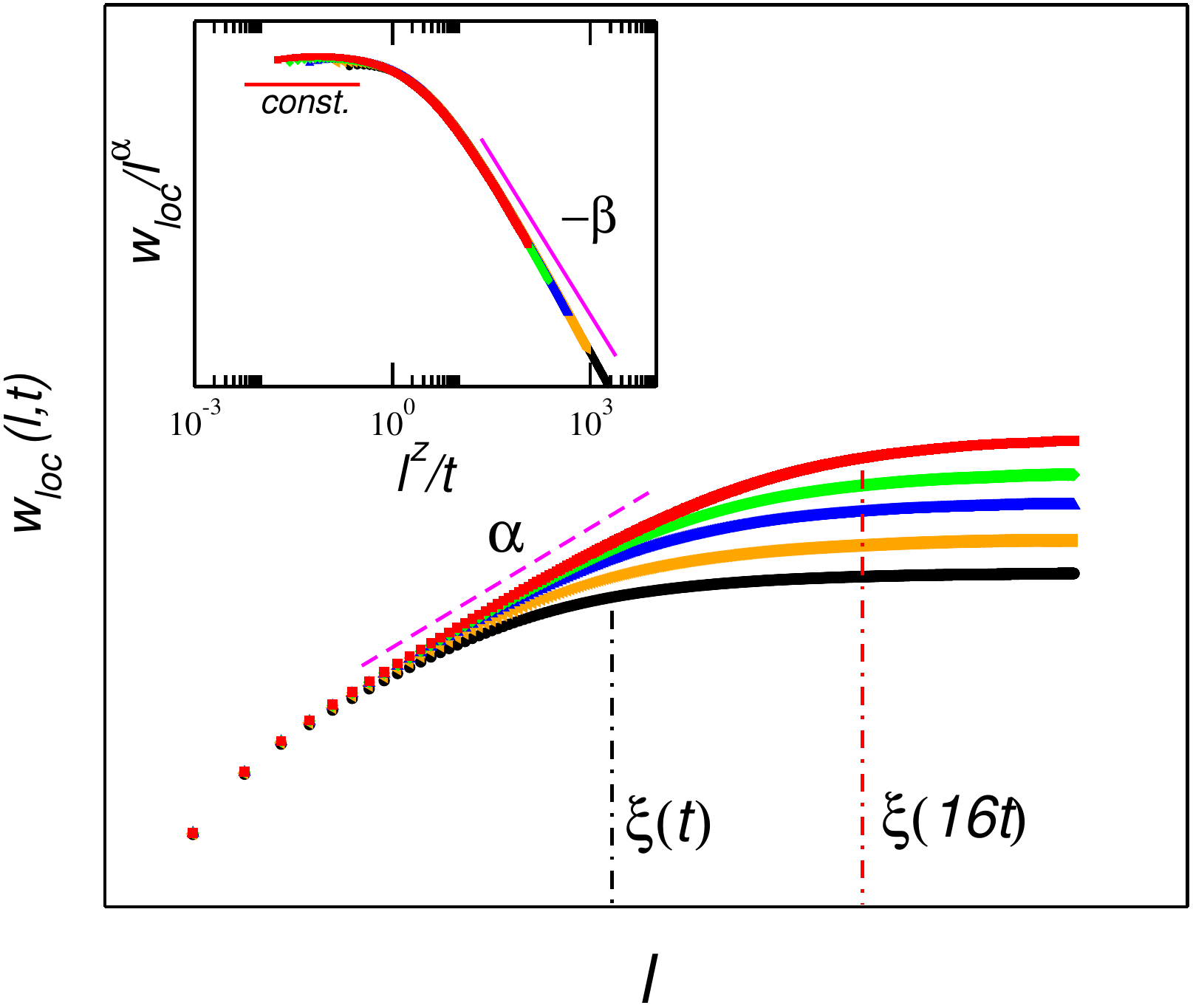}
    \caption[Typical local roughness plot (log $\times$ log) for an interface grown at time t.]{Typical local roughness plot (log $\times$ log) for an interface grown at different times.
    The curves refer to growth times: black circles (t),
    orange left triangles (2t), blue up triangles (4t), green diamonds (8t) and
    red squares (16t). Vertical dashed lines indicate the $\xi_{||}$ measured.
    Inset confirms the FV \textit{ans\"{a}tze}, described in eq. \ref{roug3} with L replaced by $l$, and
    using the appropriate values for $\alpha$ and $z$.}
    \label{wlocmodel}
    \vspace{0.5 cm}
    \end{figure}

Notice that this is the same scaling form obtained for the roughness in the eq. \ref{roug2} with $\beta$ replaced 
by $2\beta$ and $L$ by $l$. So, at first note, the example showed in the fig. \ref{wmodel} is shared by $C_h(t)$, but with 
the growth regime evolving as $t^{2\beta}$. In the same way, one can replace $L$ by $l$ in the eq. \ref{roug} to obtain
a scaling form for the local roughness [$w_{loc}(l, t)$], which is equally 
a correlation function\footnote{In the absence of anomalous roughening, the w$_{loc}$ scaling
follows the FV \textit{ans\"{a}tze} as in the eq. \ref{roug2}.}. From an experimental point of view, it is unpractical changing $L$ of a system 
in order to extract $\alpha$. Rather, one often uses local measurements, spanning boxes of lateral length $l$
in the interval $[0, L]$ and obtaining the exponent from the 
hypothesis $C_h^{1/2}(l, t) \sim w_{loc}(l, t) \sim l^{\alpha}$. Fig. \ref{wlocmodel} 
shows a typical behavior of $w_{loc}$ for interfaces in the growth regime. 

Other important CF is the \textit{spatial covariance} of heights ($C_s$), defined as:

\begin{equation}
 C_s(|\textbf{l}|, t) = \langle[ h(\textrm{\textbf{x}}, t)h(\textrm{\textbf{x}} +  \boldsymbol{l}, t) ]\rangle -\langle h \rangle^2,
\label{cov}
 \end{equation}
where it is straightforward to show that $C_h = 2w_{loc}^2 - 2C_s$. 

The covariance $C_s(l,t)$ has been computed, in particular, for one-dimensional models belonging to the 
Kardar-Parisi-Zhang (KPZ) Universality Class (UC) because they are universal and given by the covariance of Airy processes \cite{Prahofer, Takeuchi1, Tiago2d} (see fig. 5 in the ref. \cite{Takeuchi1}). 
Very recently, T. Halpin-Healy and G. Palazantzas have found numerically and confirmed experimentally the universality of the 
rescaled $C_{s(KPZ)}$ also in 2 + 1 dimensions \cite{Healy_exp}.

Likewise, one can define \textit{slope-slope} correlation functions using $\nabla$h, instead of
h, in the eq. \ref{hh_CF} and eq. \ref{cov}. Indeed, several 
experimental studies \cite{Fabio, Cordoba, Almeida} have used the slope-slope covariance (eq. \ref{slop_CF})
in order to obtain an estimative of $\xi_{||}$.

\begin{equation}
 \Gamma(|\textbf{l}|, t) = \langle[ \nabla h(\textrm{\textbf{x}}, t) \nabla h(\textrm{\textbf{x}} +  \boldsymbol{l}, t) ]\rangle.
 \label{slop_CF}
\end{equation}

\begin{figure}[!ht]
    \vspace{0.5 cm}
    \centering
    \includegraphics[width = 9.0 cm]{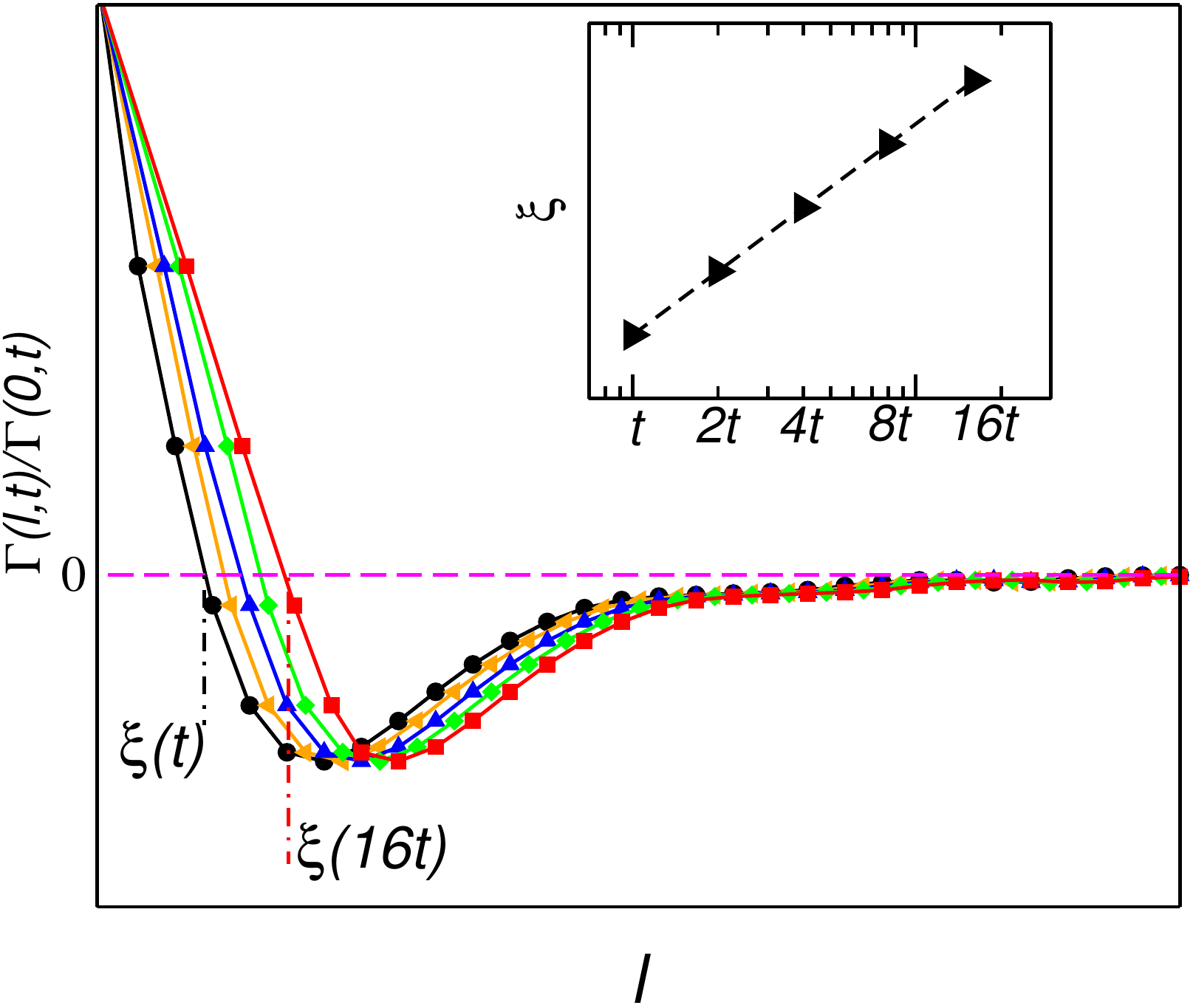}
    \caption[Example of Slope-Slope covariance function]{ Typical normalized slope-slope covariance as function of 
    the distance l and the time t in a linear $\times$ linear plot. Curve colors
    refer to growth times: black (t), orange (2t), blue (4t), green (8t) and red (16t). Magent dashed line sets $\Gamma(|\textbf{l}|, t)/\Gamma(0,t)$ equal to 0. Vertical dashed lines indicates the position of the estimated correlation length.
    The inset shows $\xi_{||}$, as function of time in a log $\times$ log plot, from where the exponent $1/z$ can be extracted.}
    \label{slopes}
    \vspace{0.5 cm}
    \end{figure}

An estimative of $\xi_{||}$ can be done measuring either the first zero or the first minimum
of the curve \cite{Siniscalco}. Figure \ref{slopes} shows a plot of a typical behavior of the slope-slope covariance
as function of the distance l and time t. The procedure used for estimating $\xi_{||}$ is also indicated, as well as
the plot of $\xi_{||}$ as function of the growth time, from where the exponent $1/z$ can be found.

 \section{Continuum Equations and Universality Classes}
 \label{section_UC}
 
 Let an interface be described by its height field \{h(\textbf{x},$t_{0}$)\}, in a appropriate coarsening
 grained level, and consider there is a known driving force, i.e., there is a far-from-equilibrium situation. An important
 question concerning on it could be: \textit{how does one can describe the height-field evolution of the interface?} 
 A priori, the answer does not appear to be reachable because it seems that there is a lack of informations such as:
 i) the \textit{kind} of interface (biological, physical, chemical reaction front, etc.) which is being dealt and 
 ii) the specific interactions ruling the dynamics at the microscopic level.
 
 Indeed, an approach to the question from this point of view makes the problem practically intractable.
 Nevertheless, looking at the macroscopic scales of evolving interfaces, one can glimpse that the \textit{collective behavior}
 of them shares many underlying similarities, which does not depend, in fact, of i) and ii). Hence, based on this one could
 perform a long-wavelength description of the system.
 Advancing in these thoughts, one can argue that, beyond of the long-wavelength hypothesis, the long-time regime is also a 
 necessary condition, since transient (finite-time) behaviors should be avoided. This particular limit of
 long wavelengths and long times is called \textit{hydrodynamic limit} by analogy to the Navier-Stokes equation 
 for a fluid of particles \cite{Kardar_book2}.
 
 Guided by these ideas, one can build general equations for describing the height-field evolution of growing surfaces, with 
 the form:
 
 \begin{equation}
 \partial_{t}h(\textbf{x},t) = F + \Theta(\textbf{x}, \{h\}, t) + \eta,
 \label{gerall}
 \end{equation}
 with F being the average number of particles per unit time arriving at the interface\footnote{Assume the particle size being 
 equal the unit, in order to satisfy dimensional analysis.}, the called driving force. Now, one must account that
 this \textit{arriving process} is stochastic, and the therm $\eta$ is inserted in the equation for capturing this feature. 
 Regarding $\eta$, one has:
 
 \begin{itemize}
  \item If the active zone (interface front) advances onto an inhomogeneous medium as a porous substrate 
  or a paper sheet, the relevant noise in the process is static and changes point to point in the medium. This is called
  of \textit{quenched noise}, once $\eta = \eta(\textbf{x}, h)$.
  If only the quenched noise is present, the growth of the interface has a \textit{deterministic} evolution. However, 
  the \textit{thermal noise}, which is always present in experiments, destroys this determinism.
  
  \item Beyond of thermal noise, when an interface grows by receiving particles from an external flux, 
  the so called \textit{shot noise} is also present and plays the crucial role on the dynamic. This noise 
  comes from the inherent randomness occurring in the deposition process. Assuming there is no 
  preferential area onto the substrate for receiving molecular flux, one has that $\langle \eta(\textbf{x}, t) \rangle = 0$
  with the spatial-temporal covariance given by:
  \begin{equation}
  \langle \eta(\textbf{x}, t)\eta(\textbf{x'}, t')\rangle = 2D\delta^{d_s}(\textbf{x} - \textbf{x'})\delta(t - t'),
  \label{noise_cov}
  \end{equation}
  where $\sqrt{D}$ is the amplitude of the white noise.
 
 \end{itemize}

 Now we turn the attention to the forms which the functional $\Theta(\textbf{x}, \{h\}, t)$ can assume supposing that the
 dynamic of the interface is ruled by \textit{local} processes. As an underlying hypothesis, descriptions of a physic fact 
 can not depend on the origin of its observation $\overleftrightarrow{S} = (t_0, \textbf{x}, h_0)$. Thus, spatial and temporal translations must be satisfied by the eq. \ref{gerall}. It rules out from the functional 
 explicit terms involving $t^n, \textbf{x}^m$ or $h^m$, where $m \in \mathbb{R^*}$. Moreover, as the growth
 does not make distinction between ``right''- and ``left-handed'', the eq. \ref{gerall} must also be invariant 
 under spatial parity transformations with respect to the $\textbf{x}$ axis. These hypothesis reduce the allowed 
 terms in $\Theta$ to combinations of even derivatives such as $(\nabla^{2n}h)(\nabla h)^{2p}$, with $n, p \in \mathbb{N}$.
 
 In a general picture, our considerations until here have led us to consider:
 
\begin{eqnarray}
\partial_{t}h(\textbf{x},t) = F + a_1\nabla^{2}h + a_2\nabla^{4}h + ... + b_1(\nabla h)^{2} + b_2(\nabla h)^{4} + ... {}
\nonumber\\
+ c_{11}(\nabla^2 h)(\nabla h)^2 + ... + c_{np}(\nabla^{2n} h)(\nabla h)^{2p} + \eta,
\label{gerall2}
\end{eqnarray}
where $a_i$, $b_i$ and $c_{np}$ are appropriate quantities making the eq. dimensionally consistent.
 
 As we are interested in the hydrodynamic limit, derivatives of higher order are irrelevant to the asymptotic scaling
 behavior (see ref. \cite{Barabasi} pag. 49). Thus, \textit{the simplest general equation} involving these terms reads:
 
  \begin{equation}
 \partial_{t}h(\textbf{x},t) = F + a_1\nabla^{2}h + b_1(\nabla h)^{2} + c_{11}(\nabla^2 h)(\nabla h)^2 + \eta,
 \label{gerall3}
 \end{equation}
 
 As we shall see, important continuum growth equations are encoded in the equation \ref{gerall3}. Nevertheless, there are
 also others important growth equations that call for a physically-motivated term which makes appear higher derivatives of $h$. In the
 following section we discuss this subject in details.
 
 
 \subsection{Edwards-Wilkinson and the Linear-MBE equation}
 \label{section_EWMH}
 The Edwards-Wilkinson (EW) equation (eq. \ref{Ew_eq}) was proposed in 1982 for describing the sedimentation 
 of granular particles \cite{EW}. The equation preserves parity symmetry in the growth direction with respect 
 to the mean height, leading the height pdf to has $S = 0$. This particular symmetry rules out 
 even powers of $h$ such as $(\nabla h)^{2n}$ and, in accordance with the general eq. \ref{gerall3}\footnote{One can show that
 the term $(\nabla^2 h)(\nabla h)^2$ under renormalization is irrelevant compared with $(\nabla^2 h)$. See Ref. \cite{Barabasi} pag. 49.}, it reads:
 
 \begin{equation}
 \partial_{t}h(\textbf{x},t) = \nu\nabla^{2}h + \eta(\textbf{x},t),
 \label{Ew_eq}
 \end{equation}
where $\nu \equiv a_1$ is a quantity of dimension $[m^2/s]$ in the S.I. convention.

 The growth average velocity of the interface $v \equiv \langle \partial_{t}h \rangle = Ft$, 
 once $\langle \nabla^{n} h \rangle$ vanishes for periodic bound conditions. The eq. above
 is written at the referential of the mean height, avoiding the explicit dependence on the flux term and setting
 $v = 0$.
 
 Physically, the laplacian term acts as a \textit{conservative smoothing mechanism}
 redistributing the irregularities on the interface, \textit{while maintaining the average 
 height unchanged} (see pag. 50 in the ref. \cite{Barabasi} for a geometric interpretation).
 
 Due to the linear character of this equation, critical exponents can be found straightforwardly by rescaling or 
 by Fourier transform methods. As a prototype case, we shall find exponent values by applying rescaling tools 
 on a general linear equation obtained replacing $2$ by $2n$ on the nabla operator in the eq. \ref{Ew_eq}. The 
 method by Fourier transform can be found in details in the pag. 173 of the Ref. \cite{Krug_advances}.
 
%
 \vspace{0.2cm}
 \textbf{Rescaling}: Suppose a rescaling as $\textbf{x} \rightarrow \varsigma\textbf{x}$, 
 $t \rightarrow \varsigma^zt$ and $h \rightarrow \varsigma^{\alpha} h$. Inserting in the eq. \ref{Ew_eq} with $2$ 
 replaced by $2n$ on the 
 nabla, one has:
 
 \begin{equation}
  \varsigma^{\alpha - z}\partial_th = \nu \varsigma^{\alpha - 2n} \nabla^{2n} h + \eta(\varsigma \textbf{x}, \varsigma^z t),
  \label{rescal_1}
 \end{equation}
where, using the noise covariance definition (eq. \ref{noise_cov}) and the delta function 
properties $\delta^{d_s}(\varsigma \textbf{x}) = \varsigma^{-d_s}\delta(\textbf{x})$, one can rewrite the rescaled 
noise in the expression above by $\varsigma^{(-z-d_s)/2} \eta(\textbf{x},t)$. Now, assuming scale invariance, 
one finds the critical exponents (valid for $n \geq 1$):

 \begin{equation}
  \alpha = \frac{2n - d_s}{2} \qquad and \qquad z = 2n.
  \label{linearexp}
 \end{equation}

\vspace{0.2cm}

The EW eq. is recovered for $n = 1$, which yields $\alpha = (2 - d_s)/2$ and $z = 2$. In particular, for $d_s = 2$, one
obtains $\alpha = 0$ and $\beta = 0$ meaning that in the eq. \ref{roug3} the roughness exhibit a \textit{logarithm} dependence
on $t$ and on $L$. The set of exponents $\alpha = \alpha(d_s)$ and $z = 2$ compose the so called 
\textit{Edwards-Wilkinson Universality Class}, well described in the hydrodynamic limit by the EW equation. A famous numerical model belonging to the EW class
is the Random Deposition with Surface Relaxation (RDSR) proposed in 1986 by Family \cite{Family_RDSR}, where
the deposited particles are allowed to diffuse at surface until reaching the local lowest height. 
Experimental evidences of surfaces belonging to the EW class, in turn, are very rare. As far as we are concerned, the EW universality has only
been found in the growth of W multilayers on Si by magnetron sputtering \cite{Salditt, Salditt2} and in the sedimentation 
process of SiO$_2$ nanospheres \cite{Salvarezza}.

Returning to eq. \ref{linearexp}, if one sets $n = 2$ one obtains the exponents for the famous growth equation known 
as \textit{linear-MBE} equation (eq. \ref{MH_eq}), firstly proposed by Wolf and Villain \cite{WV} and 
Das Sarma and Tamborenea \cite{DT} for describing the growth of surfaces in which diffusion is 
the relevant growth mechanism.

 \begin{equation}
 \partial_{t}h(\textbf{x},t) = -K_d\nabla^{4}h + \eta(\textbf{x},t),
 \label{MH_eq}
 \end{equation}
$K_d$ accounts for the strength of the diffusion and has dimension of [$m^4/s$].

A derivation of eq. \ref{MH_eq} from a conservation law is let to the appendix section \ref{section_MH}. Indeed, the deterministic form of eq. \ref{MH_eq} was known (and was solved) since a long time ago by 
Herring \cite{Herring} and Mullins \cite{Mullins} considering the effect of the scale on sintering phenomena and
the development of thermal grooves, respectively. Due this, often the \textit{stochastic} 
growth equation is called Mullins-Herring equation, instead linear-MBE. However, in this dissertation
we will refer to the eq. \ref{MH_eq} as linear-MBE equation, while its related universality class is being called
\textit{Mullins-Herring} (MH) class.

The critical exponents constituting the MH class are (from eq. \ref{linearexp}, with $n = 2$):

\begin{equation}
\alpha = \frac{4 - d_s}{2}\qquad and \qquad z = 4.
\end{equation}

\vspace{0.2cm}
Theses exponents were firstly calculated by Wolf and Villain as intention of describing the WV model \cite{WV}, although,
shortly after was found that, actually, the model does not belong to the MH class \cite{Xun}. Numerical models 
capturing the diffusion mechanism and truly belonging to the MH class carries the name of Das Sarma, Tamborenea, Ghaisas 
and J. Kim \cite{DSG, Kim_MH}. A classical review paper on the models can be found in the ref. \cite{Das}, while a discussion on the WV model is described in the chapter 15 of the ref. \cite{Barabasi}.

On the experimental side, evidences of MH class have been found in the growth of Si on Si(111) by MBE for $T = 275\,^{\circ}\mathrm{C}$ and 
deposition rate of 7 bilayers/min \cite{Yang} [scaling was obtained through STM images]; thermal evaporation of 
amorphous Si on Si(111) substrates at deposition rate of $0.8 \pm 0.2$ \AA{}$/s$ \cite{Yang2} [AFM]; the sputter-deposition 
growth of Pt on glass at deposition rate of 6 \AA{}$/s$ and with the normal substrate aligned about $45\,^{\circ}$ with 
the target surface normal \cite{Jeffries} [STM]; in fluctuations of intra-grain domains of gold electrodeposits 
grown at 100 $nm/s$ \cite{Vazquez} [STM] and in inter-grain fluctuations of $LiCoO_x$ thin films grown 
by rf sputtering, after annealing process \cite{Kleinke_2} [AFM], as well as in the circular growth 
of cultivated brain tumor \cite{Bru} [optical microscope].

\subsection{The Linear Deposition-Desorption-Diffusion equation and Grain Models from Oliveira and Reis}
\label{section_grainmodel}

At this point, we could insert together the diffusion and the desorption mechanisms in a continuum growth equation. 
This makes sense, once specific MBE conditions (very high deposition temperature and/or low supersaturation) 
can lead the growth to depend sensitively on both processes. Deposition occurs solely if there is a supersaturation ruling the vapor-to-solid
phase transition, it means that the difference of chemical potential between the vapor ($\mu_v$) and above the solid phase 
($\mu(\textbf{x},t)$) is positive. However, when the supersaturation is low [$(\mu_v - \mu) \gtrsim 0$], thermal fluctuations 
make possible, during a characteristic time, a particle to escape from the interface and to move
toward the vapor phase. In these conditions, and including diffusion, one has:

\begin{equation}
 \partial_{t}h(\textbf{x},t) = F - B[(\mu_v - \mu)] - K_d\nabla^{4}h + \eta(\textbf{x},t),
\end{equation}
with B having dimension of inverse of linear momentum.

Inserting $\mu(x,t) \propto - \nabla^2 h$, one finds the general deposition-desorption-diffusion equation:

\begin{equation}
 \partial_{t}h(\textbf{x},t) = F + \nu \nabla^2h - K_d\nabla^{4}h + B\mu_v + \eta(\textbf{x},t),
 \label{dd_eq}
\end{equation}
where the ratio $(K_d/\nu)^{1/2}$ has dimension of length and, thus, define a characteristic length in the system, $\zeta$.

A trivial rescaling in the eq. \ref{dd_eq} provides:
\begin{equation}
\varsigma^{\alpha - z}\partial_th = \nu \varsigma^{\alpha - 2} \nabla^{2} h - K_d \varsigma^{\alpha - 4} \nabla^{4} h + B \mu_v + \varsigma^{-z-d_s}\eta(\textbf{x}, t),
\end{equation}
where for long-wavelength fluctuations ($\varsigma \rightarrow \infty$, $l \gg \zeta$), the laplacian dominates the growth and
the exponents are consistent with the EW class. At short-length scales ($\varsigma \rightarrow 0$, $l \ll \zeta$), 
however, diffusion mechanism overcomes the laplacian effect and the growth is dictated by the linear-MBE equation.

This behavior has been confirmed, for instance, in copper electrodeposition in the presence of 1,3-diethyl-2-thiorea, an organic 
additive with concentration $x$ ($0.3 \leq x \leq 0.4$ $mM$), for low current density $j = 0.02$ A $cm^{-2}$ \cite{Vazquez2}.
In this work, the authors have found $\alpha$ and $\beta$ exponents for both regimes, strongly confirming the MH-EW crossover. Notwithstanding,
in many studies on the growth of thin films the local roughness is the only curve analyzed, from where $\alpha$ is usually extracted. In particular,
when $\alpha \approx 1.00$ a linear-diffusion-dominated dynamic is often suggested, but Oliveira and Reis have demonstrated that this procedure is fail in most of cases \cite{Tiago_grain1, Tiago_grain2}.

\begin{figure}[h]
    \vspace{0.5 cm}
    \centering
    \includegraphics[width = 11.0 cm, height = 6.0 cm]{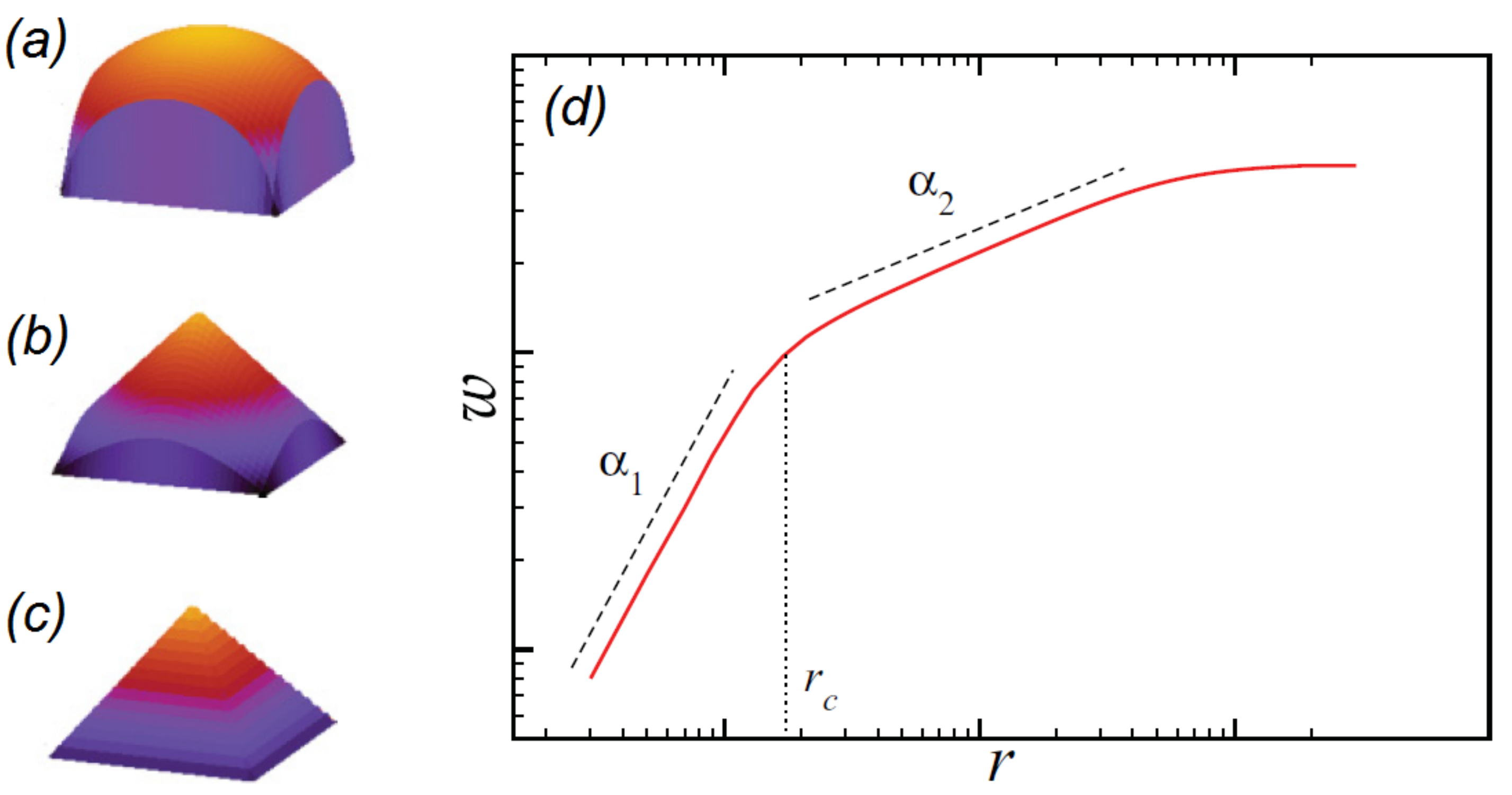}
    \caption[Grain shapes and general local roughness curve.]{Grain of (a) semi-elliptical, (b) conical and
    (c) pyramidal shapes. (d) General local roughness curve (log $\times$ log) for grainy surfaces. Fig. extracted and edited from
    Oliveira and Reis \cite{Tiago_grain1, Tiago_grain2}.}
    \label{tiago_wcurve}
    \vspace{0.5 cm}
    \end{figure}

Studying the effect of grains at surface on the local roughness scale, Oliveira and Reis have shown that the general
$w_{loc}(r,t)$\footnote{We are preserving the original notation of the paper \cite{Tiago_grain2}.} plot exhibits two crossovers - see and follow the fig. \ref{tiago_wcurve}. The first regime happens for $r \ll r_c$,
where $r_c$ is the average grain size 
and is dictated by the exponent $\alpha_1$. Length scales between grains $r_c \ll r \ll \xi$ define the second regime growing as $w_{loc} \sim r^{\alpha_2}$. The second crossover separates correlated from non-correlated
regions of the grainy surface.

Interestingly, it has been demonstrated that $\alpha_1$ has a dependence
on the grain geometric shape, on the local correlation function and on the procedure to calculate root-mean-square
averages. \textit{It implies that $\alpha_1$ can not be related to a critical exponent in the sense of capturing universal 
fluctuations at surface}. The geometric interpretation is corroborated with the local roughness curve,
where the $\alpha_1$ value can change from $0.75$ to $ 1.00$ for surfaces composed by pyramidal and flat top grains, respectively. 
The $\alpha_2$ exponent, however, keeps constant at the expected UC value related to the model. 
We recommend the reader take a look at the original papers from Olivera and Reis \cite{Tiago_grain1, Tiago_grain2}. 
Comparison with experiments has been carried out and, just to few some experimental results matching with Olivera and Reis predictions, 
one finds: the spray pyrolysis growth of ZnO films \cite{Ebothe}, which gives $0.94 \ll \alpha_1 \ll 0.97$ for high flow rates; 
the electrodeposition of cooper by Mendez \textit{et al.} \cite{Mendez}, presenting $\alpha_1 = 0.87 \pm 0.06$; 
the sputtering of niquel oxided films ($\alpha_1 = 0.70$) \cite{Cruz}; 
the growth of bilayers of poly(allylamine hydrochloride) and a side-chain-substituted azobenzene copolymer (Ma-co-DR13), 
after deposition of 10 or 20 bilayers, giving $\alpha_1 \approx 0.80$ \cite{Desouza}; 
and the surfaces of Langmuir-Blodgett films of polyaniline and neutral biphosphinic ruthenium complex, 
which yields $0.66 \ll \alpha_1 \ll 0.82$ \cite{Desouza2}.
 \chapter{The Kardar-Parisi-Zhang Universality Class: A Brief Historical and State-of-the-Art}
 \lhead{\bfseries 3. Kardar-Parisi-Zhang Universality Class}

This chapter is a natural sequence of the previous one, but due to the richness of the Kardar-Parisi-Zhang equation an entire chapter is deserved to the discussions that one follows. A basic narrative on the motivation, development
and current status of this paradigmatic continuum growth equation is presented. We do not have the intention 
to englobe in this text all what the equation has been demonstrating to offer. Instead, 
we focus on its scaling properties, on important mappings and on its universal distributions.

\section{Scaling, Mappings, Height Distributions}
\label{kpz1}
The Kardar-Parisi-Zhang (KPZ) equation was proposed in 1986 for describing growing interfaces
in which growth in the local normal direction plays a decisive role in the asymptotic dynamics \cite{kpz1986}. Based
on universality ideas, as well as on existing growth models as the Eden model for cell colony formation \cite{Eden}, and 
the ballistic deposition model for colloidal aggregates \cite{FVS, Vold}, 
KPZ proposed the simplest nonlinear stochastic equation for describing the height-field dynamic of such interfaces: 

\begin{equation}
\label{eq_kpz}
\frac{\partial h(\textbf{x},t)}{\partial t} 
= \nu \nabla^{2} h + \frac{\lambda}{2} (\nabla h)^{2} + \eta(\textbf{x},t),
\end{equation}
$\nu$ represents a ``surface tension'' due to the geometric interpretation of laplacian term (see ref. \cite{Barabasi}),
whereas the nonlinear term accounts for the growth in the \textit{local normal} direction and $\eta$ is the noise.

 \begin{figure}[ht]
    \vspace{0.5 cm}
    \centering
    \includegraphics[height = 5.0 cm ]{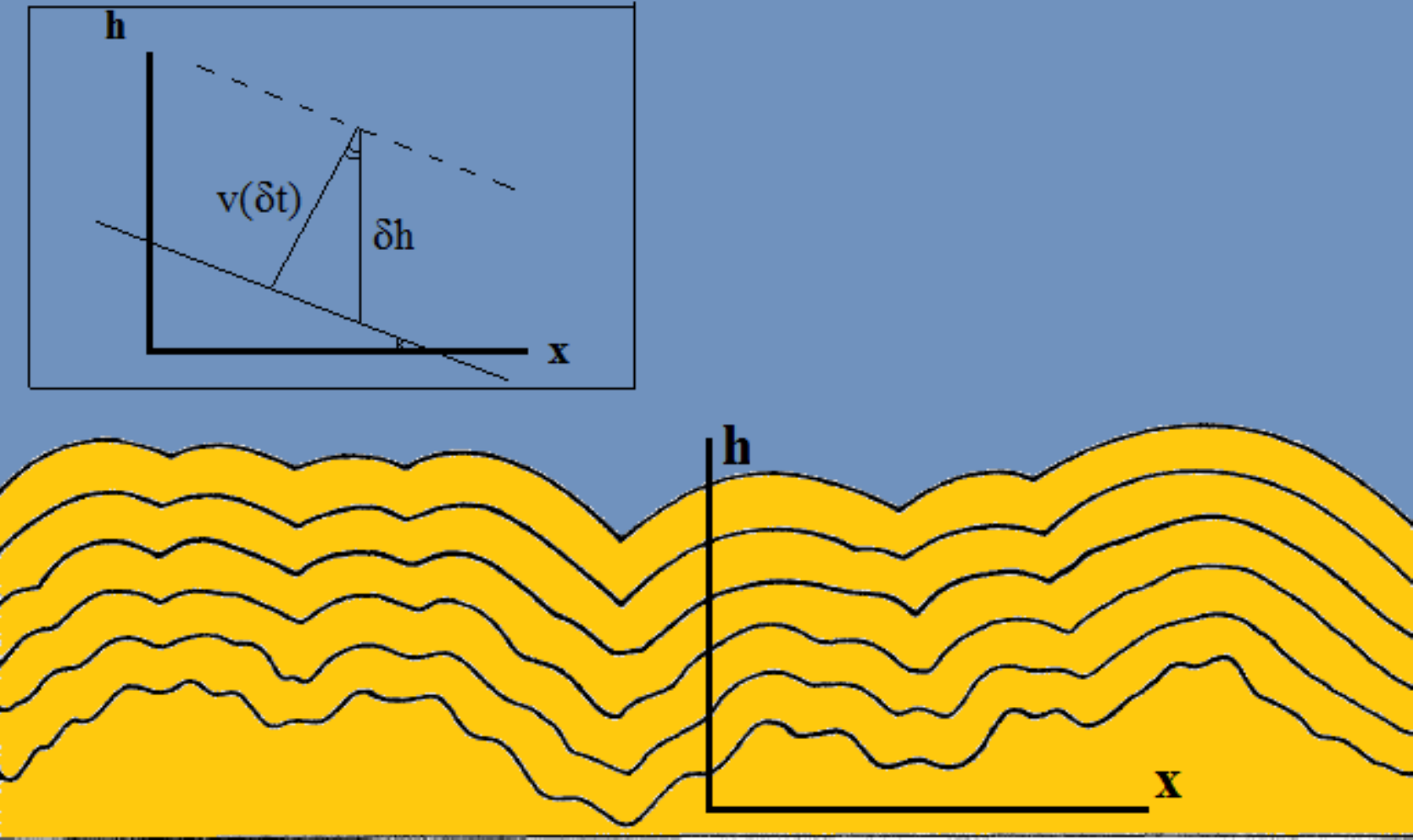}
    \caption[Successive profiles for a deterministic KPZ growth.]{Successive profiles 
    for a deterministic KPZ growth with a rough initial condition. Inset indicates how local 
    normal growth occurs. Extracted and edited from \cite{kpz1986}.}
    \label{fig_kpz}
    \vspace{0.5 cm}
    \end{figure}

As indicated in the inset of fig. \ref{fig_kpz}, when an interface grows laterally, the increment 
along the $h$ axis ($\delta h$) and the \textit{local normal velocity} $v$ are related 
by the Pythagoras' theorem: $\delta h = [(v\delta t)^2+(v \delta t \nabla h)^2]^{1/2}$. Inserting the small
slope approximation ($\nabla h \ll 1$)\footnote{This approximation is not required 
in others derivations, whose large scale limit take form of the KPZ equation \cite{Barabasi}.}, one can perform 
an expansion at the right side of the equation leading to: $\frac{\partial h}{\partial t} \simeq v + (v/2)(\nabla h)^{2} 
+ ...$ . Eq. \ref{eq_kpz} is obtained after a transformation
to the moving frame ($h \rightarrow h + vt$) and after the relaxation mechanism is 
inserted. Nonlinear higher-order terms are disconsidered because they go to zero faster
than $(\nabla h)^2$ in the hydrodynamic limit, as already discussed in chapter 2.

Unlike linear growth equations, the mean interface velocity  of the KPZ equation takes the form: 
$v = Ft + \frac{\lambda}{2} \int_0^L d^{d_s}x\langle(\nabla h)^2\rangle$, which is different from zero even when $F = 0$. $\lambda$ 
accounts for this ``excess of velocity''. In particular, if $\lambda = 0$ one obtains the EW equation (eq. \ref{Ew_eq}) and if, additionally $\nu = 0$, one finds the 
Random Growth equation (see appendix section \ref{section_RD}).

The KPZ equation can be easily solved in its deterministic form ($\eta(\textbf{x},t) = 0$), which
leads to interfaces composed by paraboloid segments resembling dendrites \cite{kpz1986}, see fig. 
\ref{fig_kpz} for a typical one-dimensional growth pattern. On the other hand, the stochastic equation (eq. \ref{eq_kpz}) has
been resisting to a complete analytical handling. Great part of the advances on the solution of the KPZ equation has moved 
on owing to the arsenal of powerful tools that mathematicians has brought to the field. For instance, analytical models that
belong to the KPZ equation as the Single-Step model \cite{Johansson} and the Poly-Nuclear-Growth (PNG) model \cite{Prahofer, Prahofer2}
have, togheter with, the Totally Asymmetric Exclusion Process (TASEP) \cite{Ferrari_Sphon} and the Directed Polymer in a Random
Potential (DPRM) \cite{Kardar_DPRM} (which can mapped onto the height field of a KPZ interface), played a crucial role within the KPZ theory.
For this last, at the beginning of this decade, analytical treatments have been achieved for d = 1 + 1 dimensions for
different geometries, which in the surface growth context refers to a curved \cite{Sasamoto, Amir, Calabrese2, Dotsenko}, 
flat \cite{Calabrese} and stationary \cite{Imamura} growth Initial Conditions (IC). Solutions of higher dimensions, however,
are in a fog of analytical hopes.

Regarding to the KPZ scaling, if one applies, naively, the trivial rescaling used in the section \ref{section_EWMH}, one obtains 
\textit{three} self-inconsistent scaling relations. Arguing that the nonlinear term should dominate the growth in the 
hydrodynamic limit and applying the rescaling again (without the laplacian term) 
one finds $\alpha = (2-d_s)/3$ and $\beta = (2-d_s)/(4+d_s)$, whose predictions are quite different from numerical results 
\cite{Healy-Review, Healy-Krug, Tiago1d, Sidiney_EPL, Sid_JSM, Kelling}. Indeed, this procedure is wrong because terms like $\nu, \lambda$ and $D$ do 
not renormalize independently - being coupled to each other \cite{Barabasi}. 
The correct prediction for KPZ exponents can be achieved using Renormalization-Group techniques and 
mappings to other problems. Through the $u = -\partial_x h$ transformation, the KPZ equation with $\lambda = 1$ is 
mapped in the Burgers' equation with noise \cite{Burger}, describing the 
vorticity-free velocity field of a stirred fluid. Additionally, the fact that $\lambda = 1$ in the Burgers' equation points out 
that we must preserve the scale invariance on the nonlinear term in the KPZ context. This reasoning provides the 
hyper-scaling relation:

\begin{equation}
\alpha + z = 2.
\label{eq_sacred}
\end{equation}
This relation is a consequence of the Burguer equation to be invariant under a \textit{galilean} transformation, 
$ u(x,t) \rightarrow u_0 + u(x-v_0t,t) $, which in turn, leads the KPZ equation to be invariant under
a \textit{tilting transformation} by an angle $\epsilon$:

\begin{equation}
h' \rightarrow h + \epsilon x; \quad x' \rightarrow x - \epsilon \lambda t; \quad t' \rightarrow t.
\end{equation}

The \textit{fluctuation-dissipation} theorem \cite{Barabasi} reveals that, at the \textit{steady state} (for $d = 1 + 1$), 
$\nabla h$ follows a Gaussian distribution likewise the position of a particle in a 
Brownian motion and brings out $\alpha = 1/2$. Together with eq. \ref{eq_sacred}, this implies
$z = 3/2$ and $\beta = 1/3$, in agreement with previous \cite{Healy-Krug, Healy-Review} and recent \cite{Tiago1d, Sidiney_EPL, 
Sid_Balistic, Healy_stationary} numerical models belonging to the \textit{KPZ universality class}.

The concept of ``universality beyond exponents'' has initiated in the classical paper from Krug, Meakin and 
Halpin-Healy, back to 1992's \cite{Healy-Krug}. In that work, by using the so-called \textit{``Krug-Meakin toolbox''} \cite{KM89, KM90}, 
it was possible to write universal amplitudes in terms of model parameters, namely, $A \equiv f(D/\nu)$\footnote{This relation is valid only for d = 1 + 1 dimensions. In higher dimensions A can also be function of $\lambda$ \cite{Healy-Krug}.} and $\lambda$, which can
be easily obtained in simulations \cite{Healy-Krug, Tiago1d, Healy_stationary, Healy_exp}. Almost 10 years after this imporant result, Johansson studied a model for which several probabilistic interpretations can be given \cite{Johansson}. Among them, an interpretation in terms of the one-dimensional TASEP (in turn mapped onto the interface fluctuations of the Single-Step model) allowed to show that the pdf of a random amplitude ($\chi$) related to the height field of the Single-Step model (eq. \ref{eq_ansatz})
is the celebrated Tracy-Widom (TW) distribution \cite{TW1, TW2}, emerging from the Random Matrix Theory context. 

\begin{equation}
h(t) \simeq v_{\infty}t + sign(\lambda)(\Gamma t)^{\beta}\chi.
\label{eq_ansatz}
\end{equation}
Here $v_{\infty}$ is the asymptotic velocity of the interface [$v_{\infty} \equiv lim_{t, L \rightarrow \infty} \langle \partial_t h \rangle$], $sign(\lambda)$ is the signal function, $\Gamma \equiv a A^{1/\alpha}|\lambda|$, with
$a$ being a constant and $A \equiv f(D/\nu)$ in d = 1 + 1 dimensions (see Ref. \cite{Healy-Krug}).

 \begin{figure}[ht]
    \vspace{0.5 cm}
    \centering
    \includegraphics[height = 6.0 cm ]{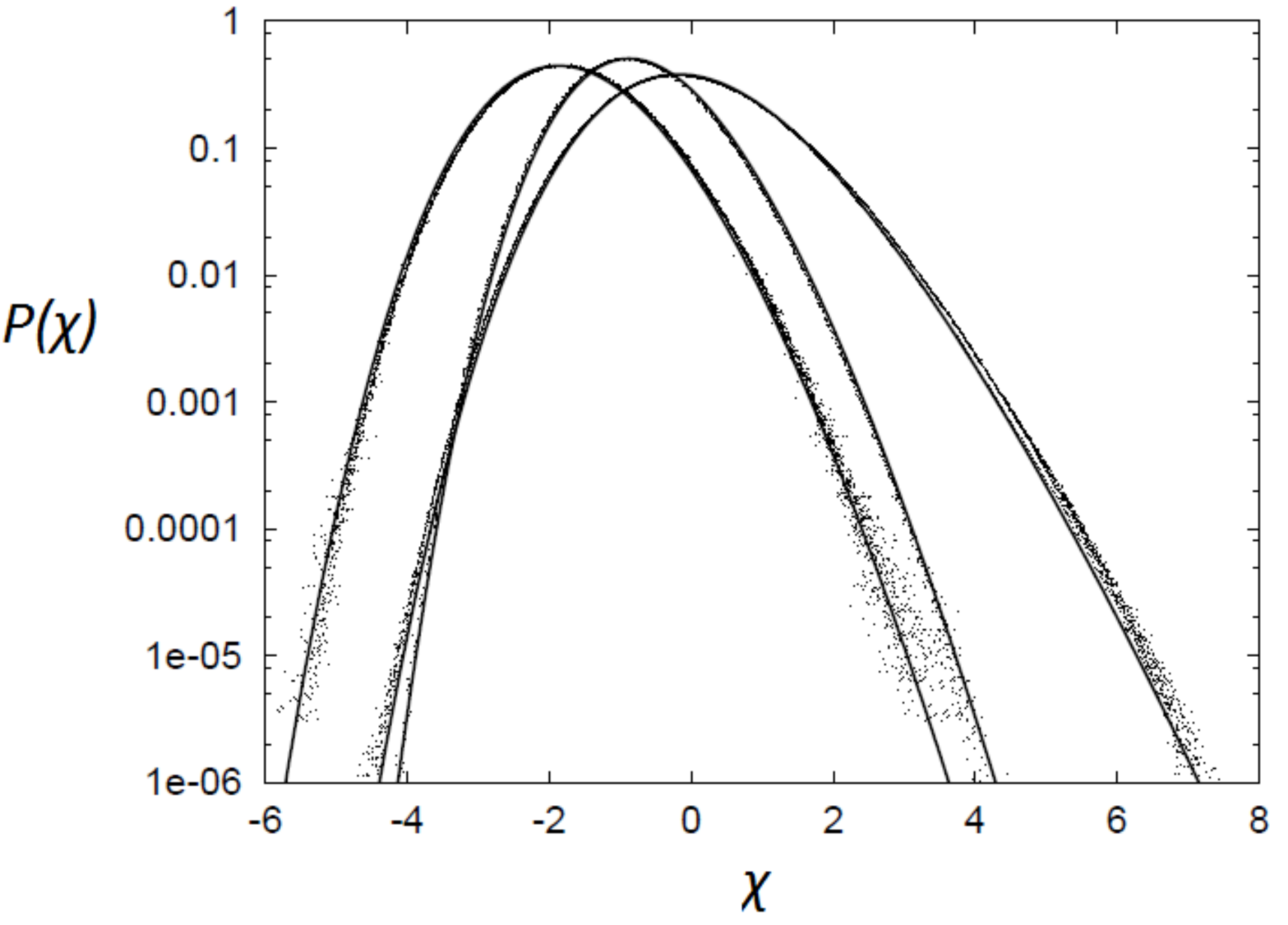}
    \caption[Universal distributions obtained by Pr\"{a}hofer and Spohn.]{Universal distributions obtained by Pr\"{a}hofer and Spohn in the Ref. \cite{Prahofer}.
    From left to rigth: universal pdfs (solid lines) for $\chi$ under curved, flat and stationary self-similar growth, respectively. Symbols refers to simulations of the PNG model. Figure extracted from the ref. \cite{Prahofer}.}
    \label{fig_praho}
    \vspace{0.5 cm}
 \end{figure}

Shortly after, Pr\"{a}hofer and Spohn \cite{Prahofer, Prahofer2} unveiled a \textit{dependence} of p($\chi$) on the
geometry (IC) of the growth. By using the mapping of the PNG model onto random 
permutations, they showed that $h(x,t)$ corresponds to the length of the longest increasing subsequence of such 
permutation, which in turn, is distributed according to: 1) the Gaussian Orthogonal Ensemble (GOE)\footnote{This means, the distribution 
of the largest eigenvalues of orthogonal matrice ensembles, whose elements are distributed from a Gaussian.} TW distribution
for a flat IC; 2) the Gaussian Unitary Ensemble (GUE)\footnote{Same as before, but now the matrices are unitary.} TW, if the growth starts from a seed and develops a curved interface; and 3) the Baik-Rains $F_{0}$ limiting distribution 
\cite{Baik} for a steady-state IC. Figure \ref{fig_praho} shows p($\chi$) for different growth geometries and for stationary initial configuration calculated from the PNG model and compared to the respectives GUE, GOE and $F_{0}$ distributions \cite{Prahofer}.
 
%

A decade later the results of Pr\"{a}hofer and Spohn \cite{Prahofer, Prahofer2}, analytical solutions on the 
one-dimensional KPZ(-DPRM) equation, already cited above, have confirmed the limiting distributions of $\chi$ as GUE for curved growth \cite{Sasamoto, Amir, Calabrese2, Dotsenko},
GOE for flat growth \cite{Calabrese} and $F_{0}$ for the stationary initial condition \cite{Imamura}. Moreover, 
finite-time corrections were found, suggesting new universal KPZ features such as a shift in the mean converging 
to GOE (GUE) values as $t^{-\beta}$. It was also demonstrated that the limiting processes, giving the one-dimensional heigth profiles of a 
KPZ interface, are dictated by the Airy$_1$ \cite{Sasamoto2, Johansson2, Borodin} and Airy$_2$ \cite{Prahofer3, Borodin} processes for flat, and curved growing, respectively.

 \begin{figure}[ht]
    \vspace{0.5 cm}
    \centering
    \includegraphics[height = 6.0 cm ]{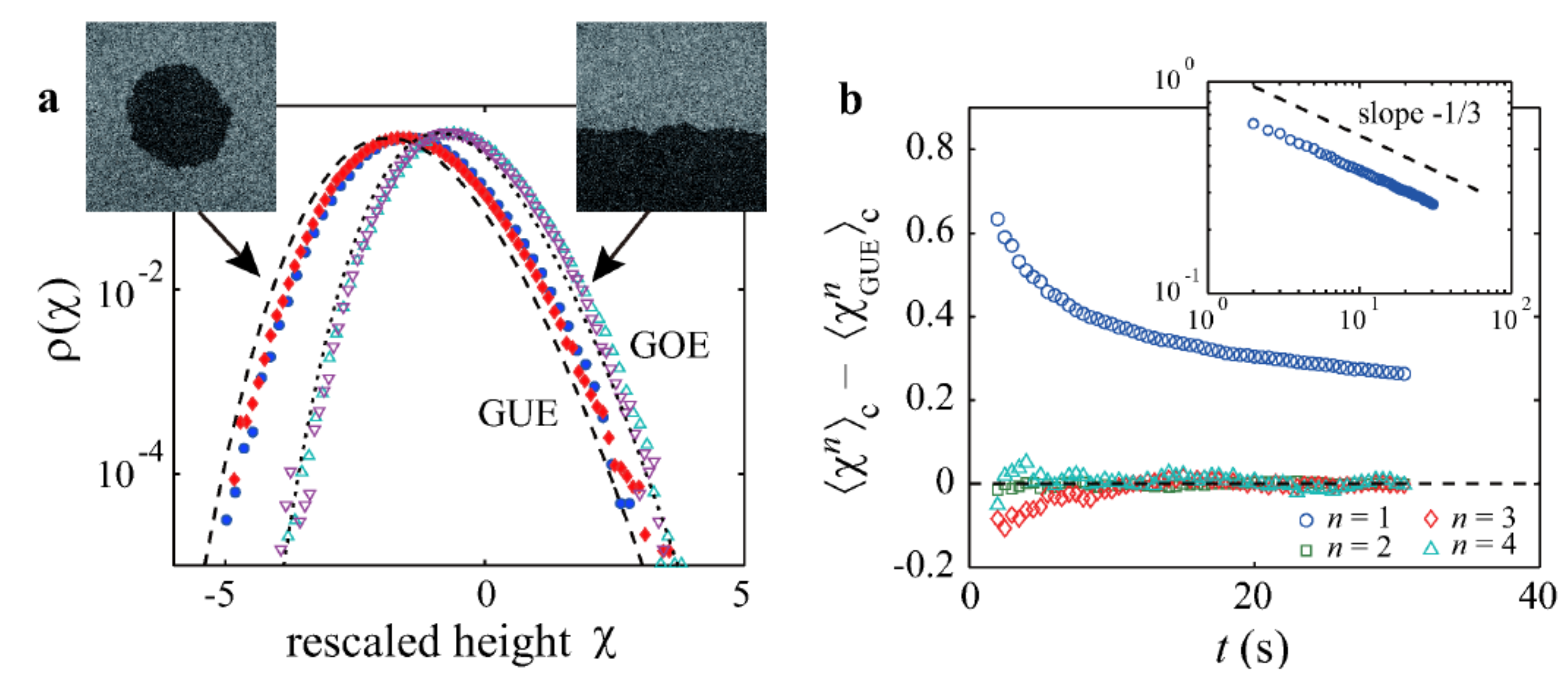}
    \caption[Results for turbulent liquid crystals.]{Results for turbulent liquid crystals extracted
    and edited from Takeuchi \cite{Takeuchi2}. (a) Histogram of rescaled local height 
    $\chi \equiv h - v_{\infty}t/(\Gamma t)^{1/3}$. The blue and red solid symbols show the 
    histograms for the circular and flat interfaces, respectively. Dashed and dotted curves show GUE and 
    GOE TW distributions. (b) How $\chi$ cumulants approach to the GUE values. Inset shows the mean 
    shift decaying as $t^{-1/3}$. The same behavior is shared by the flat case (not shown here).}
    \label{fig_takeuchi}
    \vspace{0.5 cm}
 \end{figure}

Simultaneously, experiments of unprecedented statistics on turbulent liquid crystals performed by K. Takeuchi and M. Sano \cite{Takeuchi1, Takeuchi2, Takeuchi3, Takeuchi4} have confirmed carefully great part of the predictions cited above and have given to the KPZ$_{d = 1 + 1}$ theory a reliable reality beyond mathematical, numerical, and few experimental realizations constrained to exponent results \cite{KPZ_experiments}.
By using the Krug-Meaking toolbox for unearthing non-universal parameters $A$ and $\lambda$, K. Takeuchi and M. Sano calculated
the pdf of the $\chi$ variable, defined according to the KPZ ans\"{a}tze (eq. \ref{eq_ansatz}) as $\chi \equiv (h - v_{\infty}t)/(\Gamma t)^{1/3}$.
In the fig. \ref{fig_takeuchi}(a) one can see a comparison between the experimental p($\chi$) obtained in that work for both curved and flat cases and the GUE and GOE distributions \cite{Takeuchi2}. Wonderfully, a great accordance was obtained apart from a 
slight shift at mean of the distributions. In fact, in the figure \ref{fig_takeuchi}(b) the shift at the mean is also calculated, suggesting
that it decays as $\sim t^{-1/3}$, while the shift for higher cumulants vanishes quickly.

Remarkable experiments on colloidal particles deposited at the edge of 
evaporating drops were also able to confirm the ``KPZ universality beyond exponents'', where 
curved deposits of particles slightly anisotropic following the GUE-TW distribution were found \cite{Yunker}. Analytical
and experimental results have been reinforced by numerical ones \cite{Prolhac, Sidiney_EPL, Tiago1d, Healy_stationary}, which have 
filled the last pieces toward a robust and consistent KPZ$_{d = 1 + 1}$ \textit{triumvirate}.

The KPZ$_{d = 2 + 1}$ situation is very contrasting with its one-dimensional counterpart and
almost all one knows about the most important dimension for applications has come from simulations and, very recently, from 
some remarkable experimental efforts \cite{Almeida, Healy_exp}. Best estimates for scaling exponents
indicate that \cite{Kelling, Marinari_exp, Healy_2d1}:

\begin{equation}
 \alpha \approx 0.393 \qquad \beta \approx 0.242 \qquad 1/z \approx 1.607.
 \label{kpz_exp}
\end{equation}
 
 The universality of dimensionless cumulant ratios of $h$, namely, the skewness and kurtosis (eq. \ref{skew}) were firstly calculated numerically at the \textit{stationary state}, where their values were proved to be universal (see Ref. \cite{Fabio2d} and references therein). The 
 universality in the growth regime, although glimpsed in the Ref. \cite{Paiva}, only was convincingly demonstrated in the year of 2012. Through large-scale simulations, T. Halpin-Healy \cite{Healy_2d1, Healy_2d2} and T. J. Oliveira \textit{et al.} \cite{Tiago2d} have 
 uncovered the existence of geometry-dependent KPZ \textit{universal} height distributions at the \textit{growth regime}, higher 
 dimensional GOE- and GUE-TW counterparts, lying in the heart of KPZ$_{d = 2 + 1}$ universality. Although the exact forms of these distributions are not known, Oliveira \textit{et al.} have demonstrated that rescaled height \textit{pdf's} can be well fitted by generalized Gumbel distributions \cite{Gumbel} (see the definition in the section \ref{section_MRHD}) with parameters m = 6 and 
 m = 9.5 for flat and curved cases, respectively. Moreover, as in d = 1 + 1 dimensions, results from 
 the T. J. Oliveira \textit{et al.} study \cite{Tiago2d} have supported a generalization of the KPZ$_{d = 2 + 1}$  \textit{ans\"{a}tze} (eq. \ref{eq_ansatz}) inserting appropriate finite-time corrections. The final KPZ \textit{ans\"{a}tze} reads:

 \begin{equation}
 h(t) = v_{\infty}t + sig_{\lambda}(\Gamma t)^{\beta}\chi + \eta_p + \zeta_p t^{-\gamma_p} + ...,
 \label{eq_ansatz2}
 \end{equation}
where $\eta_p$, $\zeta_p$ and $\gamma_p$ are non-universal parameters \cite{Ferrari_frings, Carrasco}. The
values for these model-dependent parameters can be found in the references \cite{Healy-Krug, Tiago1d, Sidiney_EPL, Carrasco, Healy_stationary} for d = 1 + 1, and \cite{Healy_2d1, Healy_2d2, Tiago2d, Carrasco} for d = 2 + 1. Deserving much more
attention, the universal-KPZ values for the cumulants of height distributions are grouped in the tables \ref{universalkpz}
and \ref{universalkpz2}.

\begin{table}[htb]
\centering
\setlength{\tabcolsep}{15pt}
    \begin{tabular}{c c c c c c c}
	\hline \hline
	& GOE & GUE & Baik-Rains\\
	\hline
	$\langle\chi\rangle_c$ & -0.76007 & -1.77109 & 0\\
	$\langle\chi^2\rangle_c$ & 0.63805 & 0.81320 & 1.15039\\
	|S| & 0.2935 & 0.2241 & 0.35941\\
	 K & 0.1652 & 0.09345 & 0.28916\\
	\hline\hline
    \end{tabular}
    \caption[Universal KPZ values for cumulants of HDs in d = 1 + 1]{\begin{small} Universal 
    KPZ values for cumulants of height distributions in d = 1 + 1 \cite{Prahofer}.
    \end{small}}
    \label{universalkpz}
\end{table}

\begin{table}[htb]
\centering
\setlength{\tabcolsep}{15pt}
    \begin{tabular}{c c c c c c c}
	\hline \hline
	& Flat & Curved & Flat-Growing & 1D Groove & Stationary \\
	\hline
	$\langle\chi\rangle_c$ & -0.75(5) & -2.3(1) & -2.4(2) & -1.47(2) & == \\
	$\langle\chi^2\rangle_c$ & 0.23(1) & 0.33(2) & 0.34(2) & 0.249(4) & 0.46(2) \\
	|S| & 0.423(7) & 0.33(1) & 0.33(2) & 0.396(7) & 0.244(8)\\
	K & 0.344(9) & 0.212(7) & 0.21(2)  & 0.31(2) & 0.176(4)\\
	\hline\hline
    \end{tabular}
    \caption[Universal KPZ values for cumulants of HDs in d = 2 + 1]{\begin{small} Universal KPZ values for cumulants of height  distributions in d = 2 + 1 dimensions for Flat case \cite{Healy_2d1}, Curved
    \cite{Healy_2d2}, flat-growing \cite{Carrasco}, 1D groove initial condition \cite{Healy_2d2} and Stationary
    state \cite{Healy_2d2}.
    \end{small}}
    \label{universalkpz2}
\end{table}

%
%
%
%
%
%


 As a final remark on KPZ height distributions, we point out that studying KPZ growth on enlarging flat substrates,
 Carrasco \textit{et al.} have revealed that the Tracy-Widom distributions and the Airy processes (as well as 
 their (2 + 1)-dimensional analogs) do not depend on the interface macroscopic curvature,
 but actually on the inflation of the lattice metric on the active zone \cite{Carrasco}. Furthemore, in the search
 for an upper critical dimension in KPZ class, Alves \textit{et al.} \cite{Sid_high} have confirmed that the 
 KPZ \textit{ans\"{a}tze} is valid up to d = 6 + 1 dimensions, at least for the Kim-Kosterlitz model \cite{Kim_model}.
 
\section{Universal Squared Roughness Distributions}
\label{section_SLRD}
In 1994, investigating the old problem of random-walk interfaces, F\'{o}ltin \textit{et al.} \cite{Foltin} showed
that the squared roughness \textit{pdf} [$P(w^2)$]\footnote{Notice that at the steady state, the surface roughness
fluctuates around its saturated value and $P(w^2)$ is the pdf associated to these fluctuations.} of such interface, at \textit{steady state}, behaves as:

\begin{equation}
 P({w^2}) = \frac{1}{\langle w^2 \rangle}\Phi( {w^2} / {\langle w^2 \rangle} )
\label{eq_w2}
\end{equation}
where $w^2$ is the squared global roughness (eq. \ref{roug}) of the interface, $\Phi(u)$ has a closed form and is also 
an \textit{universal} scaling function. Indeed, its universality was confirmed by numerical simulations of
EW and KPZ models, and have emphasized the power of that distribution for accessing the UC of a given 
growth process, once it presents a weak dependence on finite-size corrections.  

In the same year, Plischke \textit{et al.} \cite{PlischkeWD} extended theses studies for curvature-driven interfaces  and have 
reinforced the validity of the eq. \ref{eq_w2} as well as the universality of $\Phi$, which in this case has a different form from that for Gaussian interfaces. Following the same idea, R\'{a}cz \textit{et al.} \cite{Racz} calculated \textit{numerically} $P(w^2)$ for several UCs
in d = 2 + 1 dimensions and exposed that, unlike the lower-dimensional case, $\Phi_{EW}$ and 
$\Phi_{KPZ}$ are different, being the first a Gaussian and the last one marked by a 
slow decay in the rigth tail. $\Phi_{MH}$ and $\Phi_{VLDS}$ also exhibited remarkable differences between each other. 
In the same paper one finds a ``recipe'' of how comparing $\Phi$, for a given UC, with the ones obtained in experiments. The recipe consistes in divide the surface into boxes of lateral size $l \ll \xi_{||}$ inside which $w^2$ should be calculated to yield a large \textit{ensemble}. On the other hand, $l$ must also be larger than characteristic sizes at surface such as grains, mounds, etc.

The Squared Local Roughness Distributions (SRLDs) have been used in signal analysis \cite{Antal} and in 
issues about the upper critical KPZ dimension \cite{Marinari} because they are considered one of the 
most suitable ways for accessing the UC of a growth. For instance, in a very interesting study \cite{Aarao}, Aar\~{a}o 
Reis analyzed the rescaled distribution $\Phi$, at mean null and unitary variance, (call it $\Psi$, defined as in the
eq. \ref{Psi}) for one- and two-dimensional models belonging to the KPZ and VLDS classes.

\begin{equation}
 P(w^2) = \frac{1}{\langle w^2\rangle_c^{1/2}} \Psi \Big(\frac{w^2-\langle w^2\rangle}{\langle w^2\rangle_c^{1/2}}\Big).
 \label{Psi}
\end{equation}

\begin{figure}[ht]
    \vspace{0.5 cm}
    \centering
    \includegraphics[height = 6.0 cm ]{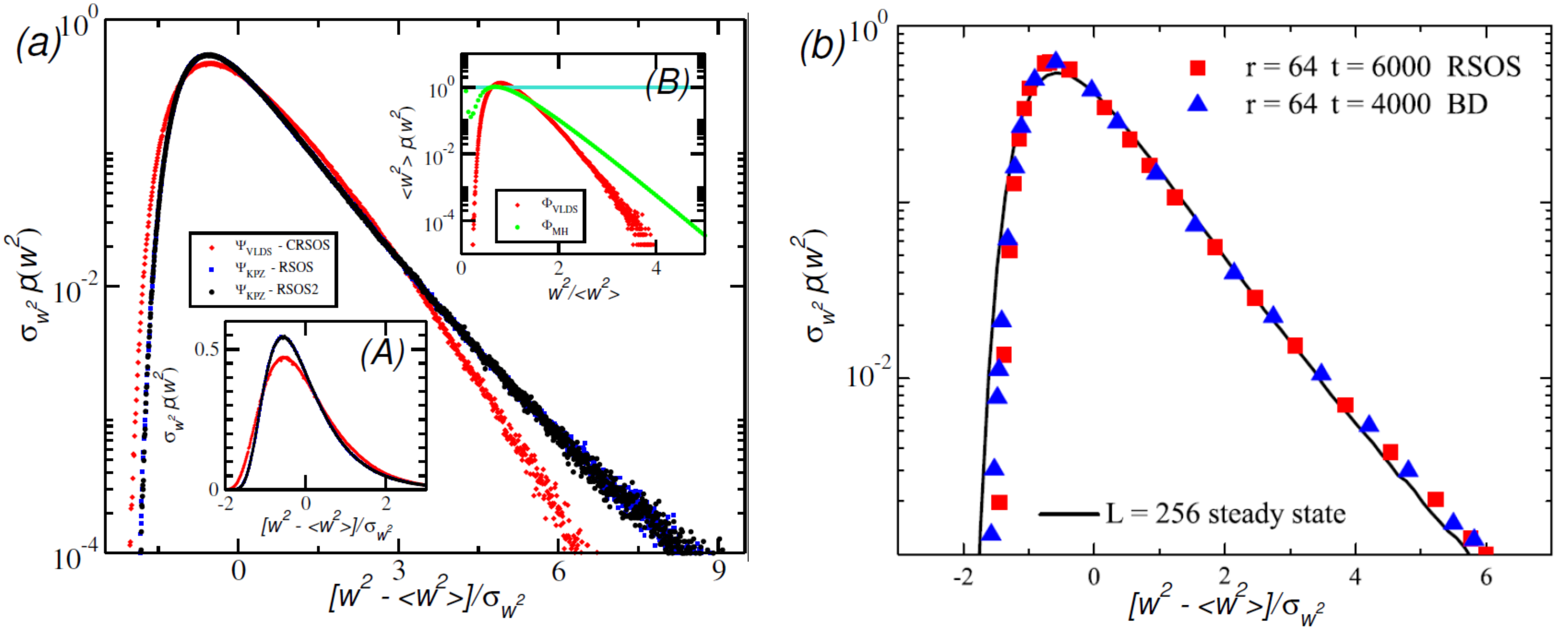}
    \caption[SLRDs for different universality classes]{(a) The main plot exhibits the contrasting form of universal $\Psi$ distributions for the VLDS and KPZ classes calculated from the CRSOS ($L = 64$), RSOS ($L = 256$) and RSOS2 ($L = 128$) models at the \textit{stationary state}, for $d = 2 + 1$, according to the ref. \cite{Aarao}. The insertion (A) reinforces the difference between $\Psi_{VLDS}$ and $\Psi_{KPZ}$ near of the peak, whereas inserts (B) shows the constrast 
    between $\Phi_{VLDS}$ and $\Phi_{MH}$. (b) Comparasion between $\Psi_{KPZ}$ calculated from the steady state (solid line) and from two-dimensional RSOS and BD models in the growth regime, for a ``box'' of lateral size $r = 64$. Results in (a) and in (b)
    can be found in the references \cite{Aarao} and \cite{Paiva}, respectively. Theses results were kindly provided 
    by Prof. Aar\~{a}o Reis.}
    \label{fabio}
    \vspace{0.5 cm}
 \end{figure}

Aar\~{a}o Reis have confirmed that, in d = 2 + 1, $\Psi_{KPZ}$ presents a \textit{stretched exponential in the rigth tail} as 
approximately $exp(-x^{0.8})$ \cite{Aarao} - see figure \ref{fabio}(a). This form is contrasted 
with those from $\Psi_{EW}$, which is Gaussian, and with the simple exponential decay of $\Psi_{VLDS}$ and $\Psi_{MH}$. Even the last ones can be easily distinguished in the $\Phi_{VLDS}$ and $\Phi_{MH}$ scaling, as exemplified in the inset (B) of fig. \ref{fabio}(a). A subsequent work from Paiva $\&$ Aar\~{a}o Reis \cite{Paiva} brought 
out $\Psi_{KPZ}$ in the \textit{growth regime} presenting a similar decay at the right tail as shown in the
fig. \ref{fabio}(b). This feature has been suggested as an universal and distinct KPZ landmark \cite{Paiva} and has been confirmed experimentally in 
the growth of CdTe \cite{Almeida} and oligomer \cite{Healy_exp} thin films.

\section{Universal Maximal Relative Height Distributions}
\label{section_MRHD}
Extreme-value statistics (EVS) play an important role in systems where rare events have drastic consequences such as floods, 
internet failures, stock market crashes as well as in important technological applications \cite{Gumbel, Raychaudhuri, Gyorgyi, Majumdar, Lee, Tiago_mahd}.
For instance, the onset of a breakdown of corroded surfaces is determined by their deepest (or weakest) point, whereas 
in batteries the highest point of the metal surface reaching the opposite metal surface is responsible by the beginning
of a short-circuit \cite{Raychaudhuri}. A well-known probability function in the EVS context is the called Gumbel's first asymptote, being the distribution of the \textit{n}th value among $N$ independent (uncorrelated) random variables \cite{Gumbel}. The Gumbel pdf, $G(X;m)$, of the variable $X$ is defined as:

\begin{equation}
 G(X;m) = \frac{m^mb}{\Gamma_f(m)}\exp[-m(z_X + \exp(-z_X))],
 \label{pdf_gumb}
\end{equation}
where $m$ is a parameter, $b = \sqrt{\psi_1(m)/\langle X^2 \rangle_{c}}$, $z_X = b(\langle X \rangle - X + s)$, $s = [ln(m) - \psi_0(m)]/b, \Gamma_f(X)$ is the gamma function, and $\psi_k(X)$ is the polygamma function or order k\cite{Gumbel, Tiago_mahd}.

Early studies on EVS applied on growing interfaces focused on the steady state of linear (EW) equations
\cite{Raychaudhuri, Gyorgyi, Majumdar}. Through maximal relative height ($m^*$) analysis, defined as the difference between the largest height minus the average height of the surface, it was shown that in the stationary regime $m^*$ scales
likewise the global roughness (this result is valid only for the one-dimensional case) \cite{Raychaudhuri} and that 
its \textit{universal} distribution, $P(m^*) = L^{-1/2}f(m^* L^{-1/2})$, is dictated by the 
Airy distribution function, $f(x)$, whether periodic boundary conditions are used \cite{Majumdar}. 

Interestingly, even for strongly correlated systems such as two-dimensional EW interfaces at steady state, it has been shown 
that $P(m^*)$ can be very well fitted by the Gumbel\footnote{See also the Oliveira \textit{et al.} results \cite{Tiago2d} about
the relationship between the Gumbel distribution and KPZ HD's$_{d = 2 + 1}$. They are already discussed in the section \ref{kpz1}.}, with a non-integer value, n = 2.6(2) in this case \cite{Lee}. A numerical work has brought theses
discussions for KPZ and VLDS two-dimensional surfaces \cite{Tiago_mahd}. It was confirmed the universality 
for either the scaled maximal- or minimal-relative height distributions (MRHDs), depending on the relevant nonlinear 
term signal. Moreover, the remarkable contrast between the right tail decay from MRHD$_{KPZ}$ (simple exponential) and 
MRHD$_{VLDS}$ (Gaussian) and the very weak finite-size effects affecting these distributions
have suggested an alternative way for accessing the UC of a given growth. Our experimental results have been
the first confirmation of such universality of this distribution in the KPZ$_{d = 2 + 1}$ context.

 \chapter{Material and Experimental Methods}
 \lhead{\bfseries 4. Material and Experimental Methods}

In this chapter a short review on the experimental techniques used along this work is given. Details on growth parameters
and on the experimental methodology is discussed in details.

\section{Hot Wall Technique}

Hot Wall Epitaxy is a well-established technique based on thermal evaporation developed at the end of 
the 1970's which has been used for growing high-quality films from II-VI, IV-VI and also III-V 
compounds \cite{Otero, SukarnoQD1}. The great difference between a simple thermal evaporation system 
and a HWE one consists the presence of a heated liner (hot wall). This liner serves as a guide for the 
vapor beam to flow from the source towards the substrate, ensuring low material loss and growth 
conditions as near as possible of the thermodynamic equilibrium.

\begin{figure}[!b]
    \vspace{0.5 cm}
    \centering
    \includegraphics[width = 8.0 cm]{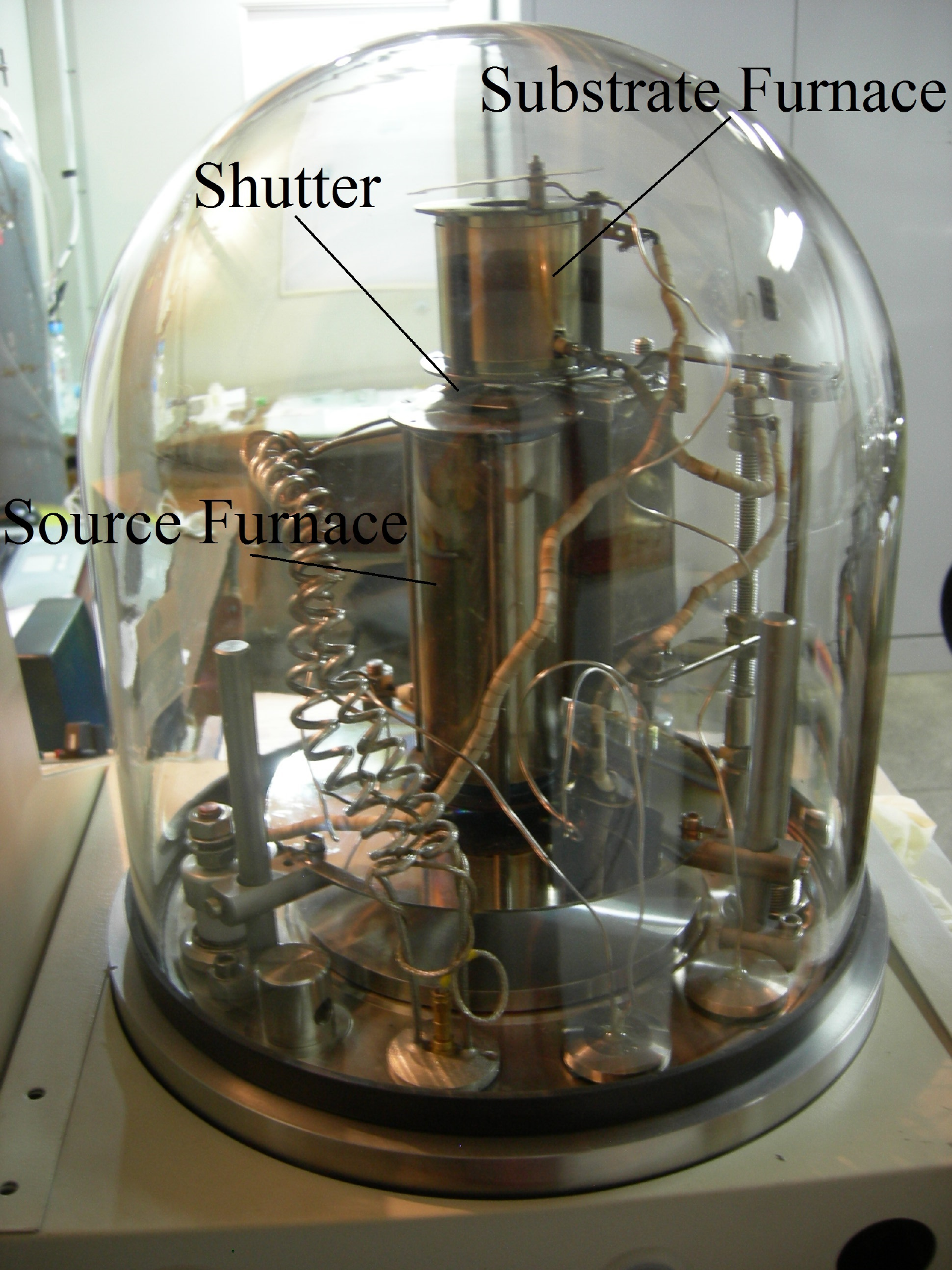}
    \caption[HWE system used in this work]{Hot Wall system of Epitaxy Laboratory, Physics Departament of
    the Universidade Federal de Vi\c{c}osa, Brazil. The legend in the figure shows two furnaces 
    (substrate and source) that work independently and a shutter which allows starting/finishing the growth.
    The pressure during the growth can reach $\sim 10^{-7}$ Torr.}
    \label{fig_HWE}
    \vspace{0.5 cm}
    \end{figure}

HWE systems share the same basic structure (see fig. \ref{fig_HWE}), but they can be modified depending on the
particular growth necessities (see \cite{Otero} for different HWE forms). In general, there are 
tree resistances windings at quartz tube to heat the substrate, the source and the wall, independently. It guarantees
that the temperature displayed at the controller is always being measured at the same 
referential\footnote{It is not the case for other systems where temperature measurements 
must be routinely calibrated, as occurs in MBE systems.}, beyond of making the system's calibration to be very reliable.
HWE works in high vacuums ($\approx 10^{-7}$ Torr), providing a relatively clean environment for the 
growth \cite{Lalev1, Lalev2, SukarnoQD1}. 
Substrate position can work as a lip for the wall or can be slightly inserted above it when a shutter is available.

As a very simple technique, HWE is unsuitable for doping, for growing compounds with very different
vapor pressures and it also does not allow \textit{in-situ} measurements. Indeed, these disadvantages turn 
the HWE use minimized. However, in situations where \textit{in-situ} measurements are not essential, where the 
compounds evaporate congruently and do not have high temperature of sublimation, HWE is absolutely one of the most 
suitable techniques to be employed due to its high reproducibility, wide growth rate range (0.01 - 10 \AA{}/s) \cite{SukarnoQD1} 
and its relative low cost (ownership and maintenance). This is, particularly, the case for the growth of CdTe
\cite{SukarnoQD1, SukarnoQD2, Lalev1, Lalev2, Fabio,Suka_Ediney2}.

\section{Atomic Force Microscopy}

\begin{figure}[!b]
    \vspace{0.5 cm}
    \centering
    \includegraphics[width = 12.0 cm]{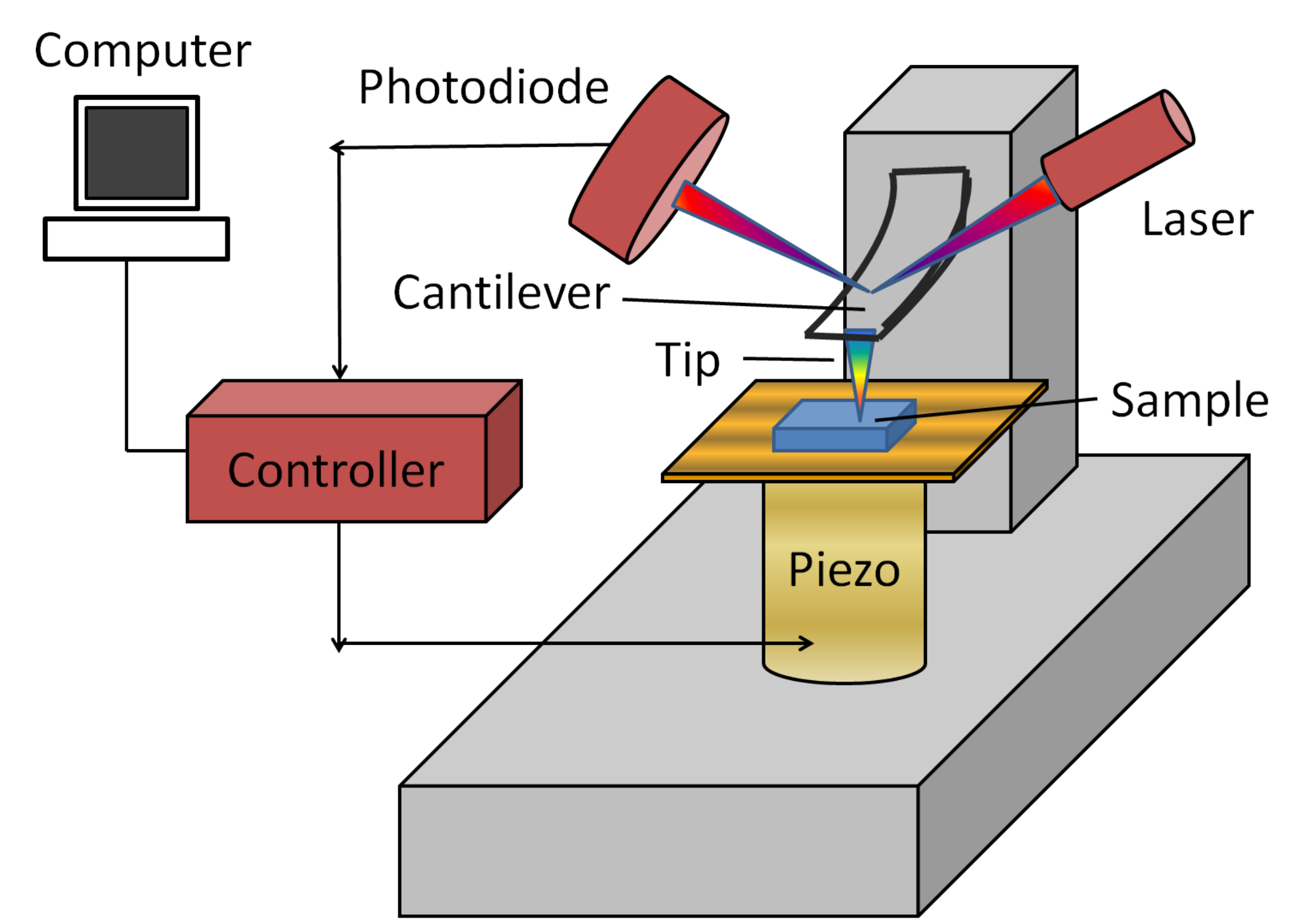}
    \caption[Basic parts of an AFM]{Basic scheme of working principles of an Atomic Force Microscope.}
    \label{fig_AFM}
    \vspace{0.5 cm}
    \end{figure}

After the 1986 year, AFM invention \cite{Binnig} has enormously assisted the surface study in
micrometer scale beyond of helping several stages of the thin-film production \cite{Pimpinelli}. As a blind man must 
use his fingers to feel topographic variations during a Braille lecture, the AFM consists
in to obtain morphological data from interactions between a very thin tip (order of $10^{-8}m$ radii) and 
the sample surface to yield a three-dimensional image. The equipment measures atomic 
forces via a spring deformation, more known as \textit{cantiveler}, under which the tip is coupled. On the 
other cantiveler face, a laser beam is focused and reflected to a photo-diode in order to send
electrical signals associated with the cantiveler's deflection to the controller. A simplified view of most 
important AFM components is shown at fig. \ref{fig_AFM}.

Once the electrical signals have arrived at controller, they are translated and sent to the computer to yield
a bidimensional profile sketched by the tip. An AFM image can have 1024 x 1024 pixels, which means that 
there are 1024 bidimensional profiles, each one of them being composed by 1024 equally spaced points. In some AFM systems, 
the sample is supported by a piezoelectric ceramic which suffers well-behaved subnanometric deformations 
to move the sample in relation to the tip. 


There are two basic AFM measurement modes, namely, \textit{contact} and \textit{tapping}. In the former, the tip
approaches to the surface until suffer a repulsion force, which causes the tip to bend up. In the tapping mode, however, the 
tip is set to oscillate near to its natural resonance frequency and gets close to the surface 
until the oscillation amplitude becomes reduced to a smaller referential value (this occurs at distance between regions of 
repulsion and attraction). For both modes, the piezo adjusts, through a feedback mechanism, the height axis (z) to hold or
the force (contact) or the amplitude (tapping) constant. Then, a surface image is obtained from those z values recorded by the piezo. 
The choice of which method should be employed depends on the aspects of the studied material (shape, structure, nature) and the 
kind of results that want to be found (local friction coefficients, phases difference, conductance variations, etc). For instance, biological
samples are most suitable for tapping mode once it avoids damage sample and frictional forces, while contact is most suitable for 
thin films and crystals in general.

Nowadays, AFM is one of the most used techniques for studying surfaces at the submicrometer level. In particular, several works
have used AFM images to perform scaling analysis of interfaces, as exemplified by works involving the growth 
of $SiO_2$ by CVD \cite{Ojeda}, the dissolution of polycrystalline pure iron \cite{Cordoba}, the amorphous Si by 
thermal evaporation \cite{Yang} and Pt sputtered on glass \cite{Jeffries}.

\section{Si surface cleaning}

Substrate cleaning is the key initial step in thin film and epitaxial growth. In order to obtain flat and contamination-free Si surfaces, several chemical and/or thermal treatments have been proposed
since 1980's \cite{Ishizaka, Takahagi, Kwabata}. Basically, the aim is to remove the silicon native oxide 
layer ($\approx 0.7$ nm of thickness) and the hydrocarbon contaminant layer ($\approx 0.2$ nm). The former 
is usually removed by low-concentration aqueous HF solution [1-10$\%$] using HF and water of high purity, 
which provide an ``ideal'' stable monolayer H-terminated Si surface, depending on the aqueous pH 
concentration \cite{Hirashita, Chabal, Higashi}. Moreover, aqueous HF solution does not etch the bare 
Si surface itself, preserving a Si smooth morphology (typically, $\approx 0.2$ nm of roughness \cite{Kwabata}).
Whereas some works stress the necessity of removing reminiscent hydrocarbon impurities \cite{Ishizaka, Takahagi}, 
X-ray photoelectron spectroscopy (XPS) measurements show that 1.5\% HF-treated Si surfaces present a very low 
concentration of $O$, $F$ and $C$ \cite{Kwabata}. Moreover, the H-terminated surface obtained by this process 
proved to be very stable against the oxidation in air. In particular, this cleaning procedure has been used 
for growing \textit{epitaxial} CdTe QD's on Si(111) \cite{SukarnoQD1}. Another approaches as the degreasing by acetone, 
ethanol and deionized (DI) water followed by repeatedly boiling in HNO$_3$, dipping in HF, rinsed with DI water 
and dried with N$_2$ also are commonly used \cite{Lalev1}.

\section{CdTe thin films: Cleaning, Growth and Characterization}

In this work, p-type Si(001) substrates of dimensions $10.0$ mm $\times$ $10.0$ mm $\times$ $0.3$ mm were dipped
2 minutes into an aqueous HF($2\%$) solution prepared with DI water. This time has been checked to be able of removing completely the Si native oxide layer \cite{Kwabata}.
At the sequence, Si surfaces were exposed to $N_2$ air jet just to remove any reminiscent droplet on the surface, once
after HF treatment the surface becomes largely H-terminated. The treated surfaces were immediately inserted
into the HWE chamber, inside which a high vacuum was performed. There was no any heating treatment before or after
the growth.

The HWE system used in this work is composed by two independent furnaces 
(source and substrate), separated by a shutter and a wall of 7 cm, as shown in the fig. \ref{fig_HWE}. 
The deposition occurs at pressures $\approx 10 ^{-7}$ Torr,
obtained by a diffusion pump system, while the source temperature can be controlled from $400$ to $520\,^{\circ}\mathrm{C}$,
producing growth rates between ($\approx 0.01$ and $2.5$ \AA{}/s). The temperature of the substrate (more known as
the deposition temperature) can be varied from 150 up to $550\,^{\circ}\mathrm{C}$.

Solid CdTe (99.999\%) has been used as source material. The temperature of the source was fixed at
$520\,^{\circ}\mathrm{C}$ yielding a deposition rate $F = 2.2 \pm 0.3$ \AA{}/s for deposition temperatures set at
150, 200, 250, and $300\,^{\circ}\mathrm{C}$. For each deposition temperature, the growth time ($t$) was varied from
15 to 240 min in a geometric progression sequence of ratio $2$, providing thicknesses ($th$) from \textit{approximately}\footnote{As we shall see in the chapters 5 and 6, the growth of CdTe in these conditions are non-conserved, this implies that the growth rate is
not constant in time.}
$0.20$ up to $3.5$ $\mu m$. The $th$ and the growth rate $F$ were determined post-growth using 
a \textbf{XP1 - AMB\'{I}OS} contact profilometer and a \textbf{ContourGT-K BRUKER} optical profilometer.

Surface characterization was performed in air by \textit{ex-situ} AFM. We have used an \textbf{Ntegra Prima SPM} working in 
contact mode. Different kinds of Si tips were used in order to check the reliability of data. All of them were used in
the statistical analysis. The frequency of the AFM scan was kept near of 1.5 lines/s during acquisition of the images, but we have also confirmed that the frequency does not affect the results as far the frequency is not set at very high 
values as $\gtrsim$ 4.0 lines/s. Surface topography of 3 to 10 different regions near of the center film of each sample was scanned producing images of $10$ $\mu m$ $\times$ $10$ $\mu m$ with $1024$ $\times$ $1024$ pixels. This size was chosen so that morphological properties in domains smaller and larger than the averaged grain size could be simultaneously investigated. However, different sizes of scanning were carried out, namely, $1$ $\mu m$ $\times$ $1$ $\mu m$, $30$ $\mu m$ $\times$ $30$ $\mu m$ and $100$ $\mu m$ $\times$ $100$ $\mu m$ to guarantee the validity of data in different scales. For all images a \textit{flatten's correction} of second 
order was performed to correct the sample misalignment and the piezo scanner error.
 \chapter{RESULTS: Uncovering the KPZ Universality in CdTe Thin Films}
 \lhead{\bfseries 5. Uncovering the KPZ universality in CdTe thin films}

The results contained in this chapter are related to films grown at a particular deposition temperature, 
namely, T = $250\,^{\circ}\mathrm{C}$. Here we develop a noveil procedure to distill the Universality Class (UC)
of a given growth, which consists in to: i) perform a visual-AFM investigation of the interfaces fluctuations focusing on the dynamic of superficial structures whose length is characteristic; ii) relate it to a deep
local scaling analysis; iii) add on the results coming from the global scaling and, finally, iv) to find and/or 
to confirm the UC of the growth by using universal distributions. As we will be seen below, through this 
scheme we were able to find the UC of CdTe thin films, in the meantime that one has been experimentally 
demonstrated the universality of KPZ$_{d = 2 + 1}$ distributions.

\section{Semi-Quantitative Morphological Analysis}

\begin{figure}
    \vspace{0.5 cm}
    \centering
    \includegraphics[width = 15.0 cm]{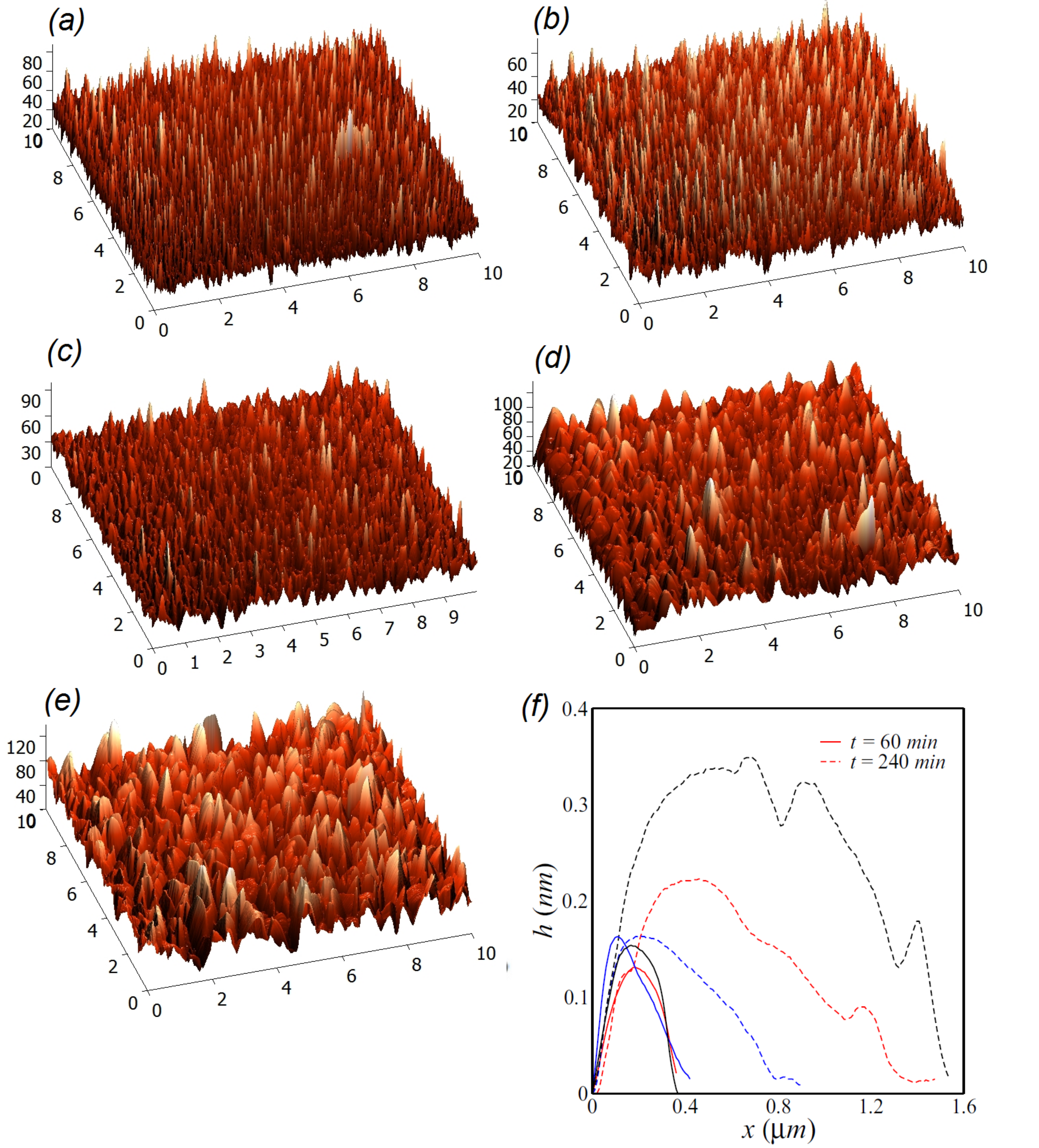}
    \caption[AFM images for CdTe thin films grown at T = $250\,^{\circ}\mathrm{C}$.]{AFM images [$10 \times 10$ $\mu m$; height scale in nanometers] for CdTe thin films 
    grown at T = $250\,^{\circ}\mathrm{C}$. (a) - (e), Interfaces grown by 15, 30, 60, 120 and 240 min, respectively.
    (f) Typical grain's shape present at the interface for t = 60 min (solid lines) and t = 240 min (dashed lines).}
    \label{T250}
    \vspace{0.5 cm}
 \end{figure}

At first, it is important to observe how the morphology evolves as function
of the growth time in a semi-quantitative fashion. Figure \ref{T250} shows $10 \times 10$ $\mu m$ AFM images for
all available times. The images depicted are those which better represent typical behavior over all regions scanned.

At initial times (figures \ref{T250} (a) and (b)), the surface is dominated by a large number of grains
with well-defined ``\textbf{sharp}''\footnote{Hereafter we consider that the grain has a sharp shape under a [nm] $\times$ [$\mu$m]
scale, which is a typical scale for analyzing the grainy morphology \cite{Ojeda, Sukarno, Fabio}.} shape, which present a very high aspect ratio ($\Omega$), defined as the ratio between the 
crystallite height and the characteristic length of its base (the average grain width - $\zeta$). Roughly, 
one observes $\zeta \ll 0.5$ $\mu m$ for $t = 15$ and $30$ min, whilst the vertical bar decreases 
from $80$ $nm$ to $60$ $nm$. Consequently, $\Omega$ seems to decrease in this interval. At the subsequent time, 
one can see a surface composed by fewer, but larger grains ($\zeta \approx 0.5$ $\mu m$)
surrounded by a crowd of smaller sharp grains. It is not difficult
to see that $\Omega$ has decreased once again. Remarkably, for the last two surfaces showed 
in the figures \ref{T250} (d) and (e), one notices the presence of grains with larger basis, in particular, 
$\zeta \lesssim 0.8$ $\mu m$, and having a smoother top, contrasting with those previous conical-like grains.
Considering that the surface (d) still presents a background of small crystallites, the aspect ratio 
in the interval of $120 - 240$ min keeps near to a constant value, once that background has grown up both 
in height and width. Another important and distinguishable feature differing surfaces 
(d)-(e) from (a)-(c) is the ubiquity of \textit{multi-peaked} grains at the interface. 
The AFM images point out that \textit{coalescence processes} have yielded these structures, at least for times 
larger than $t = 60$ min. In the fig. \ref{T250} (f) one compares typical grain's profiles at the surface for
$t = 60$ min and $t = 240$ min. It is clear that single-peaked-sharp-conical structures evolve for larger 
multi-peaked mounds with a flatter top and having a smaller aspect ratio.

%
%
%
%
%

\begin{figure}[!t]
    \vspace{0.5 cm}
    \centering
    \includegraphics[width = 12.0 cm]{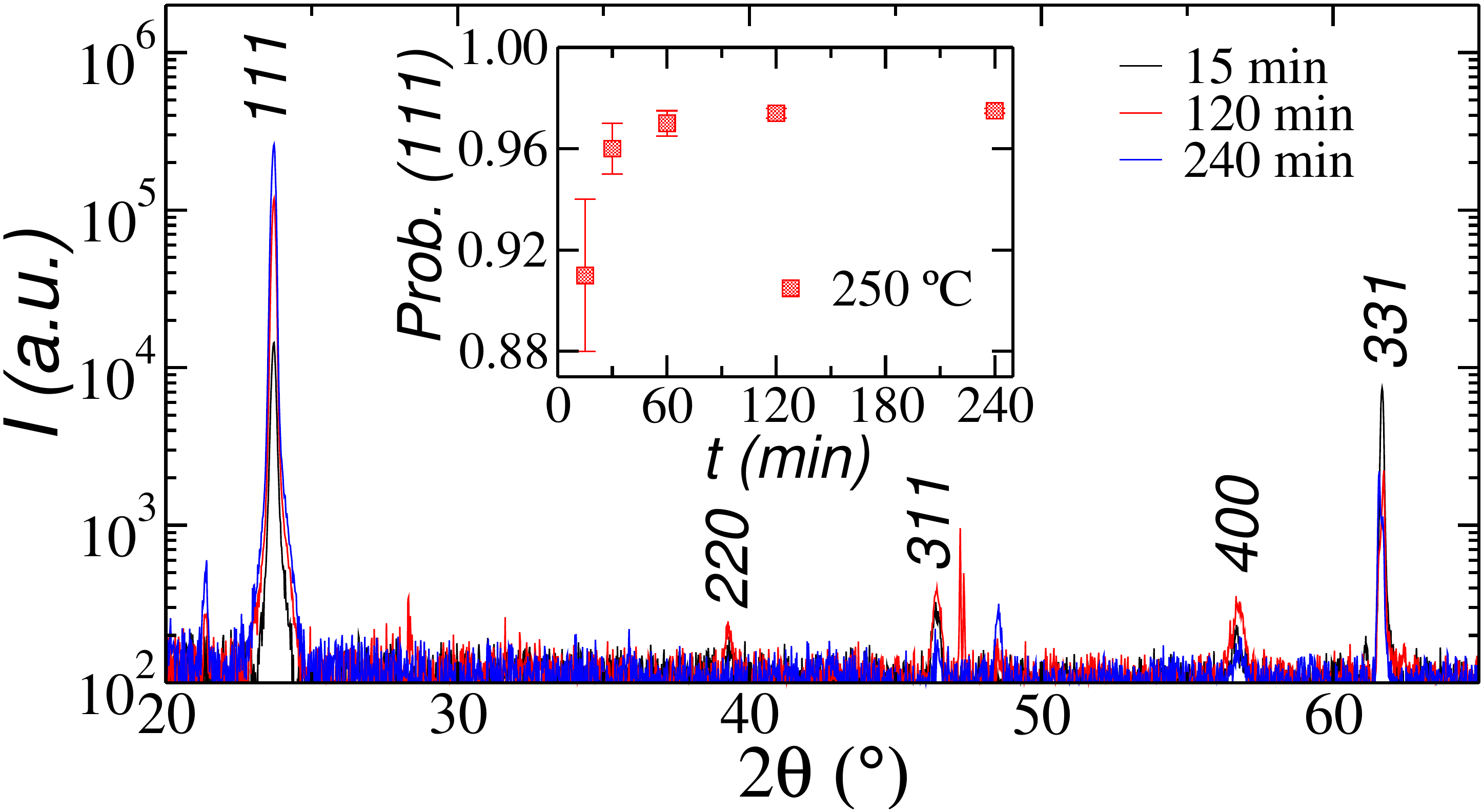}
    \caption[X-Ray spectra of CdTe polycrystalline layers]{X-Ray spectra of CdTe polycrystalline layers using $\lambda_{Cu K-\alpha}$. Spectrum for layers grown by (black) 15 min, (red) 120 min,
    and (blue) 240 min at $T = 250\,^{\circ}\mathrm{C}$. One notices the presence of grains 
    with several crystallographic orientations in reference to the substrate normal, namely: 
    (111), (220), (311), (400) and (331). However, as the time evolves, the (111) peak becomes higher. The 
    inset shows the probability of finding a grain whose [111] direction coincides with the normal vector
    of the substrate surface.}
    \label{xrd}
    \vspace{0.5 cm}
 \end{figure}

 The origin os theses results can be understood as follows. Due to the large mismatch of lattice parameters between CdTe and Si ($\approx 20.0 \%$ \cite{SukarnoQD1}), the growth of CdTe on Si hardly proceeds epitaxially, except in particular conditions reported 
 by Ferreira \textit{et al.} onto Si(111) substrates \cite{SukarnoQD1, SukarnoQD2}. Otherwise, 
 CdTe layers are usually polycrystalline presenting a \textit{strong} [111] texture \cite{Ribeiro, Suka_Ediney2, Sporken}.
 Indeed, we confirm this by performing a structural analysis of the CdTe/Si(001) films. Using a 
 \textbf{D8 - DISCOVER} (BRUCKER) X-Ray Diffractometer in the $\theta-2\theta$ coupled mode (radiation $\lambda_{Cu K_{\alpha}} 
 = 0.154056$ $nm$), the XRD spectra of films are built, as shown in the fig. \ref{xrd}. The presence of more than one crystallographic orientation is a sufficient condition to assign the polycrystalline nature of film. Nevertheless, as the deposition proceeds, the (111) peak becomes higher at the same time that the other ones are reduced. The probability of 
 finding a crystallite [111] oriented after $t$ (min) of growth, p(111, t), is defined as: 
 
 \begin{equation}
 p(111, t) \equiv \frac{[I(t)_{111}/A_{\theta-2\theta}(\theta_{111})]}{\sum_{hkl} I(t)_{hkl}/A_{\theta-2\theta}(\theta_{hkl})}
 \label{p1}
 \end{equation}
 where $I_{hkl,t}$ is the intensity of the peak (hkl) at the time t and $A_{\theta-2\theta}(\theta_{hkl})$ is the absorption
 factor for the $\theta-2\theta$ geometry dependent of the $\theta_{hkl}$ angle \cite{Birkholz_book, Triboulet}.

In the inset of the fig. \ref{xrd} one shows the probability of finding a (111) grain in the film. Thus, one can confirm a very strong [111] texture in CdTe layers grown on Si(001) substrates at $T = 250\,^{\circ}\mathrm{C}$ where for early growth times (111) crystals have already composed almost $90 \%$ of the thick layer.

 In regard to initial regime of growth, previous studies have shown that the growth of CdTe on H-terminated Si surfaces 
 follows the Volmer-Weber (VW) growth mode, where three-dimensional islands nucleate directly onto the surface without an initial wetting layer \cite{SukarnoQD1, Lalev1, Lalev2}. This result has been found in the CdTe growth on both 
 (111) \cite{SukarnoQD1, Lalev1}\footnote{See fig. \ref{cdteqd} in the appendix section \ref{section_NucleationT}.} 
 and Si(001) faces \cite{Lalev2}, by using a slow and/or a variable molecular flux during the experiment. Due to the 
 VW growth mode, conical-like grains are formed onto the template of submonolayer islands and grow by increasing their aspect ratio, since the direct contact with the substrate is unfavorable energetically. Parts of the Si remains exposed until 
 the islands start coalescing to form a continuous layer. As the environment is polycrystalline, 
 grain boundaries (GB) of colided neighborhing grains are unfavorable regions for deposition and diffusion of particles, since in these places one finds a large number of defects \cite{Tello, Gonzales, Joelma, AmarCdTe}. Defects are yielded, mainly, due the stress evolution. Many studies \cite{Tello, Gonzales} have shown that the stress is compressive prior to coalescence, becomes tensile during the coalescence (as a consequence of attractive forces between neighboring grains) and reaches a constant value as deposition proceeds. Even in the presence of the strong texture, the shared GBs of two (111) crystals also yield defects because the grains have usually different rotational orientations. This kind of defect causes shear stress (torque) which tends to reorient the crystallites \cite{Gonzales}.
 
 For long growth times (multi-layer regime), the 
 substrate-film interaction is expected to be lost \cite{Herman1} and the growth dynamic becomes dictated by the growth of CdTe on CdTe, i.e. the homoepitaxial growth \cite{Herman1, Herman2}. As the substrate temperature is high during the growth, the rate diffusion is expected to be high \cite{Barabasi}\footnote{See also section \ref{section_NucleationT} in the Appendix.}. Thus, ``defect sites'' at (an around) the GBs of colided neighboring grains are eventually covered by diffusion and deposition of particles (a kind of \textit{relaxation} process occuring at, and around, these regions), which allows
 coalescence processes becoming more operative. As consequence of this relaxation, the aspect ratio ($\Omega$) of grains
 decrease in time and the top of them becomes smoother (due to the ``filling'' process of particles at local minima), as corroborated by the AFM images (figures \ref{T250}(a)-(e)).
 
\section{Local Fluctuations: Non-Universal and Universal Scaling Exponents}

\begin{figure}[!ht]
    \vspace{0.5 cm}
    \centering
    \includegraphics[width = 7.0 cm]{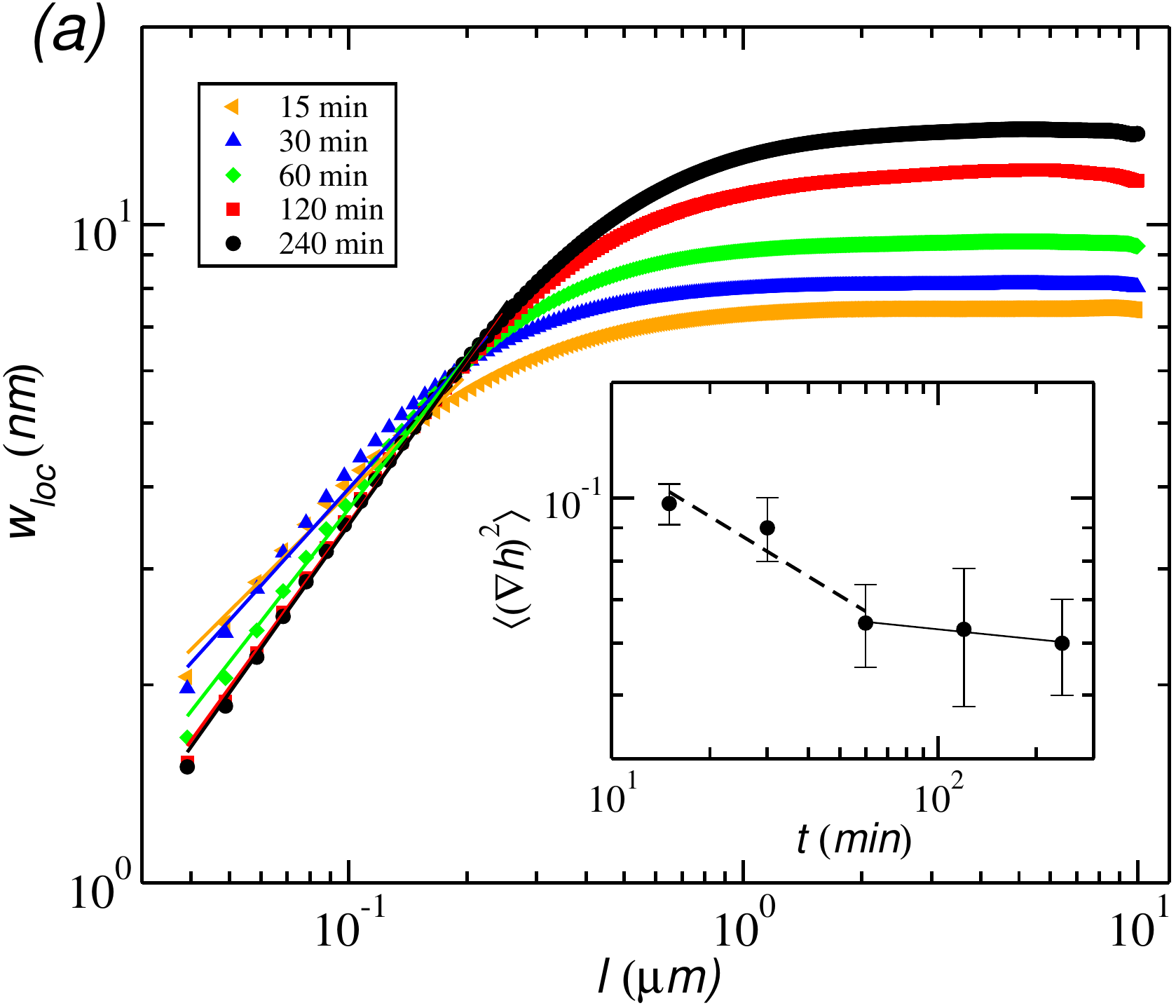}
    \includegraphics[width = 7.0 cm]{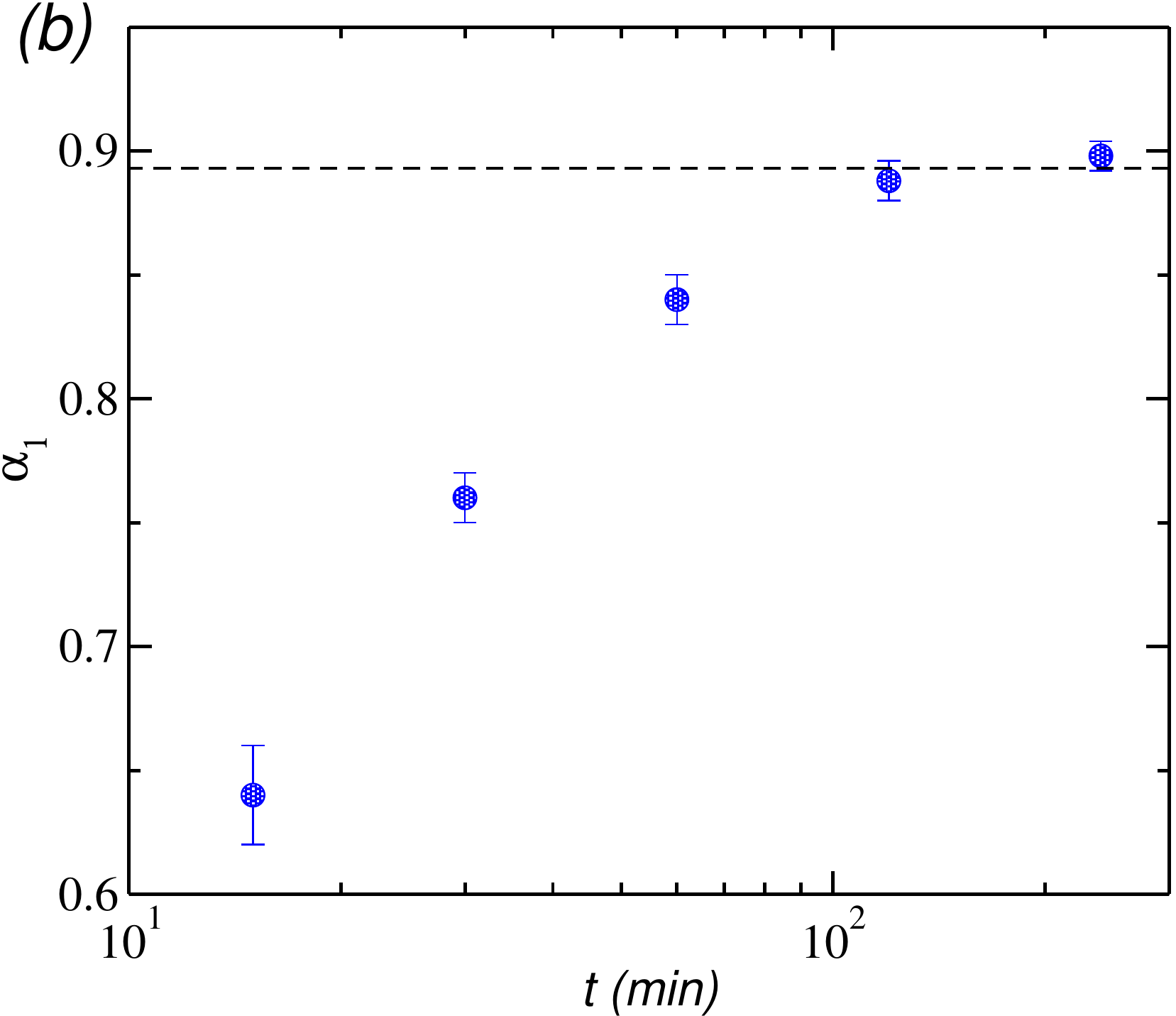}
    \caption[Local roughness scaling for CdTe thin films grown at $T = 250\,^{\circ}\mathrm{C}$]
    {(a) Local roughness scaling for CdTe thin films grown at $T = 250\,^{\circ}\mathrm{C}$
    by 15 min (left triangles), 30 min (up triangles), 60 min (diamonds), 120 min (squares) and 240 min (circles).
    Inset shows the average local inclination and (b) shows in details the non-universal (geometrical) scaling exponent
    as function of the growth time. Solid lines in the main plot of (a) represent the fits used 
    to extract $\alpha_1$ in (b).}
    \label{wloc250}
    \vspace{0.5 cm}
 \end{figure}

Now we turn our attention to local fluctuations at the interface. Figure \ref{wloc250}(a) shows the local 
roughness (defined in the section \ref{section_CF}) calculated for surfaces grown by different times.

At short length scales ($l \lesssim 10^{-1}$ $\mu m$) and early times, one can observe $w_{loc}(l,t)$ 
decreasing in time because, in this situation, $w_{loc}$ measures the intra-grain roughness of grains
evolving from a sharp to a rounded shape - i.e, from larger to smaller height fluctuations.
For comparison, this is the opposite occurring in systems that exhibits anomalous scaling, as
can be seen in the appendix section \ref{section_anomalous}. In general, the presence or absence of
anomalous roughening can be evaluate by calculating the evolution of squared slopes at the surface, $\langle(\nabla h)^2\rangle$, which is expected to scale as \cite{Lopez99}:

\begin{equation}
 \langle(\nabla h)^2\rangle \sim t^{2\kappa}.
 \label{kappa3}
\end{equation}

Positive values of $\kappa$ indicate anomalous scaling, while negative or null values confirm that the Family-Vicsek
scaling (eq. \ref{roug3}) describes the dynamic of local height fluctuations \cite{Lopez99, Ramasco00}.

The inset of the fig. \ref{wloc250}(a) shows the behavior of $\langle(\nabla h)^2\rangle$ in time, which is the same
of $w_{loc}$ for fixed box sizes $l^*$, with $l^* \lesssim 10^{-1}$ $\mu m$. In particular, one can notice that
an initial decreasing regime is followed by a ``saturated'' one. The power-law (eq. \ref{kappa3}) in this 
transient regime provides $\kappa = -0.15(5)$. This value is very close to that predicted 
by L\'{o}pez \cite{Lopez99}, namely $\kappa = -1/6$, for two-dimensional KPZ systems. Moreover, it is well-known 
that $w_{loc}(l^*,t) \sim \Omega$ \cite{Huo, Tiago_mestrado} and these results are in agreement with our previous
analysis based on AFM images. Moreover, the crossover in $w_{loc}$ occurs in $l \approx \zeta$ (compare with fig. \ref{T250}),
in accordance with numerical results \cite{Tiago_grain1, Tiago_grain2}. At the other hand, long-wavelength fluctuations ($l \gg 10^{-6}$ $\mu m$) increase in time, as expected \cite{Barabasi}.

The scaling exponent $\alpha_1$, defined as $w(l,t) \sim l^{\alpha_1}$, has its value 
\textit{changing in time} from 0.6 to 0.9, respectively, as shown in the fig. \ref{wloc250}(b).
These values, however, should not be taken as representative of universal fluctuations at the interface, as done in previous 
studies \cite{Sukarno, Fabio, Mata}. Rather than, as explained by Oliveira and Reis \cite{Tiago_grain1, Tiago_grain2} (see
section \ref{section_grainmodel}), $\alpha_1$ values are related to the grain morphology: $\alpha_1 \approx 0.6$ indicates that the grainy morphology has a sharp-conical form and, as larger is the $\alpha_1$ value, smoother is the top of typical grains' shape. Notice that these findings are in agreement with the AFM images (fig. \ref{T250}).
The true universal roughness exponent $\alpha$, however, should be found in a region where $\zeta \ll l \ll \xi$ \cite{Tiago_grain1, Tiago_grain2}.



 \begin{figure}[ht]
    \vspace{0.5 cm}
    \centering
    \includegraphics[width = 10.0 cm]{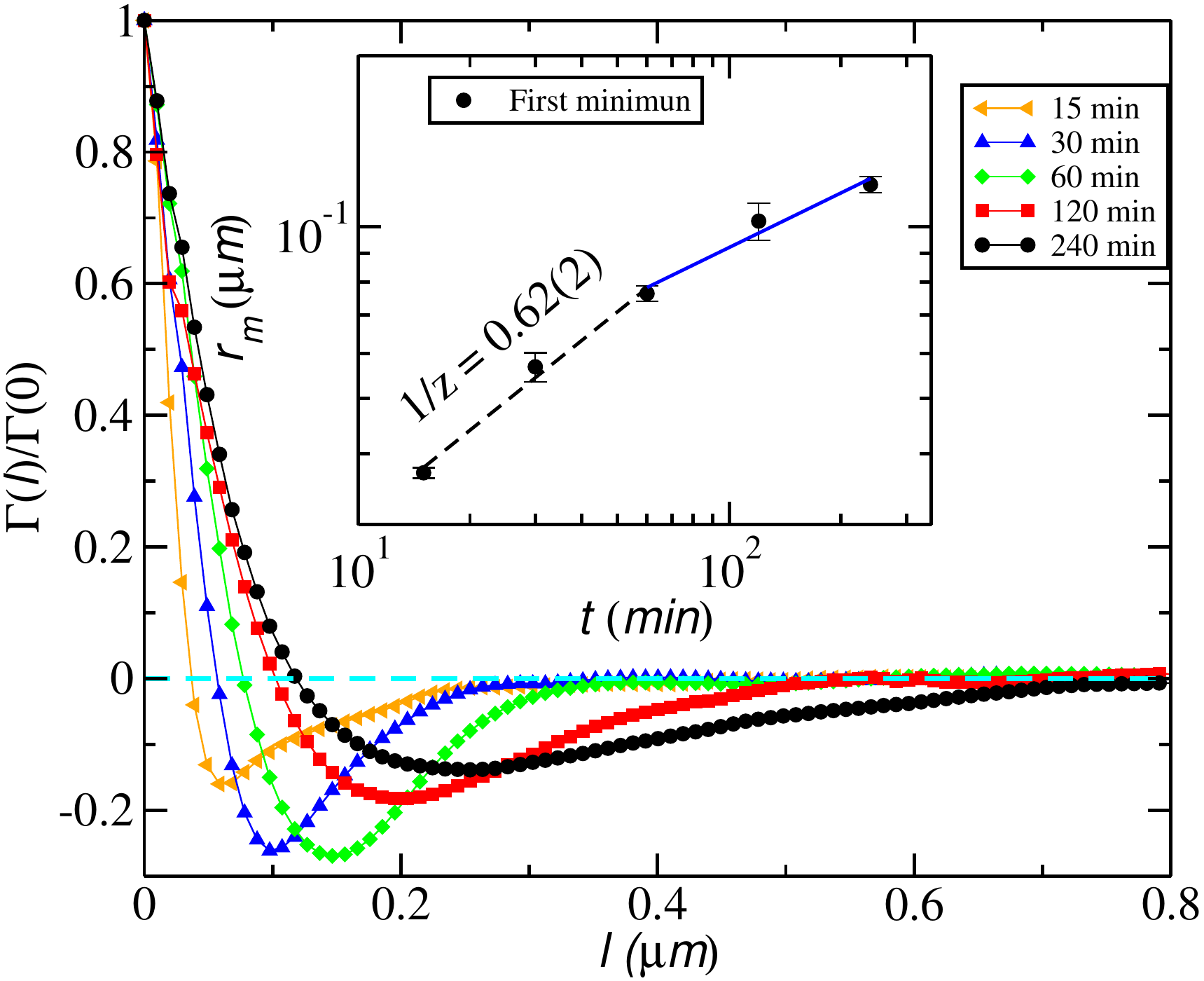}
    \caption[Normalized slope covariance for CdTe thin films grown at $T = 250\,^{\circ}\mathrm{C}$]
    {Normalized slope-slope correlation function for CdTe thin films 
    grown at $T = 250\,^{\circ}\mathrm{C}$ by 15 min (left triangles), 30 min (up triangles), 
    60 min (diamonds), 120 min (squares) and 240 min (circles). Inset shows 
    the first minimum (circles) extracted from $\Gamma(l)/\Gamma(0)$ curves as
    function of growth time.}
    \label{slope250}
    \vspace{0.5 cm}
 \end{figure}

In order to unearth the dynamic exponent, we have used the slope-slope correlation function, defined 
in the eq. \ref{slop_CF} and discussed in the section \ref{section_CF}. Figure \ref{slope250} shows
normalized $\Gamma(l)$ functions calculated for all times studied. From the first zero and/or from the first
minimum ($r_m$) of normalized $\Gamma(l)$ curves one can measure the \textit{coarsening exponent} 
$n_{coar}$, defined as $r_m \sim t^{n_{coar}}$ \cite{Fabio, Almeida, Cordoba}. For short growth times, before
the appearence of multi-peak structures at surface, one has that $n_{coar}= 1/z$. As suggested by AFM images, this 
regime in our experimental situation occurs for $t \lesssim 60$ min and provide
$n_{coar} = 1/z = 0.62(2)$ - see the dashed line in the inset of the fig. \ref{slope250}. This value is in 
excellent agreement with that expected for two-dimensional KPZ systems \cite{Tiago2d, Healy_2d1, Healy_2d2, Healy_exp, Healy_stationary, Kelling} (see eq. \ref{kpz_exp}). In turn, taking longest available growth times, 
one obtains $n_{coar} \approx 0.34$, which should not be interpreted as representative of temporal universal fluctuations because, here, peaks of reasonable size appearing on the top of a mound induce an underestimative for the average
grain size measured by the slope-slope function (eq. \ref{slop_CF}). Thus, a reliable measurement 
for $r_m$ is not achieved in this situation and, hence, one can not be validated the relation $n_{coar}= 1/z$.

\section{Global Scaling}

\begin{figure}[!h]
    \vspace{0.5 cm}
    \centering
    \includegraphics[width = 10.0 cm]{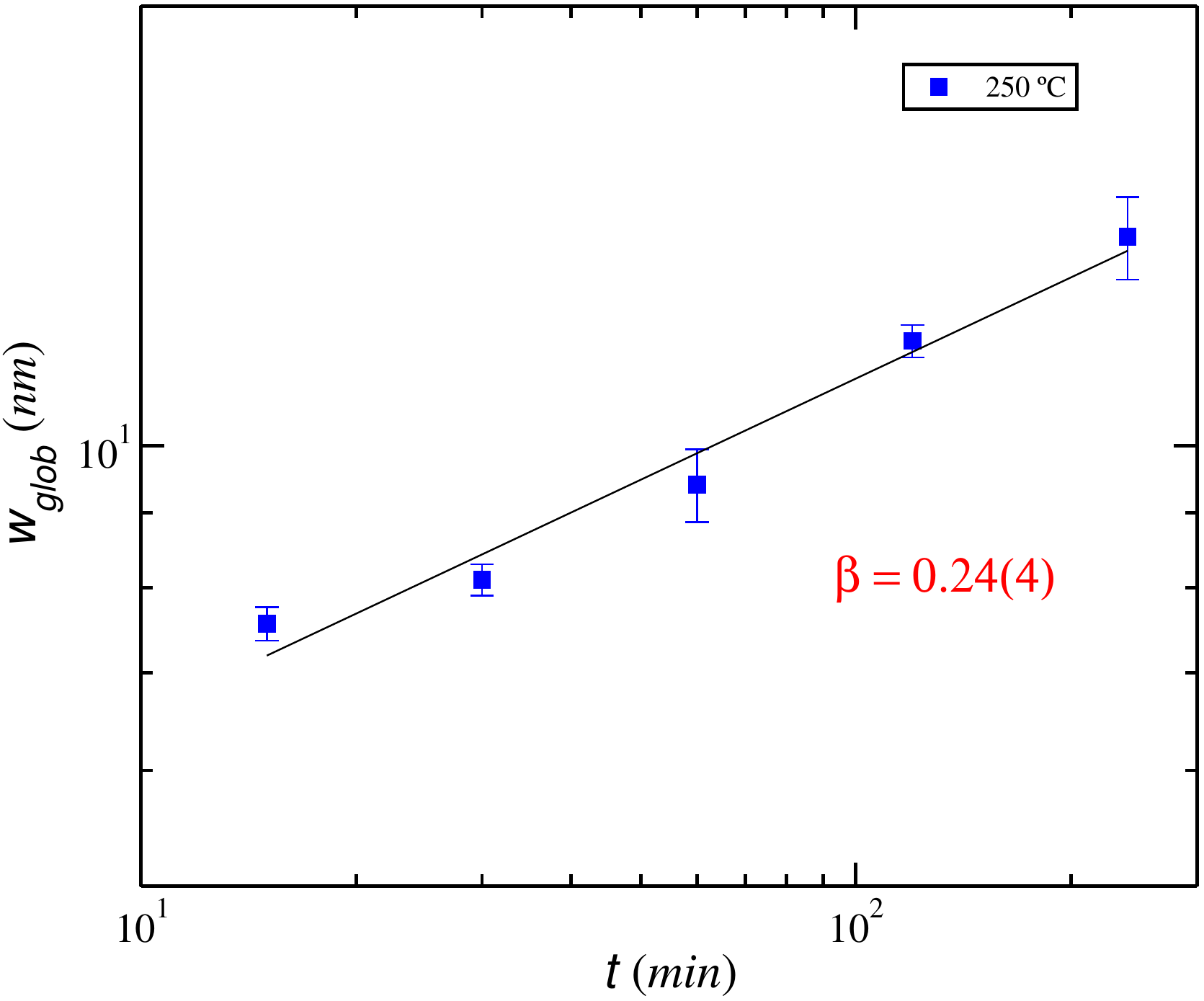}
    \caption[Global roughness of CdTe thin films as function of time]
    {Global roughness as function of time for CdTe thin films grown at $T = 250\,^{\circ}\mathrm{C}$.
    The roughness grows as a power-law with $\beta = 0.24(4)$.}
    \label{wglob250}
    \vspace{0.5 cm}
 \end{figure}

Following the Family-Vicsek \textit{ans\"{a}tze} (eq. \ref{roug3}), the global roughness scale as $t^{\beta}$. In the 
fig. \ref{wglob250} one can see the $w$ plot behaving as a power-law with exponent $\beta = 0.24(4)$. This is a
very strong evidence of KPZ$_{d = 2 + 1}$ growth, once $\beta_{KPZ} \approx 0.24$ (see eq. \ref{kpz_exp}).

At this point we have found $1/z = 0.62(2)$, $\beta = 0.24(4)$ and $\kappa = -0.15(5)$. These values are strong
indicatives of two-dimensional KPZ growth in CdTe films. If this clue is really true, we should obtain universal 
distributions of heights, local roughness, and maximal heights matching with that ones numerically predicted for the
KPZ class (see table \ref{universalkpz2} and sections \ref{section_SLRD} and \ref{section_MRHD}).

As we are dealing with a few number of available growth times, we are not able to find the 
asymptotic velocity ($v_\infty$) of the CdTe surface growth \cite{Healy-Krug}. Beyond that, by following 
the Krug-Meakin toolbox \cite{KM90, Healy-Krug}, data of CdTe interface fluctuations evolving in time
(until reaching the stationary regime) for different initial miscut angles of Si(001) substrates should be accessible.
Thus, we are not in position to validate, at this time, the two-dimensional KPZ \textit{ans\"{a}tze} (eq. \ref{eq_ansatz2}). However,
we can perform a comparison between a \textit{rescaled} height distributions and the distribution of $\chi$. 
Take the spatial average on both sides of the eq. \ref{eq_ansatz} - negleting the correction terms - one has:

\begin{equation}
\langle h \rangle = v_\infty t + sign(\lambda)(\Gamma t)^{\beta}\langle \chi \rangle.
\label{a1}
\end{equation}

One can vanish the term $v_\infty t$ subtracting eq. \ref{a1} from eq. \ref{eq_ansatz}. The result is:

\begin{equation}
h - \langle h \rangle = sign(\lambda)(\Gamma t)^{\beta}[\chi - \langle \chi \rangle].
\label{a2}
\end{equation}

Now, it easy to show that $\langle h^2 \rangle_c = (\Gamma t)^{2\beta}[\langle \chi^2 \rangle - \langle \chi \rangle^2]$. 
Hence, dividing eq. \ref{a2} by the global roughness $w = \sigma_h = \langle h^2 \rangle_c^{1/2}$ one finds:

\begin{equation}
\frac{(h - \langle h \rangle)}{\sigma_h} = sign(\lambda) \frac{(\chi - \langle \chi \rangle)}{\sigma_{\chi}},
\label{a3}
\end{equation}
where $\sigma_{\chi}$ = $\langle \chi^2 \rangle_c^{1/2}$.

 \begin{figure}[!ht]
    \vspace{0.5 cm}
    \centering
    \includegraphics[width = 12.0 cm]{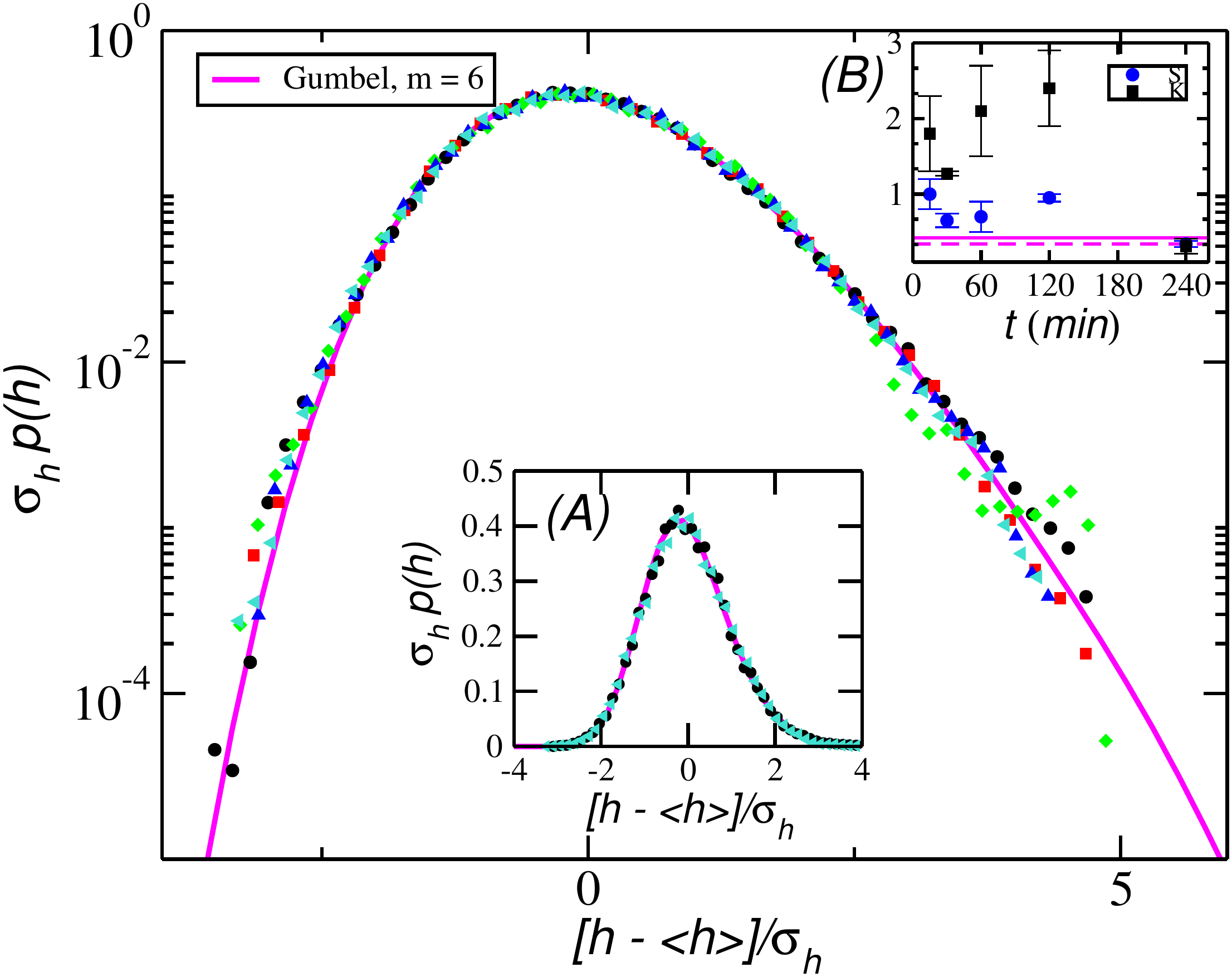}
    \caption[Rescaled HDs for several regions of the thickest CdTe layer]
    {Rescaled height distributions at mean null and unitary variance from several 
    regions (symbols) of the thickest sample grown at $T = 250\,^{\circ}\mathrm{C}$ compared with 
    the numerical KPZ curve (solid magenta line). Inset (A) highlights the very good collapse 
    between experimental and numerical data close to the peak in a linear $\times$ log plot. Insertion (B) exhibits the
    experimental S and K evolving in time. Solid and dashed magent lines refers to the numerically expected 
    KPZ$_{d = 2 + 1}$ S and K values, respectively.}
    \label{HD250}
    \vspace{0.5 cm}
 \end{figure}

Equation \ref{a3} means that to compare rescaled height distributions with mean null and unitary variance 
is equivalent to comparing the $\chi$ fluctuations rescaled in the same way. This allows us to compare these distributions even without known the non-universal KPZ parameters. Figure \ref{HD250} depicts the rescaled height distributions for 
several regions of the thickest sample. Remarkably, one notices a very good agreement between the 
experimental data and the Gumbel pdf with m = 6 (see eq. \ref{pdf_gumb}) for four decades around the peak. This astonish result
is reinforced by the data collapse close to the peak, as highlighted in the the insertion A of the fig. \ref{HD250}. 
The slight deviation observed in the left tail could be due to the fact that AFM tip does not
scan accurately deep valleys as well as it does for higher heights.

Comparing the universal KPZ$_{d = 2 + 1}$ dimensionless cumulant ratios for the flat case (table \ref{universalkpz2}), we have found, for the thickest film, $S_{t = 240} = 0.34(1)$ and $K_{t = 240} = 0.3(1)$, while numerical values are $S = 0.423(7)$ and $K = 0.344(9)$ \cite{Tiago2d, Healy_2d1, Healy_2d2, Healy_exp}. However, as shown
in the inset B of the fig. \ref{HD250}, the experimental $S$ and $K$ values approach to the 
KPZ$_{d = 2 + 1}$ ones only for $t = 240$ min. In principle, these results could be just a 
fluctuation, making the agreement of HDs in the fig. \ref{HD250} a mere coincidence. Indeed, this is not the 
case here because before to seek the universal HD distribution, \textit{critical exponents} have 
already point out to KPZ growth. Moreover, it is well know that experimental $S$ and $K$ values take a 
long time to converge to the \textit{asymptotic} values calculated numerically\cite{Takeuchi1, Takeuchi2, Yunker}. This
effect should be stronger in 2 + 1 dimensional systems.
Hence, the underestimate $S$ value can be only a \textit{finite-time} effect and results emerging from the 
fig. \ref{HD250} provide the first so-waited experimental confirmation of the 
KPZ$_{d = 2 + 1}$ universality beyond the exponents.

 \begin{figure}
    \vspace{0.5 cm}
    \centering
    \includegraphics[width = 11.0 cm]{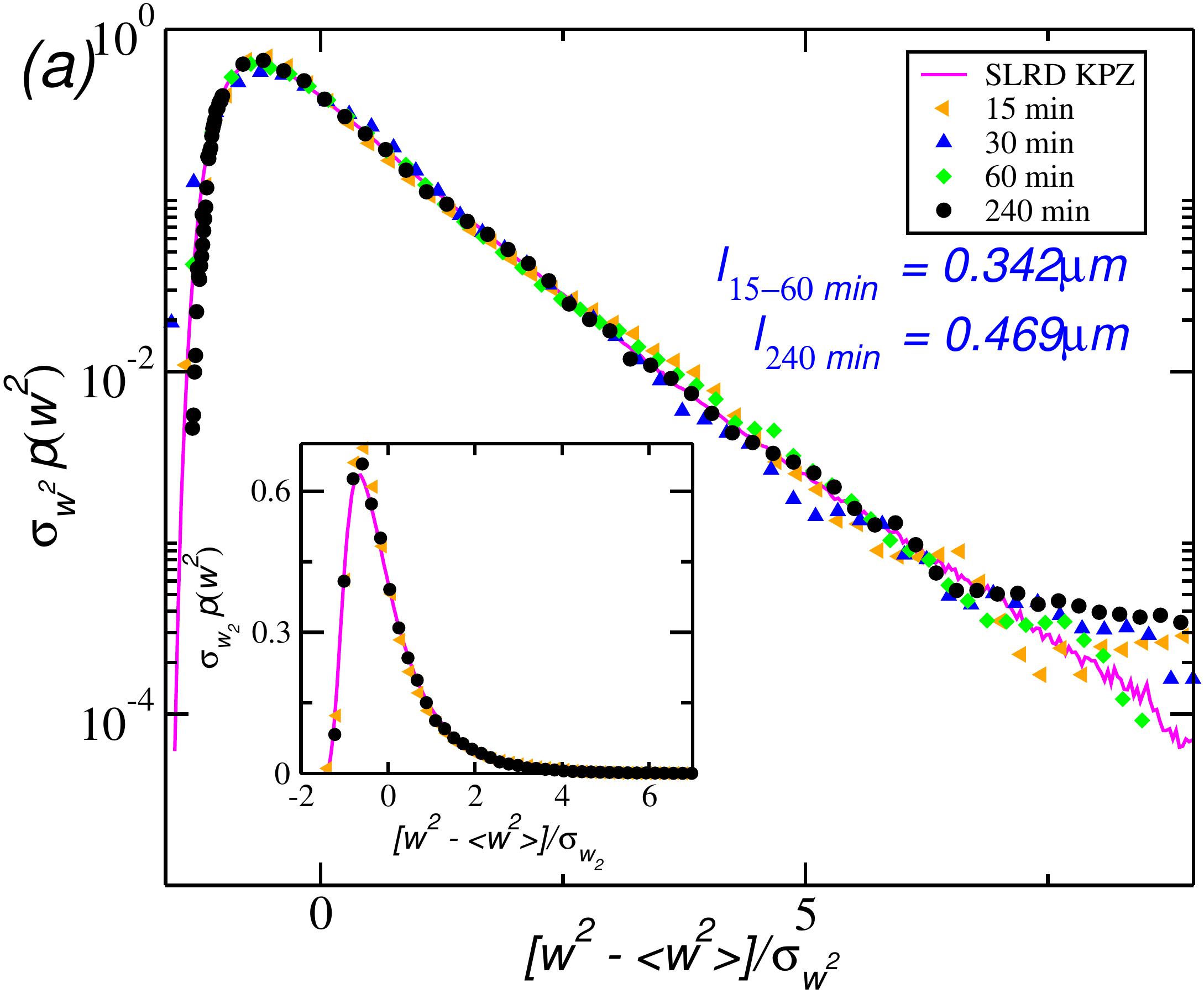}
    \includegraphics[width = 11.0 cm]{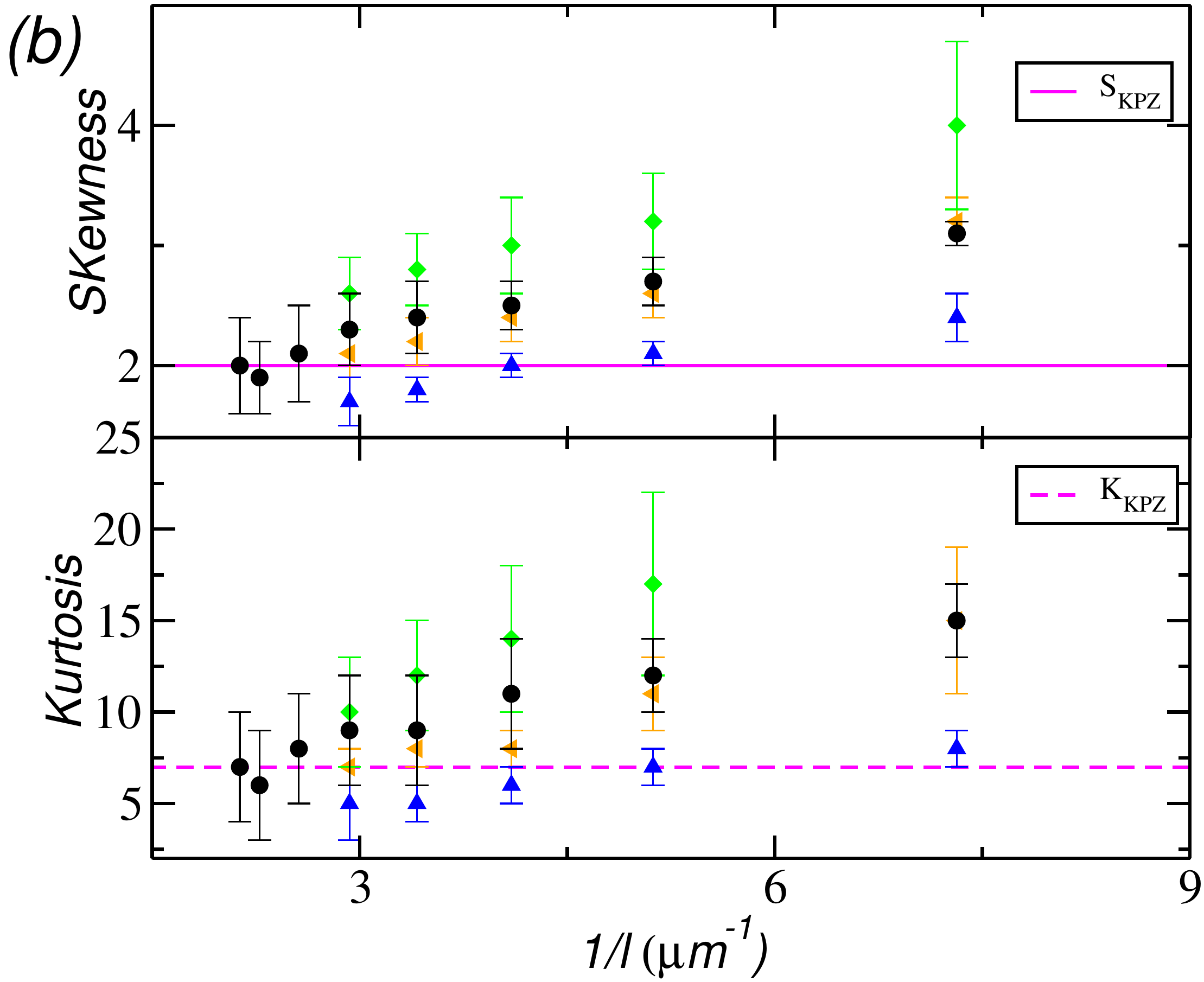}
    \caption[Rescaled SRLDs for CdTe surfaces grown at $T = 250\,^{\circ}\mathrm{C}$]
    {(a) Rescaled squared local roughness distributions for samples grown at 
    $T = 250\,^{\circ}\mathrm{C}$, by 15 min (orange left triangles), 30 min (blue up triangles),
    60 min (green diamonds) and 240 min (black circles).The curves shown are those whose the box size
    is indicated below the legend. They are different for each $t$ because the appropriate interval 
    $\zeta \ll l \ll \xi$ depends on $t$. The inset shows the good collapse around 
    the peak for the thinnest and thickest film grown. (b) Skewness and kurtosis values as 
    function of the box size $l$.}
    \label{SLRD250}
    \vspace{0.5 cm}
 \end{figure}
 \begin{figure}
    \vspace{0.5 cm}
    \centering
    \includegraphics[width = 11.0 cm]{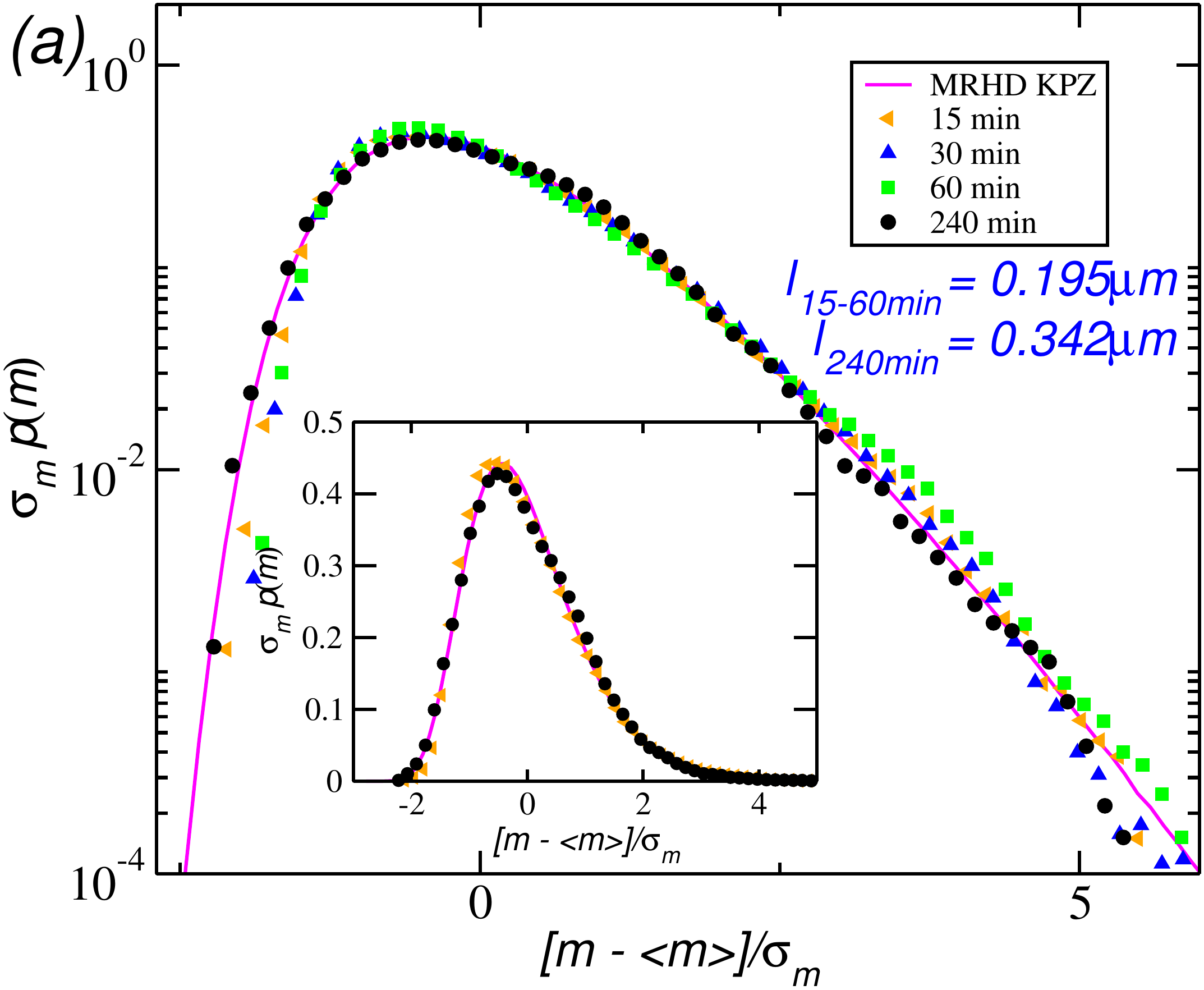}
    \includegraphics[width = 11.0 cm]{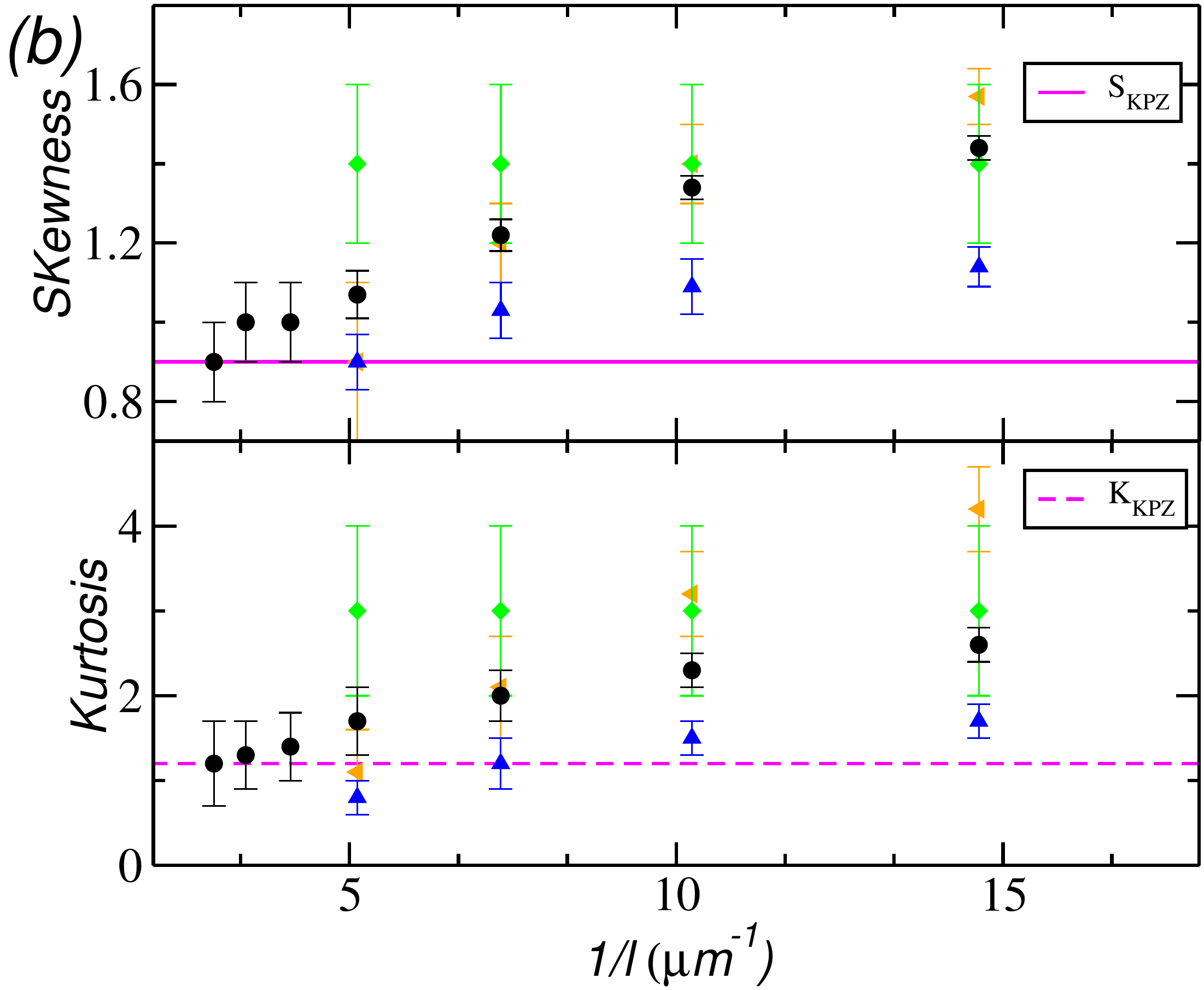}
    \caption[Rescaled MRHDs for samples grown at $T = 250\,^{\circ}\mathrm{C}$]
    {(a) Rescaled maximal relative height distributions for samples grown at 
    $T = 250\,^{\circ}\mathrm{C}$, by 15 min (orange left triangles), 30 min (blue up triangles),
    60 min (green diamonds) and 240 min (black circles). The curves shown are those whose the box size
    is indicated below the legend. They are different for each $t$ because the appropriate interval 
    $\zeta \ll l \ll \xi$ depends on $t$. The inset shows the good collapse around the peak 
    for the thinnest and thickest film grown. (b) Skewness and kurtosis values as function of the box 
    size $l$.}
    \label{MRHD250}
    \vspace{0.5 cm}
 \end{figure}

In order to prove the universality of distributions described in the sections \ref{section_SLRD} and \ref{section_MRHD}, and 
also to gain further evidence of KPZ growth, we have calculated the SLR and MRH distributions, constrained to the box size $l$ within the interval $[10 \mu m/1024] \ll l \ll \xi$. Figure \ref{SLRD250} (a) shows the SLRDs for all available growth times, whereas in fig. \ref{SLRD250} (b), the respective $S$ and $K$ values as function of $l$ are depicted. Astonishingly, a 
very good collapse is seen, in almost four decades around the peak, among the experimental and numerical KPZ SRLDs. Moreover, the stretched exponential decay at the right tail is present, giving one more (strong) evidence of KPZ growth. 

Similar results are found for the MRHDs, figures \ref{MRHD250} (a) and (b), where the nice collapse between 
the experimental and numerical data is evident. The curves showed in the main plot are those 
whose $S$ and $K$ values better represent the mean for the $l$ described in the graph.

It is very important mention why the $S$ and $K$ values change with $l$. This happens because one does not know, \textit{a priori}, what are the $l's$ satisfying the condition $\zeta \ll l \ll \xi$ (remember there is a characteristic length
at the surface). As in the CdTe/Si(100) system $\zeta \approx \xi$, the appropriate interval for $l$ is very short 
and should be identified as a convergence of $S$ and $K$ for some specific value. In the case shown
in the figures \ref{SLRD250}(b) and \ref{MRHD250}(b), these convergent values agree with those numerically calculated for 
the SLRD and MRDS of KPZ$_{d = 2 + 1}$ models \cite{Paiva, Aarao, Tiago_mahd}.

\section{The origin of the KPZ mechanism and Conclusions}

The results presented in the previous sections point out that the long-wavelength CdTe surface fluctuations, at least
for film grown at $T = 250\,^{\circ}\mathrm{C}$ and $F \approx 2.2$ \AA{}/s, evolve according 
to the KPZ equation (eq. \ref{eq_kpz}). The origin of the KPZ scaling in this experimental 
system can be understood unveiling the complex dynamic of grains, which is dictated by an interplay between 
inter-grain surface energetics, constraints yielded by the coalescence of 
neighborhing grains and filling processes occuring at (and around) the GB's of colided neighborhing grains. This leads to a 
complex packing of crystalline grains. A simple illustration is provided by the grain deposition model in the Ref. \cite{Tiago_grain1}. In the model, a new cubic grain is firmly attached to the boundary of the grains below it, but does not fill all available space in their neighborhood because its shape is constrained to be cubic. This aggregation 
mechanism has the same effect of the lateral aggregation as in the ballistic model \cite{Vold}. It generates excess velocity, which is the landmark of KPZ scaling.

In conclusion, we have found the first robust\footnote{Going beyond of the standard comparasion with exponents.} experimental $KPZ_{d = 2 + 1}$ system.
On the basis of a semi-quantitative morphological and long-wavelength fluctuation studies, we have extracted 
critical exponents agreeing with that expected for the KPZ class. Moreover, the universality of KPZ 
distributions (HDs, SLRDs and MRHDs) has been experimentally demonstrated. We have not checked 
the KPZ$_{d = 2 + 1}$ ans\"{a}tze (eq. \ref{eq_ansatz}) directly due to inherent two-dimensional experimental obstacles.
Our results were published in the Physical Review B \cite{Almeida}.

Shortly after the publication of our work \cite{Almeida}, T. Halpin-Healy and G. Palasantzas \cite{Healy_exp} have used
this noveil scheme proposed by us to confirm the KPZ universality in oligomer thin films. This clearly corroborates 
that we have given a new perspective for finding, and confirming, two-dimensional growing surfaces 
belonging to the KPZ universality class \cite{Healy_exp}.

  \chapter{RESULTS: The Effect of Temperature on CdTe Growth Dynamic}
  \lhead{\bfseries 6. Temperature Effect on CdTe Growth Dynamic}
  In this chapter one studies the effect of the deposition temperature, in the range of [150, 300]$\,^{\circ}\mathrm{C}$, 
on the mound evolution and on the long-wavelength fluctuations of CdTe surfaces grown on Si(001) substrates.

\section{Intragrain morphology and local fluctuations}

\begin{figure}[!ht]
    \vspace{0.5 cm}
    \centering
    \includegraphics[width = 15.0 cm]{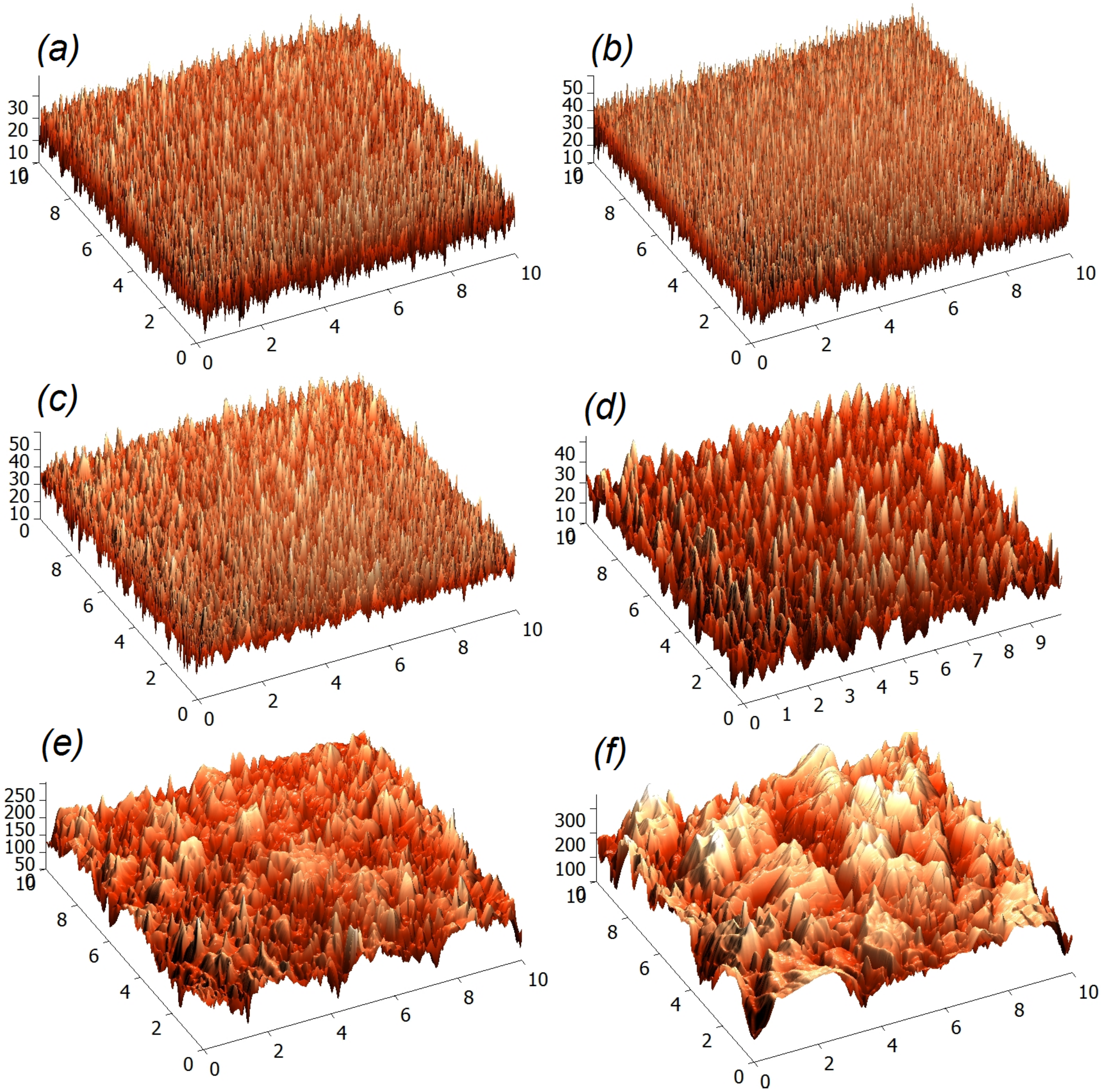}
    \caption[AFM images of CdTe thin films grown at $T = 150$, $200$ and $300\,^{\circ}\mathrm{C}$.]{AFM images ($10 \times 10$ $\mu m$, height scale in $nm$) of CdTe thin films grown at: 
    $T$ = $150\,^{\circ}\mathrm{C}$ (a) and (b), $T$ = $200\,^{\circ}\mathrm{C}$ (c) and (d), and 
    $T$ = $300\,^{\circ}\mathrm{C}$ (e) and (f) by $t = 120$ min and $240$ min, respectively.}
    \label{morph1}
 \end{figure}

\begin{figure}[h]
    \centering
    \includegraphics[width = 13.0 cm]{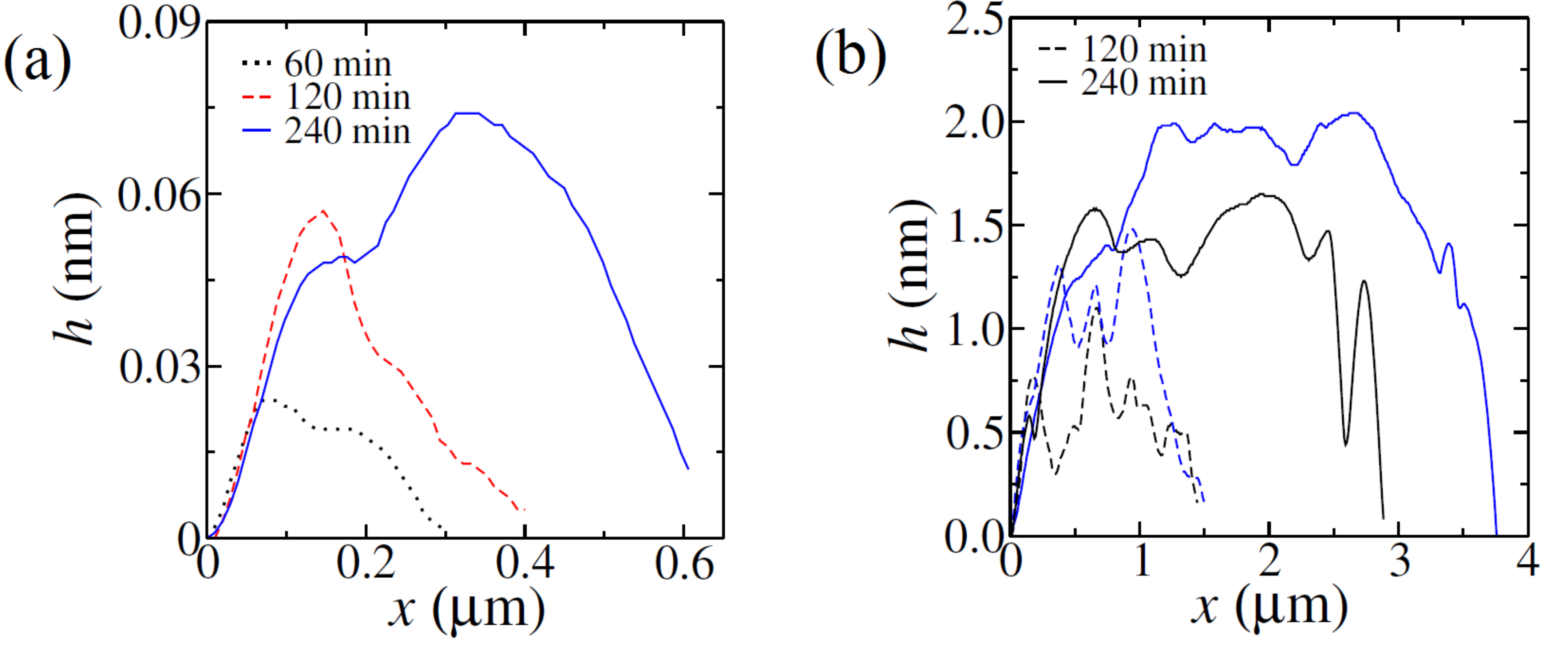}
    \caption[Typical grain/mound profiles at the surface for $T = 200\,^{\circ}\mathrm{C}$, and 
    $300\,^{\circ}\mathrm{C}$.]{Typical grain/mound profiles at the surface for (a) $T = 200\,^{\circ}\mathrm{C}$ 
    and (b) $T = 300\,^{\circ}\mathrm{C}$ for different growth times.}
    \label{morph2}
    \vspace{0.5 cm}
 \end{figure}

Figure \ref{morph1} shows typical AFM images for surfaces grown at different $T$ for the two largest available growth time, $t = 120$ min and $240$ min. For $T= 150\,^{\circ}\mathrm{C}$ (fig. \ref{morph1}(a) and (b)), one can see grains with a well-defined sharp shape  
dominating the surface, at the same time that $\Omega$ seems to increase with $t$, as suggested by the vertical bar. 
For films  grown at $T= 200\,^{\circ}\mathrm{C}$, the scenario is quite similar to that for $T= 150\,^{\circ}\mathrm{C}$, except 
for the largest growth time (fig. \ref{morph1}(d)), where $\Omega$ has decreased in the interval of $t \in$ [120, 240]min. 
Unlike the morphology of surfaces grown at $T = 250\,^{\circ}\mathrm{C}$ (fig. \ref{T250}), 
multi-peaked structures are rare in $t = 240$ min and grains with a well-defined shape still are the majority. 
See fig. \ref{morph2}(a), in which typical mound profiles are shown in order to make clear those notes. 
On the other hand, at higher temperatures, namely $T= 300\,^{\circ}\mathrm{C}$, the surfaces present 
much more complex structures, as can be noticed from the the figures \ref{morph1}(e) and (f). In fact, these 
films are composed by a mix of a few conical grains and large mounds (formed by coalesced grains) with 
$\zeta \approx 2$ $\mu m$. As the time evolves, $\zeta$ increases faster than the mound height, at least
within the interval $t \in$ [120, 240]min, to form larger mounds.


 \begin{figure}[!h]
    \vspace{0.5 cm}
    \centering
    \includegraphics[width = 8.5 cm]{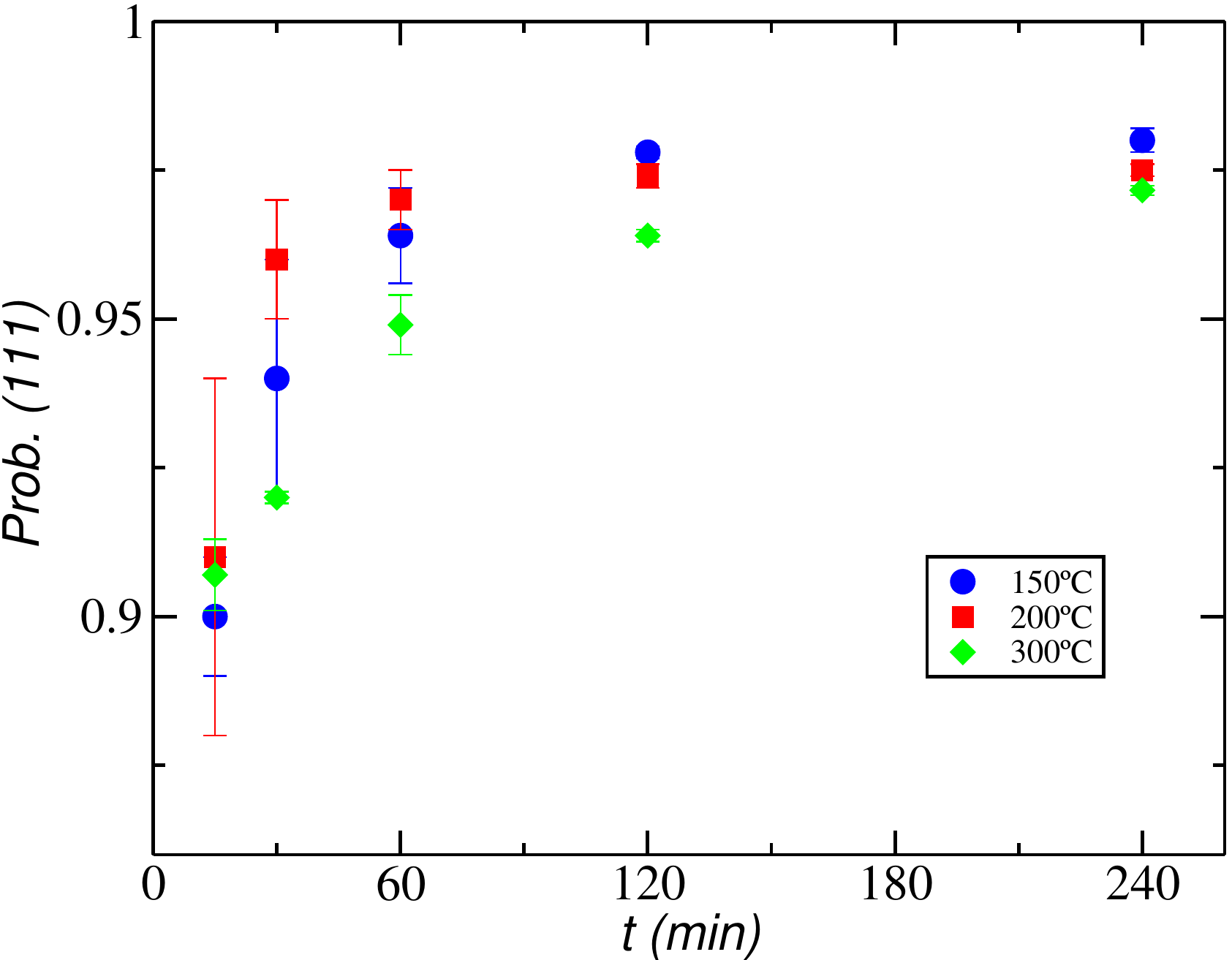}
    \caption[Texture in the (111) direction in CdTe layers grown at different $T$]{Probability of finding (111) 
    grains with respect to the substrate normal in a CdTe layer grown at $T = 150\,^{\circ}\mathrm{C}$ (circles), 
    $200\,^{\circ}\mathrm{C}$ (squares) and $300\,^{\circ}\mathrm{C}$ (diamonds).}
    \label{xrd2}
    \vspace{0.5 cm}
 \end{figure}

The orientation of grains whose crystallographic plans are parallel to the substrate surface is determined 
by $\theta-2\theta$ XRD measurements. As we already found for films grown at $T = 250\,^{\circ}\mathrm{C}$, 
a strong [111] texture has been observed also for other temperatures. Moreover, the same peaks have been seen
in the XRD espectra\footnote{The XRD spectrum is not shown because it is similar to that represented in the fig. \ref{xrd}.}. Figure \ref{xrd2} shows the probability of finding (111) grains (eq. \ref{p1}) in the CdTe layer as function of the growth time and for different temperatures. The results suggest that this texture is: 1) slightly influenced by the deposition temperature,
being this sligth deviation probably due to the fact that as higher is $T$, larger is the width of grains \cite{Evans}, including those grown in (111)-different orientions which reflects more the X-ray incident and, hence, diminhes $p(111, t, T)$. 2) independent of the substrate, once a similar [111] texture has been found by Ribeiro \textit{et al.}\cite{Ribeiro} and Ferreira \textit{et. al}\cite{SukarnoQD1, Suka_Ediney2} in the growth of CdTe on glass substrates and also by Sporken \textit{et al.} using Si(001) substrates \cite{Sporken}. Anyway, 
the important for the growth dynamic is that most coalescence process happens between (111) grains and, even this way, they give rise a large number of defects at the GBs of colided neighboring grains due to, basically, the difference between the rotational orientation of neighboring crystallites.

 \begin{figure}[!ht]
    \vspace{0.5 cm}
    \centering
    \includegraphics[width = 7.2 cm]{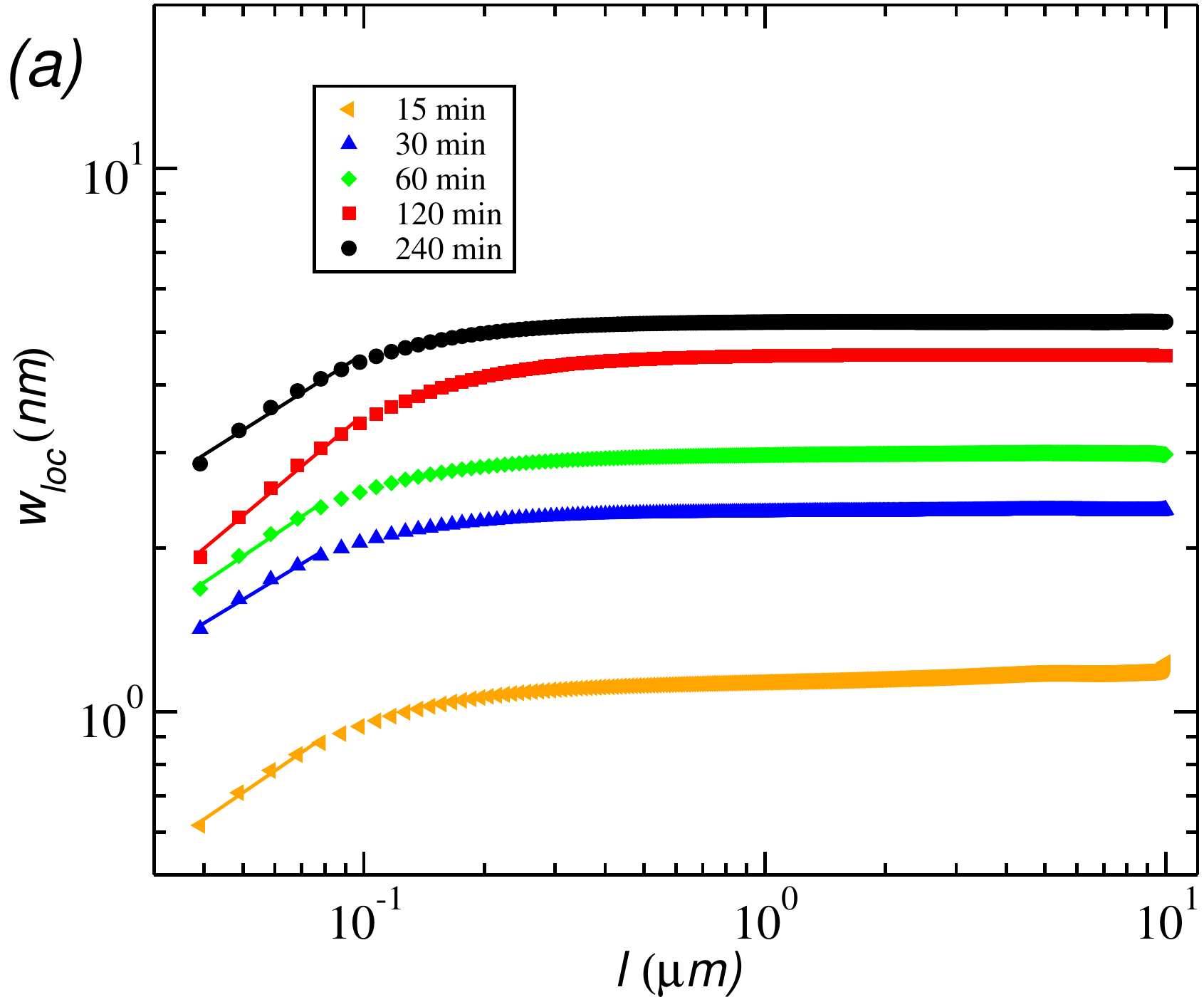}
    \includegraphics[width = 7.2 cm]{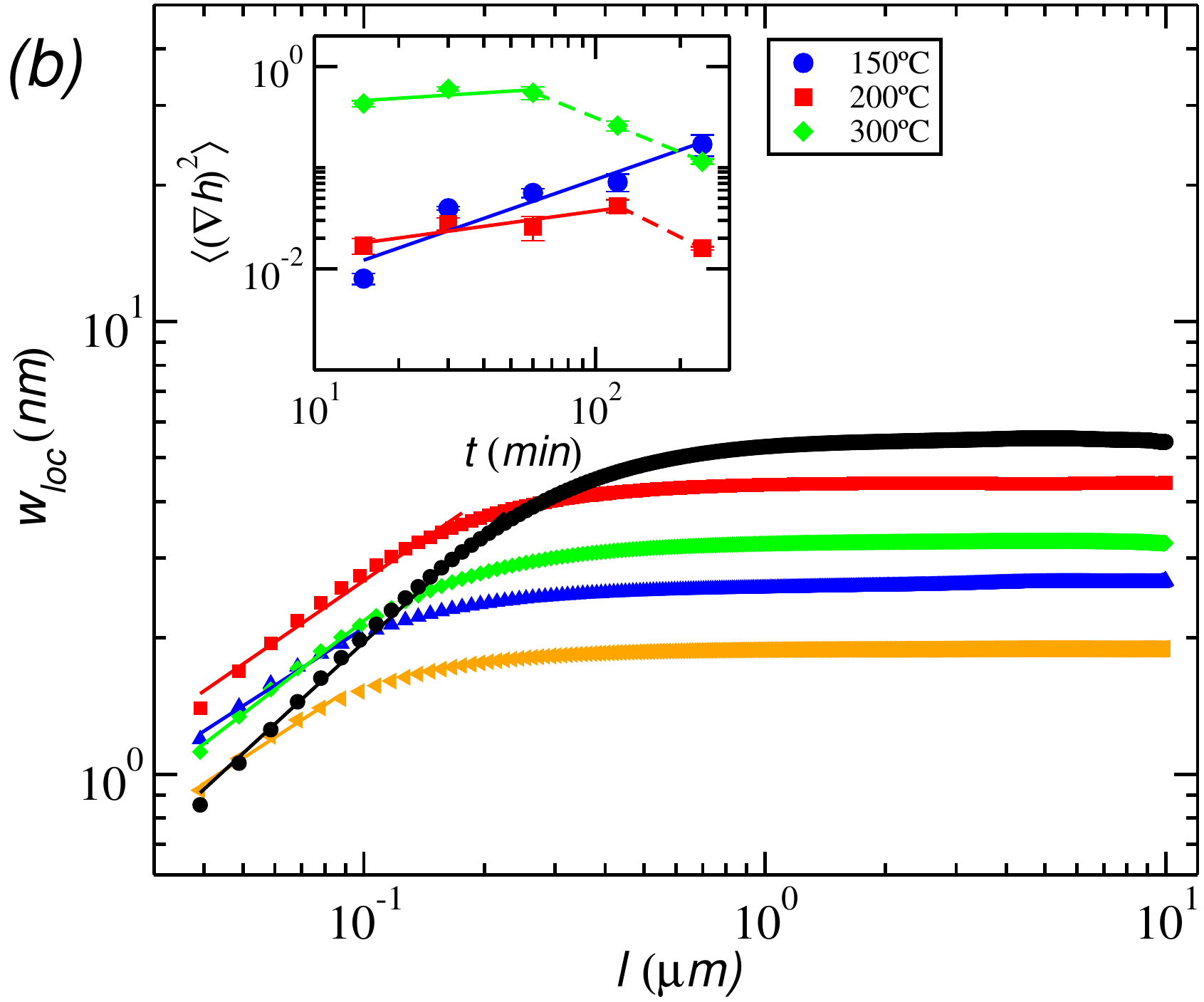}
    \includegraphics[width = 7.2 cm]{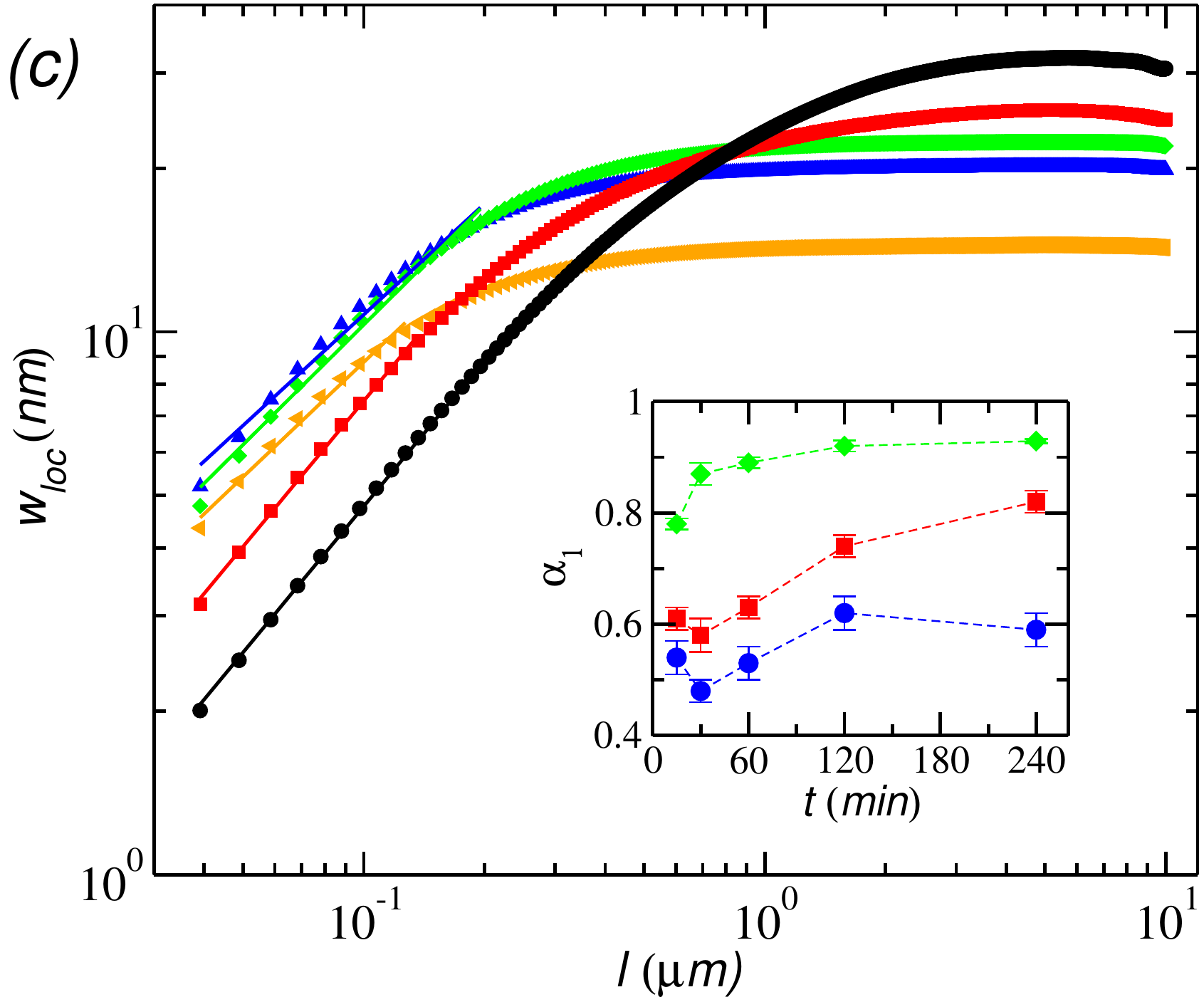}
    \caption[Local roughness for films grown at $T = 150\,^{\circ}\mathrm{C}$, $200\,^{\circ}\mathrm{C}$ and 
    $300\,^{\circ}\mathrm{C}$.]{Local roughness for films grown at (a) $T = 150\,^{\circ}\mathrm{C}$, 
    (b) $200\,^{\circ}\mathrm{C}$ and (c) $300\,^{\circ}\mathrm{C}$ by 15 min (left triangles), 30 min (up triangles), 
    60 min (diamonds), 120 min (squares) and 240 min (circles). Insertion of (b) shows the squared local inclinations 
    as function of the growth time for $T = 150\,^{\circ}\mathrm{C}$ (circles), $T = 200\,^{\circ}\mathrm{C}$ (squares), 
    $T = 300\,^{\circ}\mathrm{C}$ (diamonds). The same legend is used in the inset of the figure (c), which
    depicts the geometrical $\alpha_1$ exponent as function of $t$, extracted from the solid lines shown in the main plots.}
    \label{loc}
    \vspace{0.5 cm}
 \end{figure}

In regard to surface fluctuations at local scales, fig. \ref{loc} presents the local roughness 
calculated as function of $t$ and $T$. The first note to take look at is that, in the fig. \ref{loc}(a), the curves dislocate to 
up as the time evolves (in both short-, $l \lesssim 10^{-1}$ $\mu m$, and large-wavelengths $l \gg 10^{-1}$ $\mu m$) and, apparently, no one relaxation is observed. Indeed,
this $w_{loc}$ increasing at short-length scales is the landmark of the \textit{anomalous roughening}, discussed in the appendix section \ref{section_anomalous}, 
where, at first sight, a larger number of critical exponents must to be found for determining the UC. The $\kappa$ exponent, 
defined in the equations \ref{kappa3} and \ref{kappa2}, has been calculated for the films grown at $T = 150\,^{\circ}\mathrm{C}$ - see insertion of the fig. \ref{loc}(b). The value obtained 
is $\kappa = 0.5(1)$. Although this value is positive, indicating anomalous scaling, its particular value 
is consistent with that one obtained from an \textit{uncorrelated} interface, where the squared-local-slope fluctuations increase in time with unit power. In another words, the random growth 
is intrinsically anomalous in this sense. 

For surfaces grown at $T = 200\,^{\circ}\mathrm{C}$, the situation is more complex than that for the 
lowest $T$, as shown in the fig. \ref{loc}(b). There, one can see that the curve for $t = 240$ min has dislocated 
down at short-length scales, instead of dislocating up, as in the initial trend. Very interesting, this has been the 
same behavior presented by $\Omega$, as well as in the visual inspection of AFM images (fig. \ref{morph1}). The squared 
local slopes also show a behavior of two regimes (see inset of the fig. \ref{loc}(b)), where an 
\textit{initial anomalous regime},
characterized by $\kappa = 0.19(5)$, is followed by a second one in which the Family-Vicsek scaling 
is asymptotically recovered \cite{FVS, Lopez99} with $\kappa \approx -0.7$.

For the deposition temperature $T = 300\,^{\circ}\mathrm{C}$, the same qualitative result of two regimes
is observed - see fig. \ref{loc}(c). The transient anomalous scaling gives place 
to a ``normal'' one at $t \lesssim 60$ min, as corroborated by the local roughness at short-length scales and 
by $\langle(\nabla h)^2\rangle$ as function of $t$. The values for $\kappa$ characterizing these two regimes are, 
respectively, $0.2(2)$ and $-0.56(3)$. Notice that the first value can not guarantee 
the presence of \textit{anomaly}, since $\kappa = 0$ is a possible value within the error bar.

Now, regarding to the \textit{geometrical} scaling exponent $\alpha_1$, defined as $w_{loc} \sim l^{\alpha_1}$ 
(section \ref{section_grainmodel}), its value as function of the growth time is depicted in the inset of 
the fig. \ref{loc}(c) for different temperatures. $\alpha_1$ is close to $0.6$ for $150\,^{\circ}\mathrm{C}$, which indicates the presence of very sharp grains at surface at all times\cite{Tiago_grain1, Tiago_grain2}, according with the visual inspection of AFM images. In turn, for $T = 200\,^{\circ}\mathrm{C}$ and $300\,^{\circ}\mathrm{C}$, $\alpha_1$ changes from $0.6$ to $0.8$ and
from 0.80 to 0.95, respectively. This reveals that sharp grains give place to smoother structures at long times.
These results are totally consistent with the figures \ref{morph2}(a) and (b). A similar behavior for $\alpha_1$ occurs
for a fixed $t$, as $T$ increases. Based on these results, one can also conclude that
the behavior of $w_{loc}$ at short-length scales ($l \ll 10^{-1}$ $\mu m$) is governed by 
two contributions: the first is the grain shape, related to $\alpha_1$, which can present large (sharp grains)
or small (flat top grains) fluctuations. The second one coms from the aspect ratio, since for a fized box size $l^*$, one has
$w_{loc}(l^*,t) \sim \langle(\nabla h)^2\rangle$ \cite{Huo, Tiago_mestrado}.

The origin of the results above can be understood as follows: initially, CdTe films evolve
according to the Volmer-Weber growth mode \cite{SukarnoQD1, Lalev1, Lalev2}, and
their width becomes larger as higher is the deposition temperature \cite{Evans}
(see fig. \ref{morph1}). As the grains with well-defined shape enlarge laterally, they colide forming a continuous film
and giving origin to grains boundaries (GBs), where non-crystalline (defects) regions emerge \cite{Joelma, AmarCdTe}. These GBs fromed by colided neighboring grains are
formed mainly at the interface between (111) grains having different rotational orientations. The size of these defect 
regions is larger as higher is $T$ since large are the grain perimeters (remember that initally $\zeta$ increases with $T$) \cite{Evans}. At these GBs, hence, an additional energy barrier ($E_{GB}$) rises, which tends to repel the diffusion and the deposition of particles at, and near to, 
these sites \cite{Tello, Gonzales}. A similar barrier has also been suggested recently in the growth of CdTe/CdS films \cite{AmarCdTe}. For low $T$, where surface diffusion is slow, a small number of molecules 
can overcome the $E_{GB}$ barrier and most of them aggregates inside the grain that they have arrived. This compels 
the grain height to increase faster than its width leading $\Omega$ and $\langle(\nabla h)^2\rangle$ to increase 
as the time evolves (see figs. \ref{morph1}, \ref{loc} and inset of fig. \ref{loc}(b)). However, at long times, a 
relaxation happens because particles are eventually deposited at (and around) the GBs of colided neighboring grains. This diminishes the number of defect sites in the active zone and, consequently, enhances diffusion towards these regions.
From this moment, coalescence processes become more operative and grains with well-defined shape give place 
to large mounds, so that $\Omega$ and $\langle(\nabla h)^2\rangle$ decrease in time. In particular, 
for $T = 150\,^{\circ}\mathrm{C}$ this regime is not observed due to the experimental 
condition of \textit{finite-time} growth. For higher $T$, however, the second regime is clearly noticed and 
appears so earlier as higher is $T$, as can be checked in insets of the figs. \ref{loc}(b) and (c) and in 
the inset of fig. \ref{wloc250}(a).

\subsection{A Kinetic Monte Carlo Model}
The reliability of the above reasoning is illustrated in an one-dimensional atomistic growth model.
Since our interest is the coalescence process, the growth starts on a periodic array of pyramidal 
grains with the same width $\zeta$ and height $H$, for simplicity. The unit time is set to correspond to the deposition
of one monolayer of particles. During a deposition event, a site i is randomly 
selected and a particle is allowed diffusing at surface (starting from the site i) until finding a 
site $j$ where it aggregates permanently whether the constraint $|h_j - h_{j\pm 1}| \leq 1$ is satisfied. Thus, inside the characteristic ``grain'', aggregation follows the conservative RSOS (restricted solid-on-solid) rule \cite{Kim_CRSOS}. At the GBs of colided neighboring grains, however, an energy barrier $E_{GB}$ is present, so that a particle diffuses toward them with probability 
$P_{D}=e^{-E_{GB}/k_{B} T}$. Once a particle \textit{aggregates} at a given GB $i$, the barrier 
$E_{GB}$ at $i$ becomes null with probability $P_{R}=e^{-E_{R}/k_{B} T}$. Notice that this simple mechanism
captures the underlying feature of the relaxation process at the GBs of colided neighboring grains. 

 \begin{figure}[!h]
    \vspace{0.5 cm}
    \centering
    \includegraphics[width = 7.2 cm]{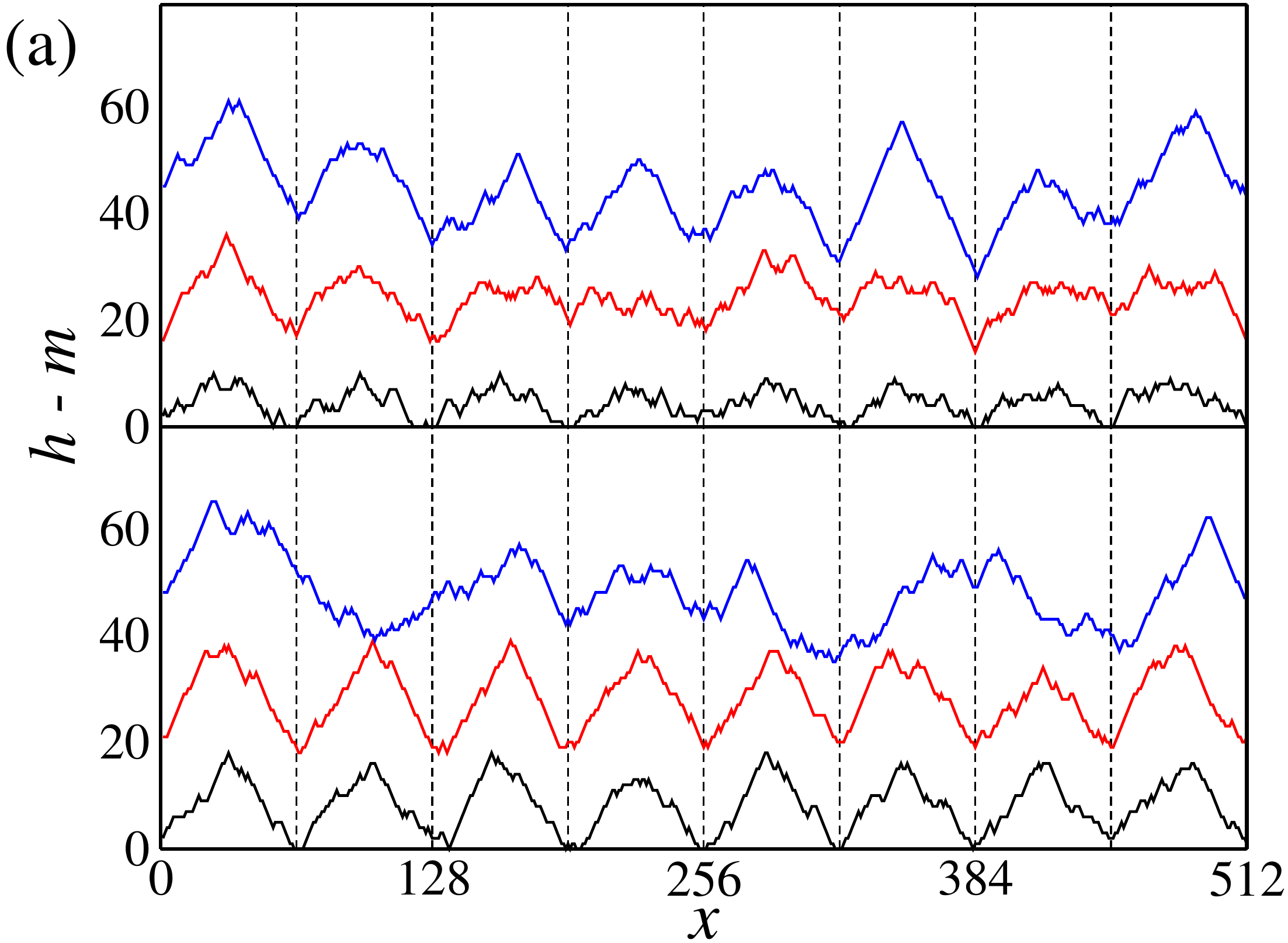}
    \includegraphics[width = 7.0 cm]{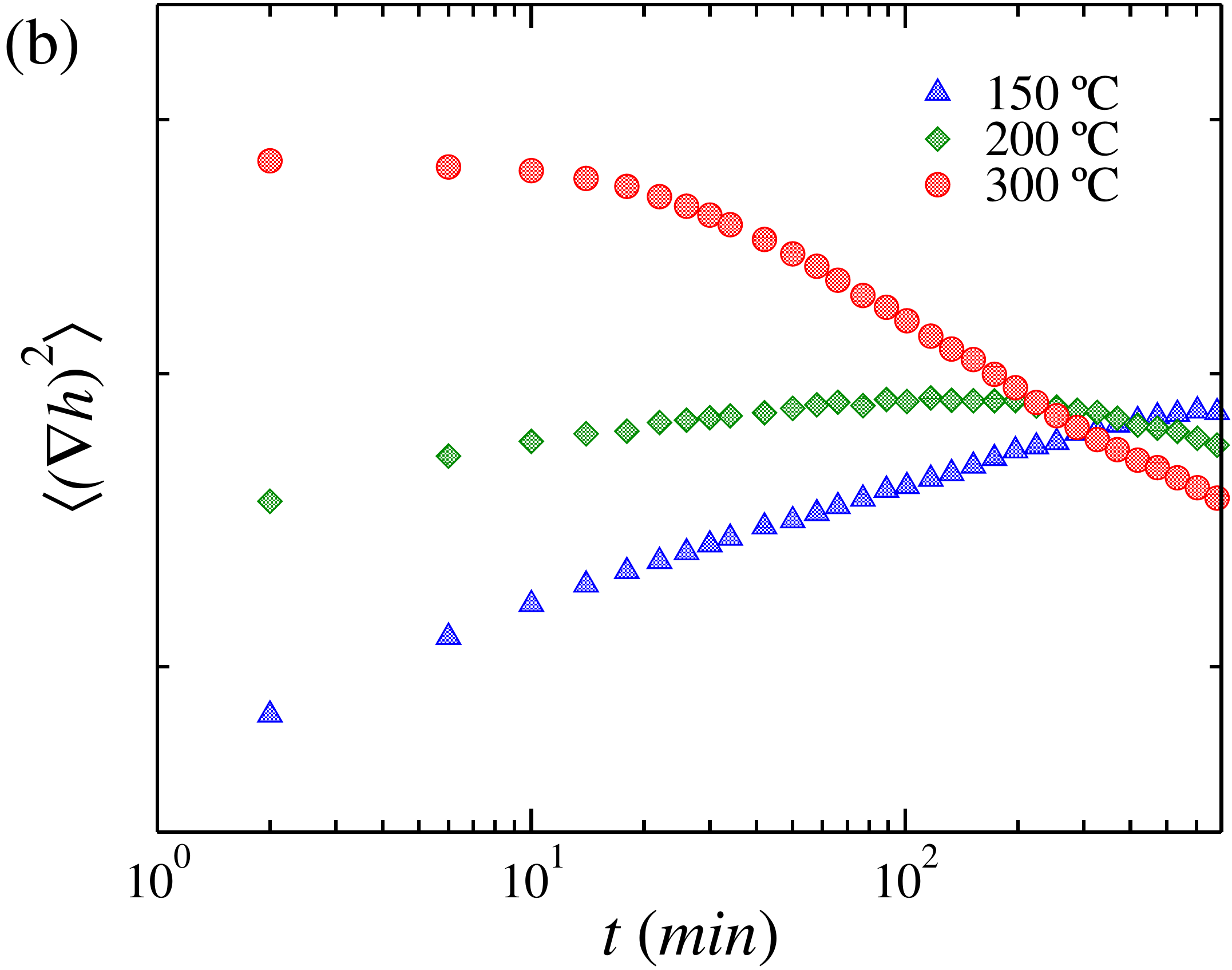}
    \caption[Results from the Kinetic Monte Carlo Model]{Results from an one-dimensional 
    Kinetic Monte Carlo model. In (a) one depicts the height profiles for $T=150\,^{\circ}\mathrm{C}$ (top)
    and $200\,^{\circ}\mathrm{C}$ (bottom) for $t = 10, 100$ and 1000, and shifted by $m = 10, 80$ and 960,
    respectively. Dashed lines represent the initial (t = 0) GB's of colided neighboring grains. (b) Average-squared-local slope $\langle(\nabla h)^2\rangle$ versus time for surfaces
    grown at $T=150\,^{\circ}\mathrm{C}$ (circles), $T=200\,^{\circ}\mathrm{C}$ (squares) and
    $T=300\,^{\circ}\mathrm{C}$ (diamonds).}
    \label{model}
    \vspace{0.5 cm}
 \end{figure}

Figure \ref{model}(a) shows typical surface evolutions for $T = 150$ and $200\,^{\circ}\mathrm{C}$. 
One has considered $E_{GB}=0.10$ eV, $E_{R}=0.30$ eV, $\zeta=64$ and $H=8$, 16 and 24 for $T=150$, 200 
and $300\,^{\circ}\mathrm{C}$, respectively. For $T=150\,^{\circ}\mathrm{C}$ one observes grains with almost 
fixed width and increasing height. A similar behavior is observed at short times for $T=200\,^{\circ}\mathrm{C}$ but, for 
large $t$, large mounds (formed by coalesced grains) appears. The same occurs for larger $T$. This qualitative agreement 
with the experiment is corroborated by the evolution of the squared local slopes displayed in 
the fig. \ref{model}(b). Comparing these results with the experimental ones (inset of fig. \ref{loc}(c)), one can confirm 
that the interplay between the relaxation process at the GBs and initial conditions (Volmer-Weber growth mode and width larger
for higher $T$), in fact, explains the CdTe/Si(001) mound evolution. Nevertheless, it is important mention that this
simple model does not reproduce (and do not have this intention) the complex dynamic taking place in the growth of CdTe films, but it gives valuable insights about underlying features which rule the coalescence process.

\section{Universal exponents}

Now, once we have got an explanation for what is going on at local scales when the deposition temperature is changed, we are prepared to turn our attention to the analysis of long-wavelength fluctuations and to the Universality Class (UC) of the growth. 

 \begin{figure}[!ht]
    \vspace{0.5 cm}
    \centering
    \includegraphics[width = 8.5 cm]{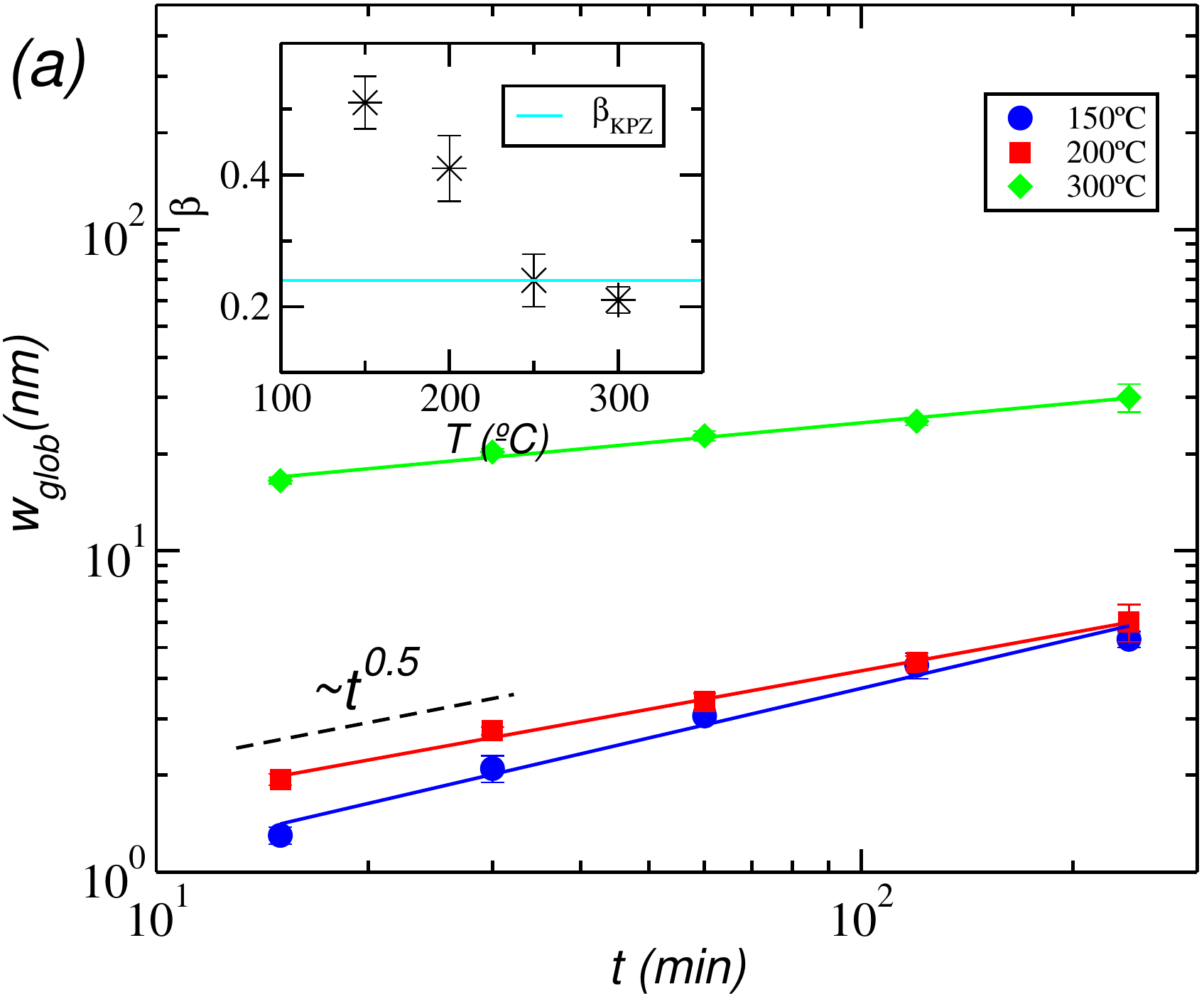}
    \includegraphics[width = 8.7 cm]{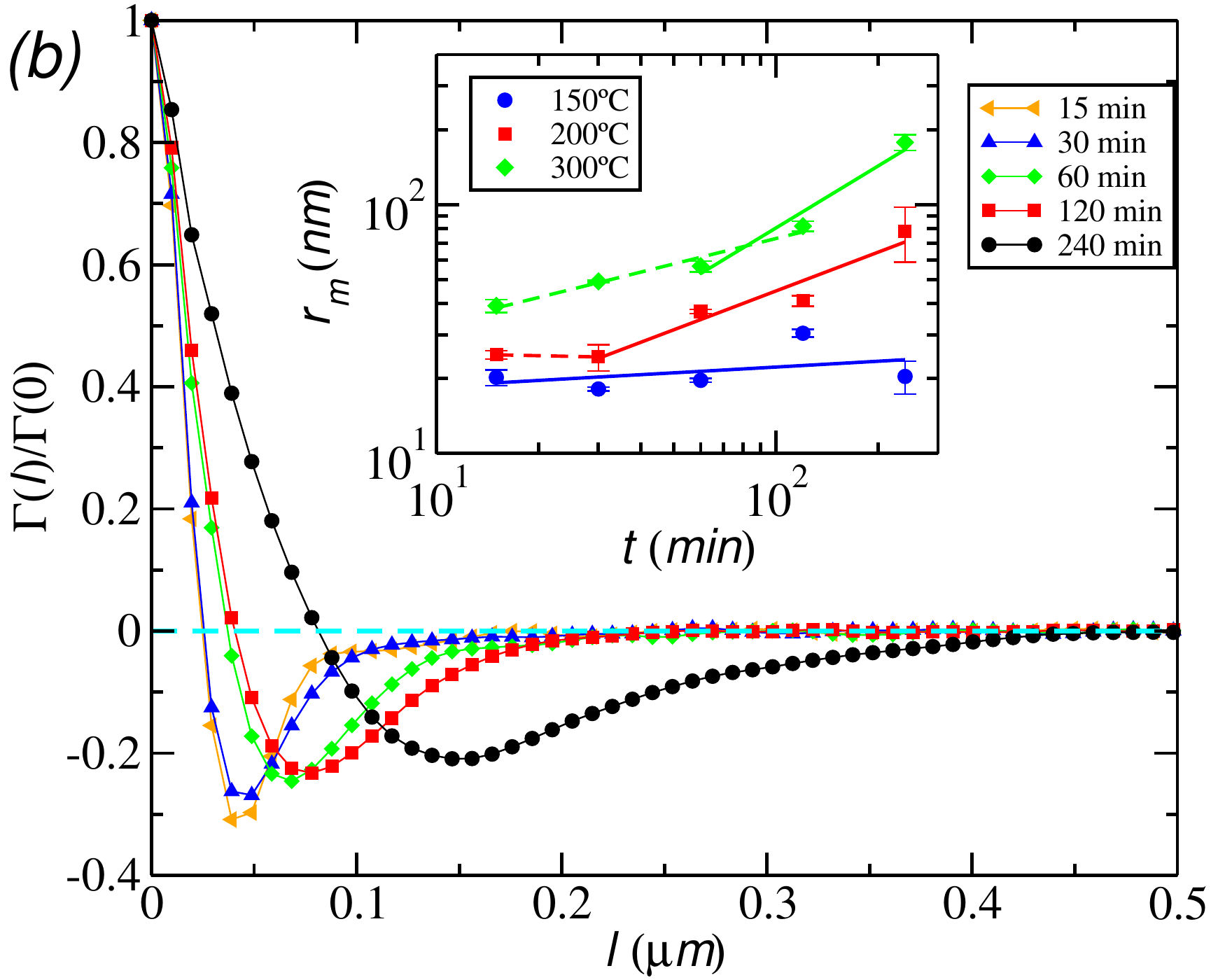}
    \caption[Global roughness curve and slope-slope covariance]{(a) Global 
    roughness as function of the growth time for different temperatures, namely, $T = 150\,^{\circ}\mathrm{C}$ (circles),
    $T = 200\,^{\circ}\mathrm{C}$ (squares) and $T = 300\,^{\circ}\mathrm{C}$ (diamonds). Solid lines are guide to 
    eyes, from which the $\beta$ exponent has been found. Inset of (a) shows how the growth exponent depends on $T$. 
    The solid line marks the value 0.239, which is consistent with $KPZ_{d = 2 + 1}$ \cite{Kelling, Healy_2d1, Healy_2d2}. (b) 
    Slope-Slope correlation function for CdTe thin films grown at $T = 200\,^{\circ}\mathrm{C}$. The oscillatory behavior
    is observed for all temperatures and the curve shown is a typical one, valid for all temperatures analyzed. 
    Insertion in (b) depicts the first zero ($r_m$) of the Slope-Slope correlation function as a function of 
    time for several $T's$. The legend is the same exhibited in the inset of (a). Dashed and solid lines 
    are guide to the eyes, from where the $n_{coar}$ and $1/z$ exponents have been determined, explained in the text.}
    \label{rug_slop}
    \vspace{0.5 cm}
 \end{figure}

Figure \ref{rug_slop}(a) shows the global roughness as function of the growth time for different temperatures. 
At first, one notices that $w(T)$ increases with $T$. In fact, as the CdTe layer
is polycrystalline, the complex competition between grains gives rise to inter-grain fluctuations that are
larger as higher is the temperature because: 1) the grains, themselves, are larger and 2) the valleys separating 
these structures are deeper as higher is $T$, as observed in the fig. \ref{morph1}. Similar results have been 
found in the growth of CdTe grown on glass substrates covered by fluorine doped with tin oxide \cite{Fabio, Sukarno}.

Regarding to the growth exponent, for $T = 150\,^{\circ}\mathrm{C}$, one obtains $\beta = 0.51(4)$ (see inset of fig. \ref{rug_slop}(a)), which in association with $\kappa = 0.5(1)$ give us a strong evidence that one has a Poissonian growth at this low temperature. For $T = 200\,^{\circ}\mathrm{C}$, however, one can
notice that for initial times ($t \lesssim 60$ min), global roughness increases with $\beta_{initial} \approx 0.5$, but
this regime is followed by a distinct one where, clearly, one finds a tendency of $\beta$ becoming 
smaller - \textbf{compare $\textrm{w(T)}$ for $\textrm{T = 150}\,^{\circ}\mathrm{C}$ and $\textrm{200}\,^{\circ}\mathrm{C}$}. Despite this fact,
one finds $\beta = 0.41(5)$ for $T = 200\,^{\circ}\mathrm{C}$. It is worth mention that this value does not match with \textit{anyone} known UC \cite{Barabasi, Healy-Review}. In particular this value has been found in homoepitaxial growth of Cu(001) at 250K \cite{Botez}, of Ag(111) and Ag(001) at 300 K \cite{Elliott}, and also in the growth of CdTe on glass 
substrates by HWE at $T = 250\,^{\circ}\mathrm{C}$ \cite{Sukarno, Fabio}.

Finally, for films grown at $T = 300\,^{\circ}\mathrm{C}$, experimental results point out $\beta = 0.21(5)$, a value 
close to the KPZ one (eq. \ref{kpz_exp}) \cite{Kelling, Healy_2d1, Healy_2d2} and also to the MH and VLDS classes - see section \ref{section_EWMH} and appendix section \ref{section_VLDS}, respectively. Indeed, diffusion should dominate the growth at very higher $T$, overcoming other mechanisms. However, with only this evidence we can not determine which growth equation 
describes the surface fluctuations of the CdTe interface.

\vspace{1cm}

Figure \ref{rug_slop}(b) shows the slope-slope correlation function (eq. \ref{slop_CF}) for CdTe thin films grown 
at $T = 200\,^{\circ}\mathrm{C}$. The same oscillatory behavior is found for samples grown at $T = 150\,^{\circ}\mathrm{C}$ and 
$300\,^{\circ}\mathrm{C}$ (not shown here). As discussed in the section \ref{section_CF}, from the first minimum 
or from first zero of $\Gamma(l,t)$, one can extract the average mound size ($r_m$), which scales
as $r_m \sim t^{n_{coar}}$ \cite{Fabio, Almeida, Cordoba, Siniscalco}. In an appropriate range of time,
one has $n_{coar} = 1/z$. The insertion in the fig. \ref{rug_slop}(b) shows the first zero ($r_m$) as function of 
the growth time, for different temperatures. For $T = 150\,^{\circ}\mathrm{C}$, multi-peaked structures do not 
appear at the surface and the equality $n_{coar} = 1/z$ must be valid for all $t$. One finds $1/z \approx 0.07$, which gives
one more prove that the growth is uncorrelated at this $T$ \cite{Barabasi}. For $T = 200\,^{\circ}\mathrm{C}$, two regimes can be reasonably
seen, they are indentified as dashed and solid lines in the inset of fig. \ref{rug_slop}(b). From the first regime, one 
has $1/z \approx -0.02$ and from the second one obtains $1/z = 0.6(1)$. The first value is consistent with random growth \cite{Barabasi}, whilst the second one agree with $KPZ_{d = 2 + 1}$ within the error bar (eq. \ref{kpz_exp}) \cite{Kelling, Healy_2d1, Healy_2d2}. Finally, for $T =300\,^{\circ}\mathrm{C}$, two short and long-time regimes are also 
observed and provide $n_{coar} = 0.32(5)$ and $n_{coar} = 0.7(1)$, respectively. The first regime 
is consistent with the 1/z value for the VLDS class \cite{Barabasi} and the second is near of the 
expected 1/z value for the KPZ class. Both values agree with numerical ones whithin the error bar. Nevertheless, at 
this temperature we must be careful in assume $n_{coar} = 1/z$, once multi-peaked structures appear since initial growth times. For times $t \lesssim 60$ min, local peaks on the top of the mounds should contribute to underestimate the 
average size of the structure. However, at large growth times, local peaks become shorter compared with the basis of very large
mounds at the surface (see fig. \ref{morph2}(b)) and, hence, they should not strongly influence in the measurement of $r_m$. Thus, the most careful action should be set $n_{coar} = 1/z$ only for \textit{long growth times}, at least 
for films grown at high deposition temperatures, as $T = 300\,^{\circ}\mathrm{C}$.

\section{Partial considerations}
At this point, we have performed a scaling analysis of CdTe surface fluctuations based also in local features
of the growth. From local roughness curves we detect transient anomalous scaling and crossover effects during
the dynamic, which have been related to the emergence of an energy barrier at the GB's of colided neighboring grains 
and to the relaxation process. A Kinetic Monte Carlo model has supported our reasonings, reproducing qualitatively the 
experimental results. From the local roughness curves it was not possible to unearth
the $\alpha$ exponent because the second (universal) regime has not been observed. However, the values found for
$\alpha_1$ are consistent with the predictions in the refs. \cite{Tiago_grain1, Tiago_grain2} as
corroborated by the AFM images - see table \ref{table_alpha}. In the sequence, one has 
found $\beta$ and $1/z$ as function of the deposition temperature. For $T = 150\,^{\circ}\mathrm{C}$, all 
results indicate a \textit{Poisson growth}, while for $T = 200\,^{\circ}\mathrm{C}$ a crossover 
from uncorrelated-to-KPZ growth seems to occur. The value $\beta = 0.41(5)$, which does not match with any 
known UC reinforces this idea. For the highest temperature studied, one has $\beta = 0.21(5)$ and $1/z = 0.7(1)$. The $\beta$ value is close to the MH, VLDS and KPZ classes \cite{Barabasi} and prevents a clear distinction of which growth 
equation describes the CdTe surface fluctuations at this temperature. However, the value found for $1/z$ at long growth times points to a KPZ growth, even at this very high deposition temperature. Moreover, we stress that based only on
results coming from exponents, one can not make clear what UC the fluctuations of CdTe surfaces (at T = $200\,^{\circ}\mathrm{C}$ and $300\,^{\circ}\mathrm{C}$) belong.

The values of all exponents found until here can be seen in the tables \ref{table_alpha}, \ref{table_kn} and \ref{tab_bz}.

\begin{table}[!h]
\centering
\setlength{\tabcolsep}{15pt}
    \begin{tabular}{c c c c c c }
	\hline \hline
	t (min) & 15 & 30 & 60 & 120 & 240\\
	\hline
	$\alpha_1 (T = 150\,^{\circ}\mathrm{C})$ & 0.54(3) & 0.48(2) & 0.53(3) & 0.62(3) & 0.59(3) \\
	$\alpha_1 (T = 200\,^{\circ}\mathrm{C})$ & 0.61(2) & 0.58(3) & 0.63(2) & 0.74(2) & 0.82(2) \\
	$\alpha_1 (T = 250\,^{\circ}\mathrm{C})$ & 0.64(2) & 0.76(1) & 0.84(1) & 0.89(1) & 0.90(1) \\
	$\alpha_1 (T = 300\,^{\circ}\mathrm{C})$ & 0.78(1) & 0.87(2) & 0.89(2) & 0.92(1) & 0.93(4) \\
	\hline\hline
    \end{tabular}
    \caption[Values for the non-universal exponent $\alpha_1(t,T)$]{\begin{small} Values for the non-universal exponent $\alpha_1(t,T)$
    coming from CdTe surfaces grown on Si(001) substrates by HWE with $F \approx 2.2$ \AA{}/s and deposition temperatures
    at $150\,^{\circ}\mathrm{C}$, $200\,^{\circ}\mathrm{C}$, $250\,^{\circ}\mathrm{C}$ and $300\,^{\circ}\mathrm{C}$.
    \end{small}}
    \label{table_alpha}
\end{table}


\begin{table}[!h]
\centering
\setlength{\tabcolsep}{15pt}
    \begin{tabular}{c c c c c c }
	\hline \hline
	$T (\,^{\circ}\mathrm{C})$ & 150 & 200 & 250 & 300 \\
	\hline
	$\kappa$ & 0.5(1) & 0.19(5)/ $\approx -0.7$ & -0.15(5)/ $\approx -0.05$ & 0.2(2)/-0.56(3) \\
	$n_{coar}$ & $\approx 0.07$ & -0.02(5)/0.6(1) & 0.62(2)/0.35(5) & 0.32(5)/0.7(1) \\
	\hline\hline
    \end{tabular}
    \caption[Values for the non-universal exponents $\kappa(t,T)$ and $n_{coar}(t,T)$]{\begin{small} 
    Values for the non-universal exponents $\kappa(t,T)$ and $n_{coar}(t,T)$ coming from CdTe surfaces grown on Si(001) 
    substrates by HWE with $F \approx 2.2$ \AA{}/s and deposition temperatures at $150\,^{\circ}\mathrm{C}$, 
    $200\,^{\circ}\mathrm{C}$, $250\,^{\circ}\mathrm{C}$ and $300\,^{\circ}\mathrm{C}$. Values separated by the symbol ($/$) 
    are valid in a appropriate range of time. These ranges of time are indicated as solid/dashed lines in the insertion of
    fig. \ref{loc}(b) for $T = 150$, 200 and $300\,^{\circ}\mathrm{C}$ and as dashed/solid lines in the inset of fig. \ref{wloc250}(a) for $T = 250\,^{\circ}\mathrm{C}$.
    \end{small}}
    \label{table_kn}
\end{table}

\begin{table}[!ht]
\centering
\setlength{\tabcolsep}{15pt}
    \begin{tabular}{c c c c c c }
	\hline \hline
	$T (\,^{\circ}\mathrm{C})$ & 150 & 200 & 250 & 300\\
	\hline
	$\beta$ & 0.51(4) & 0.41(5) & 0.24(4) & 0.21(5) \\
	$1/z$ & $\approx 0.07$ & 0.6(1) & 0.62(2) & $0.7(1)$\\
	\hline\hline
    \end{tabular}
    \caption[Values for the universal exponents $\beta(T)$ and $1/z(T)$]{\begin{small} Values for 
    the universal exponents $\beta(T)$ and $1/z(T)$ for surface fluctuations of CdTe 
    grown on Si(001) substrates by HWE with $F \approx 2.2$ \AA{}/s within the range of $15$ to $240$ min 
    and for several deposition temperatures. \end{small}}
    \label{tab_bz}
\end{table}

\section{Universal Distributions}

In this section, we supplement our previous studies with a deep analysis of height, squared roughness local and maximal
relative height distribuitons. As we shall demonstrate, this analysis, rather than a complementary one is, in some cases, \textbf{essential} for unveiling the Universality Class of a given growth.


\subsection{CdTe surface fluctuations at $T = 150\,^{\circ}\mathrm{C}$: Poissonian Growth}

\begin{figure}[!ht]
    \vspace{0.5 cm}
    \centering
    \includegraphics[width = 12.0 cm]{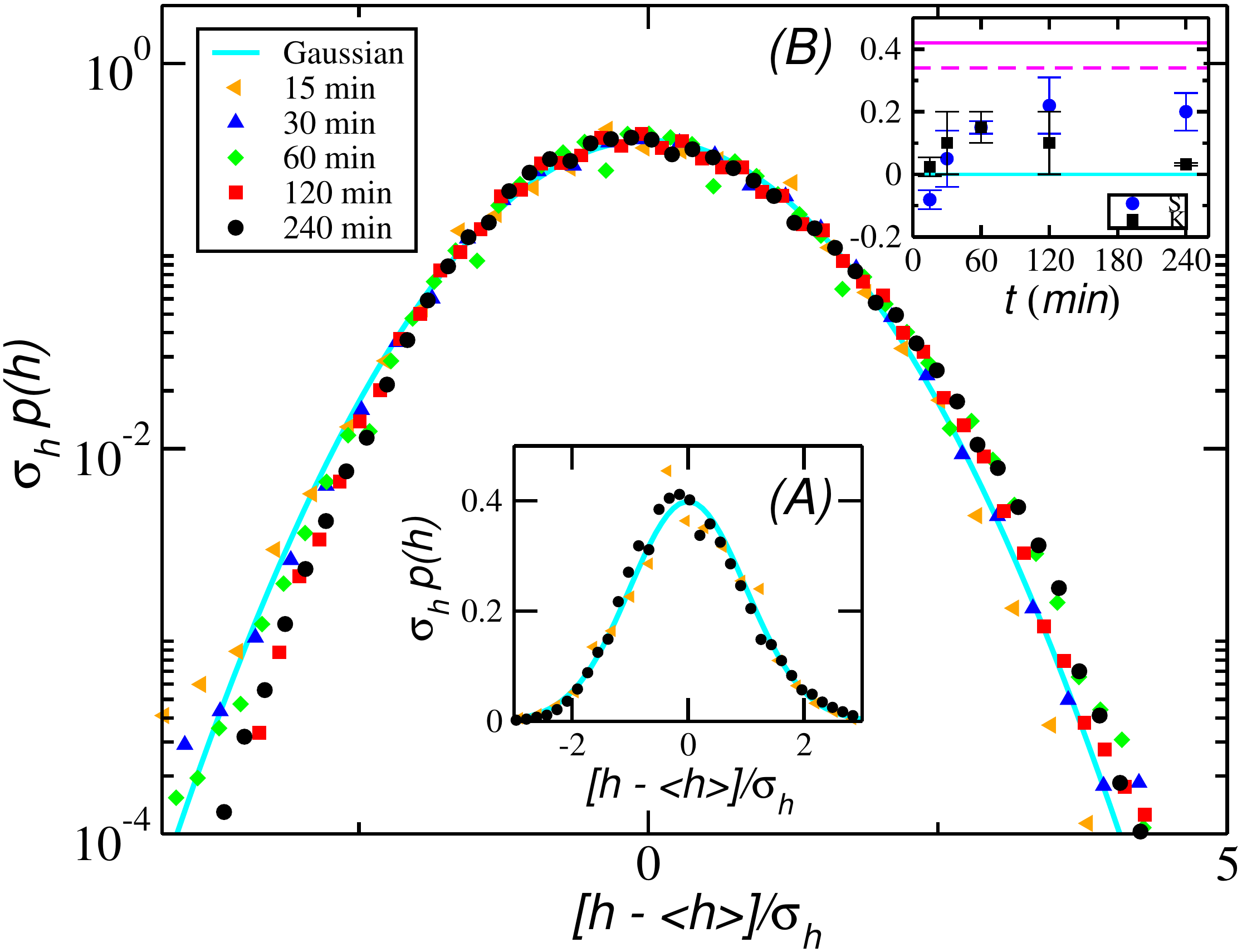}
    \caption[Rescaled HDs for surfaces grown at $T = 150\,^{\circ}\mathrm{C}$]
    {Rescaled height distributions at mean null and unitary variance for different
    growth times (symbols) from CdTe surfaces grown at $T = 150\,^{\circ}\mathrm{C}$. The experimental
    data are compared to the Gaussian (solid cyan line). Inset (A) highlights the very good collapse 
    between experimental data and the Gaussian close to the peak in a linear $\times$ log plot. 
    Insertion (B) exhibits the experimental S and K evolving in time. Solid/dashed cyan and magent lines 
    refers to the S/K values for the Gaussian-HD and KPZ-HD, respectively.}
    \label{HD150}
    \vspace{0.5 cm}
 \end{figure}

Figure \ref{HD150} shows the rescaled height distributions for different growth times for CdTe surfaces
grown at $T = 150\,^{\circ}\mathrm{C}$. The HDs are compared to the Gaussian distribution, expected for uncorrelated growing
surfaces. One notices that experimental data collapse very well with the HD-Gaussian near of the 
peak (inset (A) of the fig. \ref{HD150}) as well as at the tails - in at least four decades around the peak.

Experimental values for the Skewness ($S$) and Kurtosis ($K$) are displayed in the insertion (B) of 
the fig. \ref{HD150}, where their values seem to fluctuate around zero, as expected from interfaces having a Gaussian-HD.
Adding to this result those coming from the scaling roughness analysis (see table \ref{tab_bz}) as well as those
from the local dynamic, \textbf{one can conclude that, at this low temperature, height-field fluctuations of CdTe films are
\textit{Poissonian}} \cite{Evans, Barabasi}.

 \begin{figure}[!h]
    \vspace{0.5 cm}
    \centering
    \includegraphics[width = 7.2 cm]{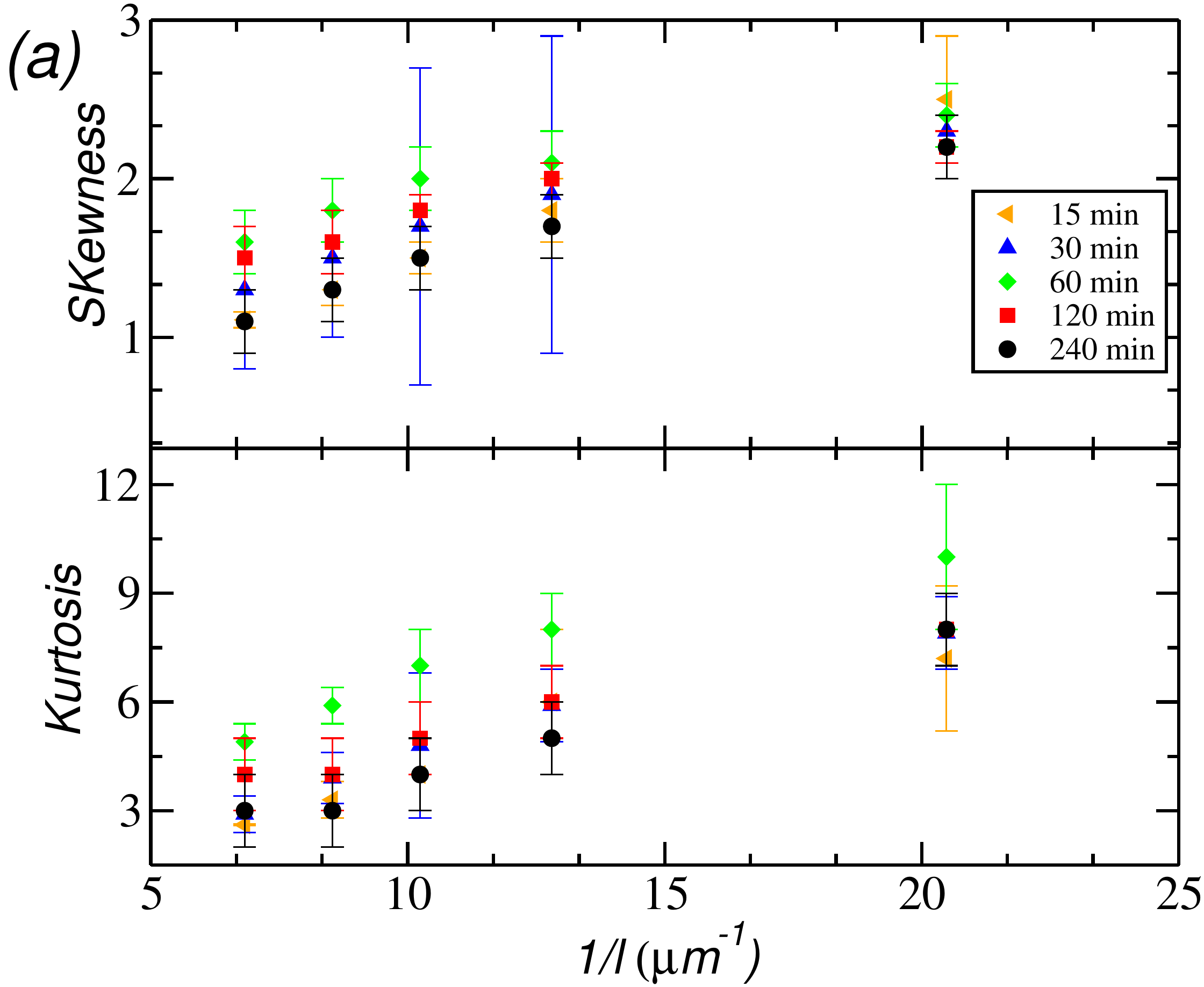}
    \includegraphics[width = 7.2 cm]{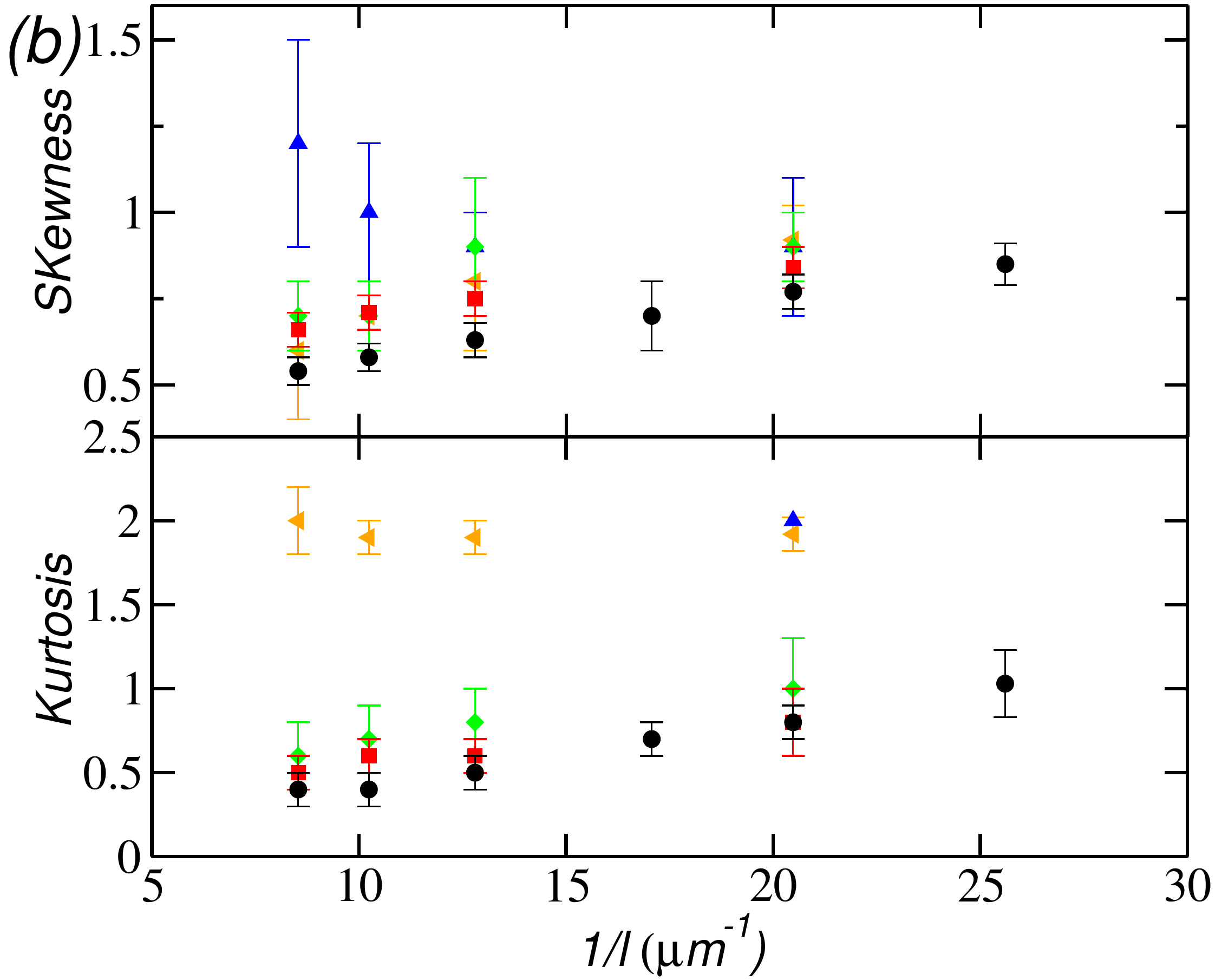}
    \caption[$S$ and $K$ values as function of the box size from SLRDs and MRHDs]{$S$ and $K$ 
    values for (a) SLRDs and (b) MRHDs from CdTe surfaces grown at $T = 150\,^{\circ}\mathrm{C}$ by 
    15 min (orange left triangles), 30 min (blue up triangles), 60 min (green diamonds), 120 min (red squares) 
    and 240 min (black circles). No one convergence for S(l) and K(l) curves are observed, as expected for 
    uncorrelated growing surfaces.}
    \label{SRLD150}
    \vspace{0.5 cm}
 \end{figure}

The $S$ and $K$ values as function of the box size ($l$) for SLRDs and MRHDs also go in favor to a uncorrelated growth - see figs. \ref{SRLD150}(a) and (b). As discussed in the sections
\ref{section_SLRD} and \ref{section_MRHD}, these ``box'' distributions must be evaluated in the range of 
$\zeta \ll l \ll \xi$. When $l$ satisfies this limit, $S$ and $K$ values become independent of $l$ and converge to 
universal values \cite{Foltin, PlischkeWD, Raychaudhuri, Majumdar}. However, when the surface is 
uncorrelated, $\xi$ keeps very small being of the same order of $\zeta$ (when there are grains at the surface). Thus, no 
one convergence of $S$ and $K$ should be observed, as can be confirmed in the figures \ref{SRLD150}(a) and \ref{SRLD150}(b).

\subsection{Random-to-KPZ Crossover in CdTe Surface Fluctuations at $T = 200\,^{\circ}\mathrm{C}$}
 \begin{figure}[ht]
    \vspace{0.5 cm}
    \centering
    \includegraphics[width = 12.0 cm]{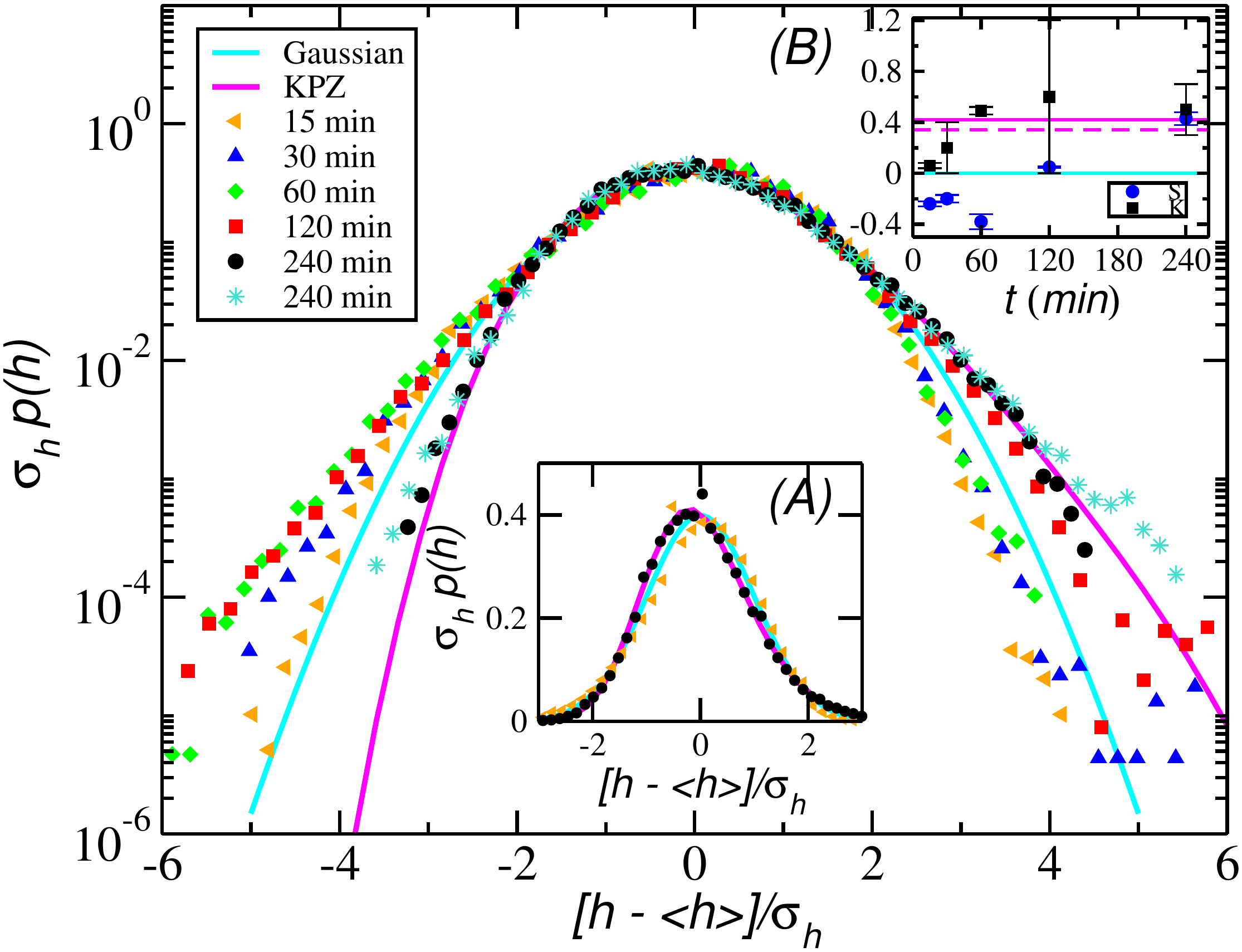}
    \caption[Rescaled HDs for surfaces grown at $T = 200\,^{\circ}\mathrm{C}$]
    {Rescaled height distributions, at mean null and unitary variance, for different
    growth times (symbols) from surfaces grown at $T = 200\,^{\circ}\mathrm{C}$. The experimental
    data are compared to the Gaussian (solid cyan line) and to the numerical KPZ curve (solid magenta line). 
    Inset (A) highlights the very good collapse between experimental and numerical data close to the peak in 
    a linear $\times$ log plot. Insertion (B) exhibits the experimental $S$ and $K$ evolving in time. Solid/dashed 
    cyan lines refers to the Gaussian S/K values, whereas the magenta line indicates the KPZ $S/K$ values.}
    \label{HD200}
    \vspace{0.5 cm}
 \end{figure}

HDs for CdTe films grown at $T = 200\,^{\circ}\mathrm{C}$ can be seen in the fig. \ref{HD200}. Very interesting, one notices
that the distributions at initial growth times are very far from the KPZ one. Instead, they are closer to
a Gaussian. This is confirmed by $S(t)$ and $K(t)$, which are closer to zero than to the KPZ values - see inset B of the fig. \ref{HD200}. At long times, however, $S$ and $K$ values seem to converge to the KPZ ones and a very good agreement with the universal KPZ-HD is observed. This disagreement/agreement with a KPZ-HD is reinforced taking into account the collapse near of the peak (see inset A of the fig. \ref{HD200}) for surfaces grown at $t = 15$ min and $t = 240$ min, respectively.

Despite of large error bars related to S and K values, which stem from low statistics, the agreement between the longest
growth time available and the expected KPZ values (see table \ref{universalkpz}) is remarkable, namely 
S$_{t = 240} = 0.43(5)$ and K$_{t = 240} = 0.5(2)$, whereas KPZ values are $S = 0.42(1)$ and $K = 0.34(2)$.
Based on our previous local and roughening scaling study, we can argue that \textbf{the HDs above show a crossover in time 
toward the KPZ regime}. In particular, this kind of crossover is widely studied numerically 
(see \cite{Juvenil, Alice1, Alice2} and ref. therein) and this experiment, as far as we know, is the first and
clean experimental evidence of such random-to-KPZ growth in $d = 2 + 1$.
  \begin{figure}[!h]
    \vspace{0.5 cm}
    \centering
    \includegraphics[width = 9.5 cm]{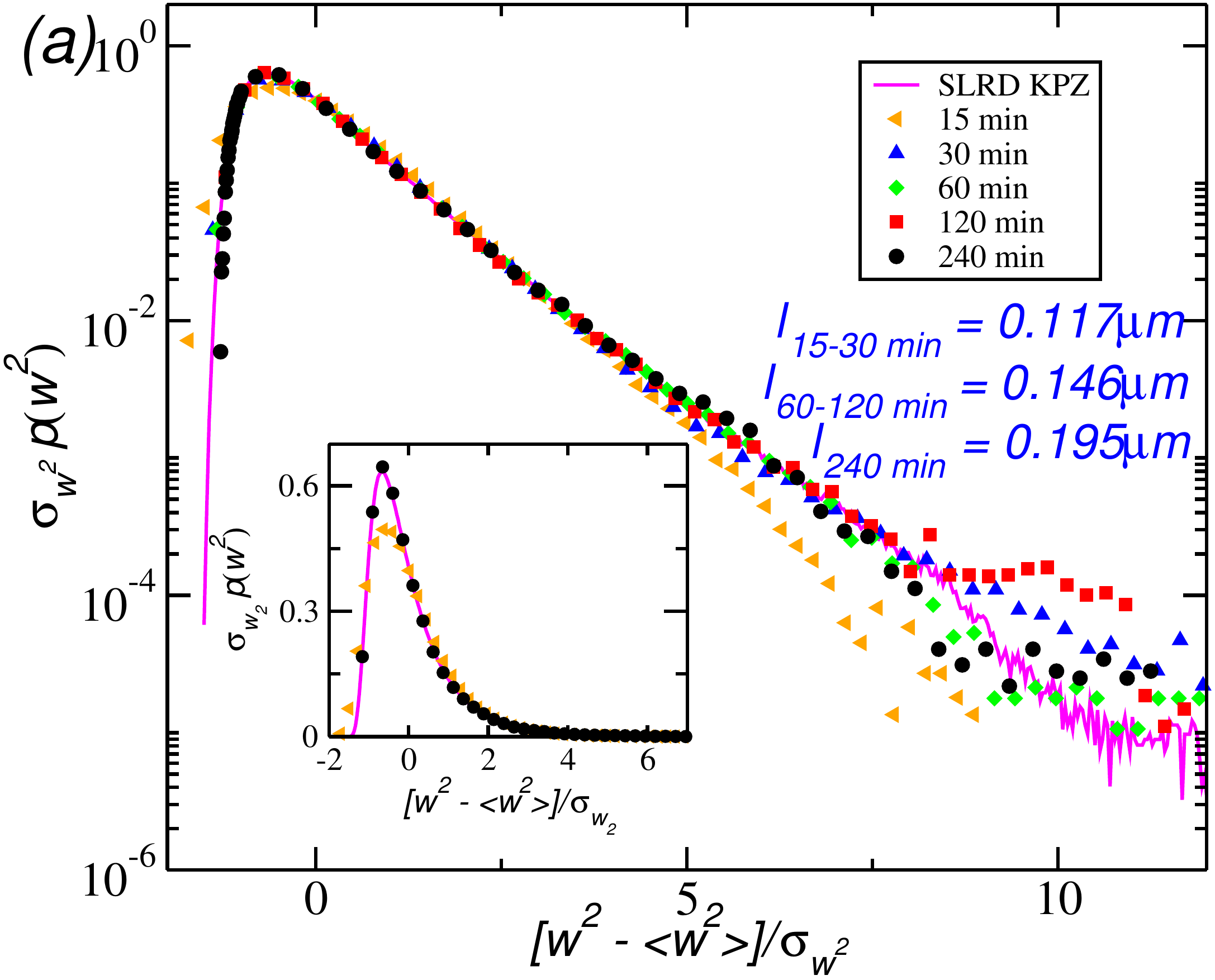}
    \includegraphics[width = 9.5 cm]{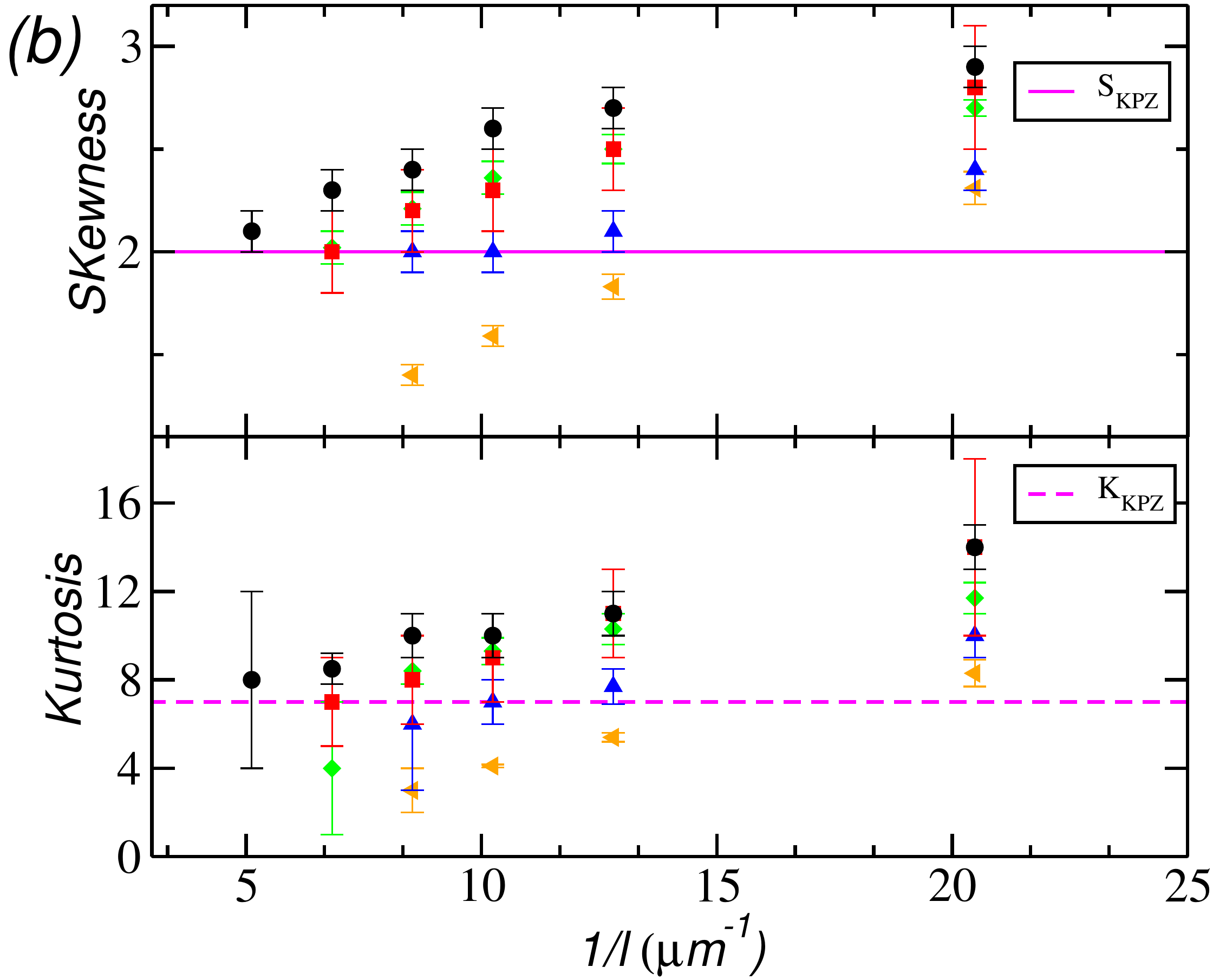}
    \caption[Rescaled SLRDs for samples grown at $T = 200\,^{\circ}\mathrm{C}$]
    {(a) Rescaled squared local roughness distributions for samples grown at 
    $T = 200\,^{\circ}\mathrm{C}$, by 15 min (orange left triangles), 30 min (blue up triangles),
    60 min (green diamonds), 120 min (red squares) and 240 min (black circles). The curves shown 
    are those whose the box size is indicated below the legend. They are different for each $t$ because 
    the appropriate interval $\zeta \ll l \ll \xi$ depends on $t$. The inset shows the poor/nice
    collapse around the peak for the thinnest/thickest film grown. (b) Skewness and Kurtosis values as 
    function of the box size $l$.}
    \label{SLRD200}
    \vspace{0.5 cm}
 \end{figure}

  \begin{figure}[!ht]
    \vspace{0.5 cm}
    \centering
    \includegraphics[width = 9.5 cm]{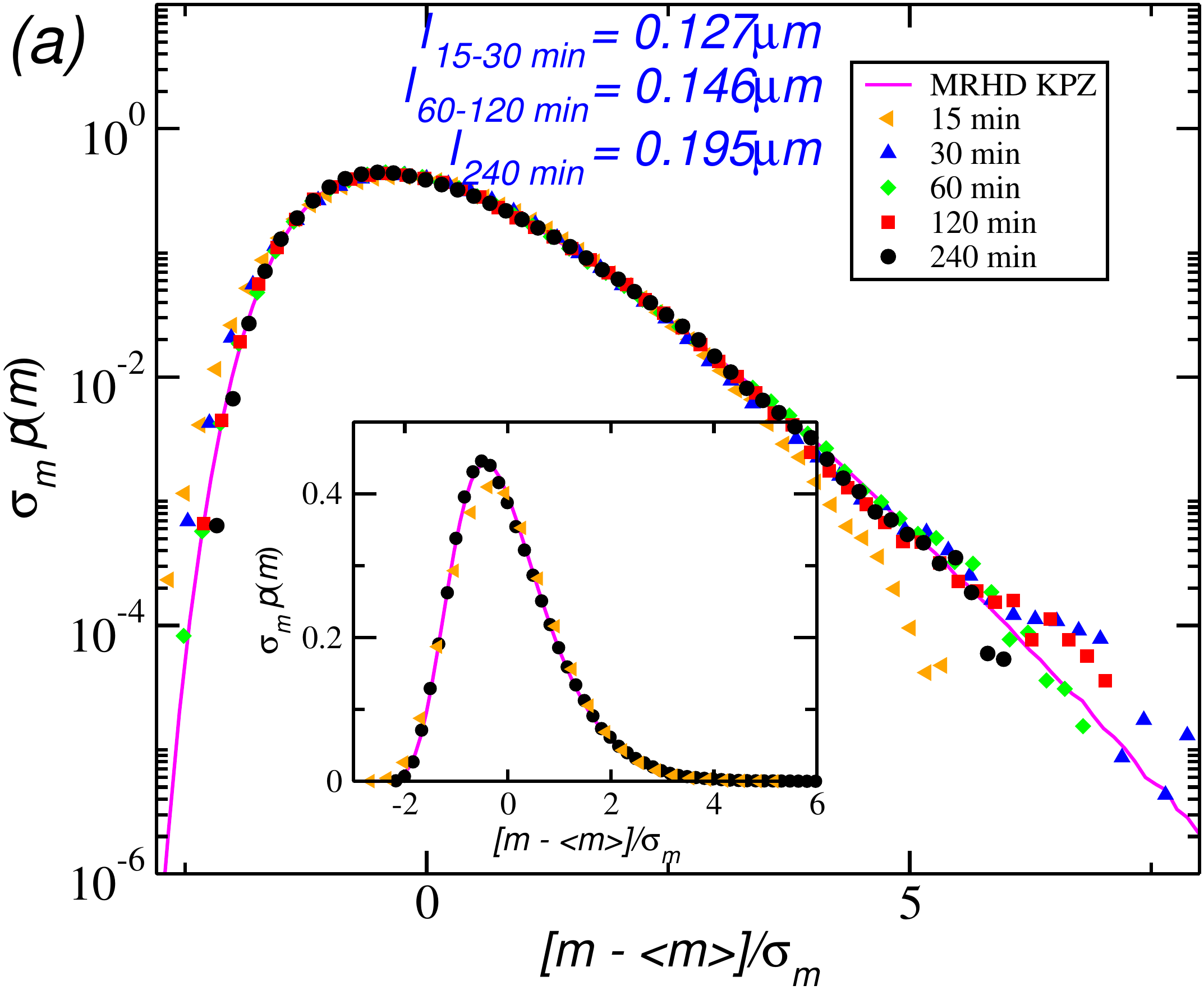}
    \includegraphics[width = 9.5 cm]{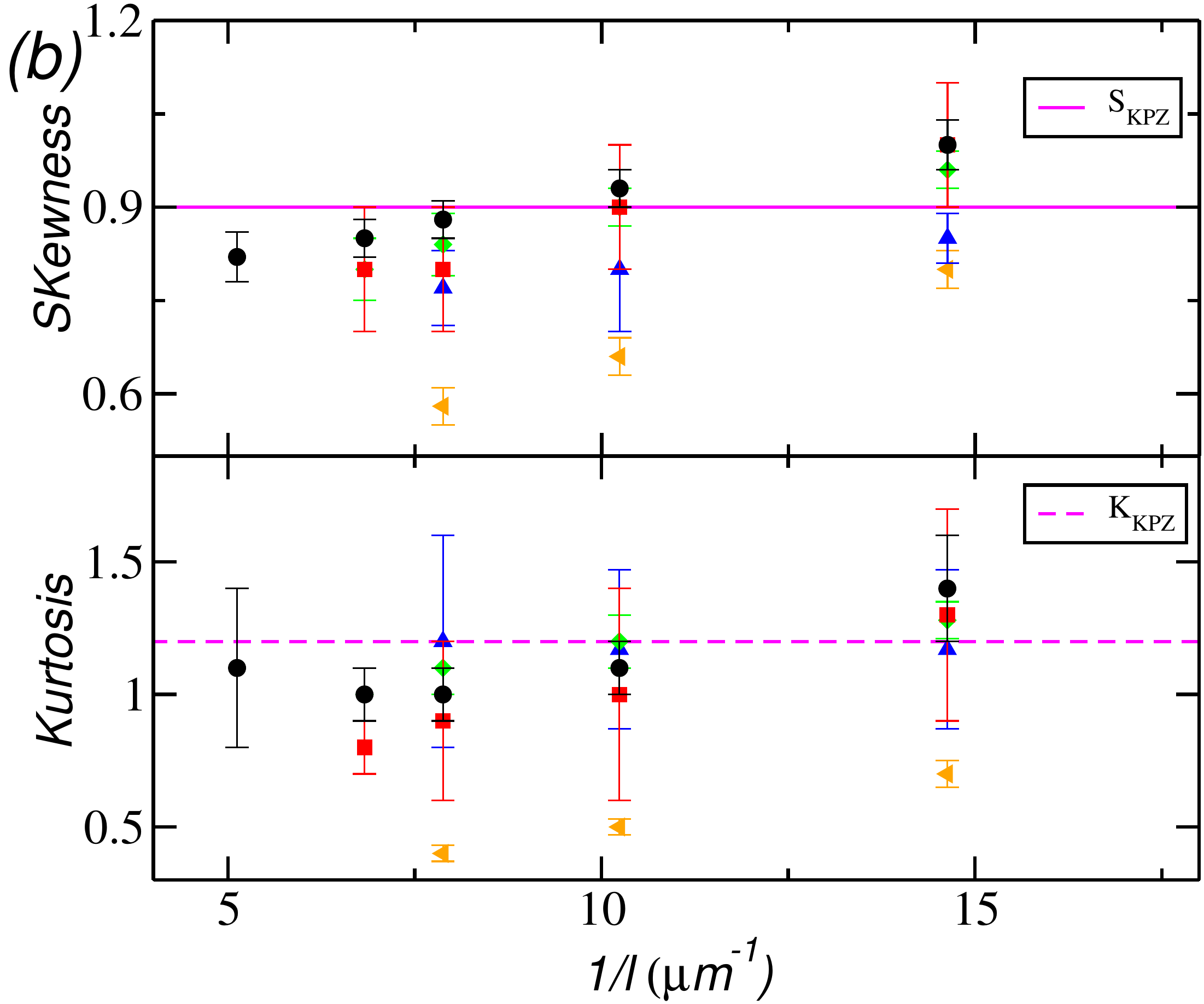}
    \caption[Rescaled MRHDs for samples grown at $T = 200\,^{\circ}\mathrm{C}$]
    {(a) Rescaled maximal relative height distributions for samples grown at 
    $T = 200\,^{\circ}\mathrm{C}$, by 15 min (orange left triangles), 30 min (blue up triangles),
    60 min (green diamonds), 120 min (red squares) and 240 min (black circles). The curves shown 
    are those whose the box size is indicated next to the legend. They are different for each $t$ 
    because the appropriate interval $\zeta \ll l \ll \xi$ depends on $t$. The inset shows the good
    collapse around the peak for the thinnest and the thickest film grown. (b) Skewness and Kurtosis values as 
    function of the box size $l$.}
    \label{MRHD200}
    \vspace{0.5 cm}
 \end{figure}
The SRLDs and MRHDs also support these findings, as exhibited in the figures \ref{SLRD200} and \ref{MRHD200}. At initial times, SLRDs curves do not present the stretched exponential decay at the right tail, which is known to be a KPZ landmark \cite{Aarao}. Moreover, $S$ and $K$ values for these times do not converge to the KPZ one as the box size is increased (see the fig. \ref{SLRD200}(b)). The comparison between the SLRD for surfaces grown at $t = 15$ min and the KPZ-SLRD
around the peak gives more evidences of such discrepancy - inset of fig. \ref{SLRD200}(a). Nevertheless, the situation is 
totally different at long times, where the agreement between the experimental data and the theoretical one 
reaches more than four decades around the peak with $S$ and $K$ values converging to the KPZ values 
as the continuous limit is approached, constrained to $\zeta \ll l \ll \xi$. Results coming from the MRHDs (fig. \ref{MRHD200}) are not so accurate as those emerging from SRLDs, but they also corroborate that, for CdTe thin-films grown at $T = 200\,^{\circ}\mathrm{C}$, a Random-to-KPZ crossover takes place.

\subsection{KPZ Growth with Deposition Refused: CdTe Surface Fluctuations at $T = 300\,^{\circ}\mathrm{C}$}

\begin{figure}[!ht]
    \vspace{0.5 cm}
    \centering
    \includegraphics[width = 12.0 cm]{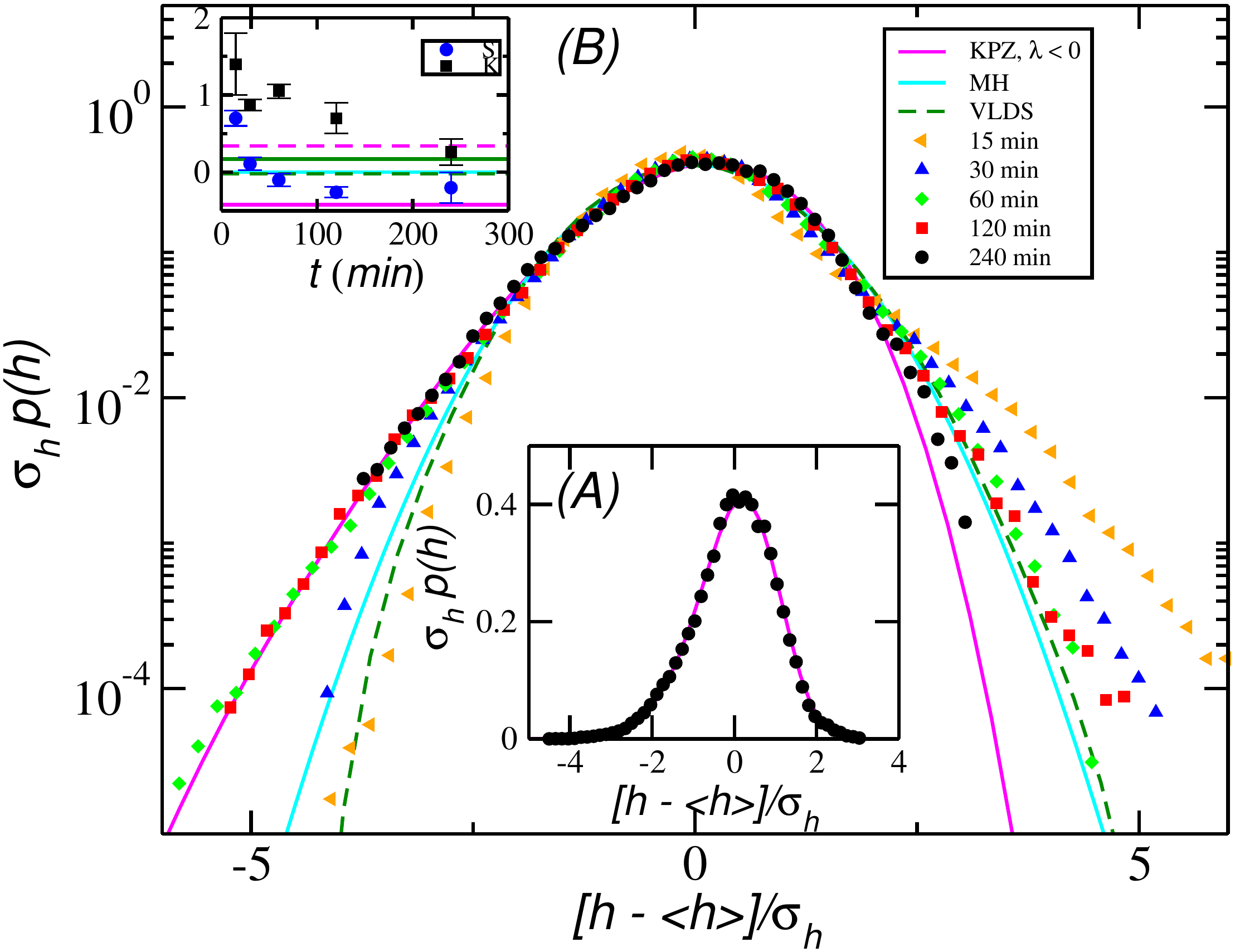}
    \caption[Rescaled HDs for surfaces grown at $T = 300\,^{\circ}\mathrm{C}$]
    {Rescaled height distributions at mean null and unitary variance for different
    growth times (symbols) from surfaces grown at $T = 300\,^{\circ}\mathrm{C}$. The experimental
    data are compared to the numerical KPZ curve (solid magenta line) considering $\lambda < 0$, to
    the universal HD-MH (solid cyan line) and to the universal HD-VLDS (dashed green line). Inset (A) highlights 
    the very good collapse between experimental and KPZ 
    numerical data close to the peak in a linear $\times$ log plot. Insertion (B) exhibits the experimental 
    S and K evolving in time. Solid/dashed lines refers to the expected S/K values.}
    \label{HD300}
    \vspace{0.5 cm}
 \end{figure}

Finally, the analysis of films grown at $T = 300\,^{\circ}\mathrm{C}$ enclose this section and bring to out the real UC of
CdTe thin films grown at high temperatures. Figure \ref{HD300} depicts the experimental HDs being compared to numerical 
KPZ, MH (linear-MBE equation) and VLDS (non-linear MBE equation) curves.

Just to mention, we have checked the universality for the HDs of 
these diffusion-dominated classes by integrating the linear equation as well as simulating several discrete models (see \cite{DSG, Das, Fabio_VLDS} and ref. therein) that belong to the MH and VLDS classes. As an additional note, the KPZ HD plotted is that one for $\lambda < 0$, which is 
the same HD plotted in the previous figures, but reflected around the origin. Anyway, one can observe that HDs for early growth times do not agree with distributions of any class, but, for long times, once again, a reasonable agreement with the KPZ-HD is observed. 
The left tails seem to converge faster than the right ones to the
KPZ curve - see HDs for $t > 30$ min. As demonstrated by the AFM images (fig. \ref{morph1}(e) and (f)) and
by the local exponent $\alpha_1$ (inset of the fig. \ref{loc}(c)), the top of grains become smoother as the time evolves. 
It leads the right tail of the HDs to move in the left direction and, consequently, to approach to the KPZ-HD.

Unlike the exponents, the experimental HDs discard the MH and VLDS classes as possibles UC of the CdTe growth.
For the largest time available, the inset A of the fig. \ref{HD300} presents a nice collapse 
around the peak between the experimental result and
the numerical one. Moreover, it is clear seen that $S(t)$ and $K(t)$ values are approaching to the KPZ ones, within the error bars -
inset B of the fig. \ref{HD300}. Particularly, for $t = 240$ min, one has S$_{t = 240} = -0.2(2)$ and K$_{t = 240} = 0.3(2)$,
which agree with the KPZ values (table \ref{universalkpz}), for $\lambda < 0$, considering the large error bars.

 \begin{figure}[!ht]
    \vspace{0.5 cm}
    \centering
    \includegraphics[width = 7.2 cm]{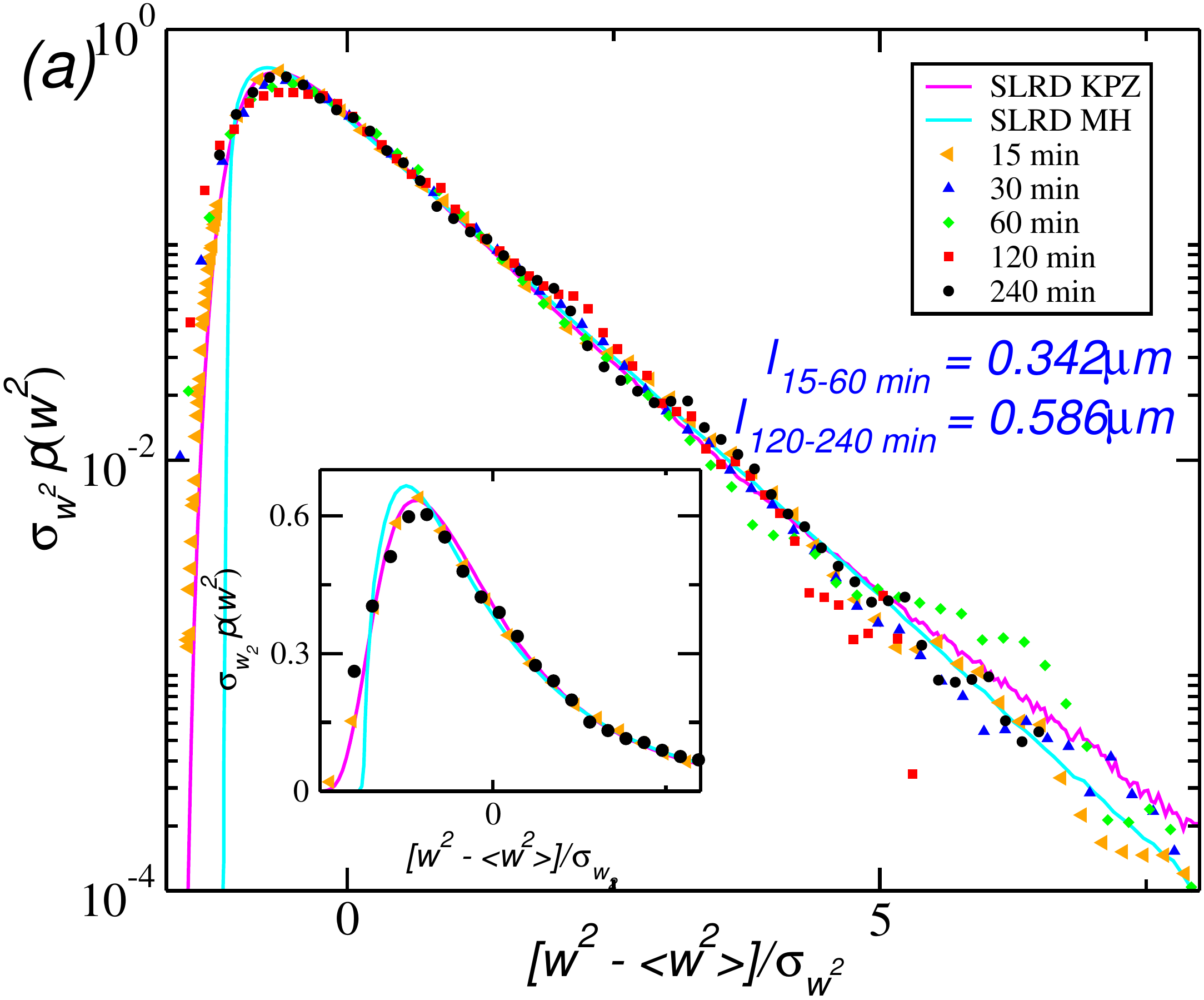}
    \includegraphics[width = 7.2 cm]{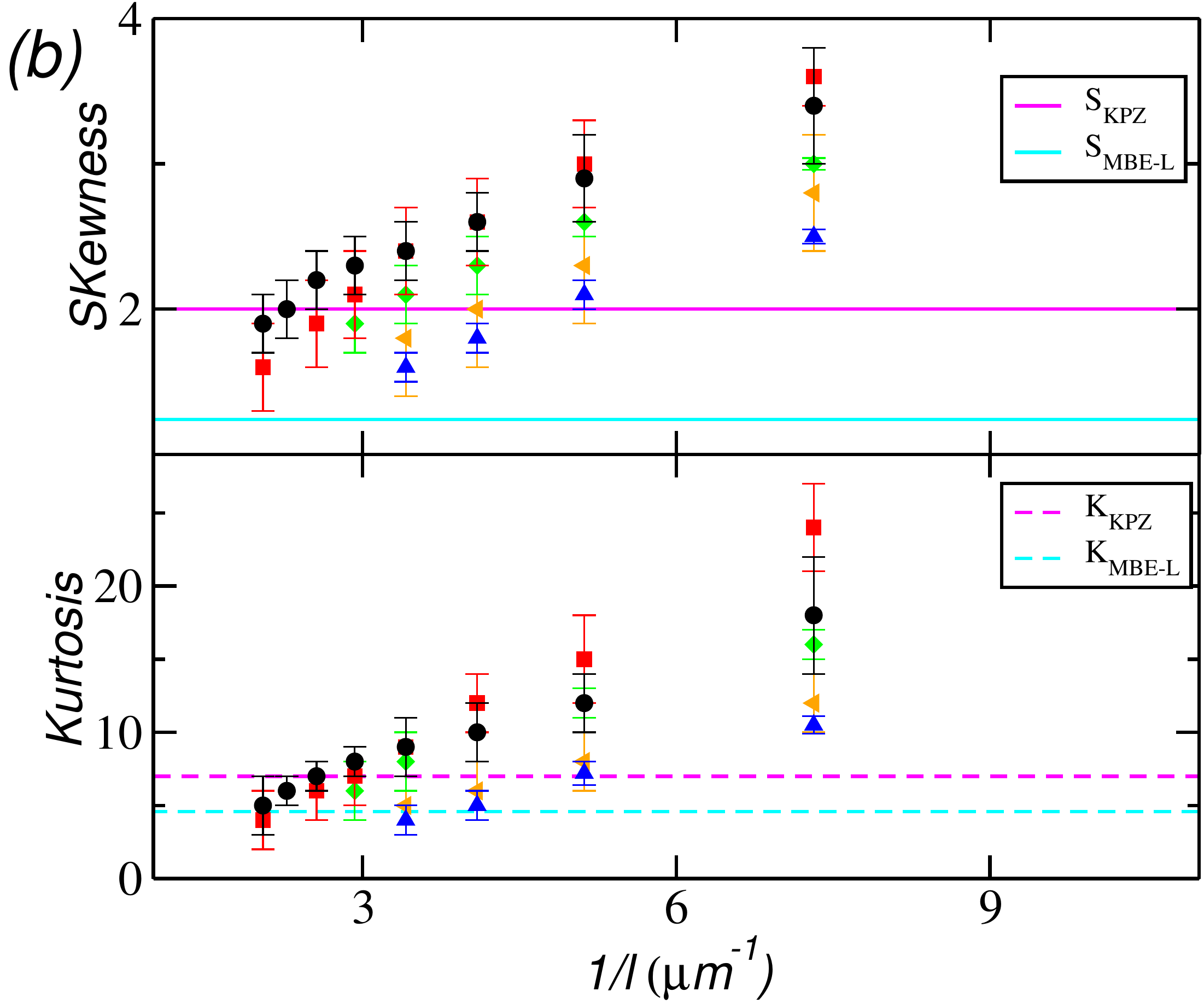}
    \caption[Rescaled SRLDs for surfaces grown at $T = 300\,^{\circ}\mathrm{C}$]
    {(a) Rescaled squared local roughness distributions for samples grown at 
    $T = 300\,^{\circ}\mathrm{C}$ by 15 min (orange left triangles), 30 min (blue up triangles),
    60 min (green diamonds), 120 min (red squares) and 240 min (black circles). The curves shown 
    are those whose the box size is indicated below the legend. They are different for each $t$ because 
    the appropriate interval $\zeta \ll l \ll \xi$ depends on $t$. The inset shows the good 
    collapse around the peak for the thinnest and the thickest film grown. (b) Skewness and Kurtosis values as 
    function of the box size $l$.}
    \label{SLRD300}
    \vspace{0.0 cm}
 \end{figure}

 \begin{figure}[!t]
    \vspace{0.0 cm}
    \centering
    \includegraphics[width = 8.0 cm]{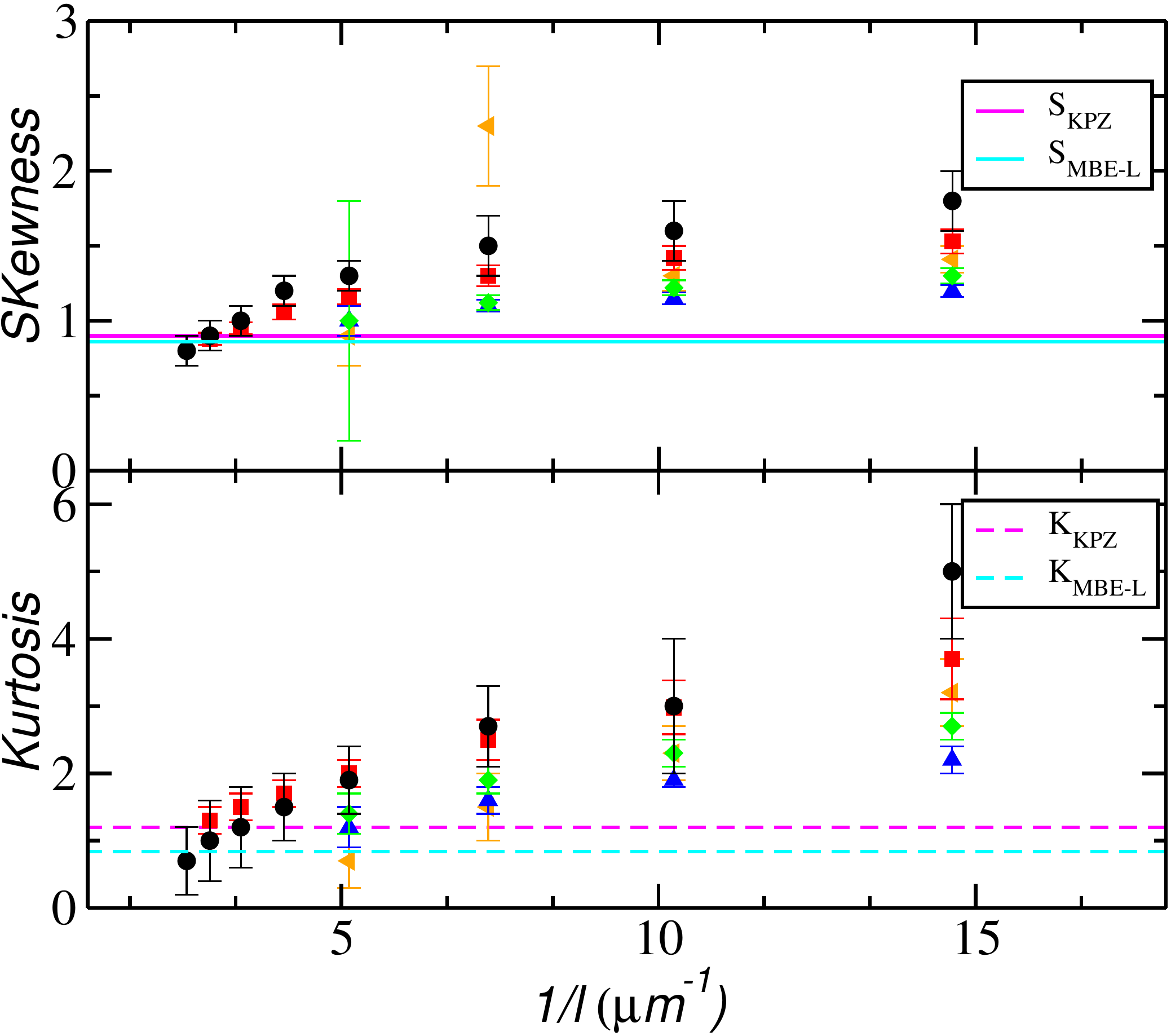}
    \caption[$S$ and $K$ as function of the box size from MRHDs for $T = 300\,^{\circ}\mathrm{C}$]
    {$S$ and $K$ as function of the box size for MRHDs for samples grown at
    $T = 300\,^{\circ}\mathrm{C}$. Solid/dashed line indicates the expected values for the $S$ and $K$
    for the KPZ (magenta line) and for the MH class (cyan line).}
    \label{MRHD300}
    \vspace{0.5 cm}
 \end{figure}

To gain further evidence of KPZ scaling, the SLRDs and MRHDs have been calculated and 
are depicted in the figures \ref{SLRD300} and \ref{MRHD300}, respectively. \textit{From the left tail and from the peak of the
SLRDs} (fig. \ref{SLRD300} and its inset), one notices that \textit{experimental results are very well described by the
KPZ curve}, whereas the decay at the right tail can not make a clear distinction between the KPZ and the MH class. 
In the fig. \ref{SLRD300}(b), $S$ values converge to the KPZ ones, in the meantime that $K$ values also can not 
be distinguished from that expected for the KPZ-MRHD and MH-MRHD. In this situation, where the exponent 
analysis point to KPZ or to MH growth, the MRHDs \textit{are not suitable} to make a separation between these two classes. As showed in the fig. \ref{MRHD300}, the expected $S$ and $K$ values for the KPZ and the MH class are very close between each other and, from the experimental point of view, one can not distill the UC based \textit{only} on this result.

These findings help us to see how hard is determining the UC of the two-dimensional experimental films. Anyhow, guided by the local-structure evolution, roughening scaling results $\&$ distributions
analysis, we can conclude that the CdTe interface for $T=300\,^{\circ}\mathrm{C}$ evolves according the 
KPZ equation (eq. \ref{eq_kpz}) with $\lambda < 0$. This indicate the existence of a mechanism of \textit{refusing} the deposition of particles depending on the local slopes as occurs in the famous RSOS model \cite{Kim_model}. A possible explanation for $\lambda<0$ is that the sticking coefficient is smaller in regions with very large slopes at surface. Indeed, one notices in the fig. \ref{loc}(b) that $\left\langle (\nabla h)^2 \right\rangle$ for $T=300\,^{\circ}\mathrm{C}$ is very larger than that for lower $T$. This can explain why this effect appears only at high $T$ \cite{Ojeda,Zhao}.

\section{Final Remarks}

A detailed study on how the deposition temperature affects the CdTe growth dynamic in a wide range of temperature
[150, 300]$\,^{\circ}\mathrm{C}$ has been performed. The local mound evolution has been explained in terms
of an interplay between a local energy barrier $E_{GB}$ that tends repelling the diffusion and the deposition of 
particles at defect sites (yielded by the colision of GBs of neighboring grains) and the effectiveness of 
the relaxation process at these GBs, which tends to eliminate defects on the active
zone of the CdTe interface. In turn, the last mechanism depends sensintively on the substrate temperature. Thus,
for $T=150\,^{\circ}\mathrm{C}$, the low diffusion at the GBs of colided neighboring grains prevents coalescences
and the propagation of correlations at interface, so that inter-grain fluctuations are described by Possianian process.
This dynamic also happens at short times for films grown at $T=200\,^{\circ}\mathrm{C}$, however the relaxation at (and around) the GB's of colided neighboring grains gives rise to coalescences and, hence, an asymptotic correlated growth. For higher $T$, where diffusion rate is higher, these processes start early \textit{as well as the KPZ scaling}.

In terms of KPZ equation (eq. \ref{eq_kpz}), the random growth at low $T$ implies $\nu \approx 0$ and $\lambda \approx 0$. For $T=200\,^{\circ}\mathrm{C}$, one expects $\lambda>0$, but small, so that growth is dominated by noise/non-linear effects initially/asymptotically. The absence of a crossover when $T=250\,^{\circ}\mathrm{C}$ \cite{Almeida} indicates 
that $\lambda$ is positively large. The same interplay also takes place in this temperature, but as the diffusion
is still more operative than that for lower $T$, coalecence processes start early enhancing the activity
the grain-packing (discussed in the previous chapter) and, hence, the strength of $\lambda > 0$. Thus, $\lambda(T)$ seems 
to be a positive increasing function in this range of $T$. However, for $T=300\,^{\circ}\mathrm{C}$, the evidences of KPZ scaling and the negative skewed HDs-KPZ reveals $\lambda < 0$, which is typical of KPZ systems where there is deposition refuse as in the RSOS model \cite{Kim_model}. A possible explanation is that the the sticking coefficient becomes
smaller in regions with very large slopes at surface. This reasoning is corroborated by $\left\langle (\nabla h)^2 \right\rangle$ for films grown $T=300\,^{\circ}\mathrm{C}$, where one confirms the presence of larger slopes than those for lower $T$. This can explain why this effect appears only at high $T$ \cite{Ojeda,Zhao}. Anyway, it is astonishing that so contrasting KPZ mechanisms can emerge in the CdTe growth system only adjusting $T$.

 \chapter{Conclusions and Perspectives}
 \lhead{\bfseries 7. Conclusions and Perspectives}
 In this work one has performed a detailed study on the growth dynamic at both short- and large-wavelength scales of CdTe thin films grown on Si(001) substrates by Hot Wall Technique. The results have provided, for the first time, a clear and robust evidence of a system in d = 2 + 1 dimensions which belongs to the Kardar-Parisi-Zhang class. Here, the phrase  ``robust evidence'' means that the prove goes beyond the comparison with critical exponents, as confirmed by the (rescaled) height distributions, squared local roughness distributions and maximal relative height distributions. In the meantime, this work demonstrates the universality of these KPZ distributions, giving them a reliable reality beyond numerical simulations. 

Along the analysis, one has found several pitfalls hampering the extraction of asymptotic scaling exponents. For 
instance, through the power-law in the $w_{loc} \times$ l plot, it was not possible to unearth the roughness exponent. 
Rather than, only the geometrical exponent, $\alpha_1$, which does not
provide any information about the Universality Class (UC) of the system, was found. Another difficulty is the relation 
between the average mound size ($\zeta$) and the correlation length ($\xi$). The validity 
of $\zeta \approx \xi$ should be taken as true only for an appropriate range of time, for which local peaks on the top of large mounds do not strongly underestimate the measurement of $\zeta$. Thus, an initial deep-visual inspection of surfaces
imaged by some microscope technique becomes important.

The KPZ mechanism occurring in CdTe films grown at $T = 250\,^{\circ}\mathrm{C}$ is related to the form in which new particles attaches to grain boundaries (GBs) of colided neighboring grains, do not filling all available space in their neighborhood and, thus, generating excess of velocity, the KPZ landmark. As will be discussed below in details, 
this packing mechanism also should be present in other temperatures.
Due to inexorable experimental obstacles it was not possible to validate the KPZ \textit{Ans\"{a}tze} in d = 2 + 1 dimensions. However, this is a natural extension of this work.

The effect of the deposition temperature ($T$) on the growth dynamic has also been distilled in broad range of $T$, namely, $T \in$ [150, 300]$\,^{\circ}\mathrm{C}$. A relation between short- and large-wavelength dynamics has been established. One has been found that the mound evolution is dictated by the interplay between the formation of defects at grain boundaries of colided neighboring grains and the relaxation process induced by diffusion and deposition of 
particles toward these regions. A simple Monte Carlo model corroborate this reasoning. This interplay leads to different scenarios at large-wavelength fluctuations as $T$ increases.
For $T=150\,^{\circ}\mathrm{C}$, the low diffusion at the GBs prevents coalescences 
and the propagation of correlations at interface, so that inter-grain fluctuations are described by a 
Poissonian process. Scaling exponents and distributions give support to these reasonings.

For $T=200\,^{\circ}\mathrm{C}$, however, a more complex scenario has been found. In the range of 
growth time studied, the relaxation process overcomes the effects of the barrier at (and around of) the GBs and 
gives rise to a crossover in the CdTe growth dynamic. Results from dynamical scaling and distributions also corroborate
with the presence of this crossover. Roughness scaling analysis, nevertheless, have not been able to, convincingly, point the UC of the CdTe growth at this deposition temperature. The supplemental study based on the distributions, in turn, have proved itself be essential for unveiling the UC of this system, once weaker finite-time effects than that in the standard roughness scaling has been found. Thus, it was possible to find a Random-to-KPZ crossover taking place 
in the CdTe growth, at the same time that the first robust experimental realization of 
such crossover in two-dimensional systems is demonstrated.

The growth of CdTe films at $T = 300\,^{\circ}\mathrm{C}$ carries also its distinct importance. Although dynamical
scaling analysis do not allow making a clear distinction among diffusion-dominated equations and the KPZ one for
describing CdTe fluctuations at this high deposition temperature, the distributions discard the formers
and point out the presence of KPZ growth with $\lambda < 0$, differently of lower temperatures.
This unanticipated result has been related to the decreasing of the sticking coefficient in regions with very large slopes at surface. The reasoning is corroborated by the squared-local slopes, which are about five times larger than those for lower temperatures, as can be noticed in the Atomic Force Microscope images. This can explain why this KPZ mechanism appears solely at high temperatures.

\begin{figure}[t]
    \vspace{0.5 cm}
    \centering
    \includegraphics[width = 12.0 cm]{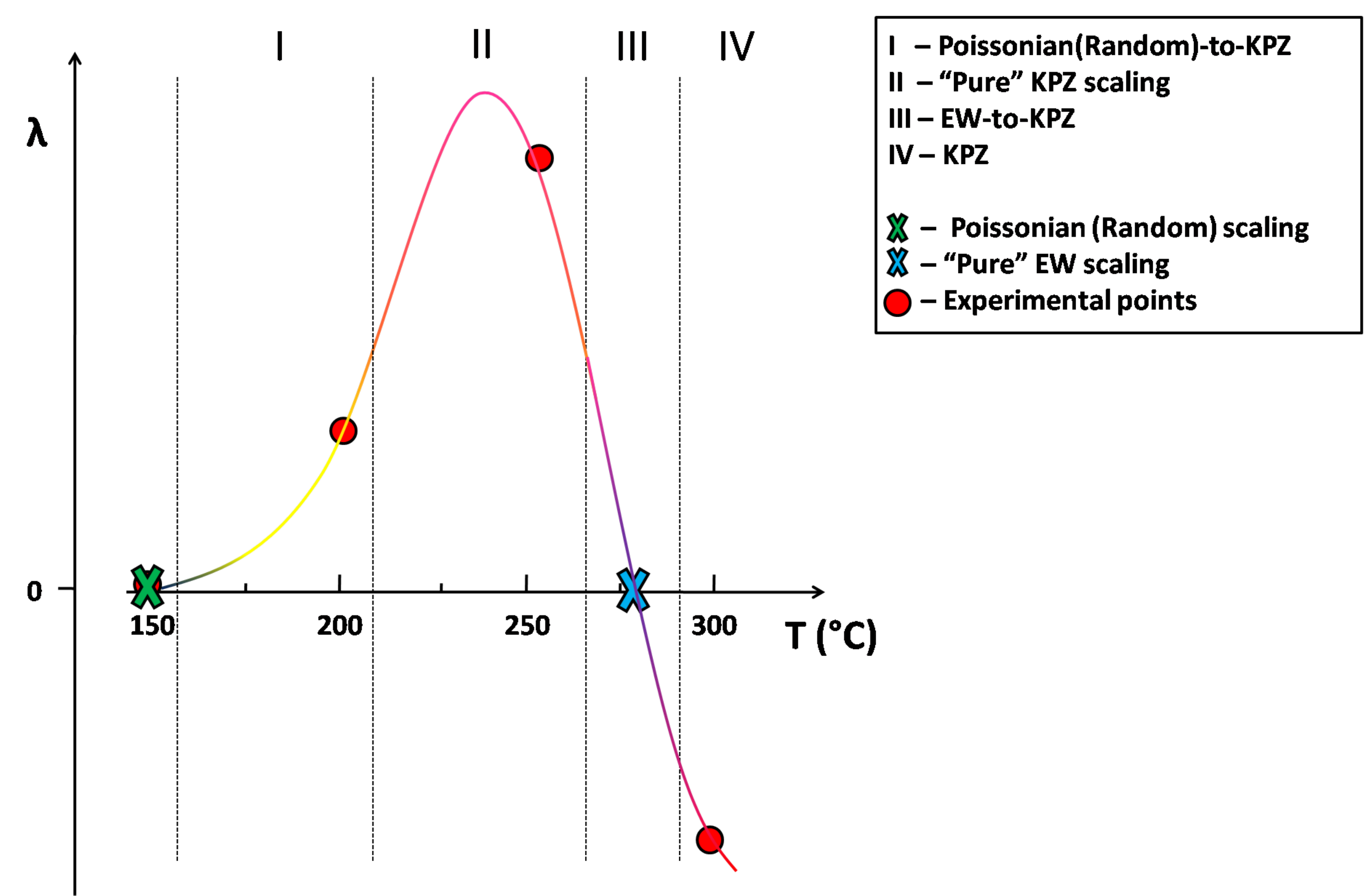}\quad
    \caption[Conjectured behavior of $\lambda$ as function of $T$]
    {Conjecture for the behavior of the excess of velocity, $\lambda$, as function of the deposition temperature $T$
    in the CdTe/Si(001) system. Circles refer to experimental points, and dashed lines indicates expected different
    regimes (I to IV) ruling the coarsening-grained dynamic. The ``$\times$'' point denotates temperatures for
    which global fluctuations are expected to belong to Random (green $\times$) and EW (blue $\times$) Classes.}
    \label{lamb_T}
    \vspace{0.5 cm}
    \end{figure}

In terms of the KPZ equation (eq. \ref{eq_kpz}), the results reveal that, 
for $T = 150\,^{\circ}\mathrm{C}$, the ``surface tension'' ($\nu$) and the excess of velocity ($\lambda$) are
very near of zero, so that noise dominates the growth. For CdTe films grown at $T= 200\,^{\circ}\mathrm{C}$, however,
the Random-to-KPZ crossover found indicates $\lambda > 0$, but small, so that non-linear effects overcome the noise
only at long growth times. In turn, a clean evidence of the KPZ scaling since inital growth times is present
in CdTe films grown at $T = 250\,^{\circ}\mathrm{C}$, in which $\lambda$ is positive and large. At the moment
that all conspire to conjecture $\lambda(T)$ as an increasing function of $T$, the results coming from
CdTe films grown at $T = 300\,^{\circ}\mathrm{C}$ show that $\lambda$ is negative. This suggests that
i) is possible to adjust $T$ at some specific $T_{EW}$, with $T \in$ ]250, 300[$\,^{\circ}\mathrm{C}$, in order 
to obtain $\lambda = 0$, i.e, to have a growth described by the Edwards-Wilkinson equation and ii) to set
$T_{(EW-KPZ)} = T_{EW} \pm \delta T$, so that EW-to-KPZ crossovers emerge into the dynamic of CdTe surface fluctuations. \textit{In summary, it is possible to adjust the KPZ non-linearity in the CdTe system only adjusting the
deposition temperature}. Figure \ref{lamb_T} summarizes theses discussions, making clear different regimes
that takes place in the growth as T is increased, namely: region I refers to the temperature interval in which
a Poissonian(Random)-to-KPZ crossover rules the dynamic. This first region is localized at the rigth border 
of temperatures close to $150\,^{\circ}\mathrm{C}$, where $\lambda \rightarrow 0$ and where 
the Poissonian(Random) growth emerges; in the region II a clean KPZ scaling emerges and precedes the thin III regime, limited by $T = T_{EW} \pm \delta T$ and symmetric in relation
to the $T_{EW}$ point, in which the pure EW growth ($\delta T = 0\,^{\circ}\mathrm{C}$) and the EW-to-KPZ crossover are expected to take place. 
Finally, at high temperatures (although expected do not be much higher than $300\,^{\circ}\mathrm{C}$), the 
KPZ scaling is recoreved with $\lambda < 0$.

We finish this work truly believing that the noveil procedure for investigating the UC of growing surfaces
open and will motivate (as exemplified by T. Halpin-Healy and G. Palasantzas \cite{Healy_exp})
a new confident perspective in the field, as well as the application of these methods in 
previously studied systems. Our system also offer the experimental possibility of a close study
of Random-to-KPZ and EW-to-KPZ crossovers.

 \appendix
\chapter{More Details About Continuum Growth Equations and Universality Classes}
\lhead{\bfseries Appendix A - Details of Continuum Growth Equations and Universality Classes}
\section{Random Growth equation}
\label{section_RD}

Concerning on EW and linear-MBE equations, one can easily see the particular case, where 
$\nu$ and $K$ are nulls respectively, resulting in the so called \textit{Random Growth} 
equation:

\begin{equation}
\partial_th = \eta(\textbf{x},t).
\label{rd}
\end{equation}

Once just the noise rules the growth dynamics, there are no correlations to be propagated through 
the system and we can argue $z \rightarrow \infty$. The $\alpha$ exponent is also ill defined, since the 
system never reaches the stationary regime. Nevertheless, we still can find as the variance of such 
random interfaces evolves with the deposition time. The result is anticipated if one looks at these surfaces 
as being composed by a set of independent Brownian motions.

Integrating the random eq. one obtains: $\int \partial_th(\textbf{x},t)dt = \int\eta(\textbf{x},t)dt$. Averaging both sides, 
one has $\langle h(\textbf{x},t) \rangle = 0$. But, if instead we square and average, one reaches $\langle h^2(\textbf{x},t) \rangle = 2Dt$. 
So, the variance grows as in the eq. \ref{w_RD}, defining the ``growth exponent''\footnote{In accordance
with our previous definition, $\beta \equiv \alpha/z$. However, the random growth is a particular case, where
one can find a power-law in time for the variance, even when the spatial- and temporal-statistical-fractality 
exponents are ill defined. In this sense, the $\beta$ exponent is fundamentally different 
from those previously cited.} $\beta = 1/2$. 

\begin{equation}
 w^2(t) = \langle h^2 \rangle - \langle h \rangle^2 = 2Dt.
 \label{w_RD}
\end{equation}

\section{A derivation for the Linear-MBE equation}
\label{section_MH}
It is well known that in MBE environments, under usual conditions, desorption\footnote{See Kinetic Effects
in the Appendix section \ref{section_Kinetic}} is a negligible process and, hence, the therm $\nabla^2 h$ in
the eq. \ref{gerall3} should not play an important role. Formation of overhangs and bulk defects are uncommon in the ideal epitaxy \cite{Herman1, Herman2} and discard the non-linear $(\nabla h)^2$ term. Our starting point for building a continuum growth equation for MBE environments is the conservation law:

\begin{equation}
 \partial_th = - \nabla \cdot \textbf{J} + \eta(\textbf{x},t),
 \label{cont_eq}
\end{equation}
where \textbf{J} represents a parallel current of particles \textit{diffusing} onto the interface.

\begin{figure}[!t]
    \vspace{0.5 cm}
    \centering
    \includegraphics[width = 4.0 cm]{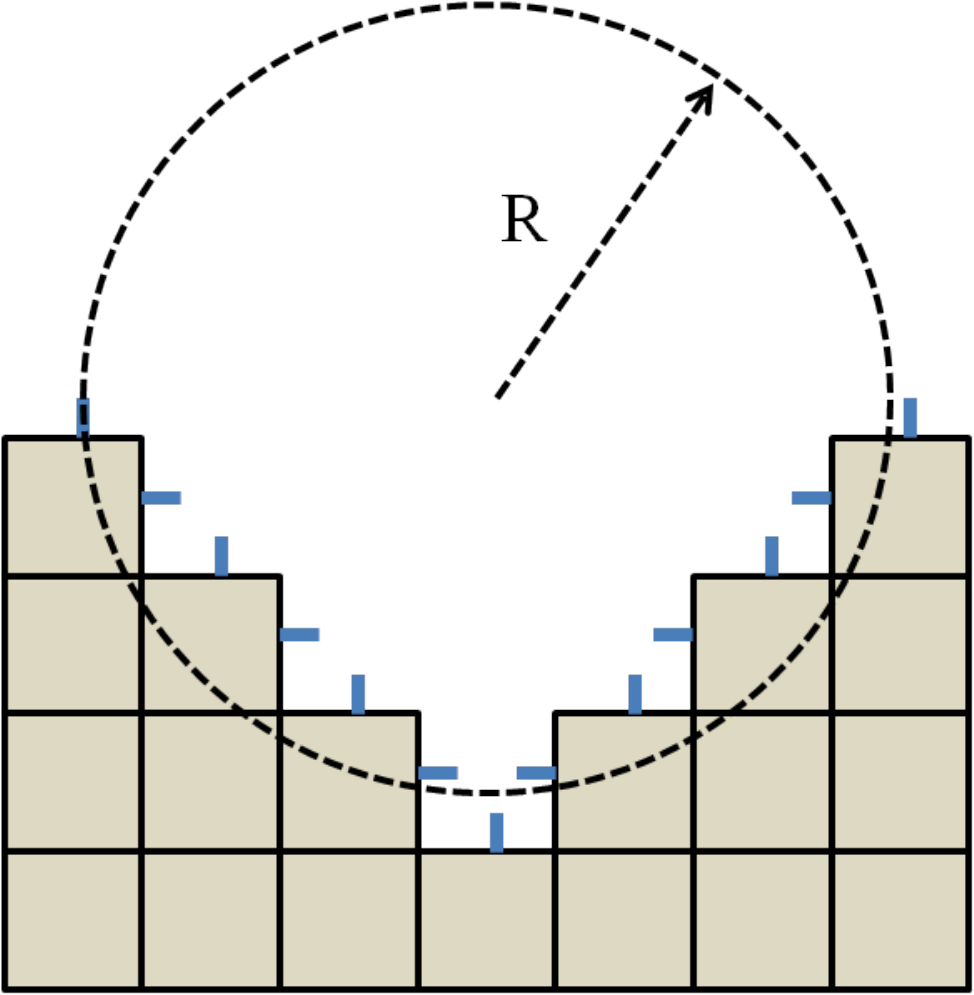}\quad
    \includegraphics[width = 4.0 cm]{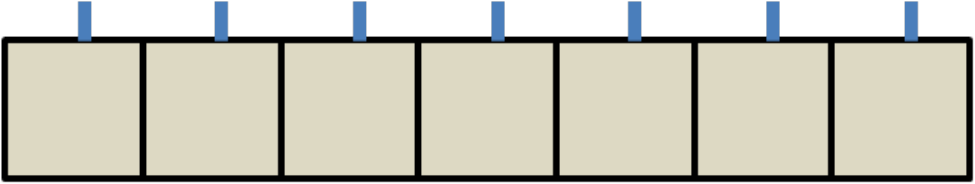}\quad
    \includegraphics[width = 4.0 cm]{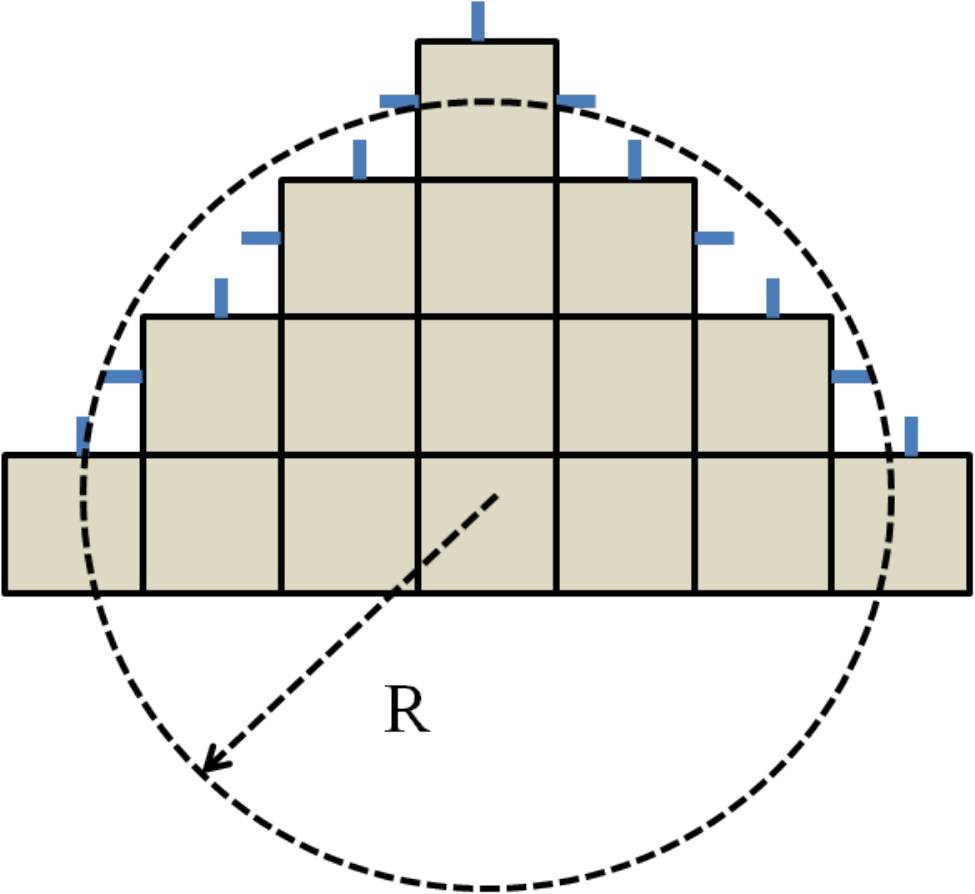}\quad
    \caption[Schematic of the chemical potential depending on the local curvature.]{Schematic of 
    the chemical potential depending on the local curvature. At left to rigth: positive curvature, 
    showing a valley, where there is a local minimum of $\mu(\textbf{x},t)$ because particles sticked at the bottom of
    the valley will not move away from there. On a terrace $R \rightarrow \infty$ and $\mu(\textbf{x},t)$ is null everywhere. 
    Negative curvature, showing an island, where $\mu(\textbf{x},t)$ has a local maximum value because particles
    on the top of the island break off their bonds and hop to a closest neighbor site where
    the free energy of the interface is decreased.}
    \label{local_bond}
    \vspace{0.5 cm}
    \end{figure}

As interfaces growing in MBE chambers are driven by chemical bond in order to minimize their surface free energy
through diffusion, $\textbf{J}$ should be $\propto -\nabla \mu$. But the simplest assumption, in turn, points out
that the chemical potential is proportional to $-1/R$, where R is the local curvature - see 
fig. \ref{local_bond}. The bottom of a valley (positive curvature, $R > 0$), for instance, is the site where a particle have major
number of neighbors and move away from there will be difficult. So, the bottom of a valley is a minimum for 
$\mu(\textbf{x},t)$. The opposite happens for the topest site of a island ($R < 0$), where there is a local maximum for $\mu(\textbf{x},t)$. On a terrace, 
all particles have the same coordination number, and locally $\mu = 0$ everywhere. 
Hence, on these considerations one can argue straightforwardly $\mu(\textbf{x},t) \propto - \nabla^2 h$. Consequently,
$\textbf{J}\propto \nabla(\nabla^2 h)$ and one reaches to the eq. \ref{MH_eq}, after eq. \ref{cont_eq}.

\section{The Non-Linear MBE equation and the VLDS class}
\label{section_VLDS}
In models with conserved dynamic, there is necessity of inserting non-linear terms, since $\alpha > 1$ breaks
the hypothesis that continuum approximations are valid in the limit of small slopes,
$\nabla h \ll L^{\alpha - 1}$ \cite{Barabasi}. As pointed out by Wolf and Villain \cite{WV}, the lowest 
non-linear term obeying conserved dynamic is the therm $\nabla^2(\nabla h)^2$.
The geometrical interpretation (see ref. \cite{LD}) indicates that particles landing at high steps (large derivatives)
relax to lower steps (small derivatives), corresponding to the ``high-temperature'' regime in MBE environments, 
where atoms break their bond and diffuse, with larger bias, to small slope kinks. The non-linear MBE equation (eq. \ref{VLDS_eq}) was proposed formally in 1991 by Lai and Das Sarma \cite{LD}.

\begin{equation}
\partial_{t}h(\textbf{x},t) = -K\nabla^{4}h + \lambda_1 \nabla^2(\nabla h)^2 + \eta(\textbf{x},t),
\label{VLDS_eq}
\end{equation}
$\lambda_1$ has the same dimension of $\nu$ and accounts for the strength of diffusion of particles at local slopes
towards small derivatives.

Critical exponents related to the non-linear MBE equation can be found by using Renormalization-Group (RG) approaches \cite{LD, Barabasi}. The method predicts:

\begin{equation}
\alpha = \frac{4-d_s}{3}; \qquad z = \frac{8 + d_s}{3}
\label{vlds_exp}
\end{equation}

Other very important result coming from RG analysis is the hyper-scaling relation valid 
for \textit{any growth with conserved dynamic and non-conversed noise}:
\begin{equation}
 z - 2\alpha - d_s = 0
 \label{hyper_dif}
\end{equation}

The set these exponents compose the \textit{Villain-Lai-Das-Sarma} (VLDS) Universality Class. There are
a small number of discrete models belonging to the VLDS class, among which \textit{possibly} are the conserved 
restricted-solid-on-solid (CRSOS) model, proposed by Kim \textit{et al.} \cite{Kim_CRSOS} and 
the Das Sarma and Tamborenea one (DT) \cite{DT}. A very good numerical study and 
descriptions of these models can be found in the Ref. \cite{Fabio_VLDS}.

\chapter{Anomalous Scaling}
\lhead{\bfseries Appendix B - Anomalous Scaling}
\label{section_anomalous}

Anomalous scaling occurs whenever a surface is not \textit{self-affine} \cite{Lopez96}. It means that
local and global fluctuations do not evolve in the same way. Initially, the therm \textit{anomalous} was
used to describe the growth of interfaces in which the exponent $\alpha$ was found to be larger (or equal) 
than 1 \cite{Leschhorn, JAmar, Schroeder, DSarma_anom}. Indeed, this comes from the fact that, 
if $\alpha \geqslant 1$, there is a divergence of fluctuations in the steady state, making the surface
\textit{super-rough}, as indicated by the Family-Vicsek ans\"{a}tze 
(eq. \ref{roug3}): $W_{sat}/L \sim L^{\alpha - 1}$. 


The current situation, however, is very different from that of almost twenty years ago. Nowadays, one knows
that several systems display anomalous scaling, independently of the $\alpha$ value (see \cite{Ramasco00, Lopez05} and 
references therein). Moreover, there are more than one possible scaling form to describe the local-fluctuations dynamic of 
a growing interface, being the FV one only a particular case where there is no difference between 
the local and global one \cite{Ramasco00}. A wide number of experimental studies have also corroborated
this conjecture, among which are the development of wood fractures \cite{Lopez_exp}, the electrodeposition of Cu \cite{Huo}, the dissolution of pure iron \cite{Cordoba} and 
the deposit of colloidal particles at the edge of evaporated drops \cite{Nicoli_coll}.

The initial clues of a break in the FV hypothesis were found in the one-dimensional Wolf-Villain 
model \cite{Schroeder, Smilauer} and, subsequently, in other linear and non-linear growth models dominated
diffusion \cite{JAmar, DSarma_anom, Smilauer, Das}. From there, analytical and numerical treatments
were strongly devoted to this issue, leading to the conclusions:

\begin{itemize}
 \item[i)] The condition $\alpha < 1$ is not sufficient to prevent the system of exhibiting the anomalous scaling \cite{Lopez96}.
 \item[ii)] There exist three types of anomaly: super-rough, intrinsic \cite{Lopez97} and faceted \cite{Ramasco00} (they are defined below).
 \item[iii)] Non-conserved growth models \textit{do not} exhibit anomalous scaling and, in particular, intrinsic 
 scaling can not occur in local growth models \cite{Lopez99, Lopez05}.
\end{itemize}

It is worth mention that the presence of the anomaly affects only \textit{local} correlation functions, keeping the
global roughness, for instance, scaling according to the standard FV scaling \cite{Ramasco00} - this is proved
in the following. For now, consider the Fourier transform of the function $h(\textbf{x},t)$ as,

\begin{equation}
 \textrm{\^{h}}(\textbf{k},t) = L^{-d_s/2} \sum_{\textbf{x}}[h(\textbf{x},t) - \langle h(t) \rangle] \exp({i\textbf{k} \cdot \textbf{x})},
 \label{FT}
\end{equation}
where $\textbf{k}$ are the wave-numbers surviving within the space bounded by the substrate.

The called structure factor [$S(k,t)$] or power spectrum (eq. \ref{PS}) is a correlation function measuring the
fluctuations in the reciprocal space.

\begin{equation}
 S(k,t) = \langle \textrm{\^{h}}(\textbf{k},t) \textrm{\^{h}}(\textbf{-k},t) \rangle = \langle |\textrm{\^{h}}(\textbf{k},t)|^2 \rangle.
 \label{PS}
\end{equation}

$S(k,t)$ can be related to the global roughness and to the height-height correlation
function ($C_h$) by the eqs. \ref{SW} and \ref{SCv}, respectively,

\begin{equation}
 W^2(L,t) = \frac{1}{L^{d_s}}\sum_{k}S(k,t) = \int \frac{d^{d_s}{k}}{(2\pi)^{d_s}}S(k,t),
 \label{SW}
\end{equation}

\begin{equation}
 C_h(l,t) \propto \int [1 - cos(\textbf{k} \cdot \textbf{x})]\frac{d^{d_s}{k}}{(2\pi)^{d_s}}S(k,t),
 \label{SCv}
\end{equation}
where the integrals run within the interval $2\pi/L \leq k \leq 2\pi/a$ for each $k$ direction, with 
$a$ being the lattice parameter.

The basic tools required to derive the different forms that anomalous scaling can appear are written in order 
in the equations above. Inserting the FV ans\"{a}tze (eq. \ref{roug3}) into the relation of eq. \ref{SW}, it is
straightforward showing that the power spectrum behaves as:

\begin{equation}
S(k,t) = k^{-(2\alpha + d_s)}s_{FV}(kt^{1/z}),
\end{equation}
where:

  \begin{equation}
  s_{FV}(u) \sim \left\{
 		\begin{array}{c}
 		const, \quad if \quad u \gg 1, \\
 		u^{2\alpha + d}, \quad if \quad u \ll 1.
 		\end{array}
 		\right.
  \label{SFV}
 \end{equation}

It means that for short-length scales ($k \gg 1/\xi$), the power-spectrum is time-independent and the curves
for all times show a power-law scaling with $k^{-(2\alpha + d_s)}$. This is the usual FV scaling in the reciprocal space.
But now, look what happens when this scaling relation is inserted into the eq. \ref{SCv}. When one sets $\alpha > 1$, 
the integral in eq. \ref{SCv} becomes divergent in the limit of $l \ll \xi$ for $L \rightarrow \infty$, and 
$a \rightarrow 0$ \cite{Lopez97}. Taking the limit  $l \ll \xi$ first, and keeping $L$ and $a$ fixed, one obtains 
\textit{a different} scaling relation for $C_h$:
  \begin{equation}
  C_h(l,t) \sim \left\{
 		\begin{array}{c}
 		l^2t^{2(\alpha -1)/z}, \quad if \quad l \ll \xi \ll L, \\
 		l^2L^{2(\alpha - 1)},  \quad if \quad l \ll L \approx \xi.
 		\end{array}
 		\right.
  \label{Canom}
 \end{equation}

\begin{figure}[h]
    \vspace{0.5 cm}
    \centering
    \includegraphics[width = 12.0 cm]{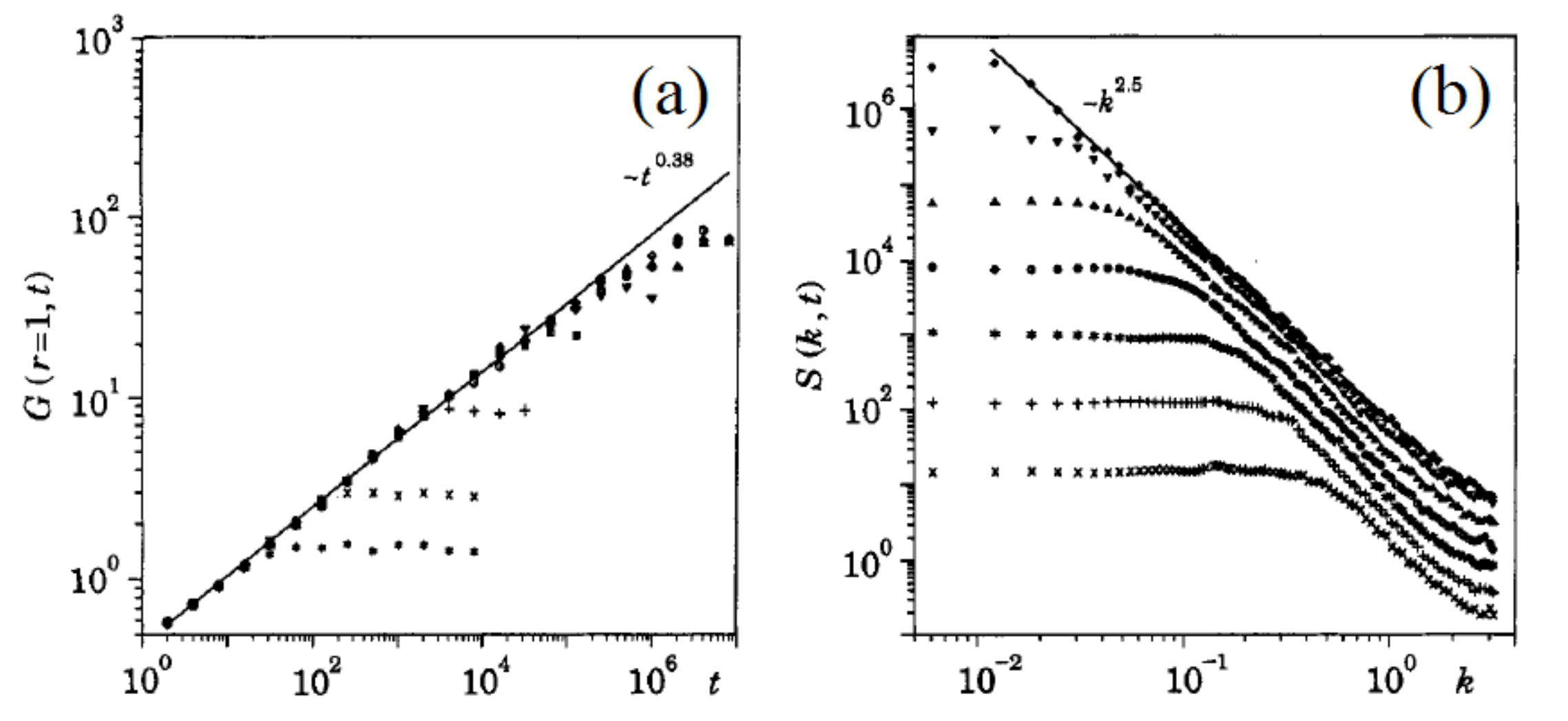}
    \caption[Example of anomalous scaling]{Example of anomalous scaling. In (a) one has the function G(1,t) (corresponding to
    the $C_h(1,t)$ function) for several systems sizes L (from the bottom o the top L = 10, 15, 30, 60, 120, 240, 800 and 1024, respectively). 
    (b) Shows the non-usual behavior for the structure factor S(k,t). From the bottom to the top
    the curves refer to growth times t = $2^{4}$, $2^{7}$, $2^{10}$, $2^{13}$, $2^{16}$, $2^{19}$, $2^{22}$, respectively.
    Figure extracted from ref. \cite{Schroeder}.}
    \label{ano_model}
    \vspace{0.5 cm}
    \end{figure}

So now a different roughness exponent has emerged and, \textit{locally}, $C_h(l,t) \sim l^{2(\alpha_{loc})}$, 
where $\alpha_{loc} \neq \alpha$ and $\alpha_{loc} = 1$. Moreover, a dependence in time for $C_h$ at 
wavelengths $l \ll \xi$ appears in the system scaling with $t^{2\kappa}$, where:

\begin{equation}
 \kappa \equiv (\alpha - \alpha_{loc})/z.
 \label{kappa}
\end{equation}
Translating this scaling for the context of the fig. \ref{wlocmodel}, one would see the curves for 
different growth times, shown in that figure, shifted to up as the time evolves for $l \ll \xi$. Figure \ref{ano_model}(a)
shows very clearly this non-usual behavior by calculating the square-local-slope evolution (eq. \ref{kappa2})
for the one-dimensional Wolf-Villain model \cite{Schroeder}. 

\begin{equation}
\langle (\nabla h)^2 \rangle \sim C_h(l = 1, t) \sim t^{2\kappa}.
\label{kappa2}
\end{equation}

In the WV model, it was found $\kappa = 0.19(1)$ during a transient regime, limited by
both finite-size and finite-time corrections \cite{Schroeder}. But this feature of non-usual behavior
of correlations functions at local scales is not the complete scenarium in which the anomaly can emerge.
Indeed, models incorporating random diffusion \cite{Lopez97} and diffusion-dominated \cite{Plischke} 
presented a non-usual behavior not only for $C_h(l,t)$, but also for the power spectrum, i.e 
[S($k \gg 1/\xi$,t)] is time dependent - see fig. \ref{ano_model}(b). Thus, it became clear that the 
FV ans\"{a}tze written in the reciprocal space (eq. \ref{SFV}) could not be the general scaling dictating
local fluctuations in the surface growth context.

The generic dynamic scaling in kinetic roughening has been developed along of many years \cite{Lopez96, Lopez97, Lopez99}, 
but a formal deduction has been achieved in the year of 2000, where Ramasco \textit{et al.} \cite{Ramasco00} have proposed
a scaling for $s(u)$ as,

 \begin{equation}
 s(u) \sim \left\{
		\begin{array}{c}
		 u^{2(\alpha - \alpha_{s})}, \quad if \quad u \gg 1, \\
		 u^{2\alpha + d}, 	      \qquad if \quad u \ll 1,
		\end{array}
		\right.
 \label{Sgeral}
 \end{equation}
where $\alpha_s$ is a new exponent called \textit{spectral roughness exponent}.

Thus, the power-spectrum, generically, scales as:

 \begin{equation}
 S(k,t) \sim \left\{
		\begin{array}{c}
		 k^{-(2\alpha_s + d_s)}, \quad  if \quad k \gg 1/t^{1/z}, \\
		 t^{\frac{2\alpha + d_s}{z}} , 	\qquad if \quad k \ll 1/t^{1/z}.
		\end{array}
		\right.
 \label{Sgeral0}
 \end{equation}

Inserting eq. \ref{Sgeral} in eq. \ref{SW}, one proves that global fluctuations are dictated by the 
standard Family-Vicsek scaling (eq. \ref{roug3}), independently of $\alpha$ and $\alpha_s$ values. Nevertheless, 
as we have shown above, the local scaling is modified, depending on the order of the several limits involved
and on values of the exponents \cite{Ramasco00}. In a general way, two major cases can be distinguished, namely,
$\alpha_s < 1$ and $\alpha_s > 1$. For the former, one obtains the same that in the eq. \ref{roug2}, but
with $L \rightarrow l$ and $\alpha \rightarrow \alpha_{s}$, i.e:

\begin{equation}
 w(l,t) = t^{\beta}f_{\alpha_{s} < 1}(l/\xi),
\end{equation}
with,

 \begin{equation}
 f_{\alpha_{s} < 1}(u) \sim \left\{
		\begin{array}{c}
		 u^{\alpha_{s}},  \qquad  if  \quad u \ll 1, \\
		 const         ,  \quad   if  \quad u \gg 1,
		\end{array}
		\right.
 \label{Fa}
 \end{equation}
implying that $\alpha_s = \alpha_{loc}$ \cite{Lopez97, Lopez96}. This is called \textit{intrinsic} anomaly: one has
the power-spectrum and $w(l,t)$ [$C_h(l,t)$] behaving non-trivially. Family-Vicsek scaling is recovered when $\alpha = \alpha_{loc}$ \cite{Ramasco00}.

Otherwise, for $\alpha_s > 1$, the integral in the eq. \ref{SCv} is divergent when $L \rightarrow \infty$. Keeping L fixed,
one finds that:

 \begin{equation}
 f_{\alpha_{s} > 1}(u) \sim \left\{
		\begin{array}{c}
		 u, \quad             \qquad  if  \quad u \ll 1, \\
		 const         ,\quad   if  \quad u \gg 1.
		\end{array}
		\right.
 \label{Fb}
 \end{equation}

This implies that $\alpha_{loc} = 1$, independently of $\alpha_s$. Additionally, whether $\alpha = \alpha_s$, so 
the power-spectrum evolves trivially in time, whereas $w(l,t)$ and $C_h(l,t)$ do not. This is the 
\textit{super-roughening} anomaly, discussed at the beginning of this section. However, when the equality 
between $\alpha$ and $\alpha_s$ is not fulfilled, so a new anomaly stems, namely, the \textit{faceted} scaling \cite{Ramasco00}.
The main characteristic of the faceted anomaly is that \textit{it can detected only by using the power spectrum} because
there is not a constrain between $\alpha$ and $\alpha_{loc}$, i.e, these exponents can be equal ($\kappa = 0$) or different ($\kappa \neq 0$).

The summary of all conditions leading to the different scaling which can appear in surface growth is, hence:

\begin{equation}
\left\{
		\begin{array}{c}
		If \quad \alpha_{s} < 1 \quad \Rightarrow \quad \alpha_{loc} = \alpha_{s} \qquad
						 \left\{
						 \begin{array}{c}
						 \alpha_{s} = \alpha	 \quad	        \Rightarrow \quad	Family-Vicsek,\\ 
						 \alpha_{s} \neq \alpha \quad	\Rightarrow \quad	Intrinsic,
						 \end{array}
						 \right.\\
		If \quad \alpha_{s} > 1 \quad \Rightarrow \quad \alpha_{loc} = 1 \qquad
						 \left\{  
 						 \begin{array}{c}
 		                                \quad \alpha_{s} = \alpha \Rightarrow Super-rough, \\
 		                                \alpha_{s} \neq \alpha \Rightarrow Faceted.
 		                                \end{array}
 		                                \right.
 		\end{array}
		\right.
 \label{spectrall2}
\end{equation}

From the experimental point of view, several studies have found the different anomalous scalings. For instance,
a transition from intrinsic to faceted scaling has been reported during dissolution of iron \cite{Cordoba}, 
whereas a faceted dynamic has been claimed in the growth of CdTe on glass substrates \cite{Fabio}. Intrinsic 
anomaly seems to appear in the growth of TiN thin films by reactive sputtering \cite{Cuerno3} and during
the deposit of colloidal particles at edge of evaporated water drops \cite{Nicoli_coll}. 
Other studies involving anomaly can be found in the Ref. \cite{Cuerno1}.

Progress has also been made in the sense of classify when a possible anomaly can appear depending on the
continuum equation considered. L\'{o}pez \textit{et al.} \cite{Lopez05} has used a 
transformation $\Upsilon = \nabla h$ to study the roughness ($W_{\Upsilon} = \langle (\nabla h)^2 \rangle$) 
of surfaces described by slopes fields. When such surfaces are rough, it implies $\kappa > 0$ and one 
has anomaly (disconsidering the faceted case). From this starting point, one can show that non-conserved
growth models, such those ruled by the Kardar-Parisi-Zhang (KPZ) class\footnote{See chapter 3.} can not
exhibit anomaly, whereas local growth models or display FV scaling or super-roughening. They can not be
exhibit intrinsic anomaly \cite{Lopez05}.

\chapter{Concepts for Surfaces in Equilibrium and for the Thin Film Growth}
\lhead{\bfseries Appendix C - Concepts for Surfaces in Equilibrium and for the Thin Film Growth}

\section{Surface Tension and Equilibrium Shape}

A system is in \textit{equilibrium} when its macroscopic properties do not change appreciably 
with the length scale of measurement time \cite{Kardar_book1}. In this regime, there is a 
suitable set of \textit{thermodynamic coordinates} describing the system, which includes 
generalized displacements $\{\textbf{x}\}$ and their related generalized 
forces $\{\textbf{Q}\}$ \cite{Kardar_book1, Callen}. Examples for the former are the volume (V) 
for a gas or the area ($A_f$) for a film, whereas their related forces are the pressure (-P) and 
the surface tension ($\gamma$), respectively.

Depending on what coordinates are fixed in an experiment, the system state can be specified 
by appropriate \textit{thermodynamic potentials}. For instance, for processes approaching to the
equilibrium isothermally [the temperature (T) is constant] without work (\dj{}$W = 0$), the Helmholtz 
free energy $F_H = F_H(T, \{\textbf{x}\})$ is well defined. Otherwise, if only chemical 
work (\dj{}$W_{che} = \sum_i \mu_i dN_i$, where $\mu_i$ is the ith-chemical potential and $N_i$
is the particle of ith-nature) is absent, 
one can use the Gibbs free energy $G = G(T, \{\textbf{Q}\})$. 
Both are minimum at the equilibrium and contains all thermodynamic information accessible to the system \cite{Kardar_book1}.

Once one dealing with surfaces (or interfaces), one shall consider that \textit{boundary atoms} are more 
energetic than those in the bulk because they have dangling bonds. Thus, for creating a new surface 
of area $dA_f$, one must provide an amount of energy $dF_H|_{T,V} = \gamma dA_f$ necessary for breaking chemical bonds. In this
situation, $\gamma$ is also interpreted as the excess of free energy per unit of area, often 
called superficial energy \cite{Venables}. Otherwise, if there are particle migrations between the bulk and 
the surface, appears a term of chemical work and one must provide $d \Xi|_{T,V} = \gamma dA_f - \Sigma_i N_i 
d\mu_i$, with $\Xi$ being the grand potential. 

Anyway, the important is: \textit{a surface minimizes its energy decreasing its area or changing $\gamma$}. The 
first case is the reason for which a water drop does not spread on the dew, and for which without gravity 
its shape is perfectly spherical\footnote{The spherical shape is that one that minimizes an area 
for a given volume, since $\gamma$ is isotropic \cite{Adamson}.}. Cappilarity phenomena and 
soap bubble dynamics are also dictated by the surface tension \cite{Pimpinelli, Venables, Adamson}. 

Typically, metals have higher $\gamma$ than oxides\footnote{This is intuitive, once metallic surface tends 
to become oxided.} and organic structures \cite{Ohring}. Water liquid-vapor interfaces 
has $\gamma = 72.94$ mN/m, whereas that ones formed by glycerine or methanol have 48.09 and 
22.50 mN/m at $20\,^{\circ}\mathrm{C}$ \cite{Adamson}. If the surface cannot decrease its area, 
$\gamma$ variations leads to modifications in the atomic arrangements and are responsible for surface reconstructions
in crystals \cite{Ohring, Venables, Pimpinelli}.

 \begin{figure}[ht]
    \vspace{0.5 cm}
    \centering
    \includegraphics[width = 12.0 cm]{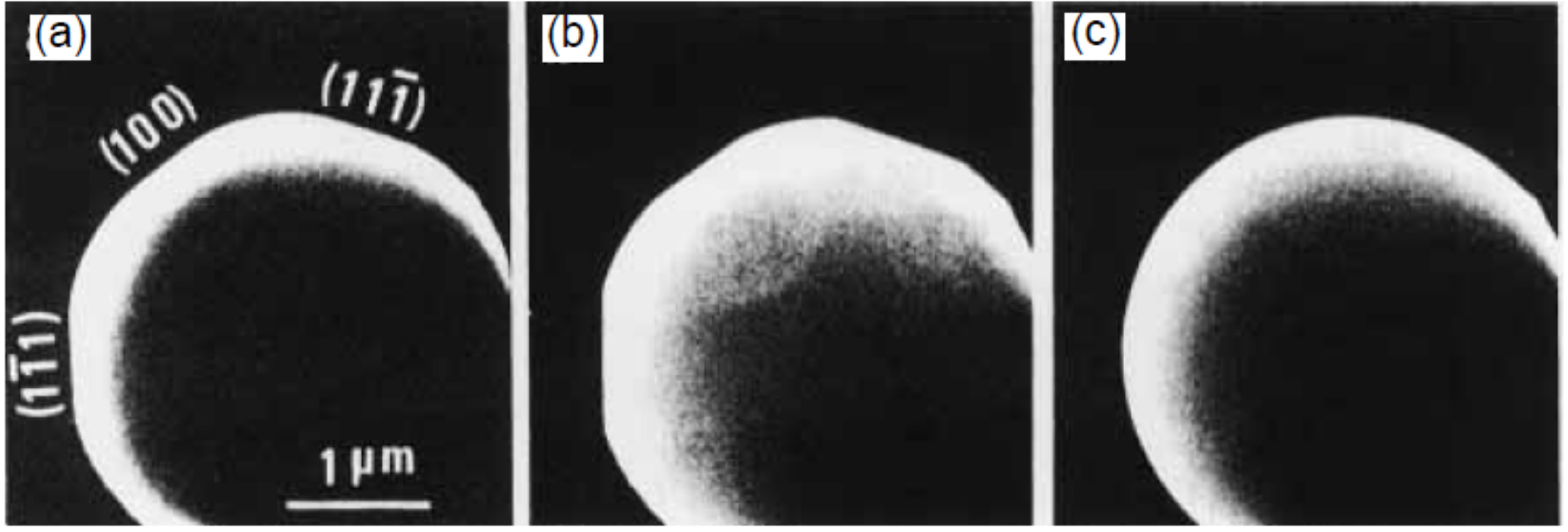}
    \caption[SEM image of Pb crystal at equilibrium]{SEM image of 
    Pb crystal at equilibrium viewed from the [011] azimuth for (a) $300\,^{\circ}\mathrm{C}$,
    (b) $320\,^{\circ}\mathrm{C}$ and (c) $327\,^{\circ}\mathrm{C}$, above the Pb melting point. Extracted
    from \cite{Venables}.}
    \label{equilibrium}
    \vspace{0.5 cm}
    \end{figure}

The shape of an object at the equilibrium can be predicted by variational principle of 
the suitable thermodynamic potential \cite{Venables}. \textit{For a crystal, the surface tension 
depends on what face is considered} and $\gamma = \gamma(hkl, T)$, where $hkl$ are the Miller's indices.
This anisotropy governs the crystal shape and, as the temperature increases, it tends to 
be minimized \cite{Pimpinelli}. Above the melting point, the anisotropy vanishes and the material must 
recover the spherical form \cite{Venables, Pimpinelli}. In the fig. \ref{equilibrium} one can seen 
equilibrium shapes for Pb crystals at different temperatures. As temperature increases the
anisotropy decreases and fewer facets are formed. Above the melting point, where Pb is liquid,
the spherical shape is achieved.


\section{Nucleation}
Nucleation is a fundamental process happening during a phase transition \cite{Ohring}. It
is important for thin films because the structures appearing in the submonolayer regime strongly affects 
the dynamic of the growth \cite{Barabasi, Evans}. Simple models for nucleation are based
on the liquid drop model, which also works for solids. The basic difference is that solids 
have two additional features: \textit{elasticity and commensurability} \cite{Pimpinelli}. These
effects will be discussed along this section. Nucleation models are important just as a 
qualitative fashion once one deals with few atoms and, hence, a continuum description 
(on thermodynamic basis) of the phenomenon becomes doubtful \cite{Ohring}.

 \begin{figure}[ht]
    \vspace{0.5 cm}
    \centering
    \includegraphics[width = 10.0 cm]{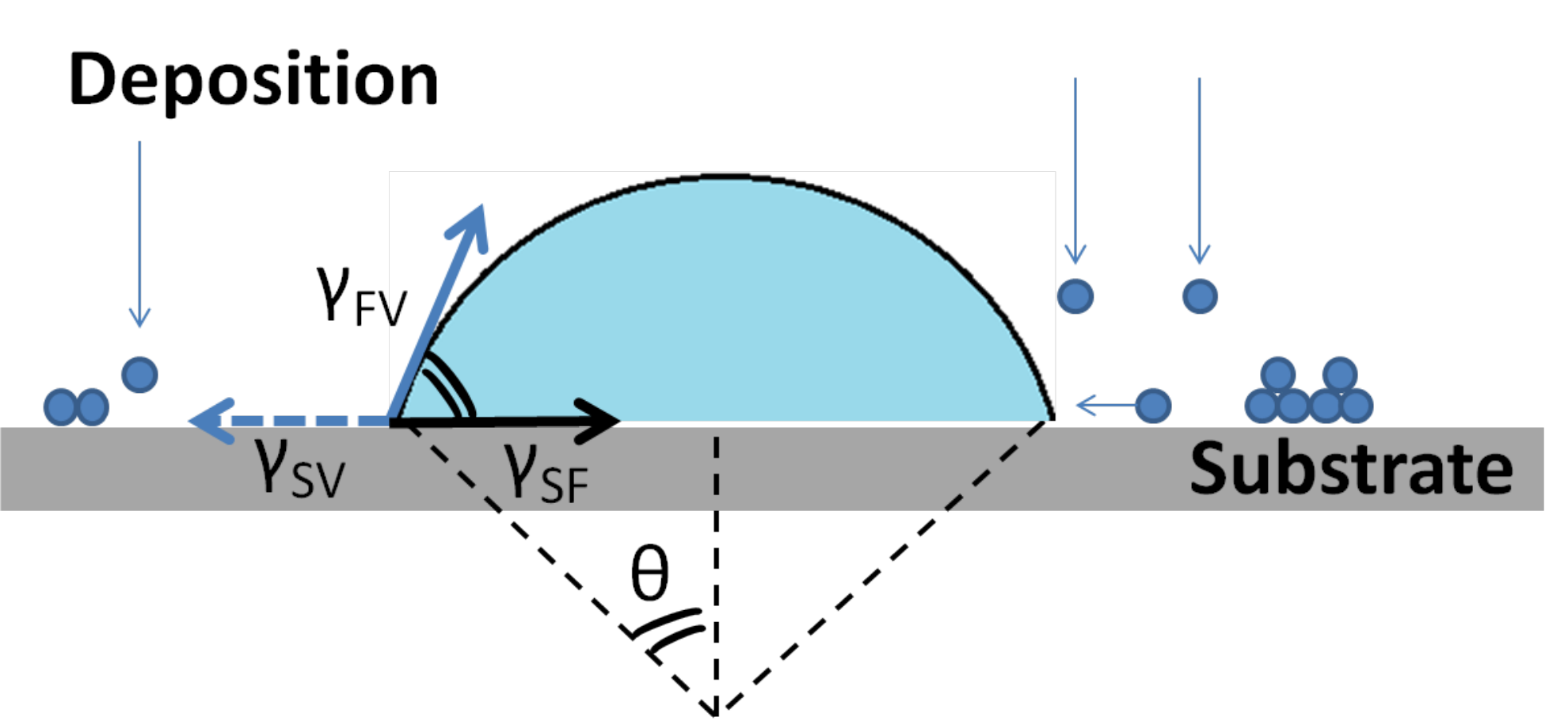}
    \caption[Schematic of nucleation processes]{Schematic of nucleation processes in a gas-to-solid
    phase transition.}
    \label{nucleation}
    \vspace{0.5 cm}
    \end{figure}

Consider a possible heterogeneous nucleation of a condensed film on a crystalline substrate 
(see a scheme in the fig. \ref{nucleation}). Deposition can occurs solely if the 
vapor is \textit{supersaturated} ($S_s > 0$). It means there exist a $\Delta P$ driving 
the atoms toward the substrate. The eq. \ref{saturation} summarizes this 
condition: 

\begin{equation}
 \Delta G_v = -\frac{K_B T}{V_a} \ln (1 + S_s),
 \label{saturation}
\end{equation}
where $\Delta G_v$ is the change in the free energy per unit volume; $K_B$ is the Boltzmann constant; 
$V_a$ the atomic volume; and $S_s \equiv \frac{P_V - P_S}{P_S}$ is the vapor supersaturation 
with $P_V - P_S$ being the difference between pressures of the vapor and above the solid, respectively.

The molecules (or atoms) impinge onto the substrate and produce, in a reasonable approximation, spherical
nucleus of mean radius $r$. The change $\Delta G$ for the substrate-film system is due to 
three contributions: (1) the first associated with all bonds to create the 
nucleus ($\sim r^3 \Delta G_v$); (2) a superficial film/vapor term owing the new nucleus 
($\sim r^2 \gamma_{fv}$); (3) and the contribution emerging from the difference between 
the energy at the old substrate/vapor surface ($\sim r^2 \gamma_{sv}$) and at the 
new substrate/film interface ($\sim r^2 \gamma_{sf}$). It reads:

\begin{equation}
\triangle G = a_{1}r^{3}\triangle G_{v} + a_{2}r^{2}\gamma_{fv} + a_{3}(r^{2}\gamma_{fs} - r^{2}\gamma_{sv}),
\label{variacaog}
\end{equation}
where $a_{1} = \frac{\pi}{3}(2 - 3 cos\theta + cos^{3}\theta)$; $a_{2} = 2\pi(1-cos\theta)$; 
$a_{3} = \pi sen^{2}\theta$; and $\theta$ is the wetting angle.

\begin{figure}[ht]
    \vspace{0.5 cm}
    \centering
    \includegraphics[width = 7.0 cm]{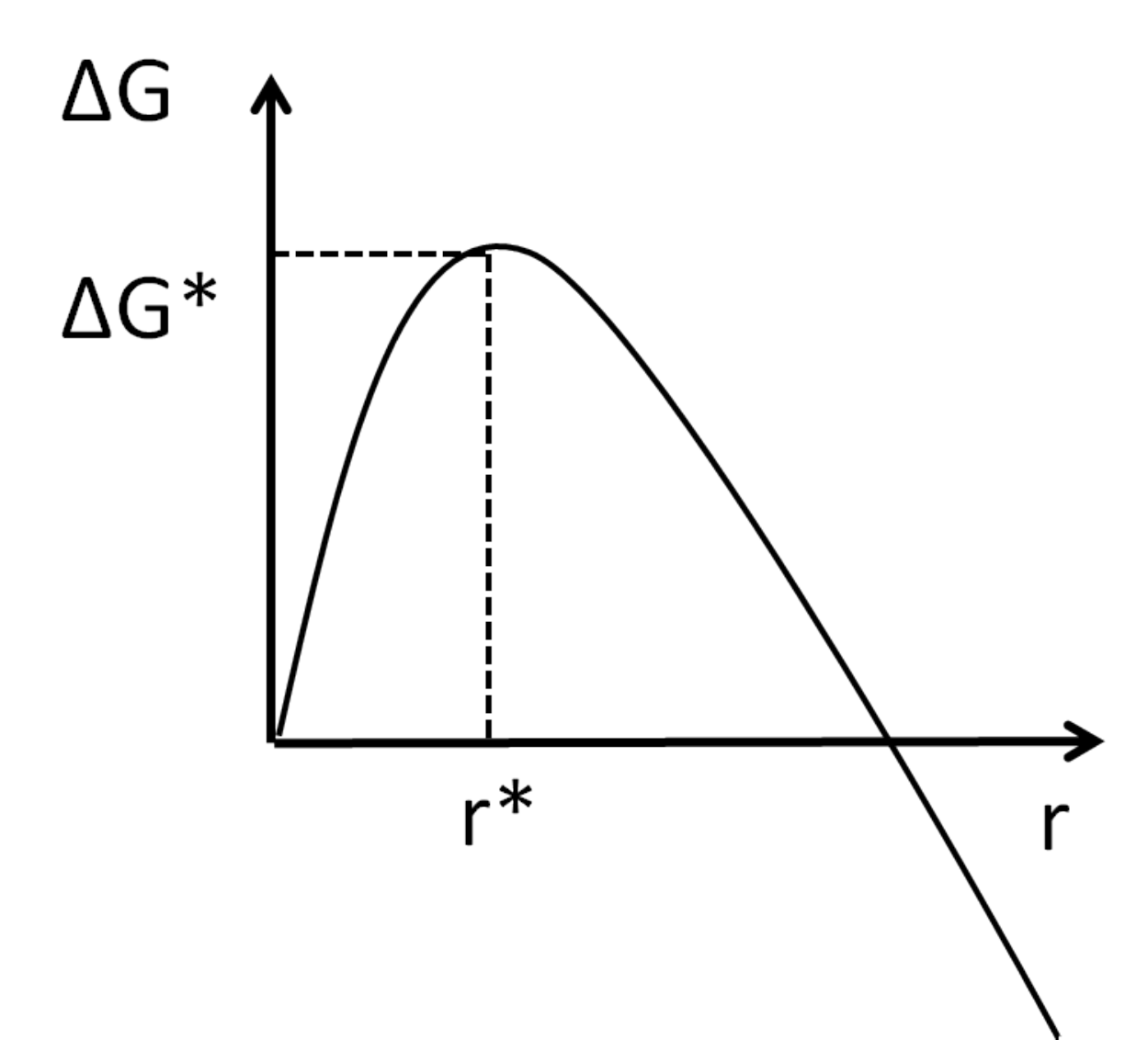}
    \caption[Change of the Gibbs free energy as function of mean nucleus size]{Qualitative change of 
    the free energy as function of the mean nucleus size during a gas-to-solid phase transition.}
    \label{barreira}
    \vspace{0.5 cm}
    \end{figure}

 In the fig. \ref{barreira} there is a plot of $\Delta G$ as function of $r$. At an \textit{instable equilibrium} 
 point, $\frac{d \Delta G}{dr} = 0$\footnote{And 
$\frac{d^2 \Delta G}{dr^2} < 0$.}, there is a critical nucleus size, $r^*$, from which island growth becomes
energetically favorable. In other words, at least an amount of $\Delta G^* = \Delta G(r^*)$ energy must 
be supplied to the system for triggering spontaneous growth, where the nucleus size grows indefinitely.

An important question raises up: \textit{How islands reach the stable size, since for $r < r^*$ they tend 
to shrink?} One must to take into account that theses processes are \textit{stochastic} and thermal fluctuations 
($K_BT \approx \Delta G^*$) are responsible to yield a density $N^*$ of stable islands per unit time. Assuming
a Boltzmann statistic, then $N^* \sim exp[-\Delta G^*/K_B T]$. Since $\Delta G^*$ must be function of T and
of the molecular flux F [atoms$/cm^2s$], one can get the feeling about how these parameters 
affect nucleation processes.

\subsection{Nucleation dependence on substrate temperature}
\label{section_NucleationT}

It is reasonable that molecular flux is directly related to $S$ such that one can replace 
$1 + S_s$ by $F/F_e$ in the eq. \ref{saturation}, with $F_e$ being the evaporation rate from the 
film at equilibrium. Inserting this in $\Delta G$ and extracting $r^*$ from the instable equilibrium
point, one can show (considering $\partial \gamma/\partial T \approx 0$ N/mK \cite{Ohring}):

  \begin{equation}
  (\partial r^*/\partial T)_{|_F} \sim [a_2\gamma_{fv} + a_3(\gamma_{fs} - \gamma_{sv})]/T^2.
  \label{restrela}
  \end{equation}

\begin{figure}[ht]
    \vspace{0.5 cm}
    \centering
    \includegraphics[width = 15.0 cm, height = 4.0 cm]{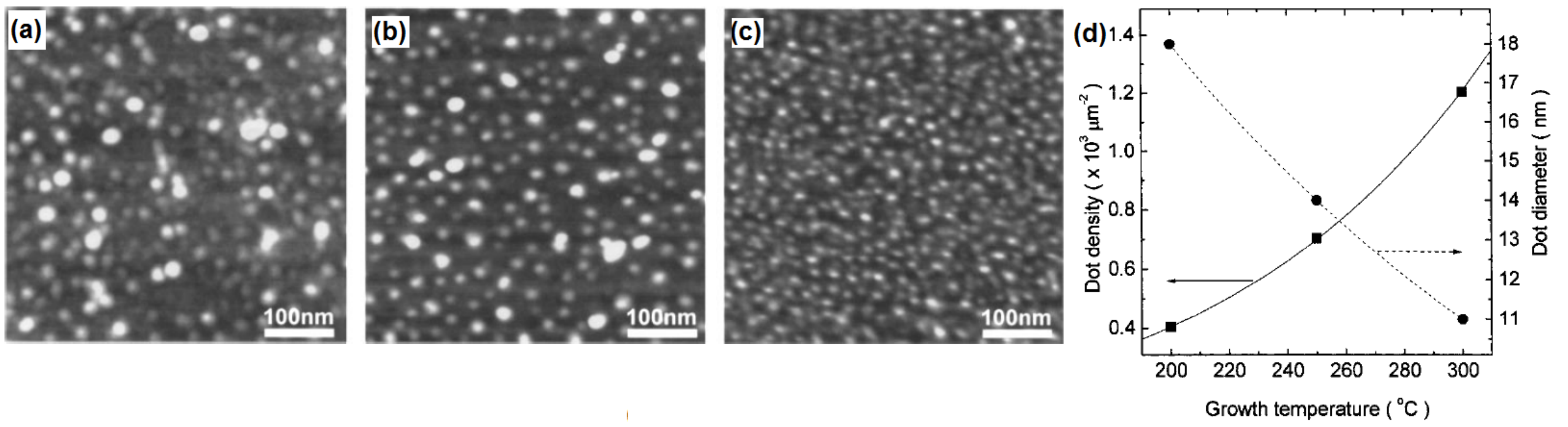}
    \caption[CdTe Quantum-Dots]{Atomic force microscopy images of CdTe Quantum-Dots grown by HWE on Si(111)
    substrates at (a) $200\,^{\circ}\mathrm{C}$, (b) $250\,^{\circ}\mathrm{C}$, and 
    (c) $300\,^{\circ}\mathrm{C}$. (d) Dot density ($\blacksquare$) and diameter (\textbullet) as function of T. 
    Courtesy of Prof. S. O. Ferreira. These results can be found in the Ref. \cite{SukarnoQD1, SukarnoQD2}.}
    \label{cdteqd}
    \vspace{0.5 cm}
    \end{figure}

At this point it is clear that the behavior of the critical size, as function of T, depends on the
relationship between the surface tensions. Indeed, this relationship is also used to describe 
growth modes of films (see next section). For instance, taking the simplest situation, where $\gamma_{fs} 
\approx \gamma_{sv}$, one has $(\partial r^*/\partial T)_{|_F} > 0$. Hence, \textit{locally}, mean critical 
size of the islands becomes larger as temperature increases (at least for a range of T, because 
the result $\sim 1/T^2$). By using the same procedure one can also find:

 \begin{equation}
  \partial (\Delta G^*)/\partial T)_{|_F} > 0,
  \label{gestrela}
 \end{equation}
 likewise $r^*$, the energy barrier $\Delta G^*$ increases as T increases. This result implies 
 that the nucleation rate is reduced exponentially in accordance with the Boltzmann statistic. Qualitatively, 
 theses results have already been found in the classical work on supercooled liquid tin and water drops, backing
 to the 1950's \cite{Vonnegut}. The crystallizations of synthetic granite and granodiorite \cite{Swanson},
 as well as the famous growth of Si on Si(001) \cite{Lagally} agree with those results. However, 
 there are also examples going in the opposite direction. S. O. Ferreira \textit{et al.} \cite{SukarnoQD1, SukarnoQD2} have 
 explored direct three-dimensional island formation in the growth of CdTe on Si(111) substrates for showing that 
 the Quantum-Dots (QD's) density increases with T in the range of 200 to $300\,^{\circ}\mathrm{C}$, whilst 
 the mean QD's size decreases (fig. \ref{cdteqd}).
 
\subsection{Nucleation dependence on molecular flux}

Once one has $r^* = r^*(T, F)$, it is straightforward showing the dependence on $F$ at T constant. Considering
the same simplest situation, one reaches to:

  \begin{equation}
  (\partial r^*/\partial F)_{|_T} < 0\qquad
  \textrm{and  }\qquad [\partial (\Delta G^*)/\partial F]_{|_T} < 0.
  \label{restrelaT}
  \end{equation}

Eq. \ref{restrelaT} tells us that nuclei size are smaller as the molecular flux increases, but at the same time,
the nucleation rate is bigger. This result agree with the intuitive feeling that when more particles arrive
onto substrate per unit time there is no time enough for the particles find an already stable island where they
can pile up. Consequently, it is likely they form dimers and other instable small islands.
These results match, for instance, with those reported for the growth of InAs/GaAs QD's on GaAs(001) 
substrates by MBE \cite{Joyce}. At $T = 490\,^{\circ}\mathrm{C}$ and $F = 0.016$ ML $s^{-1}$, 
the QD's density is about $0.2 \times 10^{11}$ $cm^{-2}$ and their diameter 
is near from 200 \AA{}. However, as $F$ is increased to $0.094$ ML $s^{-1}$, at fixed temperature, the QD's 
density also increases at least five times, whereas the QD's diameter goes to about 150 \AA{}.

Studies converging in this vein (and related to them) can be found in the cap. 9 of the Ref. \cite{Evans} as well as 
in the already cited Refs. \cite{Vonnegut, Swanson, Lagally, SukarnoQD1, SukarnoQD2, Joyce} and Refs. therein.

\section{Growth and structure of films}

\subsection{Growth Modes}

The mechanical equilibrium condition (eq. \ref{young}), predicted by the drop liquid model, provide a reasonable
way to explain three basic growth modes for solids.

\begin{equation}
 \gamma_{sv} = \gamma_{sf} + \gamma_{fv}\cos{\theta}.
 \label{young}
\end{equation}

 When the film ``wets'' the whole substrate surface, nucleating two-dimensional islands on the interface, 
 then $\theta \cong 0$ and $\gamma_{sv} \geq \gamma_{sf} + \gamma_{fv}$. This is the called 
 \textit{Frank-van der Merwe} (FM) or \textit{layer-by-layer} growth mode, often observed in 
 metal deposited on metal and homoepitaxial growth \cite{Ohring}. In this mode, ideally, the subsequent layer starts growing 
 just after the complete formation of the precedent layer. 
 
 Otherwise, when there is not the ``wetting layer'' and three-dimensional islands are nucleated directly on the 
 substrate, the \textit{Volmer-Weber} (VW) growth mode takes place. In this situation parts of the substrate
 remain exposed and the energy is minimized reducing the film/vapor interface, 
 since $\gamma_{fv} \gg \gamma_{sv} + \gamma_{sf}$. CdTe grown on Si(111) (fig. \ref{cdteqd}), for instance, 
 follows this kind of growth \cite{SukarnoQD1, SukarnoQD2}.
 
\begin{figure}[ht]
    \vspace{0.5 cm}
    \centering
    \includegraphics[width = 9.0 cm]{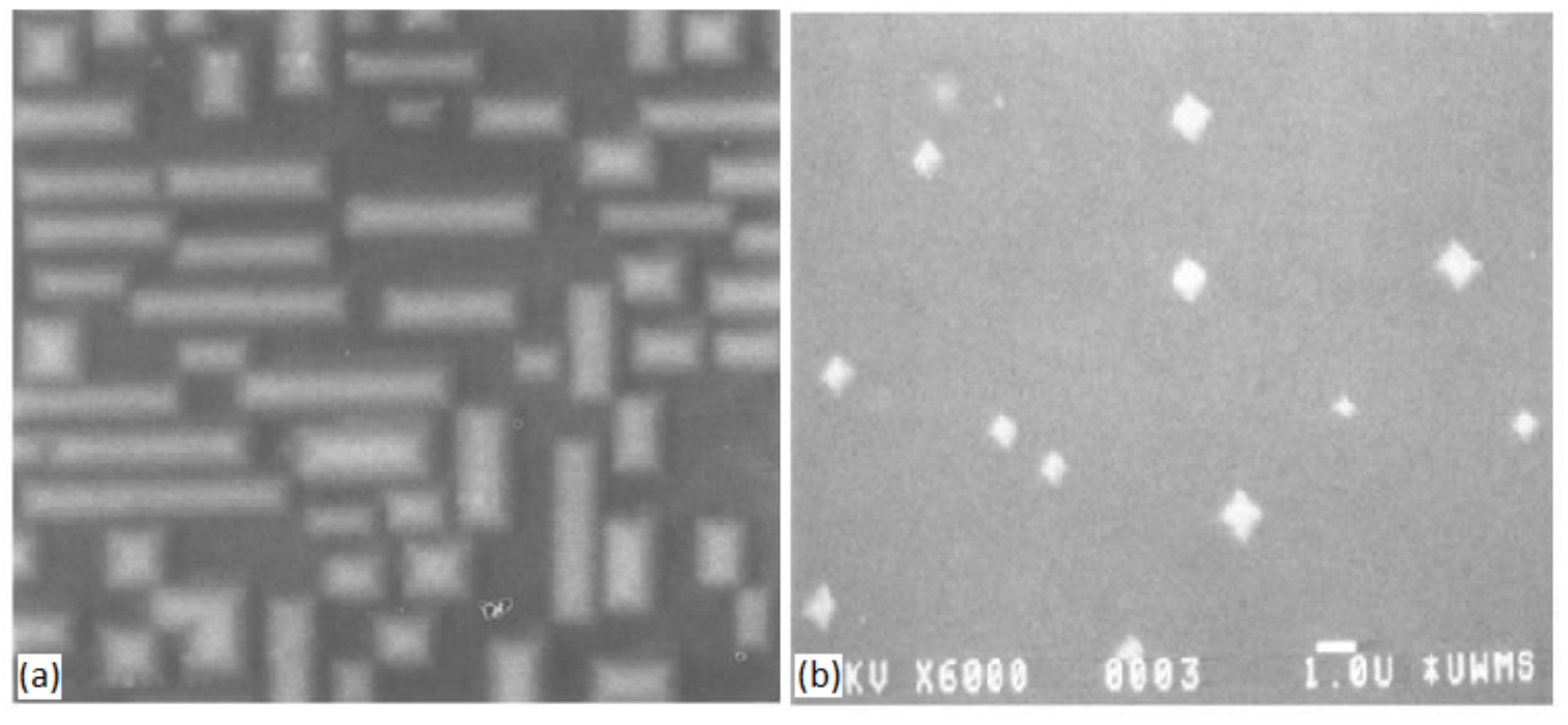}
    \caption[Ge films grown on Si(001)]{Ge films grown on Si(001) through SK mode. (a) 
    Scanning Tunneling Microscopy (STM) 2500 x 2500 \AA{} image of ``hut'' Ge clusters on Ge layers grown in the FM
    growth mode. The clusters are $\lesssim$ 1000 \AA{} long and 20-40 \AA{} high. One believes that this is a
    signature preceding macroscopic three-dimensional nucleations. (b) SEM image of the subsequent growth dominated by the 
    macroscopic three-dimensional Ge islands. These clusters are $\sim$ 250 \AA{} high and much larger than the 
    hut clusters, which can not be viewed in this scale. Images extracted and edited from \cite{Mo}.}
    \label{skge}
    \vspace{0.5 cm}
    \end{figure}

 Lastly, the \textit{Stranski-Krastanov} (SK) growth mode is understood as a transition between 
 FM to VW mode \cite{Ohring}. At the first 3-5 monolayers of deposition, $\gamma_{sv}$ dominates 
 growth making wetting. After that, direct three-dimensional islands start to nucleate on the
 subsequent layers. Any energetic disturb (defects, accumulated strain, etc.) can be the origin of this
 transition \cite{Ohring, Venables, Pimpinelli}. A typical example of SK growth mode occurs in Ge 
 films deposited on Si(001), where a metastable phase containing small clusters precedes the ``macroscopic'' 
 three-dimensional growth. See the fig. \ref{skge} \cite{Mo}.
 
 It is very important mention that the growth mode depends not only on the materials involved in the process,
 but mainly on the growth parameters, including whether impurities are present. The growth of Ag on Ag(111) 
 in the range of temperature from 175 to 575 K, for instance, changes gradually from three-dimensional (VW) 
 to step-flow growth\footnote{Explained in the section 2.3.3.}. However, if submonolayer deposits 
 of Sb are present, so the growth behavior becomes layer-by-layer (FM) over the entire 
 range of 225K to 375 K \cite{Vegt}.

\subsection{Commensurability and Polycrystallinity}

During a nucleation, under very special conditions, the film can ``copy'' the substrate crystalline structure 
\cite{Ohring}. In these cases, one says that the growth is \textit{commensurate} or \textit{epitaxial} 
\cite{Pimpinelli}. Several semiconductor devices depends on epitaxy. The most commons are optoelectronic 
ones, as LEDs and lasers based on GaInN materials, and those used in high-speed microelectronic digital 
(wireless communications), composed of Ge$_{x}$Si$_{1-x}$ \cite{Ohring}. Typically, the first step in epitaxial 
films is the homo-epitaxial Si on Si growth by CVD. The reason is that 
the \textit{epilayer} is free from defects, purer than the wafer and can be doped independently of it.

Epitaxy depends on the mismatch between the lattice parameters ($|a_f - a_s|/a_s$) of the film ($a_f$) 
and of the substrate ($a_s$), including their thermal dependence via the dilatation coefficient 
\cite{Herman1, Herman2, Birkholz_book}. It occurs whether the relative mismatch is below $ \approx 10-15\%$ \cite{Herman1, Herman2}. 
Nevertheless, exotic cases (called hard epitaxy) as the hetero-epitaxial CdTe layers 
grown on Si(111) also exist \cite{Joelma}, in spite of almost $19\%$ and $48\%$ mismatch 
of the lattice parameter and thermal coefficient, respectively. Larger mismatch implies 
larger accumulated strain (\textit{elastical} energy) which can be released as defects or 
can support a transition from FM to VW growth mode.

It is worth mention that the effect of heteroepitaxy can be included in the drop liquid model 
replacing the term ($\sim \Delta G_v$) by $[\sim (\Delta G_v + \Delta G_s)]$ in the eq. \ref{variacaog}, 
where $\Delta G_s$ inserts the extra energy accumulated in the strain form. $\Delta G^*$ increases leading
the nucleation rate to decline. Other effects as impurities, defects, chemical reactions also can be inserted 
into the model in order to get the felling about their consequences on nucleations \cite{Ohring}.
 
Unlike epitaxial single-crystals, \textit{polycrystalline} films are composed by an collection 
of \textit{grains}, each of them having its own crystallographic orientation. 
These environments lead to a complex competition between neighboring grains which are droved by the 
surface tensions, once $\gamma = \gamma(hkl)$ \cite{Ohring, Venables, Pimpinelli}. Generally, 
a preferential direction of growth (\textbf{$\Lambda$}), called \textit{texture} 
\cite{Ohring, Birkholz_book, Birkholz}, appears and evolves as deposition proceeds. Grains$_{\Lambda}$ grow faster 
than the others and, eventually, dominate the surface. Texture affects mechanical and electrical 
features of films, including their elastic modulus, yield strength, magnetic permeability, 
etching rate, diffusion rate, and others \cite{Ohring}. Standard procedure quantifying 
texture is based on X-ray diffraction techniques \cite{Birkholz, Birkholz_book}.


 \section{Kinetic Phenomena and Superficial Structures}
 \label{section_Kinetic}
 Crystal (surface) growth is a far-from-equilibrium phenomenon. During a growth from vapor phase,
 atoms impinge onto the substrate (interface) breaking and/or forming local bonds, changing the morphology and 
 hampering the system to minimize its free energy \cite{Evans}. As the substrate is heated, 
 the adosrbed atoms, or molecules, (called \textit{adatoms}) might \textit{diffuse} on the surface performing Brownian motions. 
 Commonly, these adatoms stick at positions which maximize their number of 
 coordination and \textit{deposition} occurs \cite{Ohring}. Thermal fluctuation may also induces
 \textit{desorption}, i.e, adatoms leave the surface, returning to the vapor
 \cite{Barabasi}. Deposition, desorption and aggregation mediated by diffusion are the basic mechanisms ruling a general growth. Of course,
 specific situations excludes one, maybe two of these processes, but we are considering the wide case. 
 These last two, and other cited below, are \textit{thermally activated}, namely, they 
 occur with rates given by Arrhenius' laws (eq. \ref{arrhenius}) \cite{Ohring, Venables, Pimpinelli, Barabasi}:
 
 \begin{equation}
 \tau = \tau_0 \exp{[- E_{\tau}/K_BT]},
 \label{arrhenius}
 \end{equation}
 where $\tau$ is a rate of a particular event, $\tau_0$ a constant, and $E_{\tau}$ is the energy 
 associated with the $\tau$ process.
 
 \begin{figure}[ht]
    \vspace{0.5 cm}
    \centering
    \includegraphics[width = 12.0 cm]{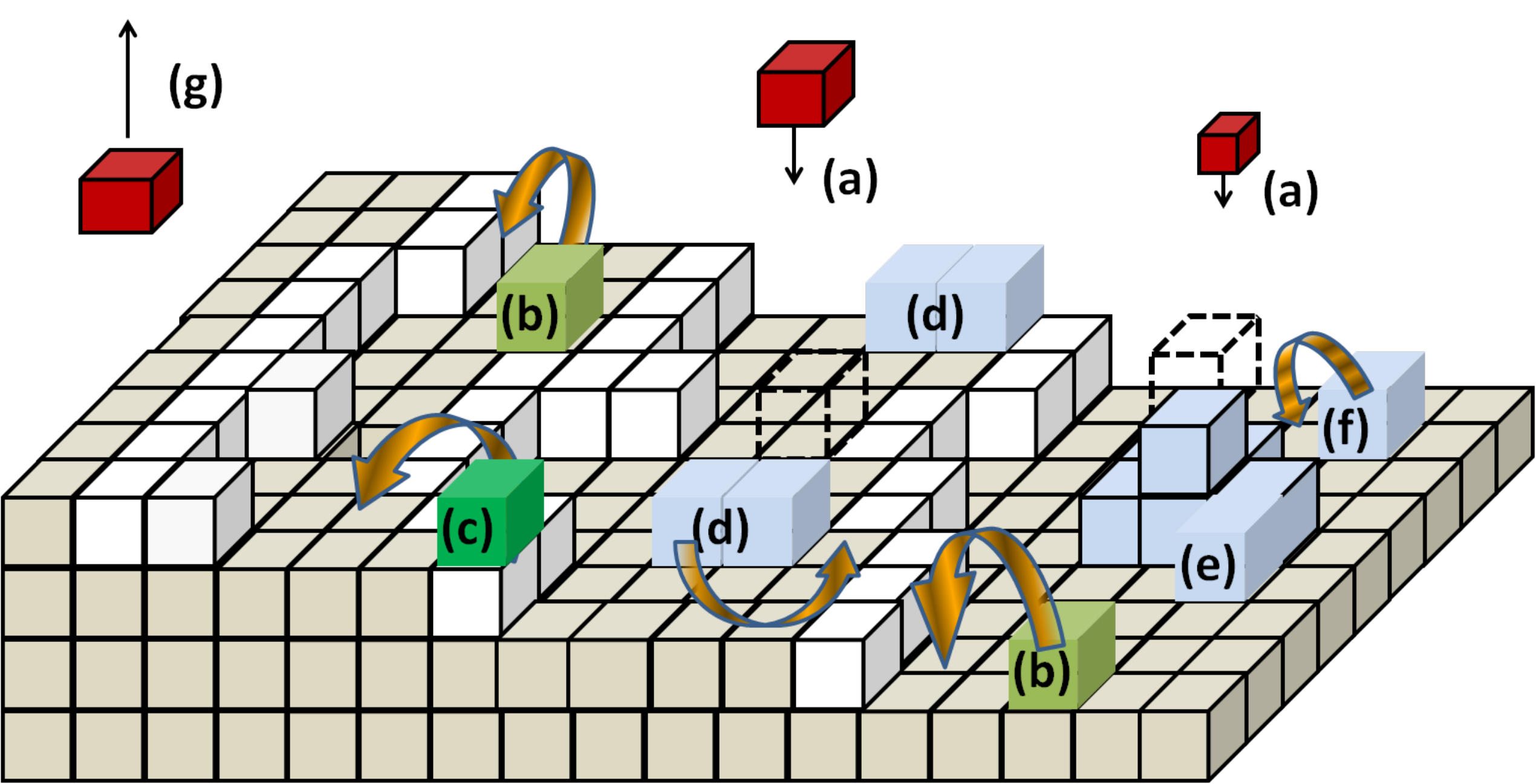}
    \caption[Schematic of kinetic processes]{Schematic of kinetic processes occurring during the growth.}
    \label{kinetic}
    \vspace{0.5 cm}
    \end{figure}

\begin{figure}
    \vspace{0.5 cm}
    \centering
    \includegraphics[width = 6.0 cm]{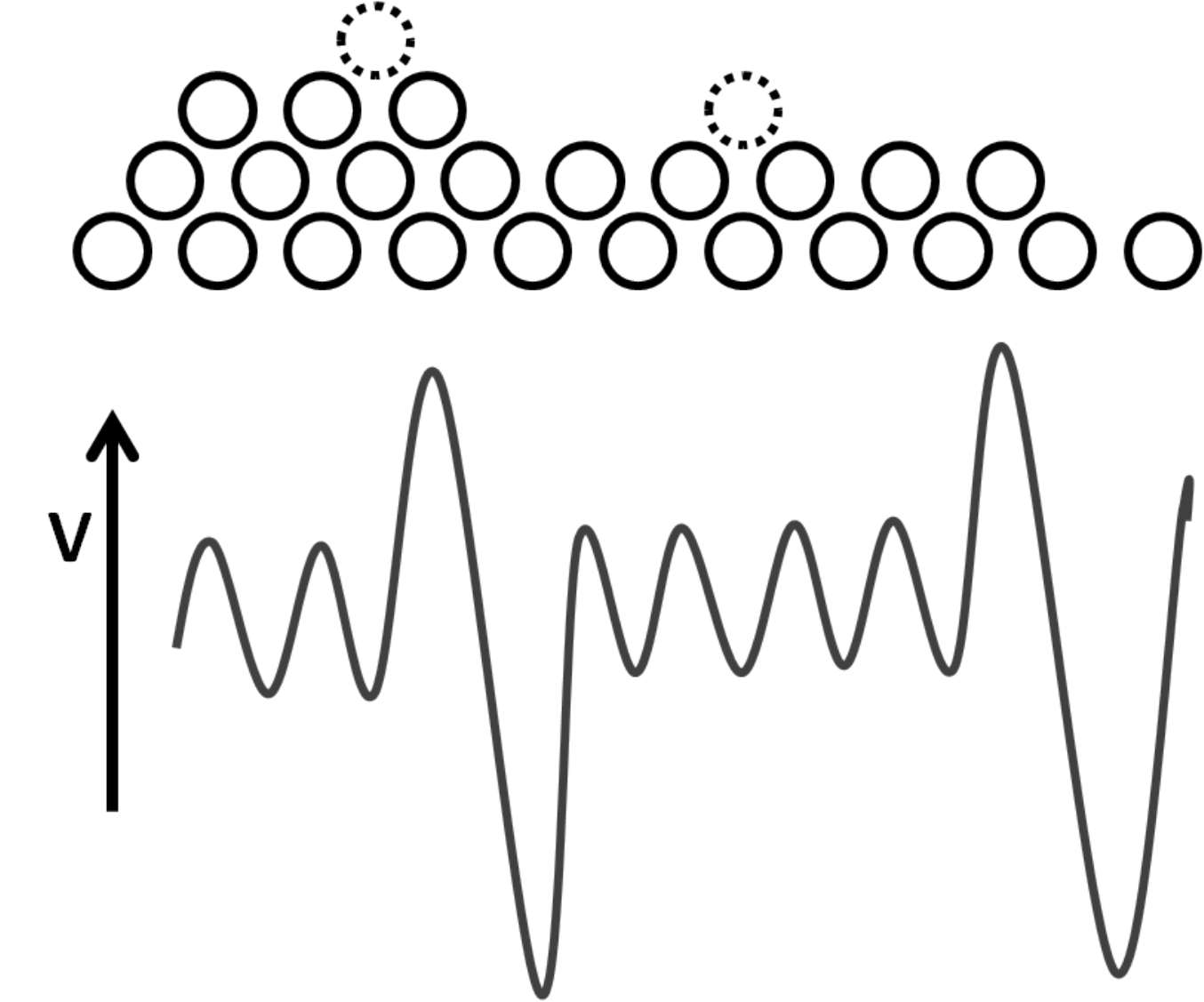}
    \caption[Schematic of the lattice potential]{Schematic of the lattice potential.}
    \label{lattice_barreira}
    \vspace{0.5 cm}
    \end{figure}

 A crystal surface is not a perfect plane. Indeed, it presents large pieces (\textit{terraces})
 separated by \textit{steps} of atomic height (or multiples of an atomic heigth)
 - see and follow the fig. \ref{kinetic}. Even theses
 steps are not perfectly straight, rather they contain straight parts separated 
 by \textit{kinks} \cite{Pimpinelli}. Particles arriving at the substrate \textbf{(a)} have an amount of
 energy relative to the vapor phase \cite{Venables}. They use this energy for breaking 
 chemical bonds already established at surface and/or for becoming adatoms \textbf{(b)}. Adatoms diffuse with a 
 rate $\tau_{diff}$, since that, for \textit{each hop}, an adatom 
 must overcome the lattice potential existing between neighboring positions 
 (see fig. \ref{lattice_barreira}). This potential depends on the nature
 of compounds involved in the growth and on the substrate face. 
%

 Preferential sites for sticking are the edge of steps and kinks, once they have a larger number of
 dangling bonds \cite{Pimpinelli}. However, adatoms reach these positions only if the temperature 
 is sufficiently high for promoting a length diffusion ($x_{diff}$) comparable to the terrace 
 separation length ($x_{terr}$) \cite{Venables}. \textit{Step-flow} growth is achieved when 
 $x_{diff} \gg x_{terr}$ \cite{Pimpinelli}. Interestingly, if an adatom diffuses on a terrace, but
 towards another one \textbf{(c)}, \textit{it tends to come back instead to jump off the terrace}. This 
 is known as \textit{Ehrlich-Schowoebel effect} (ES) \cite{Ehrlich, Evans}. In the 
 fig. \ref{lattice_barreira} the effect of the ES barrier is added. In cases where the ES barrier 
 dominates the growth (see Pt grown on Pt(111), fig. 102 in the ref. \cite{Evans}) structures 
 looking like \textit{wedding cakes} are formed \cite{Evans}. Among several ways to define an expression for the diffusion energy
 that includes the ES barrier, one usually employed is:
 
 \begin{equation}
  E_{diff} = E_0 + nE_n + E_{es},
  \label{ed}
 \end{equation}
where $n$ is the number of coordination; $E_n$ is the energy per unit bond 
and $E_{es}$ is the associated energy to the ES barrier. Typically, $E_n \sim 0.1$ eV and 
$E_{es} \sim 0.01$ eV \cite{Barabasi}. However, it is worth keep in mind that these values may change appreciably depending
on the system considered. Additionally, the eq. \ref{ed} is solely a simplified model which does not catch all source of
energies involved in real processes.
 
 Since an adatom establishes bonds, deposition happens. When an adatom meets another atom, they can form a 
 dimer \textbf{(d)}. Dimers also diffuse, albeit it is (much)less probable due to the enhance 
 of $E_{diff}$. Sometimes, dimers grow and become a two- or three-dimensional island \textbf{(e)}, 
 which tend to capture adatoms \textbf{(f)} (if $r > r^*$). At the same time, some atoms break 
 off their bonds and leave the surface \textbf{(g)}. So, desorption occurs. Desorption
 is an effect competing with deposition. Due to this effect, the molecular flux migth be different from
 the deposition rate. However, in usual MBE (and related technique) conditions, desorption is
 a neglegible process \cite{Ohring, Venables, Pimpinelli, Barabasi}. For Ga on GaAs(111) over 
 the range 860K to 960 K, the desorption energy is $E_{des} \approx 2.5$ eV \cite{Barabasi}. Assuming
 $1/\tau_{0des} = 10^{-14}$ s and $T = 850$ K in the inverse of eq. \ref{arrhenius}\footnote{The inverse
 of a given rate is the \textit{characteristic time of this process.}}, one obtains that only one Ga atom
 is desorbed from the surface in each 2 $s$. Comparing this desorption rate with the deposition rate used in MBE ($\sim 10^{16-18}$ atoms $cm^{-2}$ $s^{-1}$), it eases
 to see that desorption phenomenon is inoperative \cite{Barabasi}.
 
%

\chapter*{List of Acronyms and Symbols}
\addcontentsline{toc}{chapter}{LIST OF ACRONYMS AND SYMBOLS}
\lhead{\bfseries List of Acronyms and Symbols}
\section*{Acronyms}
\begin{longtable}[l]{p{100pt} p{500pt}} 
\textbf{Acronym} & \textbf{Description} \\
AFM & \textit{Atomic Force Microscopy} \\

BD & \textit{Ballistic Deposition} \\
CF  & \textit{Correlation Function} \\
CRSOS & \textit{Conserved Restricted Solid-On-Solid} \\
CVD & \textit{Chemical Vapor Deposition} \\

DI & \textit{Deionized} \\
DPRM & \textit{Directed Polymer in a Random Medium} \\
DT & \textit{Das-Sarma-Tamborenea} \\
EVS & \textit{Extreme-Value Statistics} \\
EW & \textit{Edwards-Wilkinson} \\

FM & \textit{Frank-van-der-Merwe} \\
FV & \textit{Family-Vicsek} \\

GBs & \textit{Grain Boundaries} \\
GOE & \textit{Gaussian Orthogonal Ensemble} \\
GPS & \textit{Global Position System} \\
GUE & \textit{Gaussian Unitary Ensemble} \\
H   & \textit{Height of ``grains'' used as parameter in a Kinetic Monte Carlo Model} \\
HDs & \textit{Height Distributions} \\
HWE & \textit{Hot Wall Epitaxy} \\

IC & \textit{Initial Conditions} \\



KPZ & \textit{Kardar-Parisi-Zhang} \\

LEDs & \textit{Light Emission Diodes} \\

MBE & \textit{Molecular Beam Epitaxy} \\
MH & \textit{Mullins-Herring} \\
ML & \textit{Mono-Layer} \\
MRHDs & \textit{Maximal Relative Height Distributions} \\



\textit{pdf} & \textit{probability density function} \\
PNG & \textit{Poly-Nuclear-Growth} \\

QDs & \textit{Quantum Dots} \\


RDSR & \textit{Random Deposition with Surface Relaxation} \\
RSOS & \textit{Restricted Solid-On-Solid} \\
SEM & \textit{Scanning Electron Microscopy} \\
SK & \textit{Stranski-Krastanov} \\
SLRDs & \textit{Squared Local Roughness Distributions} \\
STM & \textit{Scanning Tunelling Microscopy} \\

TASEP & \textit{Totally Asymetric Exclusion Process} \\
TW & \textit{Tracy-Widom} \\

UC & \textit{Universality Class} \\


VLDS & \textit{Villain-Las-Das-Sarma} \\
VW & \textit{Volmer-Weber} \\
WV & \textit{Wolf-Villain} \\

XPS & \textit{X-ray Photoelectron Spectroscopy} \\
XRD & \textit{X-Ray Diffraction} \\


\end{longtable}

\vspace{2.0cm}

\section*{Symbols}
\begin{longtable}[l]{p{100pt} p{500pt}}
\textbf{Latin Symbols} & \textbf{Description} \\

$a_f$ & \textit{Lattice parameter of the film} \\
$a_s$ & \textit{Lattice parameter of the substrate} \\
$A$ & \textit{Static amplitude in the KPZ context} \\
$A_f$ & \textit{Area of a film} \\
$A_{\theta-2\theta}$ & \textit{Absortion factor for a $\theta-2\theta$ XRD geometry} \\


$C_h$ & \textit{Heigth-difference correlation function} \\
$C_s$ & \textit{Spatial covariance} \\
$d$ & \textit{Dimension of a growth process} \\
$d_s$ & \textit{Substrate dimension} \\
$\sqrt{D}$ & \textit{Amplitude of the white noise} \\
$E_{0}$ & \textit{Enery of the lattice potential} \\
$E_{diff}$ & \textit{Characteristic energy of diffusion} \\
$E_{es}$ & \textit{Ehrlich-Showoebel barrier energy} \\
$E_{gap}$ & \textit{Energy gap} \\

$E_{n}$ & \textit{Energy per unit bond} \\
$E_{GB}$ & \textit{Energy barrier at GBs of colided neighboring grains} \\
$E_{R}$ & \textit{Relaxation energy barrier at the GBs of colided neighboring grains} \\
$E_{\tau}$ & \textit{Energy of a particular kinetic effect} \\

$F$ & \textit{Molecular Flux} \\
$F_0$ & \textit{Baik-Rains distribution} \\
$F_e$ & \textit{Evaporation rate from the film at equilibrium} \\
$F_H$ & \textit{Helmholtz free energy} \\
$G$ & \textit{Gibbs free energy} \\
$G_s$ & \textit{Gibbs free energy related to strain} \\
$G_v$ & \textit{Gibbs free energy per unit volume} \\
$G^*$ & \textit{Critical gibbs free energy} \\

$G(X;m)$ & \textit{Gumbel pdf of the X variable; m-th order} \\
$h_i$ & \textit{Surface height at the site i} \\
$h(x, t)$ & \textit{Surface height at the substrate position $x$ at time t} \\
$\langle h^n\rangle_c$ & \textit{n-th cumulant of h} \\
$\textrm{\^{h}}$ & \textit{Surface height in the reciprocal space} \\

$I_{111}$ & \textit{Intensity of the (111) peak in a XRD spectrum} \\
$j$ & \textit{Current density} \\
$J$ & \textit{Current density parallel to the surface} \\

$K$ & \textit{Kurtosis coefficient} \\
$K_d$& \textit{Strength of diffusion} \\
$K_B$& \textit{Boltzman constant } \\

$l^*$ & \textit{Characteristic box size} \\
$L$ & \textit{Lateral size of the substrate} \\

$m_n$ & \textit{nth-moment of a pdf} \\
$m^*$ & \textit{Maximal height relative to the mean height of an interface}\\

$n_{coar}$ & \textit{Coarsening exponent} \\
$N_h$ & \textit{Number of points at the surface with height h} \\
$N^*$ & \textit{Density of stable islands per unit time} \\



$p(h)$ & \textit{Density probability of the variable $h$} \\
\~{p}$(k)$  & \textit{Characteristic function} \\
$P$ & \textit{Pressure} \\
$P_D$ & \textit{Probability of particle diffusing towards a defect site} \\
$P_R$ & \textit{Probability of ocurring a relaxation process at GBs} \\
$P_S$ & \textit{Pressure above vapor the solid} \\
$P_V$ & \textit{Pressure of the vapor phase} \\
$P(h)$ & \textit{Probability of the variable $h$} \\

$Q$ & \textit{Generalized thermodynamic force} \\

$r$ & \textit{Mean radius of a quasi-spherical nucleus} \\
$r_c$ & \textit{Average grain size} \\
$r_m$ & \textit{First minimum/zero of the Slope-Slope spatial covariance} \\
$r^*$ & \textit{Critical size of an island} \\
$R$ & \textit{Local curvature} \\
$S$ & \textit{Skewness coefficient} \\
$\overleftrightarrow{S}$ & \textit{Space-time referential} \\
$S(k,t)$ & \textit{Structure factor or Power-spectrum} \\
$S_s$ & \textit{Supersaturation} \\

$t$ & \textit{Growth time} \\
$th$ & \textit{Thickness of a thin film} \\
$t_x$ & \textit{Crossover time} \\

$T$ & \textit{Temperature of the substrate} \\
$T_{EW}$ & \textit{Substrate temperature corresponding to an EW growth} \\
$T_{EW-KPZ}$ & \textit{Substrate temperature corresponding to an EW-to-KPZ crossover} \\

$v$ & \textit{Average growth velocity of an interface} \\
$v_{\infty}$ & \textit{Asymptotic growth velocity of an interface} \\

$V$ & \textit{Volume} \\
$V_a$ & \textit{Atomic volume} \\

$w_{loc}$ & \textit{Local roughness} \\
$w_{sat}$ & \textit{Saturation value for the roughness} \\
$w(L,t)$ & \textit{Global roughness} \\
$W$ & \textit{Thermodynamical work} \\

$x_{diff}$ & \textit{Characteristic length for diffusion} \\
$x_{diff}$ & \textit{Length of a terrace} \\

$z$ & \textit{Dynamic exponent} \\

\vspace{1.0cm} \\

\textbf{Greek Symbols} & \textbf{Description} \\
$\alpha$ & \textit{Roughness exponent} \\
$\alpha_1$ & \textit{Geometrical exponent} \\
$\alpha_s$ & \textit{Spectral roughness exponent} \\
$\gamma$ & \textit{Surface energy or Surface tension} \\
$\gamma_{fv}$ & \textit{Surface tension at the film-vapor interface} \\
$\gamma_p$ & \textit{Non-universal parameter in the KPZ scaling theory context} \\
$\gamma_{sf}$ & \textit{Surface tension at the substrate-film interface} \\
$\gamma_{sv}$ & \textit{Surface tension at the substrate-vapor interface} \\

$\Gamma$ & \textit{Non-universal parameter in the KPZ scaling theory context} \\
$\Gamma_f$ & \textit{Gamma function} \\
$\Gamma(l,t)$ & \textit{Slope-Slope spatial covariance} \\

$\epsilon$& \textit{Tilting angle} \\
$\zeta$ & \textit{Characteristic length in the surface} \\
$\zeta_p$ & \textit{Non-universal parameter in the KPZ scaling theory context} \\
$\eta$ & \textit{Noise} \\
$\eta_p$ & \textit{Non-universal parameter in the KPZ scaling theory context} \\
$\theta_{hkl}$ & \textit{Wetting angle} \\
$\theta_{hkl}$ & \textit{Angle of reflection for the (hkl) peak in a XRD measurement} \\
$\Theta$ & \textit{Functional of x, h and t} \\
$\kappa$& \textit{Kappa exponent} \\
$\lambda$ & \textit{Excess of velocity of an interface} \\
$\lambda_{Cu_{K\alpha}}$ & \textit{Wavelenght of the radiation in a K$_\alpha$ transition for a Cu atom} \\
$\Lambda$ & \textit{Preferential direction of growth - Texture direction} \\
$\mu$ & \textit{Chemical potential} \\
$\mu_v$ & \textit{Chemical potential in a vapor phase} \\

$\nu$ & \textit{Strength of the ``surface tension''} \\

$\xi$ & \textit{Correlation length} \\
$\xi_{||}$ & \textit{Parallel correlation length} \\
$\Xi$ & \textit{Grand potential} \\


$\sigma_{h}$ & \textit{Standard deviation of h} \\
$\sigma_{\chi}$ & \textit{Standard deviation of $\chi$} \\

$\varsigma$ & \textit{Scale factor} \\
$\tau$ & \textit{Rate of a particular kinetic effect} \\
$\tau_{0des}$ & \textit{Rate of desorption} \\
$\tau_{diff}$ & \textit{Rate of diffusion} \\
$\Phi$ & \textit{Universal scaling function} \\
$\chi$ & \textit{Random variable} \\
$\langle \chi \rangle$ & \textit{Mean of $chi$} \\
$\langle \chi^2 \rangle_c$ & \textit{Second cumulant of $\chi$} \\

$\psi_k$ & \textit{Polygamma function of kth-order} \\
$\Psi$ & \textit{Universal scaling function} \\
$\Omega$ & \textit{Aspect ratio of a grain/mound} \\

\end{longtable}


\end{document}